\documentclass[
11pt, % The default document font size, options: 10pt, 11pt, 12pt
english,  % ngerman for German
singlespacing, % Single line spacing, alternatives: singlespacing, onehalfspacing or doublespacing
parskip, % Uncomment to add space between paragraphs
headsepline, % Uncomment to get a line under the header
consistentlayout, % Uncomment to change the layout of the declaration, abstract and acknowledgements pages to match the default layout
]{MastersDoctoralThesis} % The class file specifying the document structure

\usepackage[utf8]{inputenc} % Required for inputting international characters
\usepackage[T1]{fontenc} % Output font encoding for international characters
\usepackage{lmodern} 

\usepackage{setspace} % increase interline spacing slightly
\usepackage[backend=bibtex,style=authoryear,natbib=true]{biblatex} % Use the bibtex backend with the authoryear citation style (which resembles APA)

\usepackage[autostyle=true]{csquotes} % Required to generate language-dependent quotes in the bibliography

\usepackage{amsmath}
\usepackage{amsfonts}
\usepackage{amssymb}
\usepackage{subfigure}

\newcommand{\ie}{\textit{i.e. }}
\newcommand{\eg}{\textit{e.g. }}
\newcommand{\dd}{\text{d}}
\newcommand\nnfootnote[1]{%
  \begin{NoHyper}
  \renewcommand\thefootnote{}\footnote{#1}%
  \addtocounter{footnote}{-1}%
  \end{NoHyper}
}
\newcommand\rharp[1]{\mathstrut\mkern2.5mu#1\mkern-14mu\raise1.6ex%
  \hbox{$\scriptscriptstyle\rightharpoonup$}}
\newcommand\lharp[1]{\mathstrut\mkern2.5mu#1\mkern-14mu\raise1.6ex%
  \hbox{$\scriptscriptstyle\leftharpoonup$}}
  \newcommand\smup[1]{\mathstrut\mkern2.5mu#1\mkern-13mu\raise1.6ex%
  \hbox{$\scriptscriptstyle\smallsmile$}}
\newcommand\smdown[1]{\mathstrut\mkern2.5mu#1\mkern-13mu\raise1.6ex%
  \hbox{$\scriptscriptstyle\smallfrown$}}
\newcommand\smldown[1]{\mathstrut\mkern2.5mu#1\mkern-13mu\raise2ex%
  \hbox{$\scriptscriptstyle\smallfrown$}}
  
\newcommand{\qqa}[0]{\bm{\mathsf{q}}^{\alpha}}
\newcommand{\qqf}[0]{\bm{\mathsf{q}}^{f}}
\newcommand{\qqm}[0]{\bm{\mathsf{q}}^{m}}
\newcommand{\qqh}[0]{\hat{\bm{\mathsf{q}}}}
\newcommand{\hha}[0]{\bm{\mathsf{H}}^{\alpha}}
\newcommand{\hhf}[0]{\bm{\mathsf{H}}^{f}}
\newcommand{\hhm}[0]{\bm{\mathsf{H}}^{m}}
\newcommand{\hhh}[0]{\hat{\bm{\mathsf{H}}}}

\newcommand{\mmf}[0]{\bm{\mathsf{M}}^{f}}
\newcommand{\mmm}[0]{\bm{\mathsf{M}}^{m}}
\newcommand{\mmh}[0]{\hat{\bm{\mathsf{M}}}}
\newcommand{\rra}[0]{\rho_{\alpha}}
\newcommand{\rrf}[0]{\rho_{f}}
\newcommand{\rrm}[0]{\rho_{m}}

\usepackage[bb=boondox]{mathalfa}
\usepackage{bm}

\thesistitle{A path integral approximation of conditional probability densities with application to stochastic elastic rods} 
\degree{}
\author{Giulio Corazza} 
 % Keywords for your thesis, this is not currently used anywhere in the template, print it elsewhere with \keywordnames
\university{} 
\department{}
\faculty{} 

\AtBeginDocument{
\hypersetup{pdfkeywords=\keywordnames} % Set the PDF's keywords to your keywords
\hypersetup{allcolors=black}
}

\begin{document}

\frontmatter % Use roman page numbering style (i, ii, iii, iv...) for the pre-content pages

\pagestyle{plain} % Default to the plain heading style until the thesis style is called for the body content

%----------------------------------------------------------------------------------------
%	TITLE PAGE
%----------------------------------------------------------------------------------------

\begin{titlepage}
\begin{center}
\vspace*{.06\textheight}
\vspace{1cm}
{\huge \bfseries \ttitle\par}\vspace{0.4cm} 
\vspace{4cm}
\emph{Author:}\\
{\authorname} 
\vfill
{\HRule \\[1.5cm]}  
{\large 2022}\\[4cm] % Date
\end{center}
\end{titlepage}

%----------------------------------------------------------------------------------------
%	ABSTRACT PAGE
%----------------------------------------------------------------------------------------

\begin{abstract}
\addchaptertocentry{\abstractname} % Add the abstract to the table of contents
Path integrals play a crucial role in describing the dynamics of physical systems subject to classical or quantum noise. In fact, when correctly normalized, they express the probability of transition between two states of the system. In this work, we generalise Gelfand-Yaglom-type methods in the vector case for the computation of Gaussian path integrals. The extension we propose allows to consider general second variation operators subject to different boundary conditions and to regularise the divergence in presence of zero modes. The ratio of path integrals is exploited to compute transition probabilities in the semi-classical approximation from the solutions of suitable matrix initial value problems for the Jacobi equations. Our methods are particularly useful for investigating Fokker-Planck dynamics and are applied to derive the time evolution of the $d$-dimensional Ornstein-Uhlenbeck process, and of the Van der Pol oscillator driven by white noise. The same methods are also exploited to study the statistical physics of polymers at thermodynamic equilibrium, which is the main objective of this work, where the time variable is replaced by the material parameter of a framed curve subject to stochastic fluctuations. Namely, in a continuum limit, the configuration of a polymer is a curve in the group $SE(3)$ of rigid body displacements, whose energy can be modelled via the Cosserat theory of elastic rods. It is widely known how to efficiently compute equilibria of (hyper-)elastic rods subject to prescribed two-point boundary conditions, for example using the Hamiltonian form of the Euler-Lagrange equations for the associated energy. The classification of which equilibria are also stable in the sense of realising local minima of the energy can then be carried out using conjugate point theory, specifically, counting the number of zeros of a determinant evaluated on Jacobi fields associated with the second variation of the energy. What is less well-known is that the energy of equilibria combined with the actual value of the same Jacobi field determinant can be used to estimate probability densities for the distribution of end-to-end displacements when the rod is interacting with a heat bath (as is the case for example in elastic rod models of DNA). We demonstrate how to derive such an approximate conditional probability density function governing the relative location and orientation of the two ends, first for the looping problem and second when the rod is subject to a prescribed external end-loading, in addition to external stochastic forcing from the heat bath (as happens in many single molecule experiments). Understanding looping probabilities, including the particular case of ring-closure or cyclization, of fluctuating polymers (\eg DNA) is important in many applications in molecular biology and chemistry. In this regard, Cosserat rods are a more detailed version of the classic wormlike-chain (WLC) model, which we show to be more appropriate in short-length scale, or stiff, regimes, where the contributions of extension/compression and shear deformations are not negligible and lead to high values for the cyclization probabilities (or $J$-factors). Characterizing the stochastic fluctuations about minimizers of the energy by means of Laplace expansions in a (real) path integral formulation, we develop efficient analytical approximations for the two cases of full looping, in which both end-to-end relative translation and rotation are prescribed, and of marginal looping probabilities, where only end-to-end translation is prescribed. For isotropic Cosserat rods, certain looping boundary value problems admit non-isolated families of critical points of the energy due to an associated continuous symmetry, and the standard Laplace method fails for the presence of zero modes. Taking inspiration from (imaginary) path integral techniques, we show how a quantum mechanical probabilistic treatment of Goldstone modes in statistical rod mechanics sheds light on $J$-factor computations for isotropic rods in the semi-classical context. All the results are achieved exploiting appropriate Jacobi fields arising from Gaussian path integrals, and show good agreement when compared with intense Monte Carlo simulations for the target examples. Analogously, we address the problem of computing end-to-end probabilities in the presence of a prescribed external end-loading. In this case, the partition function yields a non-trivial contribution to the final result, and its evaluation necessary undergoes a Laplace approximation leading to a Gaussian path integral with non-standard boundary conditions. Having stated the theoretical framework, we also provide an importance sampling Monte Carlo method in order to compute expectations with respect to a perturbed Gaussian distribution, as it is the case in presence of an external end-loading. We therefore compare the simulations with the predictions provided by our approximated formulas in the context of a cycling loading example, designed to highlight the role of different boundary conditions and intrinsic shapes of fluctuating elastic rods. 

\vspace{12mm}
\end{abstract}

%----------------------------------------------------------------------------------------
%	ACKNOWLEDGEMENTS
%----------------------------------------------------------------------------------------

\begin{acknowledgements}
\addchaptertocentry{\acknowledgementname} % Add the acknowledgements to the table of contents
\vspace*{3cm}
I am deeply grateful to Prof. Thomas Lessinnes for all the scientific discussions we had and his passion in research, which guided me at the beginning of this work. I thank Matteo Fadel and Raushan Singh for the scientific and non-scientific friendship, whose contributions made this paper possible.\\
\\
Thank you Harmeet Singh for all our conversations. Thank you very much to all my friends, especially ``I Pericolosi'' and the ones of ``La Repubblica del Pastatour'', two milestones in my life. I thank Axel Séguin for our friendship, which began when I started this work and will continue afterwards. To Lara the sweetest of thanks, for her unwavering support and to believe in me more than I believe in myself.

\bigskip
 
\noindent\textit{Lausanne, April 2022}
\hfill G.~C.

\nnfootnote{Ce travail a été supporté par le Fonds National Suisse (FNS) de la recherche scientifique.}
\end{acknowledgements}

%----------------------------------------------------------------------------------------
%	LIST OF CONTENTS/FIGURES/TABLES PAGES
%----------------------------------------------------------------------------------------

\tableofcontents % Prints the main table of contents

\listoffigures % Prints the list of figures

%\listoftables % Prints the list of tables

%----------------------------------------------------------------------------------------
%	DEDICATION
%----------------------------------------------------------------------------------------

\cleardoublepage
\vspace*{3cm}

\begin{raggedleft}
	Nel suo viaggio ha lasciato\\
	piccoli pezzi di lui in dono\\
	ad altre piccole nuvole\\
	giovani e ancora in venire\\
--- Tommaso\\
\end{raggedleft}

\vspace{7cm}

\begin{center}
    To Lara, and to my friends, who have always believed in me
\end{center}

%----------------------------------------------------------------------------------------
%	THESIS CONTENT - CHAPTERS
%----------------------------------------------------------------------------------------

\mainmatter % Begin numeric (1,2,3...) page numbering

\pagestyle{thesis} % Return the page headers back to the "thesis" style

% Include the chapters of the thesis as separate files from the Chapters folder
% Uncomment the lines as you write the chapters

\cleardoublepage
\chapter*{Introduction}
\markboth{Introduction}{Introduction}
\addcontentsline{toc}{chapter}{Introduction}

Path integrals are an essential tool in many branches of physics and mathematics \citep{BookPapa,BookChaichian,BookZinn,BookKleinert, BookSchulman}. Originally introduced by Wiener as a method to study Brownian motion \citep{W1,W2} in a real valued formulation, their formalism was significantly developed by Feynman for the imaginary counterpart in the context of quantum mechanics \citep{FEY,FEY0}. Since then, path integrals revealed themselves to be a powerful method for the investigation of systems subject to classical or quantum fluctuations, therefore finding a plethora of different applications.

In many relevant situations one is interested in evaluating transition (\ie conditional) probabilities. Namely, the probability for a system to be in a specific final state, given its initial state. Path integrals are precisely tailored to answer such questions, by expressing transition probabilities as an infinite weighted sum over all possible trajectories passing through both states. Typical examples where this formulation arises naturally include the dynamics of quantum particles and fields, the stochastic motion of particles in diffusion processes, and the equilibrium statistics of fluctuating polymers. In this work we focus on real valued path integrals, which implies that the first example in the latter list will not be considered, thus excluding the treatment of quantum phenomena.

In order to understand how the random component of the problem is modelled from a mathematical perspective, let us first introduce the classical deterministic description. Namely, we consider a mechanical system with configuration $\bm{q}(\tau)\in\mathbb{R}^{d}$, whose dynamics is described by the Lagrangian $\mathcal{L}(\bm{q},\dot{\bm{q}},\tau)$ and $\dot{\bm{q}}$ denotes the derivative of $\bm{q}$ with respect to $\tau\in\mathbb{R}$, denoting time. Then, given an initial value $\tau=t_0$ and a final value $\tau=t$, the action is defined as 
\begin{equation}\label{actlag}
S(\bm{q}(\tau))=\int_{t_0}^t{\mathcal{L}(\bm{q},\dot{\bm{q}},\tau)\,\dd\tau}\;,
\end{equation} 
and the variational principle of stationary action yields the equations of motion for the system in the form of differential equations. Moreover, the solutions of the equations of motion, which are paths $\bm{q}(\tau)$ satisfying prescribed boundary conditions, \eg prescribed initial and final states $\bm{q}(t_0)=\bm{q}_0$ and $\bm{q}(t)=\bm{q}$, are stationary points of the system's action functional and represent the core for characterising the behaviour of the target mechanical system.

In analogy with the latter deterministic description, let us consider now a stochastic system, \ie a system whose deterministic dynamics for some (slow) macroscopic variables is perturbed by random fluctuations arising from (fast) microscopic variables. Such a system is characterized by the fact that the equations of motion cannot be given for all its constituents, due to a complexity barrier that can only be overcome by a description of a subset of the degrees of freedom, considering the remainder as a stochastic perturbation. Informally, this is the motivation behind the Langevin equation \citep{LAN}, typically defined as a second-order stochastic differential equation of motion for the slow macroscopic variables, where the inertial Newtonian term equals the sum of forces acting on them, namely a damping term, an external potential and a random force modelling the effect of the fast microscopic variables in perturbing the dynamics of the system. Under specific hypotheses, the latter equation can be regarded as a first-order stochastic differential equation in velocity space, in phase space but also in configuration space if the inertia term is negligible compared to the damping force (Brownian or overdamped Langevin equation).

A solution of the Langevin equation is not merely a function, but a stochastic process, whose single realizations are actually functions. A particular realization associated with a solution of the Langevin equation is of no interest by itself; what is relevant, and therefore represents the core for characterising the behaviour of the target stochastic system, is the transition (or conditional) probability density function $\rho\in\mathbb{R}_{>0}$ associated with the stochastic process, and constitutes the target quantity that we aim to compute. In this way it is possible to recover a deterministic formulation, where the ``new equation of motion'' is a partial differential equation, namely the Fokker-Planck equation \citep{FOK, PLANCK} for $\rho(\bm{q},t \vert \bm{q}_0,t_0)$ (as a function of $\bm{q}$ and $t$), denoting the transition probability density of being in final state $\bm{q}(t)=\bm{q}$, given the initial state $\bm{q}(t_0)=\bm{q}_0$, and satisfying the boundary condition 
\begin{equation}\label{boudir}
\rho(\bm{q},t_0 \vert \bm{q}_0,t_0)=\delta(\bm{q}-\bm{q}_0)\;. 
\end{equation}
Furthermore, the density $\rho(\bm{q},t \vert \bm{q}_0,t_0)$ admits a path integral representation in terms of a given Onsager-Machlup function \citep{OM, OMB}, which can be understood as a Lagrangian for the trajectories of the target stochastic process. With abuse of notation, we call such a function as Lagrangian and its integral as action, denoted as in Eq.~(\ref{actlag}).

The preceding preliminary and general introduction can be considered as a common basis that serves as a thread for the various topics of the paper. In particular, we will deal with two different situations that summarise the physical applications given in the first and second parts of the work, respectively.

\subsection*{1. Path integral methods in the presence of isolated minimizers with application to Fokker-Planck dynamics}
We first consider the case of a moving particle for which the chosen physical observables satisfy the Langevin equation, and we are interested in studying the transient behaviour of the associated stochastic process. It is known that the transition probability density solves the Fokker-Planck equation and admits a path integral representation in terms of a given Onsager-Machlup function $\mathcal{L}(\bm{q},\dot{\bm{q}},\tau)$. The derivation of the latter statements is outside the scope of this work, and we regard the possibility to evaluate transition probability densities assuming the following weighting functional
\begin{equation}
\displaystyle{e^{-\int_{t_0}^t{\mathcal{L}(\bm{q},\dot{\bm{q}},\tau)\,\dd\tau}}}\;,
\end{equation}
within the path integral representation, as a starting point to introduce our methods for the computation of normalized Gaussian path integrals. 

A prescription for the calculation of the conditional probability density $\rho(\bm{q},t \vert \bm{q}_0,t_0)$ satisfying Eq.~(\ref{boudir}) can be given by a path integral formalism, which allows us to write
\begin{equation}\label{pathintbis}
\rho(\bm{q},t \vert \bm{q}_0,t_0) = \dfrac{\mathcal{K}}{\mathcal{N}} = \frac{1}{\mathcal{N}}\int\limits_{\bm{q}(t_0)=\bm{q}_0}^{\bm{q}(t)=\bm{q}}{e^{-S(\bm{q})}\,\mathcal{D}\bm{q}} \;,
\end{equation}%
where the (conditional) integration is taken over all paths with fixed end points, weighted with the action $S(\bm{q})$ and normalized by $\mathcal{N}$ to ensure
\begin{equation}\label{normbis}
\int_{\mathbb{R}^{d}}{\rho(\bm{q},t \vert \bm{q}_0,t_0)}\,\dd \bm{q} = 1 \;, \qquad \forall t\geq t_0 \;.
\end{equation}
From Eqs.~(\ref{pathintbis}) and (\ref{normbis}), we see that the normalization can be formally written as the (unconditional) path integral 
\begin{equation}\label{normexbis}
\mathcal{N} = \int\limits_{\bm{q}(t_0)=\bm{q}_0}{e^{-S(\bm{q})}\,\mathcal{D}\bm{q}} \;,
\end{equation}%
where now the integral is over all paths satisfying only the initial condition $\bm{q}(t_0)=\bm{q}_0$.

Since in applications $\rho(\bm{q},t \vert \bm{q}_0,t_0)$ is only rarely computable explicitly, the more general problem of computing marginal transition probabilities via path integrals is of interest. Therefore, we reorder the configuration variables as 
\begin{equation}
\begin{split}
&\bm{q}(\tau)=(\bm{q}_{V}(\tau),\bm{q}_{F}(\tau))\in\mathbb{R}^{d}\;,\\
&\bm{q}_{V}(\tau)=(q_1,...,q_l)(\tau)\in\mathbb{R}^l\;,\\
&\bm{q}_{F}(\tau)=(q_{l+1},...,q_d)(\tau)\in\mathbb{R}^{d-l}\;,\\
\end{split}
\end{equation}
and consider the marginals 
\begin{equation}
\rho_m(\bm{q}_{F},t \vert \bm{q}_0,t_0) = \int_{\mathbb{R}^{l}} \rho(\bm{q},t \vert \bm{q}_0,t_0)\,\dd\bm{q}_{V}\;. 
\end{equation}
The latter can be expressed as the path integral
\begin{equation}\label{pathint2bis}
\rho_m(\bm{q}_{F},t \vert \bm{q}_0,t_0) = \dfrac{\mathcal{K}_m}{\mathcal{N}} = \frac{1}{\mathcal{N}}\int\limits_{\bm{q}(t_0)=\bm{q}_0}^{\bm{q}_{F}(t)=\bm{q}_{F}}{e^{-S(\bm{q})}\,\mathcal{D}\bm{q}} \;,
\end{equation}%
where the integration for $\mathcal{K}_m$ is taken over all paths starting with $\bm{q}(t_0)=\bm{q}_0$ and with the mixed end-point conditions $\bm{q}_{F}(t)=\bm{q}_{F}$ fixed and $\bm{q}_{V}(t)=\bm{q}_{V}$ variable, further noting that the normalization term $\mathcal{N}$ remains the same as in Eq.~(\ref{normexbis}).

Within this framework, the first part of the paper is devoted to deriving approximate formulas for the evaluation of $\mathcal{K}$, $\mathcal{K}_m$ and $\mathcal{N}$ by means of a quadratic expansion about an isolated minimizer of the action, and to showing how the results obtained can be applied to solve Fokker-Planck dynamics, in particular for recovering the transition probability of the $d$-dimensional Ornstein-Uhlenbeck process \citep{OU,FALKOFF,VATI} and to investigate the non-linear Van der Pol oscillator driven by white noise, for which transition probabilities are not known analytically due to its chaotic dynamics \citep{NAESS}.

Despite their intuitive interpretation, path integrals are in general difficult to compute. Moreover, for typical cases of interest, we are often in the situation where $\mathcal{L}$ is complicated enough that closed-form solutions for $\mathcal{K}$, $\mathcal{K}_m$ and $\mathcal{N}$ do not exist. Among several different strategies to circumvent this issue and simplify part of the problem, the semi-classical (quadratic) approximation is one of the most adopted \citep{MOR, REC, BookLan}. In brief, the idea consists in approximating the weights for the paths by expanding the action to second order around an isolated minimum so that a Gaussian integral is obtained. In this work, we focus on Wiener (real valued) path integrals, for which the semi-classical approximation takes the name of Laplace expansion \citep{PIT}, and we prescribe non-standard boundary conditions for the paths entering the infinite sum, so that to extend previously known computations. 

Once the Gaussian representation is recovered, the solution is usually straightforward for conditional Dirichlet-Dirichlet path integrals ($\mathcal{K}$), where both extremal points are fixed. Nevertheless, it remains non-trivial for marginals ($\mathcal{K}_m$), where the starting point is fixed and the variables describing the end point are in part fixed and in part free, and for the unconditional Dirichlet-Neumann case of the normalization constant ($\mathcal{N}$), where only the starting point is fixed. In this regard, we point out that for special cases $\mathcal{N}$ can be evaluated analytically by means of an appropriate change of variables \citep{LUDT, LUD}, but in general we need more sophisticated methods. Addressing these remaining problems without resorting to demanding Monte-Carlo integrations, or by setting $\mathcal{N}=1$ and considering in $\mathcal{K}$ an effective (Onsager-Machlup) Lagrangian containing non-trivial additional terms that ensure normalization \citep{FALKOFF, HAKEN, GRAHAM}, is of special interest for expressing general transition probability distributions under a unified analytical and efficient approach. In summary, we represent the conditional probability density $\rho$ as a ratio of path integrals in the form of Eq.~(\ref{pathintbis}) and express its Laplace approximation taking into account a general computation for $\mathcal{N}$ in Eq.~(\ref{normexbis}). Moreover, we present an extension of the method which allows the evaluation of marginal conditional probabilities $\rho_m$ in the Laplace approximation. As a side note, without aiming at any specific application, we report the calculation of Gaussian path integrals with Neumann-Neumann boundary conditions, which differ substantially in both calculation and final results with the previously mentioned cases. 

Most of the methods presented for deriving the main results of the first part of the paper can be regarded as background material, which is organized as follows. In presence of cross terms in the second variation of $S(\bm{q}(\tau))$ for the general vector case, also the Dirichlet-Dirichlet problem $\mathcal{K}$ requires a non-trivial analysis, which was first introduced in \citep{PAP1} and that we report as background material in 1.3.1 and 1.4.2.1 of Chap. 1. Our approach is based on the generalization of such results of Papadopoulos, which allow us to evaluate both conditional and unconditional path integrals for general quadratic Lagrangians, from the solutions of the Euler-Lagrange equations and of a system of second-order non-linear differential equations (see 1.4.2.1). Furthermore, as first done in \citep{LUDT, LUD} for the Dirichlet-Dirichlet case, and illustrated in 1.4.3.1 of Chap. 1, we show that the mentioned non-linear system can be related to a simpler and well-known linear differential system of Jacobi equations. These equations, arising from the second variation of the action, together with the appropriate initial value problems, are the same exploited for detecting if an equilibrium arising from a variational principle is also a minimizer for the target functional, which is a topic with a long tradition in mathematics and physics; we give a brief summary of some useful results concerning the second variation in Subsec. 1.2.1 of Chap. 1. Finally, we mention that the Hamiltonian form of the Jacobi equations is useful for dealing with non-strictly convex problems, as described in \citep{LUDT, LUD} in the context of elastic rods and presented here as background material in 1.4.3.3 of Chap. 1.

\subsection*{2. Stochastic Cosserat rods as a model of polymers, the looping problem and path integrals in the presence of non-isolated minimizers}
If we take the limit for the time variable going to infinity, the transition probabilities for the two examples mentioned above (Ornstein-Uhlenbeck and Van der Pol oscillator driven by white noise) converge to a stationary distribution, which for the Ornstein-Uhlenbeck process is particularly simple and analytical, being Gaussian, therefore recovering Maxwell-Boltzmann statistics at thermal equilibrium. This motivates the following problem. We consider now an elastic rod in a thermal bath subject to stochastic fluctuations, whose Langevin equation can be expressed by means of a stochastic partial differential equation. In fact, in addition to the time variable $t$, a one dimensional space variable $s$ acting as a parametrization of the framed curve modelling the elastic rod is required. Since we are interested in studying the stationary behaviour in the infinite time limit at thermodynamic equilibrium, we assume to recover a Boltzmann distribution, which is not a simple function here, but rather an object with an infinite dimensional representation given by a path integral formalism on curves parametrized by the parameter $s$. 

As we can observe, unlike in the first part of the paper, the ``paths'' entering a path integral do not need to be the trajectories of a moving particle (in time), but they can also be the (static) configurations of fluctuating string-like objects at thermal equilibrium \citep{Edwards1965,Edwards1967,FR2,PapaPoly} by means of a parametrization of the curve describing their centerline and the orientation of their cross-section. In this context transition probabilities can still be defined along the parameter $s$, and take the general name of conditional probabilities. In particular, we specialize to the case of end-to-end distributions within the statistical mechanics of polymers and we are interested in conditional probabilities expressing the likelihood of finding the string's endpoints in specific positions (and possibly orientations). This observation turns out extremely useful for the study of organic and inorganic polymers at thermal equilibrium, such as chains of molecules (\eg DNA, actin filaments) that can be described by continuum elastic models \citep{Winkler94,Winkler97,Vilgis00,LUDT,LUD}.

The continuum elastic model adopted in this work is the special Cosserat theory of elastic rods \citep{COS, ANT}. A configuration of a Cosserat rod is a framed curve 
\begin{equation}
\bm{q}(s)=(\bm{R}(s),\bm{r}(s))\in SE(3)\quad\text{for each}\quad s\in [0,L]\;, 
\end{equation}
which may be bent, twisted, stretched or sheared, where the vector $\bm{r}(s)\in\mathbb{R}^3$ and the matrix $\bm{R}(s)\in SO(3)$ model respectively the rod centerline and the orientation of the material in the rod cross-section. If we consider now an elastic rod at thermodynamic equilibrium with a heat bath in absence of external forces, then we can assume without loss of generality that $\bm{q}(0)=\bm{q}_0=(\mathbb{1},\bm{0})$ and, given a prescribed $\bm{q}_L=(\bm{R}_L,\bm{r}_L)\in SE(3)$, we formulate the problem of computing a conditional probability density function for the other end of the rod to satisfy at $s=L$ either $\bm{q}(L)=\bm{q}_L$, or the weaker condition $\bm{r}(L)=\bm{r}_L$. The first case gives rise to a conditional probability density function $(\mathtt{f})$ over the space $SE(3)$, whereas the second one represents the $\mathbb{R}^3$-valued marginal $(\mathtt{m})$ over the final rotation variable, with no displacement constraint on $\bm{R}(L)$. 

We therefore assume the following weighting functional in configuration space within the path integral
\begin{equation}
\displaystyle{e^{-\beta E(\bm{q}(s))}}\;,
\end{equation}
based on the elastic potential energy $E$ of the system, with $\beta$ the inverse temperature. Analogously to what done for the first part in time, the $SE(3)$ and $\mathbb{R}^3$ densities $\rrf$ and $\rrm$ are then respectively given as the ratios of infinite dimensional Wiener integrals
\begin{equation}\label{densbis}
\rrf(\bm{q}_L,L|\bm{q}_0,0)=\frac{\mathcal{K}_f}{\mathcal{Z}}\;,\quad\rrm(\bm{r}_L,L|\bm{q}_0,0)=\frac{\mathcal{K}_m}{\mathcal{Z}}\;,
\end{equation}
\begin{equation}\label{pathint_ch2bis}
{\mathcal{K}_f}=\int\limits_{\bm{q}(0)=\bm{q}_0}^{\bm{q}(L)=\bm{q}_L}{e^{-\beta E(\bm{q})}\,\mathcal{D}\bm{q}}\;,\,\,\,{\mathcal{K}_m}=\int\limits_{\bm{q}(0)=\bm{q}_0}^{\bm{r}(L)=\bm{r}_L}{e^{-\beta E(\bm{q})}\,\mathcal{D}\bm{q}}\;,
\end{equation}
and $\mathcal{Z}$ is the partition function expressed as a path integral that guarantees the normalisation condition:
\begin{equation}\label{normex_ch2bis}
\mathcal{Z} = \int\limits_{\bm{q}(0)=\bm{q}_0}{e^{-\beta E(\bm{q})}\,\mathcal{D}\bm{q}}\;,
\end{equation}
\begin{equation}
\int_{SE(3)}{\rrf(\bm{q}_L,L|\bm{q}_0,0)}\,\dd \bm{q}_L=\int_{\mathbb{R}^3}{\rrm(\bm{r}_L,L|\bm{q}_0,0)}\,\dd \bm{r}_L=1\;.
\end{equation}

The problem denoted by $(\mathtt{f})$ arises in modelling looping in $SE(3)$, whereas problem $(\mathtt{m})$ arises in modelling looping in $\mathbb{R}^3$, where the value of $\bm{R}_L$ is a variable left free, over which one marginalises. In particular, we will specialize our analysis to the case of ring-closure or cyclization, where $\bm{r}_L=\bm{0}$ and, for problem  $(\mathtt{f})$, also $\bm{R}_L=\mathbb{1}$. On one hand, in presence of isolated minimizers, the computational methods based on Laplace approximation for the evaluation of Eqs.~(\ref{pathint_ch2bis}) and (\ref{normex_ch2bis}) are the same as the ones employed above for solving Fokker-Planck dynamics, with the difference that now we first have to deal with the rotation group $SO(3)$, being part of the configuration variable $\bm{q}(s)=(\bm{R}(s),\bm{r}(s))$, which gives rise to a manifold structure that should be treated carefully in order to recover eventually a ``flat space" formulation. On the other hand, certain looping boundary value problems admit families of non-isolated minimizers arising as a consequence of continuous symmetries (for example in the case of isotropic and/or uniform rods). In particular, we provide a theory for one symmetry parameter $\theta$ (as we want to deal with isotropic rods), but the same scheme can be suitably generalised to more symmetry parameters. The presence of a family of minimizers denoted by $\bar{\bm{q}}(s;\theta)$ translates into a zero mode $\bar{\bm{\psi}}(s;\theta)$ of the self-adjoint operator associated with the second variation of the energy. Consequently, we cannot proceed as before, for otherwise the zero eigenvalue would lead to singular results in the Laplace expansion based on Jacobi fields. We will show how to obtain approximated formulas in such cases, by integrating over variations which are orthogonal to the zero mode enforcing the constraint 
\begin{equation}
\int_0^L{\left[\left(\bm{q}-\bar{\bm{q}}\right)\cdot\frac{\bar{\bm{\psi}}}{\Vert\bar{\bm{\psi}}\Vert}\right]}\,\,\dd s=0
\end{equation}
for a particular value of $\theta$, and suitably adapting/generalizing some important results in the context of functional determinants and boundary perturbations \citep{FORM, MCK, FAL}.

Having introduced the content of the second part of the paper and explained its relationship to the first, in the following we provide a detailed introduction to this second part, motivating its scientific interest and results obtained.

It is widely known that polymers involved in biological and chemical processes are anything but static objects. In fact, they are subject to stochastic forcing from the external environment that lead to complex conformational fluctuations. One of the fundamental phenomena which is understood to perform a variety of roles is polymer looping, occurring when two sites separated by several monomers, and therefore considered far from each other, come in close proximity. A basic observation is that the interacting sites alone do not characterize the phenomenon of looping, but rather it is the whole polymeric chain that rearranges itself for this to occur. As a consequence, the length and mechanical properties of the chain, together with the thermodynamic surrounding conditions are finely tuning the likelihood of such events. There are many reasons to study this topic, which have led to a considerable literature. For instance,  looping is involved in the regulation of gene expression by mediating the binding/unbinding of DNA to proteins \citep{LOOP1, LOOP2, LOOP3}, such as the classic example of the Lac operon \citep{BIO3, LAC2}. In addition, DNA packaging (chromatin formation) \citep{PAC1}, replication and recombination \citep{PAC2, LOOP1} depend on the ability of the polymer to deform into loop configurations, as do other cellular processes. Proteins exhibit intrachain loops for organizing the folding of their polypeptide chains \citep{PROT1}, \eg antibodies use loops to bind a wide variety of potential antigens \citep{PROT2}. When dealing with a closed loop, it is usually appropriate to refer to cyclization or ring closure. In this regard, the production of DNA minicircles is being investigated for their possible therapeutic applications \citep{MINI}. Even in the context of nanotechnologies, ring closure studies have been performed for carbon nanotubes subject to thermal fluctuations \citep{NANO} and wormlike micelles \citep{MIC}.

From the modelling point of view, it is appropriate to look back at some of the historical milestones that underpin our work. In 1949, Kratky and Porod \citep{WLC} introduced the wormlike-chain (WLC) model for describing the conformations of stiff polymer chains. Soon after, the complete determination of the polymeric structure of DNA \citep{1953} guided scientists towards the application of WLC-type models in the context of DNA statistical mechanics, allowing probabilistic predictions of relevant quantities of interest. Historically, the computations have been performed in terms of Fokker-Plank equations \citep{DAN, DIFF}, but also exploiting the point of view of path integrals \citep{PIC, FR1, FR2}, a technique inherited from Wiener's work \citep{W1, W2} and quantum mechanics \citep{BookFeynman}. These ideas were largely investigated by Yamakawa \citep{YAM0, YAM00, YAM1, YAM2, YAM3, YAM4, YAM6, YAM7}, who in particular considered the problem of computing ring-closure probabilities, now ubiquitous in molecular biology \citep{BIO4, BIO5, BIO6}. Nowadays, for a homogeneous chain, the exact statistical mechanical theory of both the WLC and the helical WLC (with twist) is known \citep{WL1, WL2, WL3}, and the topic has been rigorously phrased over the special Euclidean group $SE(3)$ \citep{CHIR2}.

In parallel, in the early years of the 20th century, the Cosserat brothers Eugène and Fran\c cois formulated Kirchhoff’s rod theory using what are now known as directors \citep{COS}. However, the difficulties arising from the generality of the model, which includes the WLC as a particular constrained case, hindered its application to stochastic chains. Only quite  recently, targeting a more realistic description of DNA, the mentioned framework has been partially or fully exploited both within new analytical studies \citep{MMK, ZC, NP, LUDT, LUD, CHIR1} and intense Monte Carlo simulations \citep{ALEX1, MCDNA, ALEX2, MCD}, the latter being only a partial solution because of time and cost.

In this work, we aim to fill the gap between user-friendly but simplistic models (WLC) on the one hand, and accurate but expensive simulations (Monte Carlo) on the other, still maintaining the analytical aspect which allows one to draw conclusions of physical interest. This is achieved using \citep{LUDT, LUD} as a starting point for bridging the two historical lines of research, \ie exploiting efficient (real) path integral techniques in the semi-classical approximation \citep{2020, PAP1, BookChaichian, BookSchulman, BookWiegel, MOR} (or Laplace method \citep{PIT}), and working within the special Cosserat theory of rods in $SE(3)$, which is presented as background material in Subsec. 2.1.1 of Chap. 2, with special focus on the Hamiltonian formulation for equilibria and their stability \citep{HAM}. Namely, for studying the end-to-end relative displacements of a fluctuating polymer at thermodynamic equilibrium with a heat bath, we describe the configurations of the chain in a continuum limit by means of framed curves over the special euclidean group. Thus, from an assumed Boltzmann distribution on rod configurations, a conditional probability can be expressed as the ratio of a Boltzmann weighted integral over all paths satisfying the desired end conditions, to the analogous weighted integral over all admissible paths (partition function). Note that the latter path integral formulation in $SE(3)$ follows the work done in \citep{LUDT}, adopting a quaternion based parametrisation of $SO(3)$, which is regarded as background in Subsec. 2.2.2 of Chap. 2. The resulting path integrals are finally approximated via a quadratic (Laplace) expansion about a minimal energy configuration, for which the crucial assumption is that the energy required to deform the system is large with respect to the temperature of the heat bath. This means computing probabilities for length scales of some persistence lengths or less, which turns out to be of great relevance in biology. In particular, the contribution of local quadratic fluctuations around the minimizer is captured by an appropriate modification of a conjugate point stability analysis. In this regard, the second variation for linearly elastic rods is a standard computation and the Jacobi fields for the inextensible and unshearable Kirchhoff case are recovered as a smooth limit in Hamiltonian form as described in \citep{LUDT, LUD}. We report and elaborate the latter material in Subsec. 2.3.1 of Chap. 2.

The present study is general and is applicable to various end-to-end statistics. Due to the scientific significance of the problem, we focus on the computation of ring-closure or cyclization probabilities for elastic rods, targeting three significant aspects. The first is the possibility of systematically distinguishing between the statistics provided by end positions alone (marginal looping) and the ones provided including also end orientations (full looping) \citep{2020}, for Kirchhoff as well as for Cosserat rods. We emphasise that although Kirchhoff rod theory \citep{AN1} generalises both Euler's elastica theory to model deformations in three dimensions, and the WLC model allowing arbitrary bending, twisting and intrinsic shapes of the rod, it does not allow extension or shearing of the rod centerline. This is indeed a prerogative of the Cosserat, more general framework, where the centerline displacement and the cross-section’s rotation are considered as independent variables. We show that these additional degrees of freedom are crucial in the analysis of polymer chains in short-length scale, or stiff, regimes, in both the full and marginal cases, where the system exploits extension and shear deformations for minimizing the overall elastic energy, in the face of an increasingly penalizing bending contribution. This allows the cyclization probability density to take high values even when the WLC model (and Kirchhoff) is vanishing exponentially.

The second is addressing the ``perfect problem'' in the semi-classical context, where the symmetry of isotropy gives rise to a ``Goldstone mode'' \citep{G1} leading to a singular path integral, and requires a special treatment by suitably adapting (imaginary) quantum mechanical methods \citep{FADPOP, GHO, COL1, POLYA, JAR, BER} and functional determinant theories \citep{FORM, MCK, FAL}, which are novel in such a generality in the context of elastic rod. For simple models, an analysis in this direction is present in \citep{GUER}. The concepts of isotropy and non-isotropy can be roughly related to a circular shape rather than an elliptical shape for the cross-section of the rod, and the two cases have two different mathematical descriptions in terms of Gaussian path integrals, which we discuss in detail in the course of this paper. In particular, the effect of non-isotropy for semiflexible chain statistics has been addressed from a path integral point of view in \citep{LUD} for the planar case and in \citep{LUDT} for the three-dimensional case (and will be here taken up and simplified), but without resolving the singularity arising in the isotropic limit.

The last significant aspect included in the present work is deriving approximated solution formulas that can always be easily evaluated through straightforward numerical solution of certain systems of Hamiltonian ODE, which in some particularly simple cases can even be evaluated completely explicitly and naturally arise from the ones observed within simpler WLC models \citep{YAM00}. Versions of the solution formulas, involving evaluation of Jacobi fields at different equilibria and subject to different boundary conditions, are obtained for the two cases of full and marginal ring-closure probabilities. We recall that the case of full looping in presence of isolated minimizers has been solved in \citep{LUDT, LUD}; these background results are reported within Subsec. 2.6.2 of Chap. 2. The efficiency aspect in computing looping probabilities, maintaining the same accuracy of Monte Carlo in the biologically important range less than 1-2 persistence lengths, is fundamental. This is because Monte Carlo simulation is increasingly intractable due to the difficulty of obtaining sufficiently good sampling with decreasing polymer length, which is the limit where the approximation is increasingly accurate. Contrariwise our approximations are inaccurate in longer length regimes where good Monte Carlo sampling is easily achieved. Remarkably, the qualitative behaviour of the probability densities coming from Laplace approximation and from Monte Carlo sampling are the same regardless of the length scale, since the error does not explode but stabilizes.

We stress that the stiffness parameters expressing the physical properties of the polymer are allowed to vary along the material parameter of the curve, leading to a non-uniform rod which, in the context of DNA, would represent sequence-dependent variations. In addition, the model allows coupling between bend, twist, stretch and shear, as well as a non-straight intrinsic shape. Notwithstanding the latter generality, we prefer to illustrate our method with some basic examples of uniform and intrinsically straight rods and comparing it with a suitable Monte Carlo algorithm, in order to highlight the contributions provided by the different choices of cyclization boundary conditions in the presence of isotropy or non-isotropy, and to investigate the effect of shear and extension when moving from Kirchhoff to Cosserat rods. The results will be exposed under the hypothesis of linear elasticity, even though the theory applies to more general energy functionals.

In the same setting of the looping problem, we devote the last chapter of the paper to the study of another relevant application where elastic rods interact with both a stochastic  environment and a deterministic external input. Namely, we aim to demonstrate how to derive an approximate conditional probability density function governing the relative location and orientation of the two ends, when the rod is subject to a prescribed external end-loading, in addition to external stochastic forcing from the heat bath. The result of this work is not only an interesting analytical exploration, but is also applicable to different biological situations, \eg in the field of single molecule experiments \citep{SME1, SME2}. Moreover, for DNA, the presence of an external force can also be motivated by proteins binding the molecule in different sites. In the literature, simple WLC and helical worm-like chain models have been exploited to investigate the probability of loop formation under tension and to characterise force–extension curves varying the parameters of the model \citep{EXFOR1, EXFOR2}. It is not our goal to pursue a precise application, but rather to provide a general computational framework where the degrees of freedom of a Cosserat rod potentially include all commonly used models.

The main motivation for us will be the analysis of a system driven by a more general energy, where the non-local additional term associated with the external end-loading breaks the quadratic structure of the linearly elastic energy. The very new obstacle is therefore the evaluation of the partition function. In \citep{LUDT, LUD}, the structure of the energy functional allowed an exact computation based on a suitable change of variables, showing that the partition function was not providing any non-trivial contribution to the final result. However, in presence of end-loading the cited method can no longer be applied, and we have to look for an approximate result. We demonstrate that the theory developed for normalizing Gaussian path integrals, by means of a Laplace expansion of the denominator having Dirichlet-Neumann boundary conditions, is able to capture the relevant features of the new system. This is shown by implementing an importance sampling Monte Carlo algorithm for perturbed Gaussian distributions that is compared with our approximate predictions in the case of a cycling loading example. The latter is designed to systematically distinguish between the fluctuations about Dirichlet-Dirichlet minimizers and the ones about Dirichlet-Neumann minimizers in the presence of a non-straight intrinsic shape.

\subsection*{General form of derived approximation formulae}
All formulas for conditional probability densities $\rho$ that are derived within the present work through the Laplace approximation method, whether the minimizers are isolated or non-isolated, can be written generically in the following form 
\begin{equation}\label{formform}
\rho\approx c\,\,e^{\beta\left[S(\hat{\bm{q}})-S(\bar{\bm{q}})\right]}\sqrt{\mu\left(\bar{\bm{\psi}}\right)\det\left[\frac{{\hat{\bm{H}}}}{{\bar{\bm{H}}_{{\bm{\psi}}}}}(t_0)\right]} \;,
\end{equation}
where $c\in\mathbb{R}$ is a constant that depends on the problem, $\bar{\bm{q}}(\tau)\in\mathbb{R}^d$ is a minimizer for the computation of $\mathcal{K}$ or $\mathcal{K}_m$, $\hat{\bm{q}}(\tau)\in\mathbb{R}^d$ is a minimizer for the computation of $\mathcal{N}$, and $\bm{H}(\tau)\in\mathbb{R}^{d\times d}$ are the associated Jacobi fields, i.e. a matrix solution to the initial value problem for the Jacobi equation that we write in Hamiltonian form as 
\begin{equation}\label{formjac}
\begin{cases}
             \frac{\dd}{\dd\tau}\begin{pmatrix}
      {{\bm{H}}}  \\
     {{\bm{M}}}
     \end{pmatrix}
       =\bm{J}{{\bm{E}}}\begin{pmatrix}
       {{\bm{H}}}  \\
     {{\bm{M}}}
     \end{pmatrix}
   \\
   \\
\begin{pmatrix}
      {{\bm{H}}}  \\
     {{\bm{M}}}
     \end{pmatrix}
     (t)
       =IC
       \end{cases}\;,
\end{equation}
with $\bm{M}(\tau)\in\mathbb{R}^{d\times d}$ the conjugate variables under the Legendre transform, $\bm{J}\in\mathbb{R}^{2d\times 2d}$ the symplectic matrix, $\bm{E}(\tau)\in\mathbb{R}^{2d\times 2d}$ the symmetric matrix driving the system, and $IC$ an appropriate set of initial conditions. In particular, when Eqs.~(\ref{formform}) and (\ref{formjac}) refer to conditional probabilities for elastic rods, then we have to perform the substitutions $d\rightarrow 6$, $S\rightarrow E$, $\tau\rightarrow s$ $t_0\rightarrow 0$, $t\rightarrow L$, $\mathcal{K}\rightarrow \mathcal{K}_f$, $\mathcal{N}\rightarrow \mathcal{Z}$. Moreover, in presence of non-isolated minimizers $\bar{\bm{q}}$, Eq.~(\ref{formform}) includes the terms $\mu\left(\bar{\bm{\psi}}\right)\in\mathbb{R}$ and $\bar{\bm{H}}_{{\bm{\psi}}}$ which depend on the null-function $\bar{\bm{\psi}}$, the second denoting a suitable sub-matrix of $\bar{\bm{H}}$. If instead $\bar{\bm{q}}$ is isolated, then in Eq.~(\ref{formform}) we set $\mu\left(\bar{\bm{\psi}}\right)=1$ and $\bar{\bm{H}}_{{\bm{\psi}}}=\bar{\bm{H}}$. In summary, we present Eq.~(\ref{formform}) as the proper synthesis of zero-order and second-order contributions, arising respectively from Euler-Lagrange solutions and (regularised) Jacobi fields, which is able to capture the relevant features of general conditional probability densities under certain hypotheses related to the chosen approximation scheme.

\subsection*{Outline}
The paper is divided into two parts. Part I is devoted to path integral methods with direct application to Fokker-Plank dynamics, which include the content of \citep{2020}; Part II is about the theory of stochastic elastic rods, for which the former methods are applied, as well as new techniques that arise naturally within this topic, which include the content of \citep{2022}. More specifically, the structure of the work is as follows.

In Chap. $1$ we devote the first section to giving a summary of the problem and stating the main results, introducing the concepts of conditional probability density and of marginal conditional probability density. Then we collect some preliminary remarks, providing a background on stability analysis and introducing the Laplace method in the finite dimensional setting. The core of the chapter starts with the block diagonalization of the discretized second variation functional, which allows the evaluation of Gaussian path integrals à la Feynman. Specifically, we show how to extend Papadopoulos' method to more general boundary conditions on the paths. Finally, we recover a continuous formulation in terms of differential equations, in particular the (linear) Jacobi equations with suitable initial conditions. As a side note, the case of Neumann-Neumann boundary conditions on paths is treated for mathematical completeness. The theory is applied to solve Fokker-Planck dynamics and we consider the specific cases of the d-dimensional Ornstein–Uhlenbeck process and of the Van der Pol oscillator driven by white noise.

In Chap. $2$, first we give an overview of the statics of special Cosserat rods, with particular emphasis on equilibria and stability for the boundary value problems involved. In particular, the Hamiltonian formulation of the Euler-Lagrange and Jacobi equations provides a common theoretical framework for both Kirchhoff and Cosserat rods. Second, we present a preview of the examples that will be considered in the course of the paper, namely in the context of linear elasticity. Therefore, we characterise the minimizers of the energy, distinguishing between the non-isotropic and isotropic cases. The role of the continuous variational symmetries of isotropy and uniformity is explained. Then, we devote a section for describing the path integral formulation of fluctuating elastic rods as a model of polymers, introducing an appropriate parametrisation of the rotation group and giving the functional representations of full and marginal looping probability densities. Afterwards, the explicit approximated formulas for such densities are derived, initially in the case of isolated minimizers and thereafter in presence of non-isolation, for which a special theoretical analysis is performed. Moreover, we provide a Monte Carlo algorithm for stochastic elastic rods, exploited to benchmark our results. The examples are finally investigated from the point of view of cyclization probabilities, with special focus on shear and extension contributions for Cosserat rods in the short-length scale regimes.

In Chap. $3$ we address the problem of computing end-to-end conditional probabilities for elastic rods in presence of an external end-loading. Initially, we analyse the partition function and propose an approximate method for its evaluation based on the content of Chap. $1$. Then we develop an importance sampling Monte Carlo algorithm in order to compare our theoretical results and assess the error introduced by the Laplace approximation. To conclude, we illustrate a cyclic loading example, where the main features of the method are explained.

Further discussion and future directions follow in the ``Conclusions'' chapter.

\subsection*{Publications reported in extended version in this work}
\begin{itemize}
\item ``Normalized Gaussian path integrals''\\
      G. Corazza and M. Fadel\\
      Phys. Rev. E, 102:022135, Aug 2020
\item ``Unraveling looping efficiency of stochastic Cosserat polymers''\\
      G. Corazza and R. Singh\\
      Phys. Rev. Research, 4:013071, Jan 2022
\end{itemize}
\cleardoublepage
\part{Path integral methods}
\chapter{Normalized Gaussian path integrals}

\section{Summary of main results}
\subsection{Laplace approximation of conditional probabilities}
We briefly recall that the conditional probability density $\rho(\bm{q},t \vert \bm{q}_0,t_0)$ satisfying Eq.~(\ref{boudir}) can be given by a path integral formalism, which allows us to write
\begin{equation}\label{pathint}
\rho(\bm{q},t \vert \bm{q}_0,t_0) = \dfrac{\mathcal{K}}{\mathcal{N}} = \frac{1}{\mathcal{N}}\int\limits_{\bm{q}(t_0)=\bm{q}_0}^{\bm{q}(t)=\bm{q}}{e^{-S(\bm{q})}\,\mathcal{D}\bm{q}} \;,
\end{equation}%
where the (conditional) integration is taken over all paths with fixed end points and it is normalized by a  (unconditional) path integral over all paths satisfying only the initial condition $\bm{q}(t_0)=\bm{q}_0$
\begin{equation}\label{normex}
\mathcal{N} = \int\limits_{\bm{q}(t_0)=\bm{q}_0}{e^{-S(\bm{q})}\,\mathcal{D}\bm{q}} \;,
\end{equation}%
to get
\begin{equation}\label{norm}
\int_{\mathbb{R}^{d}}{\rho(\bm{q},t \vert \bm{q}_0,t_0)}\,\dd \bm{q} = 1 \;, \qquad \forall t\geq t_0 \;.
\end{equation}

As a starting point, in order to ensure the accuracy of the Laplace approximation, let us restrict to Lagrangian functions where the leading order term for the second variation of the action, $\bm{P}(\tau)= \frac{\partial^2 \mathcal{L}}{\partial \dot{\bm{q}}^2}\in\mathbb{R}^{d\times d}$ (symmetric), is independent of $\bm{q}, \dot{\bm{q}}$. This assumption is still general enough to include most cases of interest. On the other hand, we consider in the second variation arbitrary $\bm{Q}= \frac{\partial^2 \mathcal{L}}{\partial {\bm{q}}^2}\in\mathbb{R}^{d\times d}$ (symmetric) and cross term matrix $\bm{C}= \frac{\partial^2 \mathcal{L}}{\partial \dot{\bm{q}}\partial \bm{q}}\in\mathbb{R}^{d\times d}$, with $\bm{C}$ not necessarily symmetric.

Following the idea behind the Laplace approximation, the first step of the method consists in deriving from the Euler-Lagrange equations for $\mathcal{L}$ two solutions: 
\begin{itemize}
\item an isolated minimizer of the action $\bm{q}^{f}(\tau)$, satisfying the Dirichlet-Dirichlet boundary conditions $\bm{q}^{f}(t_0)=\bm{q}_0$ and $\bm{q}^{f}(t)=\bm{q}$,
\item an isolated minimizer $\hat{\bm{q}}(\tau)$, satisfying the Dirichlet boundary conditions $\hat{\bm{q}}(t_0)=\bm{q}_0$ and the Neumann natural boundary conditions $\frac{\partial \mathcal{L}}{\partial \dot{\bm{q}}}(t)\Big{|}_{\hat{\bm{q}}}=\bm{0}$.
\end{itemize}
Then, the second step consists in deriving two sets of solutions $\bm{H}^{f}(\tau)$, $\hat{\bm{H}}(\tau)$ of the Jacobi equations for the second variation of the action computed on the related minimum. These can be obtained from the Hamiltonian formulation of the Jacobi equations, together with the appropriate initial conditions (given at $\tau=t$), as solutions of 
\begin{equation}\label{fin3}
\begin{cases}
             \frac{\dd}{\dd\tau}\begin{pmatrix}
      {{\bm{H}^{f}}}  \\
     {{\bm{M}^{f}}}
     \end{pmatrix}
       =\bm{J}{{\bm{E}^{f}}}\begin{pmatrix}
       {{\bm{H}^{f}}}  \\
     {{\bm{M}^{f}}}
     \end{pmatrix}
   \\
   \\
\begin{pmatrix}
      {{\bm{H}^{f}}}  \\
     {{\bm{M}^{f}}}
     \end{pmatrix}
     (t)
       =\bm{J}\begin{pmatrix}
       {\mathbb{1}} \\
     \mathbb{0}
     \end{pmatrix}
       \end{cases}\;, \qquad
\begin{cases}
           \frac{\dd}{\dd\tau}\begin{pmatrix}
      {{\hat{\bm{H}}}}  \\
     {{\hat{\bm{M}}}}
     \end{pmatrix}
       =\bm{J}{{\hat{\bm{E}}}}\begin{pmatrix}
       {{\hat{\bm{H}}}}  \\
     {{\hat{\bm{M}}}}
     \end{pmatrix}
   \\
   \\
\begin{pmatrix}
      {{\hat{\bm{H}}}}  \\
     {{\hat{\bm{M}}}}
     \end{pmatrix}
     (t)
       =\bm{J}\begin{pmatrix}
     \mathbb{0} \\
     {\mathbb{1}} 
     \end{pmatrix}
       \end{cases}\;,
\end{equation}%
where $\bm{M}^{f}(\tau)$, $\hat{\bm{M}}(\tau)$ are the associated conjugate variables under the Legendre transform, 
\begin{equation}
\bm{J}=\begin{pmatrix} \mathbb{0} &\,\, \mathbb{1}\\ -\mathbb{1} &\,\, \mathbb{0}\end{pmatrix}
\end{equation}
is the symplectic matrix, and $\bm{E}^{f}(\tau)$, $\hat{\bm{E}}(\tau)$ are the related symmetric matrices driving the systems, which read
\begin{equation}\label{driv}
\bm{E} = \begin{pmatrix}
     {\bm{C}}^T{\bm{P}}^{-1}{\bm{C}}-{\bm{Q}}\,\,\,\,\,\, & -{\bm{C}}^T{\bm{P}}^{-1}\\
     \\
       -{\bm{P}}^{-1}{\bm{C}}\,\,\,\,\,\, & {\bm{P}}^{-1} 
       \end{pmatrix}\;,
\end{equation}
computed on $\bm{q}^{f}$ and $\hat{\bm{q}}$ respectively.

Finally, our first result consists in showing that we can write the Laplace approximation of the transition probability Eq.~(\ref{pathint}) as
\begin{equation}\label{fin2}
\rho(\bm{q},t \vert \bm{q}_0,t_0)\approx e^{S(\hat{\bm{q}})-S(\bm{q}^f)}\sqrt{\det\left[\frac{1}{2\pi}\frac{{\hat{\bm{H}}}}{{\bm{H}^{f}}}(t_0)\right]} \;,
\end{equation}
where $S(\bm{q}^f)$, $S(\hat{\bm{q}})$ represent the action evaluated on the related minimizer, and $\bm{H}^{f}(\tau)$, $\hat{\bm{H}}(\tau)$ are the solutions of Eq.~(\ref{fin3}).

Let us mention that in the particular case where the Lagrangian is a quadratic function of $\bm{q}$ and $\dot{\bm{q}}$, then no error is introduced by the Laplace approximation, and Eq.~(\ref{fin2}) is an equality.
In the latter case the matrix $\bm{E}$ is now independent of the particular minimum, and Eq.~(\ref{fin3}) simplifies further to
\begin{equation}\label{fin5}
\begin{cases}
              \frac{\dd}{\dd\tau}\begin{pmatrix}
      {{\bm{H}^{f}}} &  {{\hat{\bm{H}}}}\\
     {{\bm{M}^{f}}}  &  {{\hat{\bm{M}}}}
     \end{pmatrix}
       =\bm{J}{{\bm{E}}}\begin{pmatrix}
       {{\bm{H}^{f}}} &  {{\hat{\bm{H}}}}\\
     {{\bm{M}^{f}}}  &  {{\hat{\bm{M}}}}
     \end{pmatrix}
   \\
   \\
\begin{pmatrix}
    {{\bm{H}^{f}}} &  {{\hat{\bm{H}}}}\\
     {{\bm{M}^{f}}}  &  {{\hat{\bm{M}}}}
     \end{pmatrix}
     (t)
       =\bm{J}
       \end{cases} \;.
\end{equation}%

\subsection{Laplace approximation of marginal conditional probabilities}\label{marginalsub}
Another result provided in this work is the generalization of Eq.~(\ref{fin2}) that allows us to compute marginal transition probabilities. Recall that we reorder the configuration variables as 
\begin{equation}
\begin{split}
&\bm{q}(\tau)=(\bm{q}_{V}(\tau),\bm{q}_{F}(\tau))\in\mathbb{R}^{d}\;,\\
&\bm{q}_{V}(\tau)=(q_1,...,q_l)(\tau)\in\mathbb{R}^l\;,\\
&\bm{q}_{F}(\tau)=(q_{l+1},...,q_d)(\tau)\in\mathbb{R}^{d-l}\;,\\
\end{split}
\end{equation}
and consider the marginals 
\begin{equation}
\rho_m(\bm{q}_{F},t \vert \bm{q}_0,t_0) = \int \rho(\bm{q},t \vert \bm{q}_0,t_0)\,\dd\bm{q}_{V}\;. 
\end{equation}
The latter can be expressed as the path integral
\begin{equation}\label{pathint2}
\rho_m(\bm{q}_{F},t \vert \bm{q}_0,t_0) = \dfrac{\mathcal{K}_m}{\mathcal{N}} = \frac{1}{\mathcal{N}}\int\limits_{\bm{q}(t_0)=\bm{q}_0}^{\bm{q}_{F}(t)=\bm{q}_{F}}{e^{-S(\bm{q})}\,\mathcal{D}\bm{q}} \;,
\end{equation}%
where the integration for $K_m$ is taken over all paths starting with $\bm{q}(t_0)=\bm{q}_0$ and with the mixed end-point conditions $\bm{q}_{F}(t)=\bm{q}_{F}$ fixed and $\bm{q}_{V}(t)$ variable. Note that the normalization term $\mathcal{N}$ remains the same as in Eq.~(\ref{normex}).

To derive a generalization of Eq.~(\ref{fin2}) for Eq.~(\ref{pathint2}) we follow the same strategy as before, and start by computing
\begin{itemize}
\item an isolated minimizer $\bm{q}^m(\tau)$ that satisfies the Euler-Lagrange equations with Dirichlet conditions $\bm{q}^m(t_0)=\bm{q}_0$, $\bm{q}_{F}^m(t)=\bm{q}_{F}$ and with Neumann natural boundary conditions for the remaining variables $\frac{\partial \mathcal{L}}{\partial \dot{\bm{q}}_V}(t)\Big{|}_{\bm{q}^m}=\bm{0}$,
\item an isolated minimizer $\hat{\bm{q}}(\tau)$ as defined above.
\end{itemize}
As a matter of notation, the superscripts ``$f$'' used before and ``$m$'' here introduced stand for ``full'' and ``marginal'' and we will also denote the conditional probability density $\rho(\bm{q},t \vert \bm{q}_0,t_0)$ by $\rho_f(\bm{q},t \vert \bm{q}_0,t_0)$. Then, the Laplace approximation of the transition probability Eq.~(\ref{pathint2}) is 
\begin{equation}\label{fin22}
\rho_m(\bm{q}_{F},t \vert \bm{q}_0,t_0) \approx  e^{S(\hat{\bm{q}})-S(\bm{q}^m)}\sqrt{(2\pi)^{l-d}\det\left[\frac{{\hat{\bm{H}}}}{{\bm{H}^m}}(t_0)\right]} \;,
\end{equation}
with $\bm{H}^m$ obtained from
\begin{equation}\label{fin33}
\begin{cases}
           \frac{\dd}{\dd\tau}\begin{pmatrix}
      {{\bm{H}^m}}  \\
     {{\bm{M}^m}}
     \end{pmatrix}
       =\bm{J}{{\bm{E}^m}}\begin{pmatrix}
       {{\bm{H}^m}}  \\
     {{\bm{M}^m}}
     \end{pmatrix}
   \\
   \\
\begin{pmatrix}
      {{\bm{H}^m}}  \\
     {{\bm{M}^m}}
     \end{pmatrix}
     (t)
       =\begin{pmatrix}
     \mathbb{1}_{l\times l} & \mathbb{0}_{d+l\times d-l}\\
      \mathbb{0}_{2d-l\times l} & -\mathbb{1}_{d-l\times d-l}
       \end{pmatrix}
       \end{cases} \;,
\end{equation}
where $\bm{M}^m$ is the conjugate variable under the Legendre transform for the Jacobi equations in Hamiltonian form, and $\bm{E}^m$ is the symmetric matrix Eq.~(\ref{driv}) here computed on the minimizer $\bm{q}^m(\tau)$. As before, if the Lagrangian is a quadratic function of $\bm{q}$ and $\dot{\bm{q}}$, then no error is introduced by the Laplace approximation, and Eq.~(\ref{fin22}) is an equality. Note that for $l=0$ we have that Eq.~(\ref{fin22}) reduces to Eq.~(\ref{fin2}), since $\bm{q}^m(\tau)$ becomes $\bm{q}^{f}(\tau)$ for the boundary conditions of the Euler-Lagrange equations, and Eq.~(\ref{fin33}) reduces to the the first system for $\bm{H}^{f}$ in Eq.~(\ref{fin3}). On the other hand, for $l=d$ we have that $\mathcal{K}_m$ coincides with the normalization factor $\mathcal{N}$.

To summarize, our approach for computing the transition probabilities Eq.~(\ref{pathint}) and Eq.~(\ref{pathint2}) consists in taking the ratio of the conditional and unconditional Wiener path integrals in the Laplace approximation, to then express their solutions in terms of the solutions of a set of ordinary differential (Jacobi) equations. In the following we present the derivation of the mentioned results.

\section{Preliminary remarks}
\subsection{The second variation, background on stability}
We recall that the semi-classical approximation adopted here for the evaluation of the real path integrals $\mathcal{K}$, $\mathcal{K}_m$ and $\mathcal{N}$ is also known as Laplace asymptotic method. The idea is to first Taylor expand the action $S(\bm{q}(\tau))$ to second order around an isolated minimizer satisfying prescribed boundary conditions, that we denote generically as $\bar{\bm{q}}$, exploiting the fact that the first variation on a minimum is zero:
\begin{equation}
S(\bm{q})\sim S(\bar{\bm{q}})+\frac{1}{2}\delta^2S(\bm{h};\bar{\bm{q}}),\quad\bm{q}=\bar{\bm{q}}+\bm{h} \;.
\end{equation}
The existence and stability of $\bar{\bm{q}}$ are assumed. Nevertheless, we briefly discuss some known results for detecting whether an equilibrium (satisfying the Euler-Lagrange equations) is actually a minimizer of the action functional. The stability of an equilibrium arising from a variational principle can be investigated by a suitable analysis of the second variation, namely
\begin{equation}\label{secvv}
\delta^2 S(\bm{h};\bar{\bm{q}}) = \int_{t_0}^t \left( \dot{\bm{h}}^T{{\bm{P}}}\dot{\bm{h}}+2\dot{\bm{h}}^T{{\bm{C}}}{\bm{h}}+{\bm{h}}^T{{\bm{Q}}}{\bm{h}}\right)\,\dd\tau\;.
\end{equation}
The Euler-Lagrange equations for the second variation functional are better known as Jacobi equations, and the same fields appearing in Eq.~(\ref{fin3}), Eq.~(\ref{fin33}), whose volume is exploited for the computation of quadratic path integrals, are already known in the context of conjugate point theory of stability. 

In summary, given a set of boundary conditions where a full Dirichlet condition is present at least at one end, with anything self-adjoint at the other end, the classic necessary condition for an equilibrium to be a minimizer is the Legendre condition, \ie $\bm{P}\geq 0$. A sufficient condition requires (in addition to the Legendre strengthened condition $\bm{P}>0$) the absence of conjugate points or, in other words, the determinant of the matrix solution $\bm{H}(\tau)$ to the Jacobi equations presented in Eq.~(\ref{fin3}), Eq.~(\ref{fin33}) does not vanish in $[t_0,t)$. In particular, the case of Dirichlet-Dirichlet boundary conditions is standard in the literature, especially in absence of cross terms \citep{FOM}, whereas if a full Dirichlet condition is present only at one end, then a small modification of the theory is required but the underlying ideas remain unchanged. Namely, while in the Dirichlet-Dirichlet case the initial conditions for the Jacobi equations can be given either at $\tau=t_0$ or $\tau=t$, it is crucial in the other cases to shoot towards the only Dirichlet end, in our context prescribing the initial conditions at $\tau=t$. In doing so, the properties of the spectrum of the second variation operator are analogous to the Dirichlet-Dirichlet case, with positive inborn eigenvalues at $\tau=t$ and monotonically decreasing eigenvalues towards $\tau=t_0$, and the technical arguments used to prove the latter claims are recycled from the Dirichlet-Dirichlet case. Therefore, the fields presented in Eq.~(\ref{fin3}) and Eq.~(\ref{fin33}) are known in the context of stability analysis, especially in rod mechanics \citep{JHMT, JHMStab, ISO, StabMan, HEL}, where the study of minimizers of an elastic energy subject to a wide range of different boundary conditions has a long history. In addition, if the paths are not subject to full Dirichlet conditions at any end, then the conjugate point theory of stability is rather not standard and delicate. However, we will not deal with the latter situation and we simply refer to \citep{NEU, THOM} for the sake of completeness. In particular, \citep{NEU} also explains the analogy between minimization of functionals and minimization of functions, with some background on the topic.

As far as the calculation of Gaussian path integrals is concerned, the actual value of the Jacobi determinants themselves, and no longer their zeros as was the case when performing the above stability analysis, characterises the final result. On one hand, the Dirichlet-Dirichlet case leading to the first system in Eq.~(\ref{fin3}) is well-known if no cross terms are present \citep{GELYAG, MOR}, and in presence of cross terms has been derived in \citep{LUDT, LUD} on the basis of \citep{PAP1} by means of an additional change of variables closely related to the one used in the proof of sufficiency of absence of conjugate points for an extremal to be a minimum. On the other hand, the second system in Eq.~(\ref{fin3}) and the one in Eq.~(\ref{fin33}) are original to the present work if regarded as the result of Gaussian path integrals with more general boundary conditions, where a full Dirichlet condition is still present at one end.

Aware of the latter considerations, we go back to our problem and the Laplace approximation for $\mathcal{K}$, $\mathcal{K}_m$ or $\mathcal{N}$ reads
\begin{equation}\label{semicont}
e^{-S(\bar{\bm{q}})}\int{e^{-\frac{1}{2}\delta^2 S(\bm{h};\bar{\bm{q}})}\,\mathcal{D}\bm{h}} \;,
\end{equation}
where now the integral is over all admissible variations $\bm{h}(\tau)\in\mathbb{R}^{d}$ satisfying the linearised version of the boundary conditions for $\bar{\bm{q}}(\tau)$. Note that such an approximation via consideration of only quadratic fluctuations around a minimal energy configuration holds when the action required to deform the system into a different configuration is large, an hypothesis which is always true for small enough final time $t$, \ie in the short time-scale regimes. Before presenting the techniques involved in the computation of the Gaussian path integrals appearing in Eq.~(\ref{semicont}), we explore the Laplace approximation method in finite dimensions and perform some computations.

\subsection{Laplace approximation in finite dimensions, an algorithm}
First, we briefly summarise the Laplace asymptotic approximation method in one dimension. The method is useful for approximating real integrals of the form
\begin{equation}\label{1dLap}
\int_a^b{h(x)e^{-Ng(x)}}\dd x\;,
\end{equation}
where $N$ is a large positive number, $h(x)$ and $g(x)$ are twice continuously differentiable functions and $g(x)$ has a minimum at $\bar{x}\in(a,b)$. Performing a second-order Taylor expansion for $h(x)$ and $g(x)$ around the same point $\bar{x}$, we have
\begin{itemize}
\item $h(x)\sim h(\bar{x})+h'(\bar{x})(x-\bar{x})+\frac{1}{2}h''(\bar{x})(x-\bar{x})^2$ and
\item $g(x)\sim g(\bar{x})+\frac{1}{2}g''(\bar{x})(x-\bar{x})^2$,
\end{itemize}
since the derivative of $g$ at the minimum is zero.

As a consequence, the integral Eq.~(\ref{1dLap}) can be approximated by
\begin{equation}
\int_{-\infty}^{+\infty}{\left(h(\bar{x})+h'(\bar{x})(x-\bar{x})+\frac{1}{2}h''(\bar{x})(x-\bar{x})^2\right)e^{-N\left(g(\bar{x})+\frac{1}{2}g''(\bar{x})(x-\bar{x})^2\right)}}\dd x\;,
\end{equation}
where the contributions to the integral far from the minimum are arbitrarily small if we extend the interval $(a,b)$ to the whole of $\mathbb{R}$ when $N$ is arbitrarily big. Hence this is an asymptotic relation for the approximation of the integral when $N\rightarrow +\infty$. Moreover, the previous expression can be further simplified if we note that the first-order term of $h$ against the second-order term of $g$ provides a zero integral for the odd-even product. Finally we are left with
\begin{equation}
h(\bar{x})e^{-N g(\bar{x})}\int_{-\infty}^{+\infty}{e^{-\frac{N}{2}g''(\bar{x})(x-\bar{x})^2}}\dd x+\frac{1}{2}h''(\bar{x})e^{-N g(\bar{x})}\int_{-\infty}^{+\infty}{(x-\bar{x})^2e^{-\frac{N}{2}g''(\bar{x})(x-\bar{x})^2}}\dd x
\end{equation}
and the advantage of the procedure is having recovered Gaussian integrals, which can be integrated explicitly. Finally the result is
\[
\int_a^b{h(x)e^{-Ng(x)}}\dd x\sim h(\bar{x})e^{-N g(\bar{x})}\sqrt{\frac{2\pi}{N g''(\bar{x})}}+\frac{1}{2}e^{-N g(\bar{x})}\frac{h''(\bar{x})}{N g''(\bar{x})}
\]
or simply, considering $h(x)\equiv 1$
\begin{equation}\label{leading}
\int_a^b{h(x)e^{-Ng(x)}}\dd x\sim e^{-N g(\bar{x})}\sqrt{\frac{2\pi}{N g''(\bar{x})}}\;.
\end{equation}
This result can be generalized in multiple dimensions as well as in infinite dimensions, which is the case that we are going to consider for our problem.

In the light of the above considerations, let us give a ``brute force'' numerical method for the evaluation of Eq.~(\ref{pathint}) using a finite dimensional Laplace expansion. This is important for building the analogy with our infinite dimensional problem and gaining a better understanding. We recall the definition of the problem by means of path integrals
\begin{equation}
\rho(\bm{q},t \vert \bm{q}_0,t_0) = {\displaystyle{\int\limits_{\bm{q}(t_0)=\bm{q}_0}^{\bm{q}(t)=\bm{q}}{e^{-S(\bm{q})}\,\mathcal{D}\bm{q}}}}\left(\displaystyle{\,\,\int\limits_{\bm{q}(t_0)=\bm{q}_0}{e^{-S(\bm{q})}\,\mathcal{D}\bm{q}}}\right)^{-1} \;,
\end{equation}
and follow a time-slicing procedure, discretizing $\tau\in[t_0,t]$ into $n$ intervals of length $\varepsilon=\frac{t-t_0}{n}$. The subscript $j$ indicates that the associated term is evaluated in $\tau_j=t_0+j\varepsilon$ for $j=0,1,...,n$. Thus we can ``naively'' write
\begin{equation}\label{discratio}
\begin{split}
\rho(\bm{q},t \vert \bm{q}_0,t_0)= \lim_{n\rightarrow\infty} {\frac{\displaystyle{\int{e^{-\varepsilon \sum\limits_{j=0}^{n}\mathcal{L}(\bm{q}(\tau))_j\big{|}_{\bm{q}_0}^{\bm{q}_n}}}\prod\limits_{j=1}^{n-1}{\dd\bm{q}_j}}}{\displaystyle{\int{e^{-\varepsilon \sum\limits_{j=0}^{n}\mathcal{L}(\bm{q}(\tau))_j\big{|}_{\bm{q}_0}}}\prod\limits_{j=1}^{n}{\dd\bm{q}_j}}}}\;,
\end{split}
\end{equation}
using the same finite difference scheme for the discretization of numerator and denominator.
Then, for $n$ fixed, we can perform a Laplace approximation for the $d(n-1)$ dimensional integral of the multivariate function 
\begin{equation}
\displaystyle{e^{-\varepsilon \sum\limits_{j=0}^{n}\mathcal{L}(\bm{q}(\tau))_j\big{|}_{\bm{q}_0}^{\bm{q}_n}}}
\end{equation}
about the minimum ${\bm{q}}_{disc}^f$ of
\begin{equation}
{g}^f(\bm{q}_1,...,\bm{q}_{n-1})=\varepsilon \sum\limits_{j=0}^{n}\mathcal{L}(\bm{q}(\tau))_j\big{|}_{\bm{q}_0}^{\bm{q}_n}\;,
\end{equation}
and a Laplace approximation for the $d\,n$ dimensional integral of the multivariate function 
\begin{equation}
\displaystyle{e^{-\varepsilon \sum\limits_{j=0}^{n}\mathcal{L}(\bm{q}(\tau))_j\big{|}_{\bm{q}_0}}}
\end{equation}
about the minimum $\hat{\bm{q}}_{disc}$ of
\begin{equation}
\hat{g}(\bm{q}_1,...,\bm{q}_{n})=\varepsilon \sum\limits_{j=0}^{n}\mathcal{L}(\bm{q}(\tau))_j\big{|}_{\bm{q}_0}\;.
\end{equation}
The existence of such minima, together with the positive definiteness of the Hessian matrices of ${g}^f$ and $\hat{g}$ on the minima are assumed. 

Finally, the approximated computation of $\rho(\bm{q},t \vert \bm{q}_0,t_0)$ reduces to assess the limit
\begin{equation}
\lim_{n\rightarrow\infty} {\frac{\displaystyle{\int{e^{-\left[g^f(\bm{q}^f_{disc})+\frac{1}{2}(\bm{q}_1,...,\bm{q}_{n-1})^T D^2{g^f}(\bm{q}^f_{disc})(\bm{q}_1,...,\bm{q}_{n-1})\right]}}\prod\limits_{j=1}^{n-1}{\dd\bm{q}_j}}}{\displaystyle{\int{e^{-\left[\hat{g}(\hat{\bm{q}}_{disc})+\frac{1}{2}(\bm{q}_1,...,\bm{q}_{n})^T D^2{\hat{g}}(\hat{\bm{q}}_{disc})(\bm{q}_1,...,\bm{q}_{n})\right]}}\prod\limits_{j=1}^{n}{\dd\bm{q}_j}}}}\;,
\end{equation}
where $D^2{g^f}$ and $D^2{\hat{g}}$ are the Hessian matrices of the functions $g^f$ and $\hat{g}$. Note that we have already performed the substitutions 
\begin{equation}
\begin{split}
&(\bm{q}_1,...,\bm{q}_{n-1})-\bm{q}^f_{disc}\rightarrow (\bm{q}_1,...,\bm{q}_{n-1})\;,\\
&(\bm{q}_1,...,\bm{q}_{n})-\hat{\bm{q}}_{disc}\rightarrow (\bm{q}_1,...,\bm{q}_{n})\;,
\end{split}
\end{equation}
for numerator and denominator respectively, where the Jacobian factor of the transformation is one. Thus, we get
\begin{equation}\label{lapDis}
\begin{split}
\rho(\bm{q},t \vert \bm{q}_0,t_0)&\approx\lim_{n\rightarrow\infty}\frac{e^{-g^f(\bm{q}^f_{disc})}\left({2\pi}\right)^{\frac{d(n-1)}{2}}\sqrt{\det{\left[D^2{g^f}(\bm{q}^f_{disc})\right]}^{-1}}}{\displaystyle{e^{-\hat{g}(\hat{\bm{q}}_{disc})}\left({2\pi}\right)^{\frac{d\,n}{2}}\sqrt{\det{\left[D^2{\hat{g}}(\hat{\bm{q}}_{disc})\right]}^{-1}}}}\;,\\
\\
&=\lim_{n\rightarrow\infty}e^{\hat{g}(\hat{\bm{q}}_{disc})-g^f(\bm{q}^f_{disc})}\sqrt{\left(\frac{1}{2\pi}\right)^d\det{\left[\frac{D^2{\hat{g}}(\hat{\bm{q}}_{disc})}{D^2{g^f}(\bm{q}^f_{disc})}\right]}}\;,
\end{split}
\end{equation}
that must be evaluated numerically.

\subsection{A basic example}
As an example, consider the Lagrangian reported in Eq.~(1.40), p. 9 of \citep{BookLan}
\begin{equation}
\mathcal{L}(\dot{q}(\tau),q(\tau))=\frac{1}{2} \left(\dot{q}(\tau)+q(\tau)\right)^2,\quad q(\tau)\in\mathbb{R}\,\,(d=1)\;,
\end{equation}
with $t_0=q_0=0$ and $t=q=1$.
In this particular case $\rho(1,1 \vert 0,0)$ is given according to Eq.~(\ref{discratio}) as
\begin{equation}\label{discex}
\lim_{n\rightarrow\infty}{\rho_n}=\lim_{n\rightarrow\infty} {\frac{\displaystyle{\int{e^{-\frac{1}{\varepsilon} \sum\limits_{j=0}^{n-1}\left[\frac{(q_{j+1}-q_j)^2}{2}+\varepsilon q_j(q_{j+1}-q_j)+\frac{(\varepsilon q_j)^2}{2}\right]\big{|}_{q_0=0}^{q_n=1}}}\prod\limits_{j=1}^{n-1}{\dd q_j}}}{\displaystyle{\int{e^{-\frac{1}{\varepsilon} \sum\limits_{j=0}^{n-1}\left[\frac{(q_{j+1}-q_j)^2}{2}+\varepsilon q_j(q_{j+1}-q_j)+\frac{(\varepsilon q_j)^2}{2}\right]\big{|}_{q_0=0}}}\prod\limits_{j=1}^{n}{\dd q_j}}}}\;.
\end{equation}
We underline that for this quadratic example the Laplace method is going to provide the exact result without any approximation. The procedure is based on computing $\rho(1,1 \vert 0,0)$ through expression Eq.~(\ref{lapDis}) with increasing accuracy, namely by increasing $n$ at each step. Before showing the convergence of the algorithm, we compute the result predicted by the method presented in the previous section, namely we compute Eq.~(\ref{fin2}) by means of Eq.~(\ref{fin3}).

We first note that $P=C=Q=1$, therefore the Euler-Lagrange and the Jacobi equations coincide and are equal to 
\begin{equation}
-\ddot{q}(\tau)+q(\tau)=0\;.
\end{equation}
According to the boundary conditions 
\begin{equation}
q^f(0)=0,\quad q^f(1)=1,\quad H^f(1)=0,\quad M^f(1)=-1\;,
\end{equation}
we easily get 
\begin{equation}
q^f(\tau)=\frac{e^{1+\tau}-e^{1-\tau}}{e^2-1}\;,\quad S(q^f)=\frac{e^2}{e^2-1}\;,
\end{equation}
\begin{equation}
H^f(\tau)=\frac{e^{1-\tau}-e^{\tau-1}}{2}\;,\quad\displaystyle{H^f(0)=\frac{e^2-1}{2 e}}\;.
\end{equation} 
Moreover, according to the boundary conditions 
\begin{equation}
\hat{q}(0)=0,\quad\frac{\partial{\mathcal{L}}}{\partial\dot{q}}\Big{|}_{\hat{q}}(1)=\dot{\hat{q}}(1)+\hat{q}(1)=0,\quad\hat{H}(1)=1,\quad\hat{M}(1)=0\;,
\end{equation}
we also have
\begin{equation}
\hat{q}(\tau)=0\;,\quad S(\hat{q})=0\;,
\end{equation}
\begin{equation}
\hat{H}(\tau)=e^{1-\tau}\;,\quad\displaystyle{\hat{H}(0)=e}\;.
\end{equation} 
Hence, the final result is given by Eq.~(\ref{fin2}) as 
\begin{equation}\label{discexres}
\begin{split}
\rho(1,1;0,0)&=e^{S(\hat{ {q}})-S( {q}^f)}\sqrt{\det\left[\frac{1}{2\pi}\frac{{\hat{ {H}}}}{{ {H}^{f}}}(0)\right]} \;,\\
&=\displaystyle{e^{-\frac{e^2}{e^2-1}}\sqrt{\frac{e^2}{\pi(e^2-1)}}= 0.190867...}
\end{split}
\end{equation}

\begin{figure}[!ht]
\centering
\captionsetup{justification=centering,margin=1cm}
\includegraphics[width=13cm]{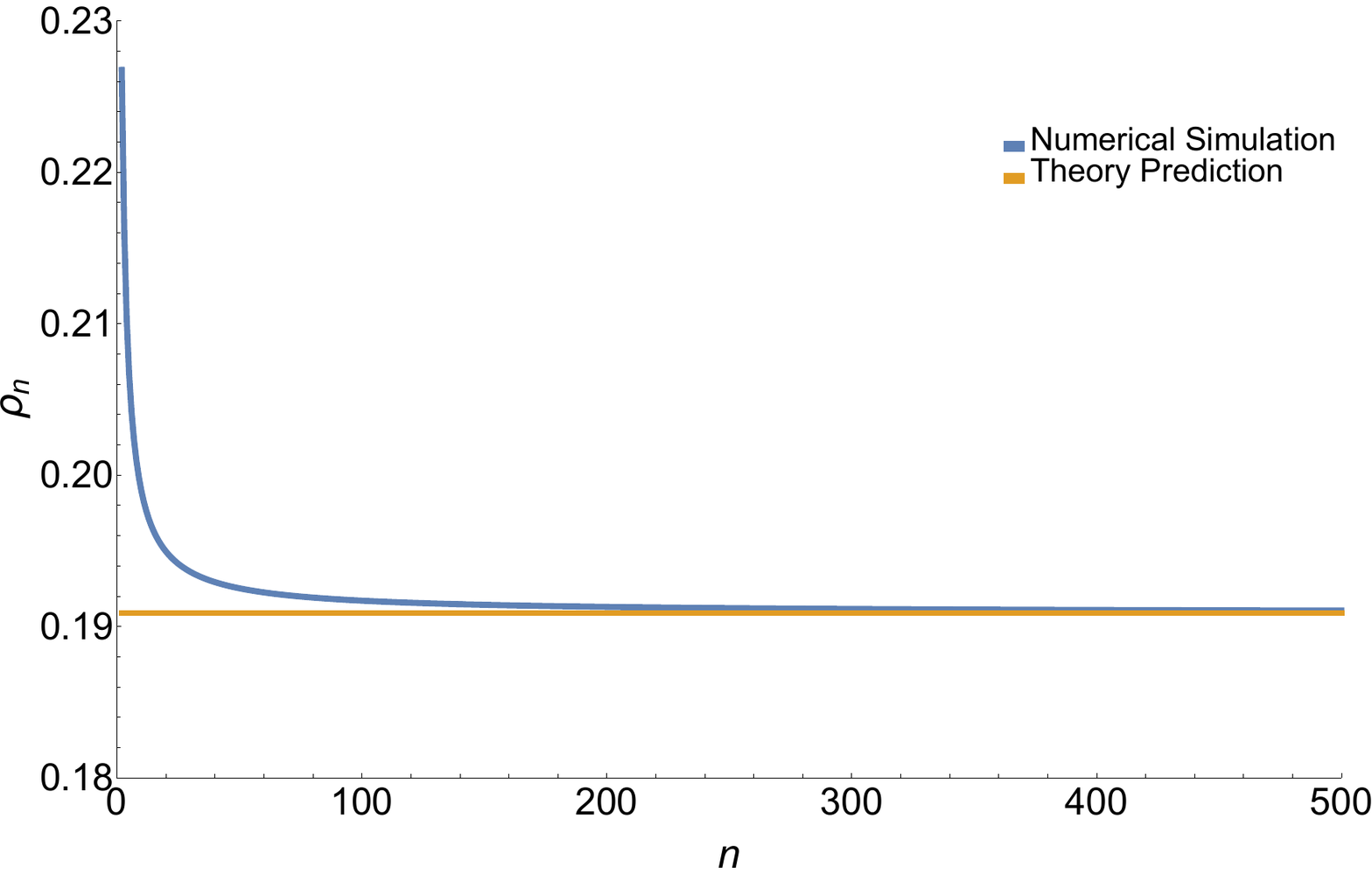}
\caption{Conditional probability density: comparison between the numerical evaluation of $\rho_n$ in Eq.~(\ref{discex}) as a function of $n$ (in blue) and the theoretical value $\rho(1,1;0,0)$ in Eq.~(\ref{discexres}) (in orange) for the simple example $\mathcal{L}(\dot{q}(\tau),q(\tau))=\frac{1}{2} \left(\dot{q}(\tau)+q(\tau)\right)^2$.}
\label{fig:conv}
\end{figure}

In Fig. \ref{fig:conv} we report the result of Eq.~(\ref{lapDis}) for Eq.~(\ref{discex}) and we compare it with our theoretical prediction Eq.~(\ref{discexres}). In the next section we start the derivation of the results stated in terms of the Jacobi equations and show how to deal with the Laplace approximation in the infinite dimensional setting, for general (matrix) second variation operators.

\section{Block diagonalization of the discretized second variation}
Following a time-slicing procedure, we discretize $\tau\in[t_0,t]$ into $n$ intervals of length $\varepsilon=\frac{t-t_0}{n}$. In this section we are interested in block-diagonalizing the finite difference approximation of the second variation Eq.~(\ref{secvv})
\begin{small}
\begin{equation}\label{secdiag}
\begin{split}
\delta^2 S(\bm{h};\bar{\bm{q}})&\approx\varepsilon\sum\limits_{j=1}^n\left[\frac{\Delta \bm{h}_j}{\varepsilon}^T\bm{P}_j\frac{\Delta \bm{h}_j}{\varepsilon}+2\frac{\Delta \bm{h}_j}{\varepsilon}^T\frac{{\bm{C}_j} \bm{h}_j+\bm{C}_{j-1}\bm{h}_{j-1}}{2}+ \bm{h}_j^T\bm{Q}_j\bm{h}_j \right] \;,\\
&=\frac{1}{\varepsilon}\sum\limits_{j=1}^n\left[{\Delta \bm{h}_j}^T\bm{P}_j{\Delta \bm{h}_j}+\varepsilon{\Delta \bm{h}_j}^T\left({\bm{C}_j} \bm{h}_j+\bm{C}_{j-1}\bm{h}_{j-1}\right)+\varepsilon^2\bm{h}_j^T\bm{Q}_j\bm{h}_j \right] \;,\\
&=\frac{1}{\varepsilon}\sum\limits_{j=1}^n\big{[}\bm{h}_j^T\left(\bm{P}_j+\varepsilon {\bm{C}_j}+\varepsilon^2\bm{Q}_j\right)\bm{h}_j-\bm{h}_{j}^T\left(\bm{P}_j-\varepsilon \bm{C}_{j-1}\right)\bm{h}_{j-1}-\bm{h}_{j-1}^T\left(\bm{P}_j+\varepsilon \bm{C}_{j}\right)\bm{h}_{j}\\
&\quad +\bm{h}_{j-1}^T\left(\bm{P}_j-\varepsilon \bm{C}_{j-1}\right)\bm{h}_{j-1}\big{]} \;,\\
&=\frac{1}{\varepsilon}\left[ \bm{h}_{0}^T\left(\bm{P}_1-\varepsilon \bm{C}_{0}\right)\bm{h}_{0}\right] \;\\
&\quad +\frac{1}{\varepsilon}\sum\limits_{j=1}^{n-1}\left[\bm{h}_j^T\left(\bm{P}_j+\bm{P}_{j+1}+\varepsilon^2\bm{Q}_j\right)\bm{h}_j-\bm{h}_j^T\left(\bm{P}_j-\varepsilon \bm{C}_{j-1}\right)\bm{h}_{j-1}-\bm{h}_{j-1}^T\left(\bm{P}_j+\varepsilon \bm{C}_{j}\right)\bm{h}_j\right]\\
&\quad +\frac{1}{\varepsilon}\left[\bm{h}_n^T\left(\bm{P}_n+\varepsilon \bm{C}_n+\varepsilon^2\bm{Q}_n\right)\bm{h}_n-\bm{h}_{n}^T\left(\bm{P}_n-\varepsilon \bm{C}_{n-1}\right)\bm{h}_{n-1}-\bm{h}_{n-1}^T\left(\bm{P}_n+\varepsilon \bm{C}_{n}\right)\bm{h}_{n}\right],
\end{split}
\end{equation}
\end{small}%
in order to obtain, for some variables ${\bm{\phi}}_j\in\mathbb{R}^{d}$ and symmetric matrices $\bm{\alpha}_j\in\mathbb{R}^{d\times d}$, an expression of the form 
\begin{equation}
\delta^2 S(\bm{h};\bar{\bm{q}})\approx\frac{1}{\varepsilon}\sum\limits_{j}{\bm{\phi}}_j^T\bm{\alpha}_j{\bm{\phi}}_j \;,
\end{equation}%
where we defined $\Delta \bm{h}_j=\bm{h}_j-\bm{h}_{j-1}$ for $j=1,...,n$, and the subscript $j$ indicates that the associated term is evaluated in $\tau_j=t_0+j\varepsilon$ for $j=0,1,...,n$, \eg $\bm{h}_j=\bm{h}(\tau_j)$.

\subsection{Background: the Dirichlet-Dirichlet case}
The Dirichlet-Dirichlet case is outlined in detail in \citep{LUDT, LUD} and the following diagonalization strategy was first presented in the work of Papadopoulos \citep{PAP1}; here we report the computation, which is used as a starting point for our extensions to the Dirichlet-Neumann and Dirichlet-Mixed (Dirichlet, Neumann) cases. Let $\bm{h}(\tau)$ be a perturbation around the Dirichlet-Dirichlet minimum $\bm{q}^f$, then $\bm{h}(t_0)=\bm{h}_0=\bm{0}$ and $\bm{h}(t)=\bm{h}_n=\bm{0}$. As a consequence of the linearised boundary conditions, Eq.~(\ref{secdiag}) becomes
\begin{small}
\begin{equation}\label{dirdisc}
\delta^2 S(\bm{h};\bm{q}^f)\approx\frac{1}{\varepsilon}\sum\limits_{j=1}^{n-1}\left[\bm{h}_j^T\left(\bm{P}_j+\bm{P}_{j+1}+\varepsilon^2\bm{Q}_j\right)\bm{h}_j-\bm{h}_j^T\left(\bm{P}_j-\varepsilon \bm{C}_{j-1}\right)\bm{h}_{j-1}-\bm{h}_{j-1}^T\left(\bm{P}_j+\varepsilon \bm{C}_{j}\right)\bm{h}_j\right]
\end{equation}
\end{small}%
Introducing now the matrices 
\begin{equation}
\bm{U}_j=\bm{P}_j+\frac{\varepsilon}{2}[\bm{C}_j^T-\bm{C}_{j-1}]\quad\text{for}\quad j=1,...,n-1\;,
\end{equation}
we have that 
\begin{itemize}
\item $\bm{h}_j^T { \bm{U}}_j \bm{h}_{j-1}=\bm{h}_j^T {\bm{P}}_j \bm{h}_{j-1}+\frac{\varepsilon}{2}\bm{h}_j^T {\bm{C}}_j^T \bm{h}_{j-1}-\frac{\varepsilon}{2}\bm{h}_j^T {\bm{C}}_{j-1} \bm{h}_{j-1}$,
\item $\bm{h}_{j-1}^T { \bm{U}}_j^T \bm{h}_{j}=\bm{h}_{j-1}^T {\bm{P}}_j \bm{h}_{j}+\frac{\varepsilon}{2}\bm{h}_{j-1}^T {\bm{C}}_j \bm{h}_{j}-\frac{\varepsilon}{2}\bm{h}_{j-1}^T {\bm{C}}_{j-1}^T \bm{h}_{j}$,
\end{itemize}
\begin{equation}
\Rightarrow\,\,-\bm{h}_j^T { \bm{U}}_j \bm{h}_{j-1}-\bm{h}_{j-1}^T { \bm{U}}_j^T \bm{h}_{j}=-\bm{h}_j^T( {\bm{P}}_j-\varepsilon {\bm{C}}_{j-1})\bm{h}_{j-1}-\bm{h}_{j-1}^T( {\bm{P}}_j+\varepsilon {\bm{C}}_{j}) \bm{h}_{j}\;.
\end{equation}
Therefore Eq.~(\ref{dirdisc}) can be written as
\begin{equation}\label{dirdisc2}
\begin{split}
\delta^2 S(\bm{h};\bm{q}^f)\approx\frac{1}{\varepsilon}\sum\limits_{j=1}^{n-1}\left[ \bm{h}_j^T\left(\bm{P}_j+\bm{P}_{j+1}+\varepsilon^2\bm{Q}_j\right) \bm{h}_j- \bm{h}_j^T \bm{U}_j \bm{h}_{j-1}-\bm{h}_{j-1}^T \bm{U}_j^T \bm{h}_j\right] \;.
\end{split}
\end{equation}

At this point, we perform a change of variables. We define the transformation with unit Jacobian 
\begin{equation}
\bm{\phi}_j=\bm{h}_j-\bm{\beta}_j \bm{h}_{j-1}\quad\text{for}\quad j=1,...,n-1\;,
\end{equation}
where the matrices $\bm{\beta}_j$ are given by the following recursive construction for the symmetric matrices $\bm{\alpha}_{j}$:
\begin{equation}
\begin{split}
&\bm{\alpha}_{n-1} = \bm{P}_{n-1}+\bm{P}_{n}+\varepsilon^2\bm{Q}_{n-1} \;,\\
&\bm{\alpha}_{j} = \bm{P}_{j}+\bm{P}_{j+1}+\varepsilon^2 \bm{Q}_{j}-\bm{\beta}_{j+1}^T\bm{\alpha}_{j+1}\bm{\beta}_{j+1} \;\qquad\text{for }\; j=n-2,...,1 \;, \\
&\bm{U}_j=\bm{\alpha}_j\bm{\beta}_j \;\hspace{58mm}\text{for }\; j=1,...,n-1 \;.
\end{split}
\end{equation}%
These expressions are motivated by the fact that they allow to express Eq.~(\ref{dirdisc2}) as a sum of quadratic forms, which is desired in view of a Gaussian integration. In fact, 
\begin{equation}
\bm{\phi}_j^T\bm{\alpha}_j\bm{\phi}_j=\bm{h}_j^T\bm{\alpha}_j \bm{h}_j-\bm{h}_j^T \bm{U}_j \bm{h}_{j-1}-\bm{h}_{j-1}^T\bm{U}_j^T \bm{h}_j+\bm{h}_{j-1}^T\bm{\beta}_j^T\bm{\alpha}_j\bm{\beta}_j \bm{h}_{j-1} \;,
\end{equation}
and recalling that $\bm{h}_0=\bm{0}$, we have
\begin{equation}\label{quaddir}
\delta^2 S(\bm{h};\bm{q}^f)\approx\frac{1}{\varepsilon}\sum\limits_{j=1}^{n-1}\bm{\phi}_j^T\bm{\alpha}_j\bm{\phi}_j \;.
\end{equation}%

\subsection{The Dirichlet-Neumann case}
Let $\bm{h}(\tau)$ be a perturbation around the Dirichlet-Neumann minimum $\hat{\bm{q}}$, then $\bm{h}(t_0)=\bm{h}_0=\bm{0}$ and no constraint is given for $\bm{h}(t)=\bm{h}_n$. As a consequence of the linearised boundary conditions, Eq.~(\ref{secdiag}) becomes
\begin{small}
\begin{equation}\label{neudisc}
\begin{split}
\delta^2 S(\bm{h};\hat{\bm{q}})&\approx\frac{1}{\varepsilon}\sum\limits_{j=1}^{n-1}\left[\bm{h}_j^T\left(\bm{P}_j+\bm{P}_{j+1}+\varepsilon^2\bm{Q}_j\right)\bm{h}_j-\bm{h}_j^T\left(\bm{P}_j-\varepsilon \bm{C}_{j-1}\right)\bm{h}_{j-1}-\bm{h}_{j-1}^T\left(\bm{P}_j+\varepsilon \bm{C}_{j}\right)\bm{h}_j\right]\\
&\quad +\frac{1}{\varepsilon}\left[\bm{h}_n^T\left(\bm{P}_n+\varepsilon \bm{C}_n+\varepsilon^2\bm{Q}_n\right)\bm{h}_n-\bm{h}_{n}^T\left(\bm{P}_n-\varepsilon \bm{C}_{n-1}\right)\bm{h}_{n-1}-\bm{h}_{n-1}^T\left(\bm{P}_n+\varepsilon \bm{C}_{n}\right)\bm{h}_{n}\right],
\end{split}
\end{equation}
\end{small}%
Introducing as before the matrices 
\begin{equation}
\bm{U}_j=\bm{P}_j+\frac{\varepsilon}{2}[\bm{C}_j^T-\bm{C}_{j-1}]\quad\text{for}\quad j=1,...,n\;,
\end{equation}
we have that Eq.~(\ref{neudisc}) can be written as
\begin{equation}\label{neudisc2}
\begin{split}
\delta^2 S(\bm{h};\hat{\bm{q}})&\approx\frac{1}{\varepsilon}\sum\limits_{j=1}^{n-1}\left[\bm{h}_j^T\left(\bm{P}_j+\bm{P}_{j+1}+\varepsilon^2\bm{Q}_j\right)\bm{h}_j-\bm{h}_j^T\bm{U}_j\bm{h}_{j-1}-\bm{h}_{j-1}^T\bm{U}_j^T\bm{h}_j\right]\\
&\quad +\frac{1}{\varepsilon}\left[\bm{h}_n^T\left(\bm{P}_n+\varepsilon \bm{C}_n+\varepsilon^2\bm{Q}_n\right)\bm{h}_n-\bm{h}_{n}^T\bm{U}_n\bm{h}_{n-1}-\bm{h}_{n-1}^T\bm{U}_n^T\bm{h}_{n}\right] \;,
\end{split}
\end{equation}

At this point, we perform a change of variables. We define the transformation with unit Jacobian 
\begin{equation}
\bm{\phi}_j=\bm{h}_j-\bm{\beta}_j \bm{h}_{j-1}\quad\text{for}\quad  j=1,...,n\;,
\end{equation}
where the matrices $\bm{\beta}_j$ are given by the following recursive construction for the symmetric matrices $\bm{\alpha}_{j}$:
\begin{equation}
\begin{split}
&\bm{\alpha}_{n}= \bm{P}_{n}+\varepsilon\frac{\bm{C}_n+\bm{C}_n^T}{2}+\varepsilon^2\bm{Q}_{n} \;,\\
&\bm{\alpha}_{j} = \bm{P}_{j}+\bm{P}_{j+1}+\varepsilon^2 \bm{Q}_{j}-\bm{\beta}_{j+1}^T\bm{\alpha}_{j+1}\bm{\beta}_{j+1} \;\qquad\text{for }\; j=n-1,...,1 \;, \\
&\bm{U}_j=\bm{\alpha}_j\bm{\beta}_j \;\hspace{58mm}\text{for }\; j=1,...,n \;.
\end{split}
\end{equation}%
These expressions are motivated by the fact that they allow to express Eq.~(\ref{neudisc2}) as a sum of quadratic forms. In fact, 
\begin{equation}
\bm{\phi}_j^T\bm{\alpha}_j\bm{\phi}_j=\bm{h}_j^T\bm{\alpha}_j \bm{h}_j-\bm{h}_j^T \bm{U}_j \bm{h}_{j-1}-\bm{h}_{j-1}^T\bm{U}_j^T \bm{h}_j+\bm{h}_{j-1}^T\bm{\beta}_j^T\bm{\alpha}_j\bm{\beta}_j \bm{h}_{j-1} \;,
\end{equation}
and recalling that $\bm{h}_0=\bm{0}$, we have
\begin{equation}\label{quadneu}
\delta^2 S(\bm{h};\hat{\bm{q}})\approx\frac{1}{\varepsilon}\sum\limits_{j=1}^{n}\bm{\phi}_j^T\bm{\alpha}_j\bm{\phi}_j \;.
\end{equation}%

\subsection{The Dirichlet-Mixed (Dirichlet, Neumann) case}
Let 
\begin{equation}
\begin{split}
&\bm{h}(\tau)=(\bm{h}_{V}(\tau),\bm{h}_{F}(\tau))\in\mathbb{R}^{d}\;,\\
&\bm{h}_{V}(\tau)=(h_1,...,h_l)(\tau)\in\mathbb{R}^l\;,\\
&\bm{h}_{F}(\tau)=(h_{l+1},...,h_d)(\tau)\in\mathbb{R}^{d-l}\;,
\end{split}
\end{equation}
be a perturbation around the Dirichlet-Mixed (Dirichlet, Neumann) minimum $\bm{q}^m$, then $\bm{h}(t_0)=\bm{h}_0=\bm{0}$, $\bm{h}_{F}(t)=\bm{h}_{Fn}=\bm{0}$ and no constraint is given for $\bm{h}_{V}(t)=\bm{h}_{Vn}$. As a consequence of the linearised boundary conditions, and setting 
\begin{equation}
\bm{U}_j=\bm{P}_j+\frac{\varepsilon}{2}[\bm{C}_j^T-\bm{C}_{j-1}]\quad\text{for}\quad j=1,...,n\;,
\end{equation}
Eq.~(\ref{secdiag}) becomes
\begin{equation}\label{dirneudisc2}
\begin{split}
\delta^2 S(\bm{h};\bm{q}^m)&\approx\frac{1}{\varepsilon}\sum\limits_{j=1}^{n-1}\left[\bm{h}_j^T\left(\bm{P}_j+\bm{P}_{j+1}+\varepsilon^2\bm{Q} _j\right)\bm{h}_j-\bm{h}_j^T \bm{U} _j \bm{h}_{j-1}-\bm{h}_{j-1}^T \bm{U} _j^T \bm{h}_{j}\right]\\
&\quad+\frac{1}{\varepsilon}\left[\bm{h}_n^T\left(\bm{P}_n+\varepsilon \bm{C} _n+\varepsilon^2\bm{Q} _n\right)\bm{h}_n-\bm{h}_{n}^T \bm{U} _n \bm{h}_{n-1}-\bm{h}_{n-1}^T \bm{U} _n^T\bm{h}_{n}\right] \;.
\end{split}
\end{equation}
To proceed further, let us introduce the following notation. For a given matrix $\bm{M}\in\mathbb{R}^{d\times d}$, we label the sub-matrices $\smup{\bm{M}}\in\mathbb{R}^{l\times l}$, $\rharp{\bm{M}}\in\mathbb{R}^{l\times d-l}$, $\lharp{\bm{M}}\in\mathbb{R}^{d-l\times l}$, $\smdown{\bm{M}}\in\mathbb{R}^{d-l\times d-l}$ and $\overline{\bm{M}}\in\mathbb{R}^{l\times d}$, such that 
\begin{equation}
\bm{M}=\begin{pmatrix} \smup{\bm{M}} & \rharp{\bm{M}}\\ \lharp{\bm{M}} & \smdown{\bm{M}}\end{pmatrix},\quad\overline{\bm{M}}=(\smup{\bm{M}}\,, \rharp{\bm{M}}\,) \;.
\end{equation}
As a consequence, since $\bm{h}_n=(\bm{h}_{Vn},\bm{0})$, we can write more precisely Eq.~(\ref{dirneudisc2}) as
\begin{equation}\label{dirneudisc3}
\begin{split}
\delta^2 S(\bm{h};\bm{q}^m)&\approx\frac{1}{\varepsilon}\sum\limits_{j=1}^{n-1}\left[\bm{h}_j^T\left(\bm{P}_j+\bm{P}_{j+1}+\varepsilon^2\bm{Q} _j\right)\bm{h}_j-\bm{h}_j^T \bm{U} _j \bm{h}_{j-1}-\bm{h}_{j-1}^T \bm{U} _j^T \bm{h}_{j}\right]\\
&\quad+\frac{1}{\varepsilon}\left[\bm{h}_{Vn}^T\left(\smup{\bm{P}}_n+\varepsilon \smup{\bm{C}} _n+\varepsilon^2\smup{\bm{Q}}_{\,\,n}\right)\bm{h}_{Vn}-\bm{h}_{Vn}^T \overline{\bm{U}} _n \bm{h}_{n-1}-\bm{h}_{n-1}^T \overline{\bm{U}} _n^T\bm{h}_{Vn}\right] \;.
\end{split}
\end{equation}%

At this point, we define the transformation with unit Jacobian 
\begin{equation}
\begin{split}
&{\bm{\phi}}_j=\bm{h}_j-\bm{\beta} _j \bm{h}_{j-1}\in\mathbb{R}^{d}\quad\text{for}\quad j=1,...,n-1\;,\\
&{\bm{\phi}}_n=\bm{h}_{Vn}-\bm{\beta} _n \bm{h}_{n-1}\in\mathbb{R}^{l}\;,
\end{split}
\end{equation}
where the matrices $\bm{\beta}_j$ are given by the following recursive construction for the symmetric matrices $\bm{\alpha}_{j}$:
\begin{align}
\bm{\alpha} _{n} &= \smup{\bm{P}}_{n}+\varepsilon\frac{\smup{\bm{C}}_{\,\,n}+\smup{\bm{C}} _{\,\,n}^{\,\,T}}{2}+\varepsilon^2\smup{\bm{Q}}_{\,\,n}\in\mathbb{R}^{l\times l} \;,&& \nonumber\\
\bm{\alpha} _{j} &= \bm{P}_{j}+\bm{P}_{j+1}+\varepsilon^2 \bm{Q} _{j}-\bm{\beta} _{j+1}^T\bm{\alpha} _{j+1}\bm{\beta} _{j+1}\in\mathbb{R}^{d\times d} \;&&\text{for }\; j=n-1,...,1 \;, \nonumber\\
\overline{\bm{U}} _n&=\bm{\alpha} _n\bm{\beta} _n\in\mathbb{R}^{l\times d}\;,\nonumber\\
\bm{U} _j&=\bm{\alpha} _j\bm{\beta} _j\in\mathbb{R}^{d\times d} \;&&\text{for }\; j=1,...,n-1 \;.
\end{align}
Then Eq.~(\ref{dirneudisc3}) reduces to a sum of quadratic forms, namely
\begin{equation}\label{quaddirneu}
\delta^2 S(\bm{h};\bm{q}^m)\approx\frac{1}{\varepsilon}\sum\limits_{j=1}^{n}{\bm{\phi}}_j^T\bm{\alpha} _j{\bm{\phi}}_j \;.
\end{equation}%

\section{Recovering a continuous formulation}
\subsection{Wiener measure}
First of all, we briefly introduce the concept of Wiener measure, as presented in \citep{BookChaichian}. This will be the key for finally recovering a continuous formulation in terms of differential equations. Given the path integral expressions
\begin{small}
\begin{equation}\label{freeK}
\begin{split}
\mathcal{K}_{W}=\int\limits_{\bm{q}(t_0)=\bm{q}_0}^{\bm{q}(t)=\bm{q}}\dd_W\bm{q}(\tau)&\stackrel{\text{def}}{=}\int\limits_{\bm{q}(t_0)=\bm{q}_0}^{\bm{q}(t)=\bm{q}}{e^{-\frac{1}{2}{\int_{t_0}^t\dot{\bm{q}}^T(\tau)\bm{P}(\tau)\dot{\bm{q}}(\tau)\,\dd\tau}}\,\mathcal{D}\bm{q}}\;,\\
&\stackrel{\text{def}}{=}\lim_{n\rightarrow\infty}\displaystyle{\int\limits_{\bm{q}(t_0)=\bm{q}_0}^{\bm{q}(t)=\bm{q}}{e^{-\frac{\varepsilon}{2} \sum\limits_{j=1}^{n}\frac{(\bm{q}_j-\bm{q}_{j-1})}{\varepsilon}^T\bm{P}_j\frac{\bm{q}_j-\bm{q}_{j-1}}{\varepsilon}}}\,\prod\limits_{j=1}^{n}\left[\frac{\det{(\bm{P}_j)}}{(2\pi\varepsilon)^{d}}\right]^{\frac{1}{2}}\prod\limits_{j=1}^{n-1}{\dd \bm{q}_j}} \;,
\end{split}
\end{equation}
\end{small}%
\begin{small}
\begin{equation}\label{freeN}
\begin{split}
\mathcal{N}_{W}=\int\limits_{\bm{q}(t_0)=\bm{q}_0}\dd_W\bm{q}(\tau)&\stackrel{\text{def}}{=}\int\limits_{\bm{q}(t_0)=\bm{q}_0}{e^{-\frac{1}{2}\int_{t_0}^t\dot{\bm{q}}^T(\tau)\bm{P}(\tau)\dot{\bm{q}}(\tau)\,\dd\tau}\,\mathcal{D}\bm{q}}\;,\\
&\stackrel{\text{def}}{=}\lim_{n\rightarrow\infty}\displaystyle{\int\limits_{\bm{q}(t_0)=\bm{q}_0}{e^{-\frac{\varepsilon}{2} \sum\limits_{j=1}^{n}\frac{(\bm{q}_j-\bm{q}_{j-1})}{\varepsilon}^T\bm{P}_j\frac{\bm{q}_j-\bm{q}_{j-1}}{\varepsilon}}}\,\prod\limits_{j=1}^{n}\left[\frac{\det{(\bm{P}_j)}}{(2\pi\varepsilon)^{d}}\right]^{\frac{1}{2}}{\dd \bm{q}_j}} \;,
\end{split}
\end{equation}
\end{small}%
we define the conditional and unconditional Wiener measure $\dd_W\bm{q}(\tau)$ equipped with the associated boundary conditions by means of Eq.~(\ref{freeK}) and Eq.~(\ref{freeN}) respectively. Moreover, setting $\bm{R}=\int_{t_0}^t{\bm{P}^{-1}(\tau)}\,\dd\tau$, the following facts hold
\begin{equation}
\mathcal{K}_{W}=\int\limits_{\bm{q}(t_0)=\bm{q}_0}^{\bm{q}(t)=\bm{q}}\dd_W\bm{q}(\tau)=\frac{e^{-\frac{1}{2}(\bm{q}-\bm{q}_0)^T\bm{R}^{-1}(\bm{q}-\bm{q}_0)}}{\sqrt{\det[2\pi\bm{R}]}}
\;,
\end{equation}
\begin{equation}
\mathcal{N}_{W}=\int\limits_{\bm{q}(t_0)=\bm{q}_0}\dd_W\bm{q}(\tau)=1\;,
\end{equation}
and therefore the multivariate Gaussian transition probability density for the $d-$ dimensional Wiener process is given by 
\begin{equation}
\rho_W(\bm{q},t \vert \bm{q}_0,t_0) = \dfrac{\mathcal{K}_W}{\mathcal{N}_W}=\frac{e^{-\frac{1}{2}(\bm{q}-\bm{q}_0)^T\bm{R}^{-1}(\bm{q}-\bm{q}_0)}}{\sqrt{\det[2\pi\bm{R}]}}\;.
\end{equation}
Using the latter formalism, the evaluation of $\mathcal{K}$ and $\mathcal{N}$ in Eq.~(\ref{pathint}) and Eq.~(\ref{normex}) is interpreted as the integral of a functional with respect to the Wiener measure.

The evaluation of $\mathcal{K}$ in the Laplace approximation is a standard textbook technique if $d=1$ and in absence of cross terms in the second variation. In general, it is remarkably more complicated. Here we adopt a time-slicing procedure for the evaluation of expressions of the type Eq.~(\ref{semicont}). From the discretisation of $\tau\in[t_0,t]$ into $n$ intervals of length $\varepsilon=\frac{t-t_0}{n}$, the Laplace approximation for $\mathcal{K}$ reads
\begin{equation}\label{appx}
\mathcal{K}\approx e^{-S(\bm{q}^f)}\int\limits_{\bm{h}(t_0)=\bm{0}}^{\bm{h}(t)=\bm{0}}{e^{-\frac{1}{2}\delta^2 S(\bm{h};\bm{q}^f)}\,\mathcal{D}\bm{h}} =e^{-S(\bm{q}^f)} \lim_{n\rightarrow\infty} I^f_n \;,
\end{equation}
\begin{equation}\label{num}
I^f_n =\displaystyle{\int\limits_{\bm{h}(t_0)=\bm{0}}^{\bm{h}(t)=\bm{0}}{e^{-\frac{\varepsilon}{2} \sum\limits_{j=0}^{n}\delta^2 S(\bm{h};\bm{q}^f)_j}}\,\prod\limits_{j=1}^{n}\left[\frac{\det{(\bm{P}_j)}}{(2\pi\varepsilon)^{d}}\right]^{\frac{1}{2}}\prod\limits_{j=1}^{n-1}{\dd \bm{h}_j}}\;,
\end{equation}%
where $\varepsilon\sum\limits_{j=0}^{n}\delta^2 S(\bm{h};\bm{q}^f)_j$ is better defined in the first line of  Eq.~(\ref{secdiag}). Moreover, the integration boundaries come from the fact that $\bm{h}(\tau)$ represents a perturbation around the minimum $\bm{q}^f(\tau)$, and as such it must satisfy null Dirichlet boundary conditions. Let us mention that the products in Eq.~(\ref{num}) form a part of the integration measure for the integral, which is here a conditional Wiener measure \citep{BookChaichian}. 

Finally, we define 
\begin{equation}
\bm{F}_j=\bm{\alpha}_j\bm{P}_j^{-1}\quad\text{for}\quad j=1,...,n-1\;,
\end{equation}
and we compute the the Gaussian integrals in Eq.~(\ref{num}) by means of the known formula
\begin{equation}
\int\limits_{\mathbb{R}^d}{e^{-\frac{\varepsilon}{2} \bm{x}^T\bm{A}\bm{x}}}\,{\dd \bm{x}}=\left(\frac{2\pi}{\varepsilon}\right)^{\frac{d}{2}}\sqrt{\det(\bm{A}^{-1})}
\end{equation}
and Eq.~(\ref{quaddir}) as:
\begin{align}\label{F1}
I^f_n &= \displaystyle{\int{e^{-\frac{1}{2\varepsilon}\sum\limits_{j=1}^{n-1}\bm{\phi}_j^T\bm{\alpha}_j\bm{\phi}_j}\,\prod\limits_{j=1}^{n}\left[\frac{\det{(\bm{P}_j)}}{(2\pi\varepsilon)^{d}}\right]^{\frac{1}{2}}\prod\limits_{j=1}^{n-1}{\dd\bm{\phi}_j}}} \;,\nonumber\\
&=\det{\left[2\pi\varepsilon\bm{P}_n^{-1}\prod\limits_{j=1}^{n-1} \bm{F}_j\right]}^{-\frac{1}{2}} \;.
\end{align}

It is easy to show that, even for simple quadratic Lagrangians, the latter procedure does not lead alone to a transition probability satisfying the normalization condition Eq.~(\ref{norm}). This can happen even if there are no cross terms (\ie $\bm{C}=\mathbb{0}$), as we will remark in the example section. To fix this issue, the condition Eq.~(\ref{norm}) is enforced by introducing the normalization factor $\mathcal{N}$, see Eq.~(\ref{pathint}). Unfortunately, computing $\mathcal{N}$ can be a non-trivial task, which we are now going to tackle.

Following the same approach as for $\mathcal{K}$, we compute the Laplace approximation for $\mathcal{N}$ as defined in Eq.~(\ref{normex}). This time we Taylor expand the action $S(\bm{q}(\tau))$ to second order around the Neumann minimum $\hat{\bm{q}}(\tau)$, since the point $\bm{q}(t)$ is unconstrained. Then, from the same discretization as before, the Laplace approximation for $\mathcal{N}$ reads
\begin{equation}\label{appx2}
\mathcal{N}\approx e^{-S(\hat{\bm{q}})}\int\limits_{\bm{h}(t_0)=\bm{0}}{e^{-\frac{1}{2}\delta^2 S(\bm{h};\hat{\bm{q}})}\,\mathcal{D}\bm{h}} =e^{-S(\hat{\bm{q}})} \lim_{n\rightarrow\infty} \hat{I}_n \;,
\end{equation}
\begin{equation}\label{den}
\hat{I}_n =\displaystyle{\int\limits_{\bm{h}(t_0)=\bm{0}}{e^{-\frac{\varepsilon}{2} \sum\limits_{j=0}^{n}\delta^2 S(\bm{h};\hat{\bm{q}})_j}}\,\prod\limits_{j=1}^{n}\left[\frac{\det{(\bm{P}_j)}}{(2\pi\varepsilon)^{d}}\right]^{\frac{1}{2}}{\dd \bm{h}_j}}\;.
\end{equation}
Here the integration boundaries come from the fact that $\bm{h}(\tau)$ represents a perturbation around the minimum $\hat{\bm{q}}(\tau)$, and as such it must satisfy only the initial null Dirichlet boundary condition. The products in Eq.~(\ref{den}) give part of the integration measure for the integral, which is here an unconditional Wiener measure \citep{BookChaichian}. In fact, note that contrary to Eq.~(\ref{num}) the product of $\dd \bm{h}_j$ runs here until $n$, which is what makes the computation of $\mathcal{N}$ in general non-trivial.

Finally, we define 
\begin{equation}
\bm{F}_j=\bm{\alpha}_j\bm{P}_j^{-1}\quad\text{for}\quad j=1,...,n\;,
\end{equation}
and we compute the Gaussian integrals as before by means of Eq.~(\ref{quadneu}):
\begin{align}\label{F2}
\hat{I}_n &= \displaystyle{\int{e^{-\frac{1}{2\varepsilon}\sum\limits_{j=1}^{n}\bm{\phi}_j^T\bm{\alpha}_j\bm{\phi}_j}\,\prod\limits_{j=1}^{n}\left[\frac{\det{(\bm{P}_j)}}{(2\pi\varepsilon)^{d}}\right]^{\frac{1}{2}}{\dd\bm{\phi}_j}}}\;, \nonumber\\
&=\det{\left[\prod\limits_{j=1}^{n} \bm{F}_j\right]}^{-\frac{1}{2}} \;. 
\end{align}

To extend this result to marginal distributions Eq.~(\ref{pathint2}) we combine the techniques involved in the computation of $\mathcal{K}$ and $\mathcal{N}$ in order to evaluate the path integral defining $\mathcal{K}_m$. The Laplace approximation for $\mathcal{K}_m$ reads
\begin{equation}\label{appxDN}
\mathcal{K}_m\approx e^{-S(\bm{q}^m)}\int\limits_{\bm{h}(t_0)=\bm{0}}^{\bm{h}_F(t)=\bm{0}}{e^{-\frac{1}{2}\delta^2 S(\bm{h};\bm{q}^m)}\,\mathcal{D}\bm{h}}=e^{-S(\bm{q}^m)} \lim_{n\rightarrow\infty} I^m_n \;,
\end{equation}
\begin{equation}\label{numDN}
I^m_n=\displaystyle{\int\limits_{\bm{h}(t_0)=\bm{0}}^{\bm{h}_F(t)=\bm{0}}{e^{-\frac{\varepsilon}{2} \sum\limits_{j=0}^{n}\delta^2 S(\bm{h};\bm{q}^m)_j}}\,\prod\limits_{j=1}^{n}\left[\frac{\det{(\bm{P}_j)}}{(2\pi\varepsilon)^{d}}\right]^{\frac{1}{2}}\prod\limits_{j=1}^{n-1}{\dd \bm{h}_j}\,\dd \bm{h}_{Vn}} \;,
\end{equation}
where $\bm{h}(\tau)=(\bm{h}_{V}(\tau),\bm{h}_{F}(\tau))$ represents a perturbation around the minimum $\bm{q}^m(\tau)$, and as such it must satisfy null Dirichlet boundary conditions corresponding to the fixed variables. Note that the integration involves only the variable part of the variation at $\tau=t$, namely $\bm{h}_{Vn}=\bm{h}_{V}(\tau_n)=\bm{h}_{V}(t)$.

Analogously to what was done previously for $\mathcal{K}$ and $\mathcal{N}$, we perform a backward integration of Eq.~(\ref{numDN}) by finding a set of positive definite matrices 
\begin{equation}
\bm{F}_j=\bm{\alpha}_j\bm{P}_j^{-1}\quad\text{for}\quad j=1,...,n-1\;,
\end{equation}
that depend on the coefficients of the second variation on $\bm{q}^m(\tau)$, \ie $\bm{P}$, $\bm{C}^m$ and $\bm{Q}^m$, and we compute the Gaussian integrals by means of Eq.~(\ref{quaddirneu}):
\begin{align}
I^m_n &= \displaystyle{\int{e^{-\frac{1}{2\varepsilon}\sum\limits_{j=1}^{n}\bm{\phi}_j^T\bm{\alpha}_j\bm{\phi}_j}\,\prod\limits_{j=1}^{n}\left[\frac{\det{(\bm{P}_j)}}{(2\pi\varepsilon)^{d}}\right]^{\frac{1}{2}}{\dd\bm{\phi}_j}}} \;,\nonumber\\
&=\left[(2\pi\varepsilon)^{d-l}\det{(\bm{\alpha}_n)}\right]^{-\frac{1}{2}}\det{\left[\bm{P}_n^{-1}\prod\limits_{j=1}^{n-1} \bm{F}_j\right]}^{-\frac{1}{2}} \;.
\end{align}
Then, we make the choice (which is part of the construction and substantially simplifies the calculations provided later on)
\begin{equation}
\bm{F} _n=\begin{pmatrix}
\bm{\alpha}_{n} & \rharp{\bm{P}}_n\\ \mathbb{0} & \varepsilon\mathbb{1}_{d-l\times d-l}\end{pmatrix}\bm{P}_n^{-1} \;,
\end{equation}
and get
\begin{equation}\label{res1DN}
{I}^m_n=(2\pi)^{\frac{l-d}{2}}\det{\left[\prod\limits_{j=1}^{n} \bm{F}_j\right]}^{-\frac{1}{2}} \;. 
\end{equation}

\subsection{The Papadopoulos equation}
\subsubsection{Background}
In \citep{PAP1}, the author showed how to recover a continuous formulation by means of a difference equation, which in the limit leads to the following (non-linear, second-order, homogeneous of order one) matrix differential equation, that we call the Papadopoulos equation following \citep{LUDT, LUD}
\begin{equation}\label{papeq}
\begin{split}
 \frac{\dd}{\dd \tau}\left[\dot{\bm{D}}{{\bm{P}}}\right]+{{\bm{D}}}\dot{\bm{C}}^{(s)}-{{\bm{D}}}\left[{{\bm{Q}}}+{{\bm{C}}}^{(a)}{{\bm{P}}}^{-1}{{\bm{C}}}^{(a)}\right]=
\dot{\bm{D}}{{\bm{C}}}^{(a)}-{{\bm{D}}}{{\bm{C}}}^{(a)}{{\bm{P}}}^{-1}{{\bm{D}}}^{-1}\dot{\bm{D}}{{\bm{P}}} \;,
\end{split}
\end{equation}
\begin{equation}
\text{with}\quad{\bm{C}}^{(s)}=\frac{\bm{C}+\bm{C}^T}{2},\quad{\bm{C}}^{(a)}=\frac{\bm{C}-\bm{C}^T}{2}
\end{equation} 
the symmetric and the anti-symmetric part of $\bm{C}$ and ${\bm{D}}(\tau)$ the unknown matrix in $\mathbb{R}^{d\times d}$. We note that non-symmetric cross terms, \ie the matrix $\bm{C}$, are accounted for within the computation. This is the crucial extension that Papadopoulos made, which was not present in the literature before. However, the Jacobi equation does not appear in this setting yet; in the next section we will refer to \citep{LUDT, LUD} in order to re-establish the connection with the Jacobi fields formulation presented at the beginning of the present paper. We further mention that Papadopoulos worked in the context of Feynman path integrals \citep{FEY, BookFeynman}, which belong to the quantum mechanical world. Nevertheless, the same results can be applied to real valued Wiener path integrals and we will describe all the machinery in the following pages, but with one main difference. In fact Papadopoulos showed that for Dirichlet-Dirichlet boundary conditions the limit for $n\rightarrow\infty$ of Eq.~(\ref{F1}) can be expressed as
\begin{equation}
\lim_{n\rightarrow\infty} I^f_n=\det\left[2\pi {\bm{D}^f}(t)\right]^{-\frac{1}{2}} \;,
\end{equation}
where ${\bm{D}^f}(\tau)$ solves Eq.~(\ref{papeq}) with initial conditions 
\begin{equation}
{\bm{D}^f}(t_0)=\mathbb{0},\quad{\dot{\bm{D}}^f}(t_0)=\bm{P}(t_0)^{-1} \;.
\end{equation}
The limitation of the latter approach lies in the direction of integration. Namely, if we aim to extend this result to different boundary conditions, e.g. where full Dirichlet at the last end is not present, it is crucial to integrate in the backward direction and the initial conditions are given at $\tau=t$. Therefore, in the following it is more sensible to present also the previously known Dirichlet-Dirichlet result in this fashion, to be consistent with the computation of the new entries $\mathcal{N}$ and $\mathcal{K}_m$, as we are going to underline.  

\subsubsection{Extension}
We are going to show that the limit for $n\rightarrow\infty$ of Eq.~(\ref{F1}), Eq.~(\ref{F2}) and Eq.~(\ref{res1DN}) can be expressed in terms of the Papadopoulos equation as follows.
\begin{equation}\label{pap}
\lim_{n\rightarrow\infty} I^f_n=\det\left[2\pi {\bm{D}^f}(t_0)\right]^{-\frac{1}{2}} \;,
\end{equation}
where ${\bm{D}^f}(\tau)$ solves Eq.~(\ref{papeq}) with initial conditions 
\begin{equation}\label{IN1}
{\bm{D}^f}(t)=\mathbb{0},\quad{\dot{\bm{D}}^f}(t)=-\bm{P}(t)^{-1} \;.
\end{equation}
\begin{equation}\label{pap2}
\lim_{n\rightarrow\infty} \hat{I}_n=\det\left[{\hat{\bm{D}}}(t_0)\right]^{-\frac{1}{2}} \;,
\end{equation}
where $\hat{\bm{D}}(\tau)$ solves Eq.~(\ref{papeq}) with initial conditions 
\begin{equation}\label{IN2}
{\hat{\bm{D}}}(t)=\mathbb{1},\quad\dot{\hat{\bm{D}}}(t)=-{\hat{\bm{C}}}^{(s)}(t)\bm{P}(t)^{-1} \;.
\end{equation}
\begin{equation}\label{pap2DN}
\lim_{n\rightarrow\infty} I^m_n=(2\pi)^{\frac{l-d}{2}}\det\left[{\bm{D}^m}(t_0)\right]^{-\frac{1}{2}} \;,
\end{equation}
where $\bm{D}^m(\tau)$ solves Eq.~(\ref{papeq}) with initial conditions
\begin{equation}\label{IN3}
\begin{split}
&\bm{D}^m(t)=\begin{pmatrix}
\mathbb{1}_{l\times l} & \mathbb{0}_{l\times d-l}\\ \mathbb{0}_{d-l\times l} & \mathbb{0}_{d-l\times d-l}\end{pmatrix}\;,\\
& \\
&\dot{\bm{D}}^m(t)=\begin{pmatrix}
\displaystyle{-\frac{\smup{\bm{C}}+\smup{\bm{C}}^{\,\,T}}{2}\smup{\bm{P}}^{-1}-\bm{W}\bm{Z}^{-1}\rharp{\bm{P}}^{\,\,T} \smup{\bm{P}}^{-1}}&\,\,\,\,\,\,\,\displaystyle{\bm{W}\bm{Z}^{-1}}\\
\\\bm{Z}^{-1}\rharp{\bm{P}}^{\,\,T}\smup{\bm{P}}^{-1}&\,\,\,\,\,-\bm{Z}^{-1}\end{pmatrix}\Big{|}_{\bm{q}^m}(t) \;,\\
& \\
&\bm{Z}=\smdown{\bm{P}}-\rharp{\bm{P}}^{\,T}\smup{\bm{P}}^{-1}\rharp{\bm{P}},\quad\bm{W}=\smup{\bm{C}}\smup{\bm{P}}^{-1}\rharp{\bm{P}}-\frac{\rharp{\bm{C}}-\lharp{\bm{C}}^{\,T}}{2} \;.
\end{split}
\end{equation}
As a consequence of the backward integration (necessary for deriving the matrices $\bm{F}_j$ for $\mathcal{N}$ and $\mathcal{K}_m$), Eq.~(\ref{papeq}) is solved in the backward direction. One of our main results is to be able to perform this passage from a modification of the method used by Papadopoulos in \citep{PAP1} as commented above. The idea is precisely that of performing a backward integration, meaning that the standard direction of discretisation $(\bm{q}_0,t_0)\rightarrow (\bm{q},t)$ is now replaced by $(\bm{q},t)\rightarrow (\bm{q}_0,t_0)$, and then exploiting proper recursion relations for ${I}_n$ and $\bm{F}_j$ in order to derive the correct initial conditions.

In conclusion, the results obtained so far allow us to express the transition probabilities Eq.~(\ref{pathint}) and Eq.~(\ref{pathint2}) in the Laplace approximation as
\begin{equation}\label{fin1}
\rho(\bm{q},t \vert \bm{q}_0,t_0)\approx e^{S(\hat{\bm{q}})-S(\bm{q}^f)}\sqrt{\det\left[\frac{1}{2\pi}\frac{{\hat{\bm{D}}}}{{\bm{D}^f}}(t_0)\right]} \;,
\end{equation}
\begin{equation}\label{finn1}
\rho_m(\bm{q}_{F},t \vert \bm{q}_0,t_0) \approx  e^{S(\hat{\bm{q}})-S(\bm{q}^m)}\sqrt{(2\pi)^{l-d}\det\left[\frac{{\hat{\bm{D}}}}{{\bm{D}^m}}(t_0)\right]} \;.
\end{equation}
Let us emphasize that when the Lagrangian is quadratic in its variables we have equality, since the second-order expansion used in the Laplace approximation does not neglect any term of higher order.

\subsubsection{Recurrence relations}
Let us start the derivation of Eq.~(\ref{pap}), Eq.~(\ref{pap2}) and Eq.~(\ref{pap2DN}). In order to derive Eq.~(\ref{pap}) we need to look for recurrence relations to express Eq.~(\ref{F1}) through a difference equation. We define 
\begin{equation}
\bm{D}_{n-k}=\varepsilon\bm{P}_n^{-1}\prod\limits_{j=1}^{k}\bm{F}_{n-j}\quad\text{for}\quad k=1,...,n-1\;, 
\end{equation}
and provide the following iterative method for $\bm{D}$ and ${\bm{\alpha}}$.
\begin{itemize}
\item Initial condition: $\bm{D}_{n-1}=\varepsilon\bm{P}_n^{-1}\bm{\alpha}_{n-1}\bm{P}_{n-1}^{-1}$\;.
\item Iteration scheme: $\bm{D}_{n-(k+1)}=\bm{D}_{n-k}\bm{\alpha}_{n-(k+1)}\bm{P}_{n-(k+1)}^{-1}$ for $\,\,k=1,...,n-2$\;.
\end{itemize}
\begin{itemize}
\item Initial condition: $\bm{\alpha}_{n-1}=\bm{P}_{n-1}+\bm{P}_{n}+\varepsilon^2 \bm{Q}_{n-1}$\;.
\item Iteration scheme: $\bm{\alpha}_{n-(k+1)}=\bm{P}_{n-(k+1)}+\bm{P}_{n-k}+\varepsilon^2 \bm{Q}_{n-(k+1)}-(\bm{\beta}_{n-k})^T\bm{\alpha}_{n-k}\bm{\beta}_{n-k}$ for $k=1,...,n-2$\;.
\end{itemize}

Regarding Eq.~(\ref{pap2}), we recover recurrence relations for Eq.~(\ref{F2}) defining 
\begin{equation}
\bm{D}_{n-k}=\prod\limits_{j=0}^{k}\bm{F}_{n-j}\quad\text{for}\quad k=0,...,n-1\;,
\end{equation}
and giving the following iterative method for $\bm{D}$ and ${\bm{\alpha}}$.
\begin{itemize}
\item Initial condition: $\bm{D}_n=\bm{\alpha}_{n}\bm{P}_{n}^{-1}$\;.
\item Iteration scheme: $\bm{D}_{n-(k+1)}=\bm{D}_{n-k}\bm{\alpha}_{n-(k+1)}\bm{P}_{n-(k+1)}^{-1}$ for $\,\,k=0,...,n-2$\;.
\end{itemize}
\begin{itemize}
\item Initial condition: $\displaystyle{\bm{\alpha}_{n}=\bm{P}_{n}+\varepsilon\frac{\bm{C}_n+\bm{C}_n^T}{2}+\varepsilon^2 \bm{Q}_{n}}$\;.
\item Iteration scheme: $\bm{\alpha}_{n-(k+1)}=\bm{P}_{n-(k+1)}+\bm{P}_{n-k}+\varepsilon^2 \bm{Q}_{n-(k+1)}-(\bm{\beta}_{n-k})^T\bm{\alpha}_{n-k}\bm{\beta}_{n-k}$ for $k=0,...,n-2$\;.
\end{itemize}

Lastly, for Eq.~(\ref{pap2DN}) we start from Eq.~(\ref{res1DN}) and define 
\begin{equation}
\bm{D}_{n-k}=\prod\limits_{j=0}^{k}\bm{F}_{n-j}\quad\text{for}\quad k=0,...,n-1\;,
\end{equation}
obtaining the following iterative method.
\begin{itemize}
\item Initial condition: $\bm{D}_n=\bm{F}_{n}$\;.
\item Iteration scheme: $\bm{D}_{n-(k+1)}=\bm{D}_{n-k}\bm{\alpha}_{n-(k+1)}\bm{P}_{n-(k+1)}^{-1}$ for $\,\,k=0,...,n-2$\;.
\end{itemize}
\begin{itemize}
\item Initial condition: $\displaystyle{\bm{\alpha}_{n}=\smup{\bm{P}}_{n}+\varepsilon\frac{\smup{\bm{C}}_n+\smup{\bm{C}}_n^{\,\,T}}{2}+\varepsilon^2\smup{\bm{Q}}_{\,\,n}}$\;.
\item Iteration scheme: $\bm{\alpha}_{n-(k+1)}=\bm{P}_{n-(k+1)}+\bm{P}_{n-k}+\varepsilon^2 \bm{Q}_{n-(k+1)}-(\bm{\beta}_{n-k})^T\bm{\alpha}_{n-k}\bm{\beta}_{n-k}$ for $k=0,...,n-2$\;.
\end{itemize}

Since the recurrence relations for the three cases are the same except for the conditions at the slice $n-1$ or $n$, the following facts hold in general. Recalling that
$\bm{\beta}_{n-k}=\bm{\alpha}_{n-k}^{-1}\bm{U}_{n-k}$, and that $\bm{U}_{n-k}=\bm{P}_{n-k}+\frac{\varepsilon}{2}[\bm{C}_{n-k}^T-\bm{C}_{n-(k+1)}]$,
it is possible to give the explicit recurrence relation for $\bm{\alpha}_{n-(k+1)}$ as\\
\\
\begin{equation}\label{rec2}
\begin{split}
\bm{\alpha}_{n-(k+1)}&=\bm{P}_{n-(k+1)}+\bm{P}_{n-k}+\varepsilon^2 \bm{Q}_{n-(k+1)}-\bm{P}_{n-k}\bm{\alpha}_{n-k}^{-1}\bm{P}_{n-k}\\
&\quad-\varepsilon\left[\frac{\bm{C}_{n-k}-\bm{C}_{n-(k+1)}^T}{2}\right]\bm{\alpha}_{n-k}^{-1}\bm{P}_{n-k}-\varepsilon \bm{P}_{n-k}\bm{\alpha}_{n-k}^{-1}\left[\frac{\bm{C}_{n-k}^T-\bm{C}_{n-(k+1)}}{2}\right]\\
&\quad-\varepsilon^2\left[\frac{\bm{C}_{n-k}-\bm{C}_{n-(k+1)}^T}{2}\right]\bm{\alpha}_{n-k}^{-1}\left[\frac{\bm{C}_{n-k}^T-\bm{C}_{n-(k+1)}}{2}\right] \;.
\end{split}
\end{equation}
Moreover, the recurrence formula for $\bm{D}$ provides the additional useful relations
\begin{align}\label{Drec}
\bm{\alpha}_{n-(k+1)} &= \bm{D}_{n-k}^{-1}\bm{D}_{n-(k+1)}\bm{P}_{n-(k+1)} \;,\nonumber\\
\nonumber\\
\bm{\alpha}_{n-k}^{-1} &= \bm{P}_{n-k}^{-1}\bm{D}_{n-k}^{-1}\bm{D}_{n-(k-1)}\;.
\end{align}%
Finally, substituting Eq.~(\ref{Drec}) in Eq.~(\ref{rec2}), and multiplying to the left both sides by $\bm{D}_{n-k}$, we get the full difference equation for the matrix $\bm{D}$, in terms of $\bm{P}$, ${\bm{C}}$ and ${\bm{Q}}$:\\
\\
\begin{small}
\begin{equation}\label{diff}
\begin{split}
\bm{D}_{n-(k+1)}\bm{P}_{n-(k+1)}&=\bm{D}_{n-k}\bm{P}_{n-(k+1)}+\bm{D}_{n-k}\bm{P}_{n-k}-\bm{D}_{n-(k-1)}\bm{P}_{n-k}+\varepsilon^2 \bm{D}_{n-k}\bm{Q}_{n-(k+1)}\\
&\, -\varepsilon \bm{D}_{n-k}\left[\frac{\bm{C}_{n-k}-\bm{C}_{n-(k+1)}^T}{2}\right]\bm{P}_{n-k}^{-1}\bm{D}_{n-k}^{-1}\bm{D}_{n-(k-1)}\bm{P}_{n-k}\\
&\, -\varepsilon \bm{D}_{n-(k-1)}\left[\frac{\bm{C}_{n-k}^T-\bm{C}_{n-(k+1)}}{2}\right]\\
&\, -\varepsilon^2 \bm{D}_{n-k}\left[\frac{\bm{C}_{n-k}-\bm{C}_{n-(k+1)}^T}{2}\right]\bm{P}_{n-k}^{-1}\bm{D}_{n-k}^{-1}\bm{D}_{n-(k-1)}\left[\frac{\bm{C}_{n-k}^T-\bm{C}_{n-(k+1)}}{2}\right].
\end{split}
\end{equation}
\end{small}%
Our goal is now to take the continuous limit ($n\rightarrow\infty$, $\epsilon\rightarrow 0$) for the latter expression, in order to obtain a differential equation for the unknown $\bm{D}$. To this end, recall that \eg $\bm{D}_{n-(k+1)}$ stands for $\bm{D}(t-s_{k+1})$ with $s_{k+1}=(k+1)\varepsilon=s_k+\varepsilon$, and that similar expressions hold for all other terms. We can therefore Taylor expand each $\bm{D}$ around $s_k$ to second order in $\varepsilon$, and each other coefficient to first order. Then, dividing everything by $\varepsilon^2$ we obtain
\begin{align}
& \frac{\dd}{\dd s}\left[ \frac{\dd}{\dd s}\left[\bm{D}(t-s)\right]\bm{P}(t-s)\right]-\bm{D}(t-s) \frac{\dd}{\dd s}\left[\bm{C}^{(s)}(t-s)\right]\nonumber\\
&\quad -\bm{D}(t-s)\left[\bm{Q}(t-s)+\bm{C}^{(a)}(t-s)\bm{P}^{-1}(t-s)\bm{C}^{(a)}(t-s)\right]= \nonumber\\
&=- \frac{\dd}{\dd s}\left[\bm{D}(t-s)\right]\bm{C}^{(a)}(t-s)\nonumber\\
&\quad +\bm{D}(t-s)\bm{C}^{(a)}(t-s)\bm{P}^{-1}(t-s)\bm{D}^{-1}(t-s) \frac{\dd}{\dd s}\left[\bm{D}(t-s)\right]\bm{P}(t-s) \;. \label{sys.2}
\end{align}

We provide the details for the derivation of Eq.~(\ref{sys.2}). Namely, in Eq.~(\ref{diff}) we simplify the notation in the following way:
\begin{itemize}
\item $\bm{f}(x)= {\bm{D}}_{n-k}$, $\bm{a}(x)= {\bm{P}}_{n-k}$, $\bm{b}(x)= {\bm{Q}}_{n-k}$, $\bm{c}(x)= {\bm{C}}_{n-k}$ and $\bm{c}^t(x)= {\bm{C}}_{n-k}^t$\;,
\item $\bm{f}(x+\varepsilon)= {\bm{D}}_{n-(k+1)}$, $\bm{a}(x+\varepsilon)= {\bm{P}}_{n-(k+1)}$, $\bm{b}(x+\varepsilon)= {\bm{Q}}_{n-(k+1)}$, $\bm{c}(x+\varepsilon)= {\bm{C}}_{n-(k+1)}$ and $\bm{c}^t(x+\varepsilon)= {\bm{C}}_{n-(k+1)}^t$\;,
\item $\bm{f}(x-\varepsilon)= {\bm{D}}_{n-(k-1)}$\;.
\end{itemize}
Now we can Taylor expand $\bm{f}(x+\varepsilon)$, $\bm{f}(x-\varepsilon)$, $\bm{a}(x+\varepsilon)$, $\bm{b}(x+\varepsilon)$, $\bm{c}(x+\varepsilon)$ and $\bm{c}^t(x+\varepsilon)$ to the desired order, but for this particular case of a second-order difference equation in the variable $\bm{f}(x)$ and coefficients $\bm{a}(x)$, $\bm{b}(x)$, $\bm{c}(x)$, $\bm{c}^t(x)$, it is convenient to expand the unknown to the second order and the coefficients just to the first order. Forgetting the rest of the Taylor expansion and the dependence on $x$ for notational convenience, we can rewrite Eq.~(\ref{diff}) as follows.
\begin{small}
\begin{equation}
\begin{split}
 \left[\bm{f}+\bm{f}'\varepsilon+\frac{1}{2}\bm{f}''\varepsilon^2 \right] \left[\bm{a}+\bm{a}'\varepsilon \right]&=\bm{f} \left[\bm{a}+\bm{a}'\varepsilon \right]+\bm{f}\bm{a}- \left[\bm{f}-\bm{f}'\varepsilon+\frac{1}{2}\bm{f}''\varepsilon^2 \right]\bm{a}+\varepsilon^2 \bm{f} \left[\bm{b}+\bm{b}'\varepsilon \right]\\
&\,-\varepsilon \bm{f} \left[\frac{\bm{c}-\bm{c}^t-{\bm{c}^t}'\varepsilon}{2}\right]\bm{a}^{-1}\bm{f}^{-1} \left[\bm{f}-\bm{f}'\varepsilon+\frac{1}{2}\bm{f}''\varepsilon^2 \right]\bm{a}\\
&\,-\varepsilon \left[\bm{f}-\bm{f}'\varepsilon+\frac{1}{2}\bm{f}''\varepsilon^2 \right]\left[\frac{\bm{c}^t-\bm{c}-\bm{c}'\varepsilon}{2}\right]\\
&\,-\varepsilon^2 \bm{f}\left[\frac{\bm{c}-\bm{c}^t-{\bm{c}^t}'\varepsilon}{2} \right]\bm{a}^{-1}\bm{f}^{-1} \left[\bm{f}-\bm{f}'\varepsilon+\frac{1}{2}\bm{f}''\varepsilon^2 \right]\left[\frac{\bm{c}^t-\bm{c}-\bm{c}'\varepsilon}{2}\right],
\end{split}
\end{equation}
\end{small}%
where the prime sign here denotes the derivative in the backward direction with respect to $s$.
Multiplying out all the terms and neglecting the elements with power of $\varepsilon$ greater than $2$ (which comes from the fact that the equation is of order $2$), we find that 
\begin{equation}
\begin{split}
\bm{\mathbb{0}}&=\bm{f}\bm{a}+\bm{f}'\bm{a}\varepsilon+\frac{1}{2}\bm{f}''\bm{a}\varepsilon^2+\bm{f}\bm{a}'\varepsilon+\bm{f}'\bm{a}'\varepsilon^2-\bm{f}\bm{a}-\bm{f}\bm{a}'\varepsilon-\bm{f}\bm{a}+\bm{f}\bm{a}-\bm{f}'\bm{a}\varepsilon+\frac{1}{2}\bm{f}''\bm{a}\varepsilon^2\\
&\quad-\bm{f}\bm{b}\varepsilon^2+\frac{1}{2}\bm{f}\bm{c}\varepsilon-\frac{1}{2}\bm{f}\bm{c}^t\varepsilon-\frac{1}{2}\bm{f}{\bm{c}^t}'\varepsilon^2-\frac{1}{2}\bm{f}\bm{c}\bm{a}^{-1}\bm{f}^{-1}\bm{f}'\bm{a}\varepsilon^2+\frac{1}{2}\bm{f}\bm{c}^t\bm{a}^{-1}\bm{f}^{-1}\bm{f}'\bm{a}\varepsilon^2\\
&\quad+\frac{1}{2}\bm{f}\bm{c}^t\varepsilon-\frac{1}{2}\bm{f}\bm{c}\varepsilon-\frac{1}{2}\bm{f}\bm{c}'\varepsilon^2-\frac{1}{2}\bm{f}'\bm{c}^t\varepsilon^2+\frac{1}{2}\bm{f}'\bm{c}\varepsilon^2+\frac{1}{4} \bm{f}\bm{c}\bm{a}^{-1}\bm{c}^t\varepsilon^2-\frac{1}{4}\bm{f}\bm{c}\bm{a}^{-1}\bm{c}\varepsilon^2\\
&\quad -\frac{1}{4}\bm{f}\bm{c}^t\bm{a}^{-1}\bm{c}^t\varepsilon^2+\frac{1}{4} \bm{f}\bm{c}^t\bm{a}^{-1}\bm{c}\varepsilon^2 \;,
\end{split}
\end{equation}
and dividing by $\varepsilon^2$ after simplifying we get exactly Eq.~(\ref{sys.2}):
\begin{equation}
\begin{split}
\bm{\mathbb{0}}&=(\bm{f}'\bm{a})'-\bm{f}\left(\frac{\bm{c}^t+\bm{c}}{2}\right)'-\bm{f}\left[\bm{b}+\left(\frac{\bm{c}-\bm{c}^t}{2}\right)\bm{a}^{-1}\left(\frac{\bm{c}-\bm{c}^t}{2}\right)\right]\\
&\quad +\bm{f}'\left(\frac{\bm{c}-\bm{c}^t}{2}\right)-\bm{f}\left(\frac{\bm{c}-\bm{c}^t}{2}\right)\bm{a}^{-1}\bm{f}^{-1}\bm{f}'\bm{a} \;.
\end{split}
\end{equation}

\subsubsection{Initial conditions}
The initial conditions are a consequence of the recurrence relations for $\bm{D}$ and $\bm{\alpha}$, and they are derived as it follows. 

$\bullet$ For $\bm{D}^f$ we have
\begin{equation}
\begin{split}
&\bm{\alpha}_{n-1}=\bm{P}_{n-1}+\bm{P}_{n}+\varepsilon^2 \bm{Q}_{n-1}\sim 2\bm{P}_n \,\;\text{as}\;\, \varepsilon\rightarrow 0\\
&\Rightarrow\quad \bm{D}_{n-1}=\varepsilon\bm{P}_n^{-1}\bm{\alpha}_{n-1}\bm{P}_{n-1}^{-1}\sim\bm{\mathbb{0}} \,\;\text{as}\;\, \varepsilon \rightarrow  0  \;,
\end{split}
\end{equation}
\begin{equation}
\begin{split}
&\bm{\alpha}_{n-2}=\bm{P}_{n-2}+\bm{P}_{n-1}+\varepsilon^2 \bm{Q}_{n-2}-(\bm{\beta}_{n-1})^T\bm{\alpha}_{n-1}\bm{\beta}_{n-1}\sim \frac{3}{2}\bm{P}_n \,\;\text{as}\;\, \varepsilon\rightarrow 0\,\,\\
&\text{because of }\text{Eq}.~(\ref{rec2})\;,\\
&\Rightarrow\quad \frac{\bm{D}_{n-2}-\bm{D}_{n-1}}{\varepsilon}=\frac{\bm{D}_{n-1}\left(\bm{\alpha}_{n-2}\bm{P}_{n-2}^{-1}-{\mathbb{1}}\right)}{\varepsilon}\sim\frac{\bm{P}_n^{-1}\bm{\alpha}_{n-1}\bm{P}_{n-1}^{-1}}{2}\sim\bm{P}_n^{-1}\,\,\text{as}\,\,\varepsilon\rightarrow 0 \;,
\end{split}
\end{equation}
leading to 
\begin{equation}\label{ICID}
\bm{D}^f(t-s)\big{|}_{s=0}=\bm{\mathbb{0}},\quad \frac{\dd}{\dd s}\left[\bm{D}^f(t-s)\right]\big{|}_{s=0}=\bm{P}(t)^{-1} \;.
\end{equation}

$\bullet$ For $\hat{\bm{D}}$ we have
\begin{equation}\label{boun}
\begin{split}
&\bm{\alpha}_{n}=\bm{P}_{n}+\varepsilon\frac{\bm{C}_n+\bm{C}_n^T}{2} + \varepsilon^2\bm{Q}_{n}\sim \bm{P}_{n} \,\;\text{as}\;\, \varepsilon\rightarrow 0\\
&\Rightarrow\quad \bm{D}_{n}  =\bm{\alpha}_{n}\bm{P}_{n}^{-1}\sim\mathbb{1} \,\;\text{as}\;\, \varepsilon \rightarrow  0  \;.
\end{split}
\end{equation}
Moreover, note that
\begin{equation}\label{bounD}
\frac{\bm{D}_{n-1}-\bm{D}_{n}}{\varepsilon}=\frac{\bm{D}_{n}\left(\bm{\alpha}_{n-1}\bm{P}_{n-1}^{-1}-{\mathbb{1}}\right)}{\varepsilon}\sim\frac{\left(\bm{\alpha}_{n-1}\bm{P}_{n-1}^{-1}-{\mathbb{1}}\right)}{\varepsilon}\,\,\text{as}\,\,\varepsilon\rightarrow 0 \;,
\end{equation}
and, because of Eq.~(\ref{rec2}) and the fact that 
\begin{equation}
\left[\frac{\bm{C}_{n}-(\bm{C}_{n-1})^T+(\bm{C}_{n})^T-\bm{C}_{n-1}}{2}\right]\bm{P}_{n-1}^{-1}\rightarrow 0\,\,\text{as}\,\,\varepsilon\rightarrow 0\;,
\end{equation}
this can also be written as
\begin{equation}\label{Neum}
\begin{split}
\frac{\bm{\alpha}_{n-1}\bm{P}_{n-1}^{-1}-{\mathbb{1}}}{\varepsilon}&=\frac{1}{\varepsilon}\big{[}\bm{P}_{n}\bm{P}_{n-1}^{-1}+\varepsilon^2 \bm{Q}_{n-1}\bm{P}_{n-1}^{-1}-\bm{P}_{n}\bm{\alpha}_{n}^{-1}\bm{P}_{n}\bm{P}_{n-1}^{-1}\\
&\qquad-\varepsilon\left[\frac{\bm{C}_{n}-\bm{C}_{n-1}^T}{2}\right]\bm{\alpha}_{n}^{-1}\bm{P}_{n}\bm{P}_{n-1}^{-1}-\varepsilon \bm{P}_{n}\bm{\alpha}_{n}^{-1}\left[\frac{\bm{C}_{n}^T-\bm{C}_{n-1}}{2}\right]\bm{P}_{n-1}^{-1}\\
&\qquad-\varepsilon^2\left[\frac{\bm{C}_{n}-\bm{C}_{n-1}^T}{2}\right]\bm{\alpha}_{n}^{-1}\left[\frac{\bm{C}_{n}^T-\bm{C}_{n-1}}{2}\right]\bm{P}_{n-1}^{-1}\big{]}\;,\\
&\sim \frac{1}{\varepsilon}\left[\mathbb{1}-\bm{P}_{n}\bm{\alpha}_{n}^{-1}\right]\,\,\text{as}\,\,\varepsilon\rightarrow 0 \;.
\end{split}
\end{equation}
Inserting now the definition for $\bm{\alpha}_{n}$ we have
\begin{equation}\label{bouns}
\frac{1}{\varepsilon}\left[\mathbb{1}-\bm{P}_{n}\bm{\alpha}_{n}^{-1}\right]=\frac{1}{\varepsilon}\left[\mathbb{1}-\left(\mathbb{1}+\varepsilon\frac{\bm{C}_n+\bm{C}_n^T}{2}\bm{P}_{n}^{-1}+\varepsilon^2\bm{Q}_{n}\bm{P}_{n}^{-1}\right)^{-1}\right] \;,
\end{equation}
which, exploiting the Neumann series 
\begin{equation}
({\mathbb{1}}+\bm{\Lambda})^{-1}={\mathbb{1}}-\bm{\Lambda}+\bm{\Lambda}^2-\bm{\Lambda}^3+...
\end{equation}
with 
\begin{equation}
\displaystyle{\bm{\Lambda}=\varepsilon\frac{\bm{C}_n+\bm{C}_n^T}{2}\bm{P}_{n}^{-1}+\varepsilon^2\bm{Q}_{n}\bm{P}_{n}^{-1}}\;,
\end{equation}
gives
\begin{equation}\label{finbou}
\frac{\bm{D}_{n-1}-\bm{D}_{n}}{\varepsilon}\sim \bm{C}_n^{(s)}\bm{P}_{n}^{-1}\,\,\text{as}\,\,\varepsilon\rightarrow 0 \;,
\end{equation}
leading to 
\begin{equation}\label{ICIN}
\hat{\bm{D}}(t-s)\big{|}_{s=0}={\mathbb{1}},\quad \frac{\dd}{\dd s}\left[\hat{\bm{D}}(t-s)\right]\big{|}_{s=0}=\hat{\bm{C}}^{(s)}(t)\bm{P}(t)^{-1} \;.
\end{equation}

$\bullet$ For $\bm{D}^m$ we have
\begin{equation}
\bm{\alpha}_{n}=\smup{\bm{P}}_{n}+\varepsilon\frac{\smup{\bm{C}}_n+\smup{\bm{C}}_n^{\,\,T}}{2}+\varepsilon^2\smup{\bm{Q}}_{\,\,n}\rightarrow\smup{\bm{P}}_n\quad\text{for}\quad\varepsilon\rightarrow 0
\end{equation}
and, using the block matrix inversion formula
\begin{equation}
\begin{split}
&\bm{P}_n^{-1}=\begin{pmatrix}
\smup{\bm{P}}_n & \rharp{\bm{P}}_n\\ \rharp{\bm{P}}_n^{\,T} & \smdown{\bm{P}}_n
\end{pmatrix}^{-1}=\begin{pmatrix}
\smup{\bm{P}}_n^{-1}+\smup{\bm{P}}_n^{-1}\rharp{\bm{P}}_n \bm{Z}_n^{-1}\rharp{\bm{P}}_n^{\,T} \smup{\bm{P}}_n^{-1}&\,\,\,\, -\smup{\bm{P}}_n^{-1}\rharp{\bm{P}}_n\bm{Z}_n^{-1}\\ \\ -\bm{Z}_n^{-1}\rharp{\bm{P}}_n^{\,T}\smup{\bm{P}}_n^{-1} &\,\,\,\, \bm{Z}_n^{-1}
\end{pmatrix} \;,\\
&\bm{Z}=\smdown{\bm{P}}-\rharp{\bm{P}}^{\,T}\smup{\bm{P}}^{-1}\rharp{\bm{P}} \;,
\end{split}
\end{equation}
the following expression is easily obtained 
\begin{equation}\label{boDN}
\bm{D}_n=\begin{pmatrix}
\bm{\alpha}_{n} & \rharp{\bm{P}}_n\\ \mathbb{0} & \varepsilon\mathbb{1}_{d-l\times d-l}\end{pmatrix}\bm{P}_n^{-1}\rightarrow\begin{pmatrix}
\mathbb{1}_{l\times l} & \mathbb{0}_{l\times d-l}\\ \mathbb{0}_{d-l\times l} & \mathbb{0}_{d-l\times d-l}\end{pmatrix}\quad\text{for}\quad\varepsilon\rightarrow 0 \;.
\end{equation}
Moreover, the derivative at the boundary is discretised as
\begin{equation}
\frac{\bm{D}_{n-1}-\bm{D}_{n}}{\varepsilon}=\frac{\bm{D}_{n}\left(\bm{\alpha}_{n-1}\bm{P}_{n-1}^{-1}-{\mathbb{1}}\right)}{\varepsilon} \;,
\end{equation}
where the term
\begin{equation}
\varepsilon^{-1}\left(\bm{\alpha}_{n-1}\bm{P}_{n-1}^{-1}-{\mathbb{1}}\right)\sim\varepsilon^{-1}\left(\bm{P}_{n}-\bm{\beta}_{n}^T\bm{\alpha}_{n}\bm{\beta}_{n}\right)\bm{P}_{n-1}^{-1}
\end{equation}
is evaluated by computing
\begin{equation}
\begin{split}
\bm{\beta}_{n}^T\bm{\alpha}_{n}\bm{\beta}_{n}&=\overline{\bm{U}}_n^T\bm{\alpha}_{n}^{-1}\overline{\bm{U}}_n\;,\\
&=\left(\overline{\bm{P}}_n+\frac{\varepsilon}{2}\left[\overline{\bm{C}_n^T}-\overline{\bm{C}} _{n-1}\right]\right)^T\bm{\alpha}_{n}^{-1}\left(\overline{\bm{P}}_n+\frac{\varepsilon}{2}\left[\overline{\bm{C}_n^T}-\overline{\bm{C}} _{n-1}\right]\right)\;,\\
&\sim \bm{P}^{\bm{\alpha}}_n+\varepsilon\frac{\bm{T}_n+\bm{T}_n^T}{2} \;,
\end{split}
\end{equation}
with
\begin{equation}
\begin{split}
&\bm{P}^{\bm{\alpha}}_n=\begin{pmatrix}
\smup{\bm{P}}_n\bm{\alpha}_{n}^{-1}\smup{\bm{P}}_n &\,\,\, \smup{\bm{P}}_n\bm{\alpha}_{n}^{-1}\rharp{\bm{P}}_n\\
\\ \rharp{\bm{P}}_n^T\bm{\alpha}_{n}^{-1}\smup{\bm{P}}_n &\,\,\, \rharp{\bm{P}}_n^T\bm{\alpha}_{n}^{-1}\rharp{\bm{P}}_n\end{pmatrix}\;,\\
& \\
&\bm{T}_n=\begin{pmatrix}
\left(\smup{\bm{C}}_n-\smup{\bm{C}}_{n-1}^{\,T}\right)\bm{\alpha}_{n}^{-1}\smup{\bm{P}}_n &\,\,\,\,\, \left(\smup{\bm{C}}_n-\smup{\bm{C}}_{n-1}^{\,T}\right)\bm{\alpha}_{n}^{-1}\rharp{\bm{P}}_n\\ \\ \left(\lharp{\bm{C}}_n-\rharp{\bm{C}}_{n-1}^{\,T}\right)\bm{\alpha}_{n}^{-1}\smup{\bm{P}}_n &\,\,\,\,\, \left(\lharp{\bm{C}}_n-\rharp{\bm{C}}_{n-1}^{\,T}\right)\bm{\alpha}_{n}^{-1}\rharp{\bm{P}}_n\end{pmatrix} \;.
\end{split}
\end{equation}
To conclude, we have
\begin{equation}
\frac{\bm{D}_{n-1}-\bm{D}_{n}}{\varepsilon}\sim \bm{D}_{n}\frac{\left(\bm{P}_{n}-\bm{P}^{\bm{\alpha}}_{n}\right)}{\varepsilon}\bm{P}_{n-1}^{-1}-\bm{D}_{n}\frac{\bm{T}_n+\bm{T}_n^T}{2}\bm{P}_{n-1}^{-1} \;.
\end{equation}
From the definition $\bm{T}(t)=\lim\limits{_{\varepsilon\rightarrow 0}}{\bm{T}_n}$, and noting that 
\begin{equation}
\frac{\mathbb{1}-\smup{\bm{P}}_n\bm{\alpha}_{n}^{-1}}{\varepsilon}\rightarrow\frac{\smup{\bm{C}}+\smup{\bm{C}}^{\,\,T}}{2}\smup{\bm{P}}^{-1}(t)\quad\text{for}\quad\varepsilon\rightarrow 0 \;,
\end{equation}
we perform the necessary computations, leading to 
\begin{small}
\begin{equation}\label{ICIDN}
\begin{split}
&\bm{D}^m(t-s)\big{|}_{s=0}=\begin{pmatrix}
\mathbb{1}_{l\times l} & \mathbb{0}_{l\times d-l}\\ \mathbb{0}_{d-l\times l} & \mathbb{0}_{d-l\times d-l}\end{pmatrix}\;,\\
& \\
& \frac{\dd}{\dd s}\left[\bm{D}^m(t-s)\right]\big{|}_{s=0}=\begin{pmatrix}
\displaystyle{\frac{\smup{\bm{C}}+\smup{\bm{C}}^{\,\,T}}{2}\left[\smup{\bm{P}}^{-1}+\smup{\bm{P}}^{-1}\rharp{\bm{P}} \bm{Z}^{-1}\rharp{\bm{P}}^{\,\,T} \smup{\bm{P}}^{-1}\right]}&\,\displaystyle{-\frac{\smup{\bm{C}}+\smup{\bm{C}}^{\,\,T}}{2}\smup{\bm{P}}^{-1}\rharp{\bm{P}} \bm{Z}^{-1}}\\
\\-\bm{Z}^{-1}\rharp{\bm{P}}^{\,\,T}\smup{\bm{P}}^{-1}&\,\,\,\,\,\bm{Z}^{-1}\end{pmatrix}\Big{|}_{\bm{q}^m}(t)\\
& \\
&\hspace{38mm}-\begin{pmatrix}
\,\displaystyle{\overline{\frac{\bm{T}+\bm{T}^T}{2}\bm{P}^{-1}}}\,\,\\
\mathbb{0}
\end{pmatrix}\Big{|}_{\bm{q}^m}(t)\; .
\end{split}
\end{equation}
\end{small}

To summarize, setting $\tau=t-s$, then Eq.~(\ref{sys.2}) becomes Eq.~(\ref{papeq}) and the initial conditions Eq.~(\ref{ICID}), Eq.~(\ref{ICIN}) and Eq.~(\ref{ICIDN}) for Eq.~(\ref{sys.2}) are respectively transformed into Eq.~(\ref{IN1}), Eq.~(\ref{IN2}) and Eq.~(\ref{IN3}), which are the initial conditions for Eq.~(\ref{papeq}).

\subsubsection{Observation on different discretization choices}
We remark that the results Eq.~(\ref{pap}), Eq.~(\ref{pap2}) and Eq.~(\ref{pap2DN}) are specific to the (Stratonovich-type) discretisation prescription adopted in \citep{PAP1} for the cross terms 
\begin{equation}
2\dot{\bm{h}}(\tau)^T{{\bm{C}}}(\tau){\bm{h}}(\tau)\approx\frac{1}{\varepsilon}(\bm{h}_{j}-\bm{h}_{j-1})^T(\bm{C}_{j}\bm{h}_{j}+\bm{C}_{j-1}\bm{h}_{j-1})\;.
\end{equation}
In fact, there is in general a one-parameter family of discretisations
\begin{equation}\label{disc}
\frac{2}{\varepsilon}(\bm{h}_{j}-\bm{h}_{j-1})^T(\gamma \bm{C}_{j}\bm{h}_{j}+(1-\gamma)\bm{C}_{j-1}\bm{h}_{j-1})\;,\,\,\,\gamma\in[0,1]\;,
\end{equation}
which allows to rewrite Eq.~(\ref{secdiag}) as a function of $\gamma\in[0,1]$ as
\begin{equation}\label{startgam}
\begin{split}
\delta^2 S(\bm{h};\bar{\bm{q}})\approx\frac{1}{\varepsilon}\sum\limits_{j=1}^n\left[\Delta { \bm{h}}_j^T\bm{P}_j\Delta { \bm{h}}_j+2\varepsilon \Delta { \bm{h}}_j^T\left(\gamma{ \bm{C}_j} { \bm{h}}_j+(1-\gamma) \bm{C}_{j-1}{ \bm{h}}_{j-1}\right)+\varepsilon^2 { \bm{h}}_j^T \bm{Q}_j{ \bm{h}}_j \right] \;.\\
\end{split}
\end{equation}
From Eq.~(\ref{startgam}), we can repeat all the steps followed in the present and former sections to see that the difference equation Eq.~(\ref{diff}) now becomes
\begin{small}
\begin{equation}\label{diff.gam}
\begin{split}
 \bm{D}_{n-(k+1)}\bm{P}_{n-(k+1)}=& \bm{D}_{n-k}\bm{P}_{n-(k+1)}+ \bm{D}_{n-k}\bm{P}_{n-k}- \bm{D}_{n-(k-1)}\bm{P}_{n-k}+\varepsilon^2  \bm{D}_{n-k} \bm{Q}_{n-(k+1)}\\
&\;+2\varepsilon(2\gamma-1) \bm{D}_{n-k} \bm{C}_{n-(k+1)}\\
&\;-\varepsilon  \bm{D}_{n-k}\left[\gamma{ \bm{C}_{n-k}-(1-\gamma)\bm{C}_{n-(k+1)}^T}\right]\bm{P}_{n-k}^{-1} \bm{D}_{n-k}^{-1} \bm{D}_{n-(k-1)}\bm{P}_{n-k}\\
&\;-\varepsilon  \bm{D}_{n-(k-1)}\left[\gamma{\bm{C}_{n-k}^T-(1-\gamma) \bm{C}_{n-(k+1)}}\right]\\
&\;-\varepsilon^2  \bm{D}_{n-k}\left[\gamma{ \bm{C}_{n-k}-(1-\gamma)\bm{C}_{n-(k+1)}^T}\right]\bm{P}_{n-k}^{-1}\bm{D}_{n-k}^{-1} \bm{D}_{n-(k-1)}\\
&\;\;\;\,\left[\gamma{\bm{C}_{n-k}^T-(1-\gamma) \bm{C}_{n-(k+1)}}\right]\;.
\end{split}
\end{equation}
\end{small}%

If $\bm{C}$ is not symmetric, it is possible to derive a differential equation from Eq.~(\ref{diff.gam}) through Taylor expansion, giving a finite result, only for the mid-point rule, \ie for $\gamma=\frac{1}{2}$, which gives Eq.~(\ref{papeq}). This is easy to check by performing the calculation. 

On the other hand, when $\bm{C}$ is symmetric, in principle, the cross terms could be integrated away from the second variation since 
\begin{equation}
2\dot{\bm{h}}^T\bm{C}\bm{h}=\frac{\dd}{\dd\tau}\left(\bm{h}^T\bm{C}\bm{h}\right)-\bm{h}^T\dot{\bm{C}}\bm{h}\;,
\end{equation} 
and the values of $\mathcal{K}$, $\mathcal{K}_m$ and $\mathcal{N}$ depends on the particular choice of $\gamma\in[0,1]$, see \citep{LANG1, LANG2} for $\mathcal{K}$. However, the consistency of the formulation is recovered in the ratio, \eg $\mathcal{K}$ over $\mathcal{N}$ for $\rho(\bm{q},t|\bm{q}_0,t_0)$. Namely, if $\bm{C}^f$ and $\hat{\bm{C}}$ are symmetric, and if we adopt the same discretisation prescription for both $\mathcal{K}$ and $\mathcal{N}$, then there is a one parameter family of different equations
\begin{equation}\label{sys.new.gam}
\frac{\dd}{\dd\tau}\left[\dot{ \bm{D}}{{\bm{P}}}\right]+2\gamma {{ \bm{D}}}\dot{ \bm{C}}-{{ \bm{D}}}[{{ \bm{Q}}}-(1-2\gamma)^2{{ \bm{C}}}{{\bm{P}}}^{-1} \bm{C}]=
(1-2\gamma)\dot{ \bm{D}}{{ \bm{C}}}+(1-2\gamma){{ \bm{D}}}{{ \bm{C}}}{{\bm{P}}}^{-1} \bm{D}^{-1}\dot{ \bm{D}}\bm{P},
\end{equation}
with initial conditions
\begin{equation}
\begin{split}
&\bm{D}^f(t)=\mathbb{0},\quad{\dot{\bm{D}}^f}(t)=-\bm{P}(t)^{-1}\quad\text{and}\\
&{\hat{\bm{D}}}(t)=\mathbb{1},\quad{\dot{\hat{\bm{D}}}}(t)=-2\gamma  \hat{\bm{C}}(t) \bm{P}(t)^{-1}\;,
\end{split}
\end{equation}
providing the same normalized result for the transition probability density.

\subsection{The Jacobi equation}
\subsubsection{Background}
Despite the apparent simplicity of Eq.~(\ref{fin1}) and Eq.~(\ref{finn1}), the $\bm{D}$ matrices have to be found by solving second-order non-linear differential equations of the form of Eq.~(\ref{papeq}). However, we expect a connection with the (linear) Jacobi equation if we observe that the techniques adopted here for the evaluation of Gaussian path integrals fall in the class of Gelfand-Yaglom-type methods \citep{GELYAG}. Hence it turns out that there exists a relation between the matrix differential equation (\ref{papeq}) and the Jacobi equation for a vector field ${\bm{h}}\in\mathbb{R}^d$
\begin{equation}\label{jac}
\frac{\dd}{\dd \tau}\left[{\bm{P}}\dot{\bm{h}}+{\bm{C}}{\bm{h}}\right]-{\bm{C}}^T \dot{\bm{h}}-{\bm{Q}}{\bm{h}}=\bm{0} \;
\end{equation}
as observed for the first time in \citep{LUDT, LUD}. This is the last piece of the puzzle: the approach of Gelfand and Yaglom showed how to compute Gaussian path integrals in terms of the Jacobi equation in absence of cross terms, then Papadopoulos managed to derive a non-linear differential equation for stating the result in the general vector case with non-symmetric cross terms, and finally in \citep{LUDT, LUD} it is reported how to compute Gaussian path integrals in terms of the Jacobi equation in presence of cross terms. In fact, if $\bm{H}=\bm{L}^T$ is a matrix whose columns $\bm{h}$ are solutions of the Jacobi equation Eq.~(\ref{jac}), then the solutions of Eq.~(\ref{papeq}) and the ones of
\begin{equation}\label{jactr}
 \frac{\dd}{\dd \tau}\left[\dot{\bm{L}}{\bm{P}}+{\bm{L}}{\bm{C}}^T\right]-\dot{\bm{L}}{\bm{C}}-{\bm{L}}{\bm{Q}}={\mathbb{0}}
\end{equation}
are related by the non-linear transformation
\begin{equation}\label{tran}
\bm{L}^{-1}\dot{\bm{L}}=\bm{D}^{-1}\dot{\bm{D}}+{{\bm{C}}}^{(a)}\bm{P}^{-1} \;.
\end{equation}
This is true because Eq.~(\ref{papeq}) is associated with a matrix Riccati equation and as such is known to correspond, or to be locally equivalent, to a second-order linear ordinary equation, which is the Jacobi equation Eq.~(\ref{jac}) \citep{LUD}.

We remark that the mentioned background work has been performed exclusively for the case of Dirichlet-Dirichlet boundary conditions and we are not aware of any systematic extension of this general approach in the literature. The aim of this chapter is to describe how the same machinery can be suitably adapted to more general boundary conditions and derive analogous results differing only in the initial conditions for the system of Jacobi equations, as it is shown in the following.

\subsubsection{Initial conditions}
Imposing the condition $\det{(\bm{L})}=\det{(\bm{D})}$, which also equals $\det{(\bm{H})}$, we ensure the uniqueness of the change of variables and find the associated initial conditions for the Jacobi equations. These read 
\begin{equation}\label{par1}
{\bm{H}^f}(t)=\mathbb{0},\quad{\dot{\bm{H}}^f}(t)=-\bm{P}(t)^{-1}\quad \text{for}\quad\mathcal{K} \;,
\end{equation}
\begin{equation}\label{par2}
{\hat{\bm{H}}}(t)=\mathbb{1},\quad{\dot{\hat{\bm{H}}}}(t)=-\bm{P}(t)^{-1}{\hat{\bm{C}}}(t)\quad\text{for}\quad\mathcal{N} \;,
\end{equation}
\begin{align}\label{par3}
{\bm{H}^m}(t)=\begin{pmatrix}
\mathbb{1}_{l\times l} & \mathbb{0}_{l\times d-l}\\ \mathbb{0}_{d-l\times l} & \mathbb{0}_{d-l\times d-l}\end{pmatrix}\;,\qquad\qquad\qquad\qquad\nonumber\\
\\
{\dot{\bm{H}}^m}(t)=\begin{pmatrix}
-\smup{\bm{P}}^{-1}\smup{\bm{C}}-\smup{\bm{P}}^{-1}\rharp{\bm{P}}\bm{Z}^{-1}{\bm{W}^*}^T&\,\,\,\,\,\,\,\,\,\,\smup{\bm{P}}^{-1}\rharp{\bm{P}}\bm{Z}^{-1}\\
\\
\bm{Z}^{-1}{\bm{W}^*}^T&\,\,\,\,\,\,\,\,\,\,-\bm{Z}^{-1}
\end{pmatrix}\Big{|}_{\bm{q}^m}(t)\quad\text{for}\quad\mathcal{K}_m \;,\nonumber\\
\nonumber\\
\bm{Z}=\smdown{\bm{P}}-\rharp{\bm{P}}^{\,T}\smup{\bm{P}}^{-1}\rharp{\bm{P}},\quad\bm{W}^*=\smup{\bm{C}}^T\smup{\bm{P}}^{-1}\rharp{\bm{P}}-\lharp{\bm{C}}^T \;.\,\,\,\,\,\quad\qquad\qquad\qquad\qquad\qquad\nonumber
\end{align}
In the following we present in detail how the initial conditions for $\bm{D}$ translate into initial conditions for $\bm{L}$, in the context of the backward integration procedure.

$\bullet$ First, let us consider the case for $\mathcal{K}$. Eq.~(\ref{tran}) gives us a mapping between $\bm{L}$ and $\bm{D}$, as far as they are invertible. If we assume $\bm{L}$ and $\bm{D}$ invertible for all $\tau\neq t$ (no conjugate points) the transformation is valid except for $\tau=t$, where $\bm{D}(t)=\mathbb{0}$ because of the initial conditions in the backward direction Eq.~(\ref{IN1}). To derive the initial conditions for $\bm{L}$ in $\tau=t$ from the initial conditions for $\bm{D}$ in $\tau=t$, we consider the following reasoning. For $\tau\neq t$ we can write
\begin{equation}
\dot{\bm{L}}=\bm{L}\bm{D}^{-1}\dot{\bm{D}}+\bm{L}\bm{C}^{(a)}{\bm{P}}^{-1}\;,
\end{equation}
and we know that $\bm{D}\rightarrow \mathbb{0}$, $\dot{\bm{D}}\rightarrow -{\bm{P}}(t)^{-1}$ as $\tau\rightarrow t$, for continuity of $\bm{D}$ and $\dot{\bm{D}}$. As a consequence, in order to obtain a finite initial condition for $\dot{\bm{L}}$, we necessarily want $\bm{L}\bm{D}^{-1}\rightarrow {\bm{X}}$ as $\tau\rightarrow t$, where ${\bm{X}}$ is a finite valued matrix. This implies that $\bm{L}(t)=\lim_{\tau\rightarrow t}{\bm{L}}=\mathbb{0}$, for continuity of $\bm{L}$. Furthermore, having $\bm{L}$ invertible for $\tau\neq t$ implies $\dot{\bm{L}}(t)$ is not singular, meaning that the matrix ${\bm{X}}$ is not singular as well, since $\bm{L}(t)=\mathbb{0}$. To summarize, we have that Eq.~(\ref{jactr}) is subject to the initial conditions
\begin{equation}\label{parp1}
\bm{L}^f(t)=\mathbb{0},\quad\dot{\bm{L}}^f(t)=-{\bm{X}^f} {\bm{P}}(t)^{-1}\;.
\end{equation}

$\bullet$ In the same way, we now consider the case for $\mathcal{N}$. Assuming $\bm{L}$ and $\bm{D}$ to be non-singular also for $\tau=t$ (note $\bm{D}(t)=\bm{\mathbb{1}}$, $\dot{\hat{\bm{D}}}(t)=-{\hat{\bm{C}}}^{(s)}(t)\bm{P}(t)^{-1}$ from Eq.~(\ref{IN2})), we have
\begin{align}
\bm{L}(t)^{-1}\dot{\bm{L}}(t) &= -\bm{C}^{(s)}(t){\bm{P}}(t)^{-1}+\bm{C}^{(a)}(t){\bm{P}}(t)^{-1} \nonumber\\
&= -\bm{C}^T(t){\bm{P}}(t)^{-1} \;. \label{ttr}
\end{align}
In addition, as we want $\bm{L}$ to be invertible in $\tau=t$, then ${\bm{Y}}=\bm{L}(t)$ must be a non-singular matrix. To summarize, we have that Eq.~(\ref{jactr}) is subject to the initial conditions
\begin{equation}\label{parp2}
\hat{\bm{L}}(t)=\hat{\bm{Y}},\quad\dot{\hat{\bm{L}}}(t)=-\hat{\bm{Y}}\hat{\bm{C}}^T(t){\bm{P}}(t)^{-1}\;.
\end{equation}

$\bullet$ Here we explain in detail how the initial conditions for $\bm{D}$ translate into initial conditions for $\bm{L}$ for the general problem of marginal distributions and therefore we consider the case for $\mathcal{K}_m$. Eq.~(\ref{tran}) gives us a mapping between $\bm{L}$ and $\bm{D}$, as far as they are invertible. If we assume $\bm{L}$ and $\bm{D}$ invertible for all $\tau\neq t$ (no conjugate points) the transformation is valid except for $\tau=t$, where $\bm{D}$ is singular because of the initial conditions Eq.~(\ref{IN3}). Therefore, the first step is to find the Taylor expansion with singular term of $\bm{D}^{-1}(\tau)$ around $\tau=t$:
\begin{equation}
\bm{D}^{-1}(\tau)=\bm{A}\frac{1}{\tau-t}+\bm{B}+O(\tau-t)\;.
\end{equation}
Since the condition $\bm{D}(\tau)\bm{D}^{-1}(\tau)=\bm{\mathbb{1}}$ must be satisfied, with 
\begin{equation}
\bm{D}(\tau)=\bm{D}(t)+\dot{\bm{D}}(t)(\tau-t)+O(\tau-t)^2\;,
\end{equation}
then we necessarily have 
\begin{equation}
\begin{cases}
\bm{D}(t)\bm{A}=\bm{\mathbb{0}}\;,\\
\bm{D}(t)\bm{B}+\dot{\bm{D}}(t)\bm{A}=\bm{\mathbb{1}}\;.
\end{cases}
\end{equation}
In the light of the initial conditions for $\bm{D}$ Eq.~(\ref{IN3}), the computation leads to
\begin{equation}\label{dm1}
\bm{A}=\begin{pmatrix}
\smup{\bm{A}} & \rharp{\bm{A}}\\ \lharp{\bm{A}} & \smdown{\bm{A}}
\end{pmatrix}=\begin{pmatrix}
\mathbb{0} & \mathbb{0}\\ \mathbb{0} & \smldown{\dot{\bm{D}}}\,\,(t)^{-1}=-\bm{Z}(t)
\end{pmatrix},\,\,\,\bm{B}=\begin{pmatrix}
\smup{\bm{B}} & \rharp{\bm{B}}\\ \lharp{\bm{B}} & \smdown{\bm{B}}
\end{pmatrix}=\begin{pmatrix}
\mathbb{1} & *\\ * & *
\end{pmatrix}\;,
\end{equation}
where the entries denoted by ``$*$" are unnecessary for the derivation of the results.

Furthermore, Eq.~(\ref{tran}) allows us to write
$\dot{\bm{L}}=\bm{L}\bm{D}^{-1}\dot{\bm{D}}+\bm{L}\bm{C}^{(a)}{\bm{P}}^{-1}$ for $\tau\neq t$. As a consequence, in order to obtain a finite initial condition for $\dot{\bm{L}}$, we necessarily want $\bm{L}\bm{D}^{-1}\rightarrow {\bm{X}}$ as $\tau\rightarrow t$, where ${\bm{X}}$ is a finite valued matrix. Since around $\tau=t$ we can also write $\bm{L}(\tau)=\bm{L}(t)+\dot{\bm{L}}(t)(\tau-t)+O(\tau-t)^2$, then we necessarily want $\bm{L}(t)=\bm{Y}$, with 
\begin{equation}
\begin{cases}
\bm{Y}\bm{A}=\bm{\mathbb{0}}\;,\\
\bm{Y}\bm{B}+\dot{\bm{L}}(t)\bm{A}=\bm{X}\;,
\end{cases}
\end{equation}
leading to $\rharp{\bm{Y}}=\mathbb{0}$, $\smdown{\bm{Y}}=\mathbb{0}$, $\smup{\bm{X}}=\smup{\bm{Y}}$ and $\lharp{\bm{X}}=\lharp{\bm{Y}}$, so that
\begin{equation}\label{inter}
\lim\limits{_{\tau\rightarrow t}}\,\,{\bm{L}\bm{D}^{-1}}=\begin{pmatrix}
\smup{\bm{Y}} & \rharp{\bm{X}}\\ \lharp{\bm{Y}} & \smdown{\bm{X}}
\end{pmatrix}
\end{equation}
and we have that Eq.~(\ref{jactr}) is subject to the initial conditions
\begin{equation}\label{parp3}
\bm{L}^{m}(t)=\begin{pmatrix}
\smup{\bm{Y}}^{\,\,m} & \bm{\mathbb{0}}\\ \lharp{\bm{Y}}^{\,\,m} & \bm{\mathbb{0}}
\end{pmatrix},\quad\dot{\bm{L}}^{m}(t)=\begin{pmatrix}
\smup{\bm{Y}}^{\,\,m} & \rharp{\bm{X}}^{\,\,m}\\ \lharp{\bm{Y}}^{\,\,m} & \smdown{\bm{X}}^{\,\,m}\end{pmatrix}\dot{\bm{D}}^{m}(t)+\begin{pmatrix}
\smup{\bm{Y}}^{\,\,m} & \bm{\mathbb{0}}\\ \lharp{\bm{Y}}^{\,\,m} & \bm{\mathbb{0}}
\end{pmatrix}{\bm{C}^{(a)}}^{\,m}{\bm{P}}^{-1}(t)\;.
\end{equation}

At this point, we use the observation that $\det{\bm{L}}=c\det{{\bm{D}}}$ for all $\tau$,
where $c$ is a constant \citep{LUD}. In order to make the transformation unique (up to invertible matrices sharing the same determinant), we impose $c=1$ for both $\bm{L}^f$, $\bm{D}^f$ and $\hat{\bm{L}}$, $\hat{\bm{D}}$, which allows us to fix the matrices $\bm{X}^f$ and $\hat{\bm{Y}}$. Namely,
\begin{align}
\lim\limits_{\tau\rightarrow t}{\det\left[{\bm{L}^f(\bm{D}^f)^{-1}(\tau)}\right]} &= \det({\bm{X}^f}) \;,\\
\lim\limits_{\tau\rightarrow t}{\det\left[{\hat{\bm{L}}\hat{\bm{D}}^{-1}(\tau)}\right]} &= \det(\hat{\bm{Y}})
\end{align}
so that we can set 
\begin{equation}
{\bm{X}}^f=\hat{\bm{Y}}={\mathbb{1}}\;.
\end{equation}
In this way, recalling that $\bm{H}=\bm{L}^T$, then the initial conditions Eq.~(\ref{parp1}) and Eq.~(\ref{parp2}) for  Eq.~(\ref{jactr}) become Eq.~(\ref{par1}) and Eq.~(\ref{par2}) for the Jacobi equations.

For the marginal problem, we use the same observation and we impose $c=1$ by choosing 
\begin{equation}
\smup{\bm{Y}}=\mathbb{1},\quad\smdown{\bm{X}}=\mathbb{1},\quad\lharp{\bm{Y}}=\mathbb{0},\quad\displaystyle{\rharp{\bm{X}}=\left[\lharp{\bm{C}}^{\,\,T}+\frac{\smup{\bm{C}}-\smup{\bm{C}}^{\,\,T}}{2}\smup{\bm{P}}^{-1}\rharp{\bm{P}}\right](t)}\;,
\end{equation}
leading to (according to Eq.~(\ref{inter}))
\begin{equation}
\lim\limits_{\tau\rightarrow t}{\det\left[{\bm{L}^m(\bm{D}^m)^{-1}(\tau)}\right]}= \det\begin{pmatrix}
\smup{\bm{Y}} & \rharp{\bm{X}}\\ \lharp{\bm{Y}} & \smdown{\bm{X}}
\end{pmatrix}=1 \;.
\end{equation}
The latter choice will be the key element for deriving the initial conditions appearing in Eq.~(\ref{fin33}).
We obtain $\bm{L}^m(t)=\bigl( \begin{smallmatrix} \mathbb{1} & \mathbb{0}\\ \mathbb{0} & \mathbb{0}\end{smallmatrix}\bigr)$ and, after some algebra, from Eq.~(\ref{parp3}) and Eq.~(\ref{IN3}), we arrive at
\begin{equation}
\begin{split}
\dot{\bm{L}}^m(t)&=\begin{pmatrix}
\mathbb{1} &\,\,\,\,\,\,\, \displaystyle{\lharp{\bm{C}}^{\,\,T}+\frac{\smup{\bm{C}}-\smup{\bm{C}}^{\,\,T}}{2}\smup{\bm{P}}^{-1}\rharp{\bm{P}}}\\ 
\\
\mathbb{0} &\,\,\,\,\,\,\, \mathbb{1}
\end{pmatrix}\Big{|}_{\bm{q}^m}\dot{\bm{D}}^m(t)+\begin{pmatrix}
\,\overline{\bm{C}^{(a)}\bm{P}^{-1}}\,\,\\
\mathbb{0}
\end{pmatrix}\Big{|}_{\bm{q}^m}(t)\;,\\
\\
&=\begin{pmatrix}
-\smup{\bm{C}}^T\smup{\bm{P}}^{-1}-\bm{W}^*\bm{Z}^{-1}\rharp{\bm{P}}^T \smup{\bm{P}}^{-1}&\,\,\,\,\,\,\,\,\,\,\bm{W}^*\bm{Z}^{-1}\\
\\
\bm{Z}^{-1}\rharp{\bm{P}}^T \smup{\bm{P}}^{-1}&\,\,\,\,\,\,\,\,\,\,-\bm{Z}^{-1}
\end{pmatrix}\Big{|}_{\bm{q}^m}(t) \;,\\
& \\
&\bm{Z}=\smdown{\bm{P}}-\rharp{\bm{P}}^{\,T}\smup{\bm{P}}^{-1}\rharp{\bm{P}},\quad\bm{W}^*=\smup{\bm{C}}^T\smup{\bm{P}}^{-1}\rharp{\bm{P}}-\lharp{\bm{C}}^T \;.
\end{split}
\end{equation}
In view of the latter computation, and $\bm{H}=\bm{L}^T$, the initial conditions Eq.~(\ref{parp3}) for  Eq.~(\ref{jactr}) become Eq.~(\ref{par3}) for the Jacobi equations.

\subsubsection{Background on Hamiltonian formulation}
The last step is reducing the Jacobi (linear second-order) equations to a first-order linear system. Since the Jacobi equations are the Euler-Lagrange equations for the second variation of the action, we can easily provide the Hamiltonian formulation given in Eq.~(\ref{fin3}) and Eq.~(\ref{fin33}). This is a standard procedure which is of no real advantage in the case of strictly convex problems, except to gain elegance of representation and to improve computational efficiency. As far as this work is concerned, we call convex problems those that are represented by a Lagrangian that is a quadratic form in certain variables. For example, the Lagrangian involved in the path integral representation associated with solutions of the Langevin equation (see next section), or the linearly elastic potential energy for a fluctuating rod (see next chapter) are both quadratic forms in their respective moments. In the first case the leading matrix of the quadratic form is called diffusion matrix, whereas in the second case it takes the name of compliance matrix. The problem is strictly convex if the leading matrix is positive definite. The interesting case where the latter matrix is positive semi-definite has been considered in \citep{LUDT, LUD} for enforcing the constraint of inextensibility and unshearability of a fluctuating elastic rod. The idea is to approximate the positive semi-definite matrix by the limit of a sequence of positive definite matrices, for then compute the second variation and observe that the limit of the associated Jacobi fields is smooth in the Hamiltonian representation (the same is not true in the Lagrangian formalism). As a side comment, in \citep{ISO, LUDT, LUD} the authors further observe that the resulting Jacobi fields are exactly those entering the Bolza conjugate point theory  of stability \citep{BOLZA} for constrained calculus of variation. In the present work we show that the limiting procedure is not affected by non-Dirichlet boundary conditions and we provide an analogous strategy in the context of Fokker-Planck dynamics within the next section.

Having said that, we move on with the actual computation that leads to the Hamiltonian form of the Jacobi equations. Namely, from Eq.~(\ref{secvv}), we consider the Euler-Lagrange equations associated with the Lagrangian
\begin{equation}\label{seclan}
\frac{1}{2}\left[ \dot{\bm{h}}^T{{\bm{P}}}\dot{\bm{h}}+2\dot{\bm{h}}^T{{\bm{C}}}{\bm{h}}+{\bm{h}}^T{{\bm{Q}}}{\bm{h}}\right]\;,
\end{equation}
leading to the Jacobi equations Eq.~(\ref{jac}).
Now we switch to the Hamiltonian formulation of the problem, deriving the Hamilton equations from the Legendre transform in $\dot{\bm{h}}$ of Eq.~(\ref{seclan}). The Hamiltonian related to Eq.~(\ref{seclan}) is
\begin{equation}
H(\bm{h},\bm{m})=\bm{m}^T \dot{\bm{h}}-\frac{1}{2}\left[\dot{\bm{h}}^T \bm{P} \dot{\bm{h}}+2 \dot{\bm{h}}^T \bm{C} \bm{h}+\bm{h}^T \bm{Q}\bm{h}\right]\;,\,\,\,\text{where}
\end{equation}
\begin{itemize}
\item $\bm{m}=\frac{1}{2}\frac{\partial}{\partial \dot{\bm{h}}}\left[ \dot{\bm{h}}^T{{\bm{P}}}\dot{\bm{h}}+2\dot{\bm{h}}^T{{\bm{C}}}{\bm{h}}+{\bm{h}}^T{{\bm{Q}}}{\bm{h}}\right]=\bm{P}\dot{\bm{h}}+\bm{C}\bm{h}$ (conjugate variable)\;,
\item $\dot{\bm{h}}=-\bm{P}^{-1}\bm{C}\bm{h}+\bm{P}^{-1}\bm{m}\;.$
\end{itemize}
Hence 
\begin{equation}
\begin{split}
H(\bm{h},\bm{m})&=\bm{m}^T\bm{P}^{-1}\bm{m}-\bm{m}^T\bm{P}^{-1}\bm{C}\bm{h}-\frac{1}{2}\big[\bm{m}^T\bm{P}^{-1}\bm{m}-2\bm{m}^T\bm{P}^{-1}\bm{C}\bm{h}+\bm{h}^T\bm{C}^T\bm{P}^{-1}\bm{C}\bm{h}\\
&\,\,\,\,\,\,+2\bm{m}^T\bm{P}^{-1}\bm{C}\bm{h}-2\bm{h}^T\bm{C}^T\bm{P}^{-1}\bm{C}\bm{h}+\bm{h}^T\bm{Q}\bm{h}\big]\;,\\
&=\frac{1}{2}\left[\bm{m}^T\bm{P}^{-1}\bm{m}-2\bm{m}^T\bm{P}^{-1}\bm{C}\bm{h}+\bm{h}^T(\bm{C}^T\bm{P}^{-1}\bm{C}-\bm{Q})\bm{h}\right]\;,
\end{split}
\end{equation}
and the Hamilton equations read as
\begin{itemize}
\item $\dot{\bm{h}}=\frac{\partial}{\partial \bm{m}}H(\bm{h},\bm{m})=-\bm{P}^{-1}\bm{C}\bm{h}+\bm{P}^{-1}\bm{m}\;,$
\item $\dot{\bm{m}}=-\frac{\partial}{\partial \bm{h}}H(\bm{h},\bm{m})=(\bm{Q}-\bm{C}^T\bm{P}^{-1}\bm{C})\bm{h}+\bm{C}^T\bm{P}^{-1}\bm{m}\;.$
\end{itemize}
In matrix form the linear system is
\begin{equation}
     \begin{pmatrix}
      \dot{\bm{h}}  \\
     \dot{\bm{m}}
     \end{pmatrix}
     =\begin{pmatrix}
       \mathbf{0} & \bm{  \mathbb{1}}_{d\times d}  \\
      -\bm{  \mathbb{1}}_{d\times d} & \mathbf{0}
     \end{pmatrix}
     \begin{pmatrix}
      \frac{\partial}{\partial \bm{h}}H  \\
    \frac{\partial}{\partial \bm{m}}H
     \end{pmatrix}=
     \bm{J}\bm{E}\begin{pmatrix}
     \bm{h}  \\
     \bm{m}
     \end{pmatrix}
     =\begin{pmatrix}
       -\bm{P}^{-1}\bm{C} & \bm{P}^{-1}  \\
       \\
      \bm{Q}-\bm{C}^T\bm{P}^{-1}\bm{C} & \bm{C}^T\bm{P}^{-1}
     \end{pmatrix}
     \begin{pmatrix}
     \bm{h}  \\
     \bm{m}
     \end{pmatrix}\;,
\end{equation}
with $\bm{E}$ given in Eq.~(\ref{driv}).

Finally, if 
\begin{equation}\label{conjmom}
\bm{M}=\bm{P}\dot{\bm{H}}+\bm{C}\bm{H}
\end{equation}
is the (matrix) conjugate variable under the Legendre transform of the second variation with respect to $\dot{\bm{H}}$, we can express Eq.~(\ref{fin2}) in terms of the solutions $\bm{H}^f$ and $\hat{\bm{H}}$ of the Jacobi equations in Hamiltonian form subject to the transformed final initial conditions, \ie Eq.~(\ref{fin3}). In fact Eq.~(\ref{par1}) and Eq.~(\ref{par2}) give 
\begin{equation}
\begin{split}
&\bm{M}^f(t)=\bm{P}\dot{\bm{H}}^f(t)+\bm{C}^f\bm{H}^f(t)=-\bm{\mathbb{1}}\;,\\
\\
&\hat{\bm{M}}(t)=\bm{P}\dot{\hat{\bm{H}}}(t)+\hat{\bm{C}}\hat{\bm{H}}(t)=\bm{\mathbb{0}}\;.
\end{split}
\end{equation}
In the same way we can express Eq.~(\ref{fin22}) in terms of the solutions $\bm{H}^m$ and $\hat{\bm{H}}$ of the Jacobi equations in Hamiltonian form subject to the transformed final initial conditions, \ie Eq.~(\ref{fin33}). In particular, Eq.~(\ref{par3}) gives
\begin{equation}
\begin{split}
\bm{M}^m(t)&=\bm{P}\dot{\bm{H}}^m(t)+\bm{C}^m\bm{H}^m(t) \;,\\
& \\
&=\begin{pmatrix}
-\smup{\bm{C}}&\,\,\,\,\,\,\,\,\,\,\bm{\mathbb{0}}_{l\times d-l}\\
-\lharp{\bm{C}}&\,\,\,\,\,\,\,\,\,\,-\bm{\mathbb{1}}_{d-l\times d-l}
\end{pmatrix}\Big{|}_{\bm{q}^m}(t)+\begin{pmatrix}
\smup{\bm{C}}&\,\,\,\,\,\,\,\,\,\,\bm{\mathbb{0}}_{l\times d-l}\\
\lharp{\bm{C}}&\,\,\,\,\,\,\,\,\,\,\bm{\mathbb{0}}_{d-l\times d-l}
\end{pmatrix}\Big{|}_{\bm{q}^m}(t)\;,\\
& \\
&=\begin{pmatrix}
\mathbb{0}_{l\times l}&\mathbb{0}_{l\times d-l}\\
\mathbb{0}_{d-l\times l}&-\mathbb{1}_{d-l\times d-l}
\end{pmatrix} \;.
\end{split}
\end{equation}

In conclusion, we have shown how to compute the path integrals appearing in Eq.~(\ref{pathint}) and Eq.~(\ref{pathint2}) in the Laplace approximation, from the solutions of the Euler-Lagrange equations and of the systems of linear differential equations Eq.~(\ref{fin3}) and Eq.~(\ref{fin33}).

\section{The Neumann-Neumann case}
In this section we provide the calculation of Gaussian path integrals with Neumann-Neumann boundary conditions on paths. Although this investigation will not cover any physical application in the course of the paper, we consider it an important mathematical extension of the previously presented cases, completing the picture with an exhaustive range of boundary conditions. In particular, the final result shows a substantial difference with the other cases. In fact, the determinant is no longer that of certain Jacobi fields with prescribed initial conditions, but the determinant of their conjugate moments, introducing the derivative of the fields within the result. In part, this section can be regarded by the reader as a useful summary of the various steps involved in the method used above.

Let us start again with the block diagonalization of the discretized second variation computed on the Neumann-Neumann minimizer denoted by $\bar{\bm{q}}$ as in Eq.~(\ref{secdiag}).
\begin{small}
\begin{equation}\label{NNsec}
\begin{split}
\delta^2 S(\bm{h};\bar{\bm{q}})&\approx\varepsilon\sum\limits_{j=1}^n\left[\frac{\Delta \bm{h}_j}{\varepsilon}^T\bm{P}_j\frac{\Delta \bm{h}_j}{\varepsilon}+2\frac{\Delta \bm{h}_j}{\varepsilon}^T\frac{{\bm{C}_j} \bm{h}_j+\bm{C}_{j-1}\bm{h}_{j-1}}{2}+ \bm{h}_j^T\bm{Q}_j\bm{h}_j \right] \;,\\
&=\frac{1}{\varepsilon}\sum\limits_{j=1}^n\left[{\Delta \bm{h}_j}^T\bm{P}_j{\Delta \bm{h}_j}+\varepsilon{\Delta \bm{h}_j}^T\left({\bm{C}_j} \bm{h}_j+\bm{C}_{j-1}\bm{h}_{j-1}\right)+\varepsilon^2\bm{h}_j^T\bm{Q}_j\bm{h}_j \right] \;,\\
&=\frac{1}{\varepsilon}\sum\limits_{j=1}^n\big{[}\bm{h}_j^T\left(\bm{P}_j+\varepsilon {\bm{C}_j}+\varepsilon^2\bm{Q}_j\right)\bm{h}_j-\bm{h}_{j}^T\left(\bm{P}_j-\varepsilon \bm{C}_{j-1}\right)\bm{h}_{j-1}-\bm{h}_{j-1}^T\left(\bm{P}_j+\varepsilon \bm{C}_{j}\right)\bm{h}_{j}\\
&\quad +\bm{h}_{j-1}^T\left(\bm{P}_j-\varepsilon \bm{C}_{j-1}\right)\bm{h}_{j-1}\big{]} \;,\\
&=\frac{1}{\varepsilon}\left[ \bm{h}_{0}^T\left(\bm{P}_1-\varepsilon \bm{C}_{0}\right)\bm{h}_{0}\right] \;\\
&\quad +\frac{1}{\varepsilon}\sum\limits_{j=1}^{n-1}\left[\bm{h}_j^T\left(\bm{P}_j+\bm{P}_{j+1}+\varepsilon^2\bm{Q}_j\right)\bm{h}_j-\bm{h}_j^T\left(\bm{P}_j-\varepsilon \bm{C}_{j-1}\right)\bm{h}_{j-1}-\bm{h}_{j-1}^T\left(\bm{P}_j+\varepsilon \bm{C}_{j}\right)\bm{h}_j\right]\\
&\quad +\frac{1}{\varepsilon}\left[\bm{h}_n^T\left(\bm{P}_n+\varepsilon \bm{C}_n+\varepsilon^2\bm{Q}_n\right)\bm{h}_n-\bm{h}_{n}^T\left(\bm{P}_n-\varepsilon \bm{C}_{n-1}\right)\bm{h}_{n-1}-\bm{h}_{n-1}^T\left(\bm{P}_n+\varepsilon \bm{C}_{n}\right)\bm{h}_{n}\right],
\end{split}
\end{equation}
\end{small}%
Let $\bm{h}(\tau)$ be a perturbation around the Neumann-Neumann minimum $\bar{\bm{q}}$, then no constraints are given for $\bm{h}(t_0)=\bm{h}_0$ and $\bm{h}(t)=\bm{h}_n$. Introducing now the matrices 
\begin{equation}
\bm{U}_j=\bm{P}_j+\frac{\varepsilon}{2}[\bm{C}_j^T-\bm{C}_{j-1}]\quad\text{for}\quad j=1,...,n\;,
\end{equation}
we have that 
\begin{itemize}
\item $\bm{h}_j^T { \bm{U}}_j \bm{h}_{j-1}=\bm{h}_j^T {\bm{P}}_j \bm{h}_{j-1}+\frac{\varepsilon}{2}\bm{h}_j^T {\bm{C}}_j^T \bm{h}_{j-1}-\frac{\varepsilon}{2}\bm{h}_j^T {\bm{C}}_{j-1} \bm{h}_{j-1}$,
\item $\bm{h}_{j-1}^T { \bm{U}}_j^T \bm{h}_{j}=\bm{h}_{j-1}^T {\bm{P}}_j \bm{h}_{j}+\frac{\varepsilon}{2}\bm{h}_{j-1}^T {\bm{C}}_j \bm{h}_{j}-\frac{\varepsilon}{2}\bm{h}_{j-1}^T {\bm{C}}_{j-1}^T \bm{h}_{j}$,
\end{itemize}
\begin{equation}
\Rightarrow\,\,-\bm{h}_j^T { \bm{U}}_j \bm{h}_{j-1}-\bm{h}_{j-1}^T { \bm{U}}_j^T \bm{h}_{j}=-\bm{h}_j^T( {\bm{P}}_j-\varepsilon {\bm{C}}_{j-1})\bm{h}_{j-1}-\bm{h}_{j-1}^T( {\bm{P}}_j+\varepsilon {\bm{C}}_{j}) \bm{h}_{j}\;.
\end{equation}
Therefore Eq.~(\ref{NNsec}) can be written as
\begin{small}
\begin{equation}\label{NNsec2}
\begin{split}
\delta^2 S(\bm{h};\bar{\bm{q}})&\approx\frac{1}{\varepsilon}\left[ \bm{h}_{0}^T\left(\bm{P}_1-\varepsilon \bm{C}_{0}\right)\bm{h}_{0}\right] \;\\
&\quad +\frac{1}{\varepsilon}\sum\limits_{j=1}^{n-1}\left[ \bm{h}_j^T\left(\bm{P}_j+\bm{P}_{j+1}+\varepsilon^2\bm{Q}_j\right) \bm{h}_j- \bm{h}_j^T \bm{U}_j \bm{h}_{j-1}-\bm{h}_{j-1}^T \bm{U}_j^T \bm{h}_j\right]\\
&\quad +\frac{1}{\varepsilon}\left[\bm{h}_n^T\left(\bm{P}_n+\varepsilon \bm{C}_n+\varepsilon^2\bm{Q}_n\right)\bm{h}_n-\bm{h}_{n}^T\bm{U}_n\bm{h}_{n-1}-\bm{h}_{n-1}^T\bm{U}_n^T\bm{h}_{n}\right] \;.
\end{split}
\end{equation}
\end{small}%
At this point, we perform a change of variables. We define the transformation with unit Jacobian 
\begin{equation}
\bm{\phi}_0=\bm{h}_0\;,\quad\bm{\phi}_j=\bm{h}_j-\bm{\beta}_j \bm{h}_{j-1}\quad\text{for}\quad  j=1,...,n\;,
\end{equation}
where the matrices $\bm{\beta}_j$ are given by the following recursive construction for the symmetric matrices $\bm{\alpha}_{j}$:
\begin{equation}
\begin{split}
&\bm{\alpha}_{n}= \bm{P}_{n}+\varepsilon\frac{\bm{C}_n+\bm{C}_n^T}{2}+\varepsilon^2\bm{Q}_{n}\;,\\
&\bm{\alpha}_{j} = \bm{P}_{j}+\bm{P}_{j+1}+\varepsilon^2 \bm{Q}_{j}-\bm{\beta}_{j+1}^T\bm{\alpha}_{j+1}\bm{\beta}_{j+1} \;\qquad\text{for }\; j=n-1,...,1 \;, \\
&\bm{\alpha}_{0}= \bm{P}_{1}-\varepsilon\frac{\bm{C}_0+\bm{C}_0^T}{2}-\bm{\beta}_{1}^T\bm{\alpha}_{1}\bm{\beta}_{1} \;, \\
&\bm{U}_j=\bm{\alpha}_j\bm{\beta}_j \;\hspace{58mm}\text{for }\; j=1,...,n \;.
\end{split}
\end{equation}%
These expressions are motivated by the fact that they allow to express the discretized second variation Eq.~(\ref{NNsec2}) as a sum of quadratic forms. In fact, 
\begin{equation}
\bm{\phi}_0^T\bm{\alpha}_0\bm{\phi}_0=\bm{h}_0^T\bm{\alpha}_0\bm{h}_0,\,\,\bm{\phi}_j^T\bm{\alpha}_j\bm{\phi}_j=\bm{h}_j^T\bm{\alpha}_j \bm{h}_j-\bm{h}_j^T \bm{U}_j \bm{h}_{j-1}-\bm{h}_{j-1}^T\bm{U}_j^T \bm{h}_j+\bm{h}_{j-1}^T\bm{\beta}_j^T\bm{\alpha}_j\bm{\beta}_j \bm{h}_{j-1},
\end{equation}
and we have
\begin{equation}
\delta^2 S(\bm{h};\bar{\bm{q}})\approx\frac{1}{\varepsilon}\sum\limits_{j=0}^{n}\bm{\phi}_j^T\bm{\alpha}_j\bm{\phi}_j \;.
\end{equation}

With the aim of computing 
\begin{equation}
\int{e^{-\frac{1}{2}\delta^2 S(\bm{h};\bar{\bm{q}})}\,\mathcal{D}\bm{h}} = \lim_{n\rightarrow\infty} \bar{I}_n \;,
\end{equation}
\begin{equation}
\bar{I}_n =\displaystyle{\int{e^{-\frac{\varepsilon}{2} \sum\limits_{j=0}^{n}\delta^2 S(\bm{h};\bar{\bm{q}})_j}}\,\prod\limits_{j=1}^{n}\left[\frac{\det{(\bm{P}_j)}}{(2\pi\varepsilon)^{d}}\right]^{\frac{1}{2}}{\dd \bm{h}_j}}\;,
\end{equation}
we define 
\begin{equation}
\bm{F}_j=\bm{\alpha}_j\bm{P}_j^{-1}\quad\text{for}\quad j=1,...,n\;,
\end{equation}
and we evaluate the Gaussian integrals:
\begin{align}
\bar{I}_n &= \displaystyle{\int{e^{-\frac{1}{2\varepsilon}\sum\limits_{j=0}^{n}\bm{\phi}_j^T\bm{\alpha}_j\bm{\phi}_j}\,\prod\limits_{j=1}^{n}\left[\frac{\det{(\bm{P}_j)}}{(2\pi\varepsilon)^{d}}\right]^{\frac{1}{2}}{\dd\bm{\phi}_j}}}\;, \nonumber\\
&=\det{\left[\frac{1}{2\pi\varepsilon}\bm{\alpha}_0\prod\limits_{j=1}^{n} \bm{F}_j\right]}^{-\frac{1}{2}} \;. 
\end{align}
Moreover, defining 
\begin{equation}
\bm{D}_{n-k}=\prod\limits_{j=0}^{k}\bm{F}_{n-j}\quad\text{for}\quad k=0,...,n-1\;,
\end{equation}
we recover the following recurrence relations for $\bm{D}$ and ${\bm{\alpha}}$.
\begin{itemize}
\item Initial condition: $\bm{D}_n=\bm{\alpha}_{n}\bm{P}_{n}^{-1}$\;.
\item Iteration scheme: $\bm{D}_{n-(k+1)}=\bm{D}_{n-k}\bm{\alpha}_{n-(k+1)}\bm{P}_{n-(k+1)}^{-1}$ for $\,\,k=0,...,n-2$\;.
\end{itemize}
\begin{itemize}
\item Initial condition: $\displaystyle{\bm{\alpha}_{n}=\bm{P}_{n}+\varepsilon\frac{\bm{C}_n+\bm{C}_n^T}{2}+\varepsilon^2 \bm{Q}_{n}}$\;.
\item Iteration scheme: $\bm{\alpha}_{n-(k+1)}=\bm{P}_{n-(k+1)}+\bm{P}_{n-k}+\varepsilon^2 \bm{Q}_{n-(k+1)}-(\bm{\beta}_{n-k})^T\bm{\alpha}_{n-k}\bm{\beta}_{n-k}$ for $k=0,...,n-2$\;.
\end{itemize}

Now we are ready to start the main computation for the Neumann-Neumann problem, which deviates significantly from the other cases. Recalling the formulas for the determinant of a block matrix
\begin{equation}
\begin{split}
\det\begin{pmatrix}
      \bm{A} &  \bm{B} \\
     \bm{C} & \bm{D}
     \end{pmatrix}&=\det{(\bm{A})}\det{(\bm{D}-\bm{C}\bm{A}^{-1}\bm{B})}\;,\\
     &=\det{(\bm{D})}\det{(\bm{A}-\bm{B}\bm{D}^{-1}\bm{C})}\;
     \end{split}
\end{equation}
for given $\bm{{A}}$, $\bm{{B}}$, $\bm{{C}}$, $\bm{{D}}$ $\in\mathbb{R}^{d\times d}$, and defining $\bm{\alpha}_0^*=\bm{P}_{1}-\varepsilon\frac{\bm{C}_0+\bm{C}_0^T}{2}$, we get
\begin{equation}
\begin{split}
\det(\bm{\alpha}_0)&=\det\left(\bm{\alpha}_0^*-\bm{\beta}_{1}^T\bm{\alpha}_{1}\bm{\beta}_{1}\right)=\det
\begin{pmatrix}
      \bm{\mathbb{1}} &  \bm{\alpha}_{1}\bm{\beta}_{1} \\
     \bm{\beta}_{1}^T & \bm{\alpha}_0^*
     \end{pmatrix}\;,\\
&=\det\left(\bm{\alpha}_0^*\right)\det\left[\bm{\mathbb{1}}-\bm{\alpha}_{1}\bm{\beta}_{1}(\bm{\alpha}_0^*)^{-1}\bm{\beta}_{1}^T\right]
\end{split}
\end{equation}
leading to
\begin{align}
\det{\left[\frac{1}{2\pi\varepsilon}\bm{\alpha}_0\prod\limits_{j=1}^{n} \bm{F}_j\right]}&=\det\left(\frac{1}{2\pi\varepsilon}\bm{\alpha}_0^*\right)\det\left[\bm{\mathbb{1}}-\bm{\alpha}_{1}\bm{\beta}_{1}(\bm{\alpha}_0^*)^{-1}\bm{\beta}_{1}^T\right]\det{\left[\left(\prod\limits_{j=2}^{n} \bm{F}_j\right)\bm{\alpha}_1\bm{P}_1^{-1}\right]}\;, \nonumber\\
&=\det\left(\frac{1}{2\pi\varepsilon}\bm{\alpha}_0^*\right)\det\left[\bm{D}_2\bm{\alpha}_1\bm{P}_1^{-1}-\bm{D}_2\bm{\alpha}_{1}\bm{\beta}_{1}(\bm{\alpha}_0^*)^{-1}\bm{\beta}_{1}^T\bm{\alpha}_1\bm{P}_1^{-1}\right]\;,\nonumber\\
&=\det\left[\frac{1}{2\pi\varepsilon}\bm{\alpha}_0^*\left(\bm{D}_1-\bm{D}_2\bm{U}_1(\bm{\alpha}_0^*)^{-1}\bm{U}_1^T\bm{P}_1^{-1}\right)\right]\;.
\end{align}
Furthermore, $(\bm{\alpha}_0^*)^{-1}=\left[\bm{\mathbb{1}}-\frac{\varepsilon}{2}\bm{P}_1^{-1}\left(\bm{C}_0+\bm{C}_0^T\right)\right]^{-1}\bm{P}_1^{-1}\sim\left[\bm{\mathbb{1}}+\frac{\varepsilon}{2}\bm{P}_1^{-1}\left(\bm{C}_0+\bm{C}_0^T\right)\right]\bm{P}_1^{-1}$,
\begin{small}
\begin{equation}
\begin{split}
\bm{U}_1(\bm{\alpha}_0^*)^{-1}\bm{U}_1^T\bm{P}_1^{-1}&\sim\left[\bm{P}_1+\frac{\varepsilon}{2}(\bm{C}_1^T-\bm{C}_{0})\right]\left[\bm{\mathbb{1}}+\frac{\varepsilon}{2}\bm{P}_1^{-1}\left(\bm{C}_0+\bm{C}_0^T\right)\right]\bm{P}_1^{-1}\left[\bm{P}_1+\frac{\varepsilon}{2}(\bm{C}_1-\bm{C}_{0}^T)\right]\bm{P}_1^{-1}\;,\\
&\sim\left[\bm{\mathbb{1}}+\frac{\varepsilon}{2}(\bm{C}_1^T-\bm{C}_{0})\bm{P}_1^{-1}+\frac{\varepsilon}{2}(\bm{C}_0+\bm{C}_{0}^T)\bm{P}_1^{-1}\right]\left[\bm{P}_1+\frac{\varepsilon}{2}(\bm{C}_1-\bm{C}_{0}^T)\right]\bm{P}_1^{-1}\;,\\
&\sim\bm{\mathbb{1}}+\frac{\varepsilon}{2}(\bm{C}_1^T-\bm{C}_{0})\bm{P}_1^{-1}+\frac{\varepsilon}{2}(\bm{C}_0+\bm{C}_{0}^T)\bm{P}_1^{-1}+\frac{\varepsilon}{2}(\bm{C}_1-\bm{C}_{0}^T)\bm{P}_1^{-1}\;,\\
&=\bm{\mathbb{1}}+\frac{\varepsilon}{2}(\bm{C}_1+\bm{C}_{1}^T)\bm{P}_1^{-1}\;,
\end{split}
\end{equation}
\end{small}%
and consequently
\begin{small}
\begin{equation}
\begin{split}
\det{\left[\frac{1}{2\pi\varepsilon}\bm{\alpha}_0\prod\limits_{j=1}^{n} \bm{F}_j\right]}&\sim\det\left[\frac{1}{2\pi\varepsilon}\left(\bm{P}_{1}-\varepsilon\frac{\bm{C}_0+\bm{C}_0^T}{2}\right)\left(\bm{D}_1-\bm{D}_2\left(\bm{\mathbb{1}}+\frac{\varepsilon}{2}(\bm{C}_1+\bm{C}_{1}^T)\bm{P}_1^{-1}\right)\right)\right]\;,\\
&\sim\det\left[\frac{1}{2\pi\varepsilon}\left(\bm{P}_{1}-\varepsilon\frac{\bm{C}_0+\bm{C}_0^T}{2}\right)(\bm{D}_1-\bm{D}_2)-\frac{1}{2\pi}\bm{P}_{1}\bm{D}_2\frac{\bm{C}_1+\bm{C}_{1}^T}{2}\bm{P}_1^{-1}\right]\;,\\
&\sim\det\left[-\frac{1}{2\pi}\bm{P}_{1}\left(\frac{\bm{D}_2-\bm{D}_1}{\varepsilon}+\bm{D}_2\bm{C}_1^{(s)}\bm{P}_1^{-1}\right)\right]\;,\\
&\overset{\tiny{\varepsilon\rightarrow 0}}{\longrightarrow}\,\,\det\left[-\frac{1}{2\pi}\bm{P}\left(\dot{\bm{D}}+\bm{D}\bm{C}^{(s)}\bm{P}^{-1}\right)(t_0)\right]\;.
\end{split}
\end{equation}
\end{small}

We note that the matrix $\bm{D}$ solves the Papadopoulos equation Eq.~(\ref{papeq}) with the same initial conditions of the Dirichlet-Neumann case
\begin{equation}
\bm{D}(t)=\bm{\mathbb{1}}\;,\quad\dot{{\bm{D}}}(t)=-{{\bm{C}}}^{(s)}(t)\bm{P}(t)^{-1}
\end{equation}
since the recurrence relations and initial data are the same in both cases. Moreover, if $\bm{H}=\bm{L}^T$ is a matrix whose columns $\bm{h}$ are solutions of the Jacobi equation Eq.~(\ref{jac}), then the solutions of the Papadopoulos equation and the ones of
\begin{equation}
 \frac{\dd}{\dd \tau}\left[\dot{\bm{L}}{\bm{P}}+{\bm{L}}{\bm{C}}^T\right]-\dot{\bm{L}}{\bm{C}}-{\bm{L}}{\bm{Q}}={\mathbb{0}}
\end{equation}
are related by the non-linear transformation
\begin{equation}
\bm{L}^{-1}\dot{\bm{L}}=\bm{D}^{-1}\dot{\bm{D}}+{{\bm{C}}}^{(a)}\bm{P}^{-1} \;.
\end{equation}
Then, making appropriate assumptions we can write the following expressions in $\tau=t_0$
\begin{align}
\dot{\bm{L}}&=\bm{L}\bm{D}^{-1}\dot{\bm{D}}+\bm{L}{{\bm{C}}}^{(a)}\bm{P}^{-1}\;,\nonumber\\
\dot{\bm{L}}-\bm{L}{{\bm{C}}}^{(a)}\bm{P}^{-1}+\bm{L}{{\bm{C}}}^{(s)}\bm{P}^{-1}&=\bm{L}\bm{D}^{-1}\dot{\bm{D}}+\bm{L}\bm{D}^{-1}\bm{D}{{\bm{C}}}^{(s)}\bm{P}^{-1}\;,\nonumber\\
\bm{P}\left(\dot{\bm{L}}-\bm{L}{{\bm{C}}}^{(a)}\bm{P}^{-1}+\bm{L}{{\bm{C}}}^{(s)}\bm{P}^{-1}\right)&=\bm{P}\bm{L}\bm{D}^{-1}\left(\dot{\bm{D}}+\bm{D}{{\bm{C}}}^{(s)}\bm{P}^{-1}\right)\;,
\end{align}
and recalling that the initial conditions in $\tau=t$ (which coincide with Eq.~(\ref{par2})) have been chosen such that $\det\bm{L}=\det\bm{D}$, we finally get
\begin{equation}
\begin{split}
\det\left[\bm{P}\left(\dot{\bm{D}}+\bm{D}{{\bm{C}}}^{(s)}\bm{P}^{-1}\right)\right]&=\det\left(\dot{\bm{L}}\bm{P}-\bm{L}{{\bm{C}}}^{(a)}+\bm{L}{{\bm{C}}}^{(s)}\right)\;,\\
&=\det\left(\bm{P}\dot{\bm{L}}^T+{{\bm{C}}}^{(a)}\bm{L}^T+{{\bm{C}}}^{(s)}\bm{L}^T\right)\;,\\
&=\det\left(\bm{P}\dot{\bm{H}}+{{\bm{C}}}\bm{H}\right)=\det\bm{M}\;
\end{split}
\end{equation}
according to Eq.~(\ref{conjmom}).

Thus the final result for Neumann-Neumann boundary conditions is 
\begin{equation}
\int{e^{-\frac{1}{2}\delta^2 S(\bm{h};\bar{\bm{q}})}\,\mathcal{D}\bm{h}} = \lim_{n\rightarrow\infty} \bar{I}_n=\det\left[-\frac{1}{2\pi}\bm{M}(t_0)\right]^{-\frac{1}{2}}\;
\end{equation}
with $\bm{M}$ that solves the initial value problem 
\begin{equation}
\begin{cases}
           \frac{\dd}{\dd\tau}\begin{pmatrix}
      {{{\bm{H}}}}  \\
     {{{\bm{M}}}}
     \end{pmatrix}
       =\bm{J}{{{\bm{E}}}}\begin{pmatrix}
       {{{\bm{H}}}}  \\
     {{{\bm{M}}}}
     \end{pmatrix}
   \\
   \\
\begin{pmatrix}
      {{{\bm{H}}}}  \\
     {{{\bm{M}}}}
     \end{pmatrix}
     (t)
       =\bm{J}\begin{pmatrix}
     \mathbb{0} \\
     {\mathbb{1}} 
     \end{pmatrix}
       \end{cases}\;.
\end{equation}

\section{Examples}
\subsection{Langevin, Fokker-Planck and path integral representation}\label{LanHam}
To show how our approach applies to a number of relevant problems in Fokker-Planck dynamics, we present here three illustrative examples, the last representing a small theoretical investigation.

To begin, let us summarize briefly the relations between the Langevin and the Fokker-Planck equations, with the associated path integral formulation \citep{GRAHAM, FALKOFF, HAKEN}. We consider the following Langevin equation in first-order form for the process $\bm{\mathfrak{q}}(t)\in\mathbb{R}^d$ representing the physical quantity of interest
\begin{equation}\label{lang}
\dd \bm{\mathfrak{q}}(t)=\bm{\mu}(\bm{\mathfrak{q}}(t),t)\dd t+\bm{\sigma}(t)\dd \bm{B}(t)\;,
\end{equation}
with $\bm{\mu}(\bm{\mathfrak{q}}(t),t)\in\mathbb{R}^d$, $\bm{\sigma}(t)\in\mathbb{R}^{d\times l}$ and $\bm{B}(t)$ an $l-$dimensional standard Wiener process. It is known that the transition probability $\rho(\bm{q},t \vert \bm{q}_0,t_0)$ for the continuous Markovian process $\bm{\mathfrak{q}}(t)$ is the fundamental solution of the Fokker-Planck equation
\begin{equation}\label{fok}
\begin{cases}
\displaystyle{\frac{\partial}{\partial t}\rho=-\frac{\partial}{\partial q_i}\left[\mu_i(\bm{q},t)\rho\right]+\frac{1}{2}\Sigma_{ij}(t)\frac{\partial^2}{\partial q_i\partial q_j}\rho}\\
\\
\rho(\bm{q},t_0 \vert \bm{q}_0,t_0)=\delta(\bm{q}-\bm{q}_0)\;,\vspace{2mm}
\end{cases}
\end{equation}
where $\bm{\Sigma}(t)=\bm{\sigma}(t)\bm{\sigma}(t)^T\in\mathbb{R}^{d\times d}$ is the diffusion matrix, $\bm{\mu}(\bm{q},t)$ the drift vector, and the Einstein summation convention is adopted for $i,j=1,...,d$. In particular, if $\bm{\Sigma}$ is constant and $\bm{\mu}(\bm{q})$ is a function of the configuration, $\rho(\bm{q},t \vert \bm{q}_0,t_0)$ has the path integral representation Eq.~(\ref{pathint}), where the Lagrangian is given by the Onsager-Machlup function \citep{OM}
\begin{equation}\label{OM}
\mathcal{L}(\bm{q}(\tau),\dot{\bm{q}}(\tau))=\frac{1}{2}\left[\left(\dot{\bm{q}}-\bm{\mu}(\bm{q})\right)^T\bm{\Sigma}^{-1}\left(\dot{\bm{q}}-\bm{\mu}(\bm{q})\right)\right] \;.
\end{equation}
In the literature, however, an additional factor $+\frac{1}{2}\text{div}\left({\bm{\mu}(\bm{q})}\right)$ usually appears in Eq.~(\ref{OM}). As mentioned before, we observe that this correction is necessary for providing a normalized result when the path integral expression for the transition probability is only defined by $\mathcal{K}$. Our method offers an alternative approach, where we avoid the problem of finding an effective Lagrangian for every application by introducing  explicitly the normalization constant $\mathcal{N}$, see Eq.~(\ref{pathint}).

In the situation where $\bm{\Sigma}$ is not strictly positive definite, even if the Onsager-Machlup Lagrangian Eq.~(\ref{OM}) is ill-defined, its Hamiltonian form is well defined by 
\begin{equation}
\mathcal{L}(\bm{p}(\tau))=\frac{1}{2} \bm{p}^T\bm{\Sigma}\,\bm{p},\quad\text{where}\quad\dot{\bm{q}}=\bm{\Sigma} \,\bm{p}+\bm{\mu}(\bm{q})\;,
\end{equation}
and $\bm{p}$ is the conjugate variable of $\dot{\bm{q}}$ under the Legendre transform. This procedure is justified by taking the limit for a sequence of strictly positive definite matrices converging to $\bm{\Sigma}$.
At the same time the Hamilton and Jacobi equations for the minima and the fluctuations are also well defined. In particular, the Hamilton equations
\begin{equation}\label{Ham}
\displaystyle{\frac{\dd}{\dd\tau}\begin{pmatrix}
      \bm{q}  \\
     \bm{p}
     \end{pmatrix}
       =\bm{J}\begin{pmatrix}
       \dd[\bm{\mu}(\bm{q})]^T\,\bm{p}  \\
     \bm{\Sigma} \,\bm{p}+\bm{\mu}(\bm{q})
     \end{pmatrix}}
\end{equation}
are subject to the boundary conditions $\bm{q}^f(t_0)=\bm{q}_0$, $\bm{q}^f(t)=\bm{q}$ for the Dirichlet-Dirichlet minimum $\bm{q}^f(\tau)$, $\hat{\bm{q}}(t_0)=\bm{q}_0$, $\hat{\bm{p}}(t)=\bm{0}$ for the Dirichlet-Neumann minimum $\hat{\bm{q}}(\tau)$, and $\bm{q}^m(t_0)=\bm{q}_0$, $\bm{q}_{F}^m(t)=\bm{q}_{F}$, $\bm{p}_{V}^m(t)=\bm{0}$ for the Dirichlet-mixed minimizer $\bm{q}^m(\tau)$. On the other hand, the Jacobi equations in Hamiltonian form presented in Eqs.~(\ref{fin3},\ref{fin33}) are driven by the matrix
\[
\bm{E}=\begin{pmatrix}
     \dd^2[\bm{\mu}(\bm{q})]^T\,\bm{p}\,\,\, \,\,& \dd[\bm{\mu}(\bm{q})]^T\\
     \\
       \dd[\bm{\mu}(\bm{q})]\,\,\,\,\, & \bm{\Sigma}
       \end{pmatrix}\;.
\]
Here and in Eq.~(\ref{Ham}), $\dd[\bm{\mu}(\bm{q})]$ and $\dd^2[\bm{\mu}(\bm{q})]$ denote respectively the rank-2 and rank-3 tensors of the first and second derivatives in $\bm{q}$ of the vector field $\bm{\mu}(\bm{q})$.

\subsection{Ornstein–Uhlenbeck process}
As a first application of our method, we consider the d-dimensional Ornstein-Uhlenbeck process \citep{OU,FALKOFF,VATI}, which is described by the Fokker-Planck equation Eq.~(\ref{fok}) where $\bm{\Sigma}\in\mathbb{R}^{d\times d}$ is a constant symmetric diffusion matrix and where $\bm{\mu}(q)=-\bm{\Theta}\, q$, $\bm{\Theta}\in\mathbb{R}^{d\times d}$, defines the drift.
It is easy to see that the system is exactly characterised by the same linear Hamilton and Jacobi equations
\begin{equation}\label{OUeq}
\displaystyle{\frac{\dd}{\dd\tau}\begin{pmatrix}
      \bm{h}  \\
     \bm{m}
     \end{pmatrix}
       =\begin{pmatrix}
       -\bm{\Theta} &\,\,\,\bm{\Sigma}  \\
      \mathbb{0}  &\,\,\,\bm{\Theta}^T
     \end{pmatrix}}
     \begin{pmatrix}
      \bm{h}  \\
     \bm{m}
     \end{pmatrix} \;.
\end{equation}
The analytical solution is given by
\begin{align}\label{analy}
&\displaystyle{\bm{h}(\tau)=e^{-\bm{\Theta}\tau}\left[\int_0^{\tau}{e^{\bm{\Theta} s}\,\bm{\Sigma}\, e^{\bm{\Theta}^T s}C_1}\,\dd s+C_2\right]} \;,\nonumber\\
\nonumber\\
&\displaystyle{\bm{m}(\tau)=e^{\bm{\Theta}^T\tau}C_1} \;,
\end{align}
where $C_1$ and $C_2$ are determined from the appropriate boundary conditions. In particular, setting $t_0=0$, we find that 
\begin{align}
&S(\hat{\bm{q}})=0\;,\nonumber\\
&S(\bm{q}^f) = \frac{1}{2} \tilde{\bm{q}}^T\left[\int_0^{t}{e^{\bm{\Theta} (\tau-t)}\,\bm{\Sigma}\, e^{\bm{\Theta}^T (\tau-t)}}\,\dd \tau\right]^{-1} \tilde{\bm{q}} \;,\quad\tilde{\bm{q}} = \bm{q}-e^{-\bm{\Theta} t}\bm{q}_0\;.
\end{align}
The Jacobi fields lead to the factors
\begin{equation}
\displaystyle{\hat{\bm{H}}(0)=e^{\bm{\Theta} t}\;,\quad\bm{H}^f(0)=\int_0^{t}{e^{\bm{\Theta} \tau}\,\bm{\Sigma}\, e^{\bm{\Theta}^T (\tau-t)}}\,\dd \tau} \;.
\end{equation}
Finally, inserting these quantities in Eq.~(\ref{fin2}), we recover the Gaussian transition probability
\begin{equation}\label{OUGauss}
\displaystyle{\rho(\bm{q},t \vert \bm{q}_0,0)=\frac{\exp{[-\frac{1}{2} \tilde{\bm{q}}^T\bm{Co}^{-1}(t) \tilde{\bm{q}}]}}{\sqrt{\det[2\pi\,\bm{Co}(t)]}}} \;,
\end{equation}
with mean and covariance matrix
\begin{equation}\label{gaussCov}
\displaystyle{\bm{Av}(\bm{q}_0,t)=e^{-\bm{\Theta} t}\bm{q}_0\;,\quad
\bm{Co}(t)=\int_0^{t}{e^{\bm{\Theta} (\tau-t)}\,\bm{\Sigma}\, e^{\bm{\Theta}^T (\tau-t)}}\,\dd \tau} \;.
\end{equation}

In addition, note that the marginal probability density Eq.~(\ref{pathint2}) for the Ornstein-Uhlenbeck process can be derived analytically by means of Eq.~(\ref{fin22}). In fact, simple algebra gives us 
\begin{equation}
S(\bm{q}^m)=\frac{1}{2}\tilde{\bm{q}}_{F}^T\,\bm{Co}_m^{-1}(t)\,\tilde{\bm{q}}_{F}\;,\quad\tilde{\bm{q}}_{F}=\bm{q}_{F}-\bm{Av}_m(\bm{q}_0,t)\;,
\end{equation}
\begin{equation}\label{Wast}
\hat{\bm{H}}^{-1}\bm{H}^m(0)=\begin{pmatrix}
\mathbb{1}_{l\times l} & *\\
\mathbb{0}_{d-l\times l} & \bm{Co}_m(t)
\end{pmatrix} \;,
\end{equation}
with 
\begin{align}
&\bm{Av}_m(\bm{q}_0,t)=(\mathbb{0}_{d-l\times l},\mathbb{1}_{d-l\times d-l})\bm{Av}(\bm{q}_0,t)\;,\nonumber\\
\nonumber\\
&\bm{Co}_m(t)=\begin{pmatrix}
\mathbb{0}_{d-l\times l} & \mathbb{1}_{d-l\times d-l}
\end{pmatrix}\bm{Co}(t)\begin{pmatrix}
\mathbb{0}_{l\times d-l}\\
\mathbb{1}_{d-l\times d-l}
\end{pmatrix} \;.
\end{align}
Since in Eq.~(\ref{Wast}) the entry denoted by ``$*$" is not relevant, Eq.~(\ref{fin22}) automatically implies that
\begin{equation}
\displaystyle{\rho_m(\bm{q}_{F},t \vert \bm{q}_0,0) = \frac{\exp{[-\frac{1}{2}\tilde{\bm{q}}_{F}^T\,\bm{Co}_m^{-1}(t)\,\tilde{\bm{q}}_{F}]}}{\sqrt{\det[2\pi\,\bm{Co}_m(t)]}}} \;,
\end{equation}
which is indeed the (Gaussian) marginal of the full transition probability distribution Eq.~(\ref{OUGauss}).

\subsection{Van der Pol oscillator}
\begin{figure}[!ht]
\centering
\captionsetup{justification=centering,margin=1cm}
\includegraphics[width=13cm]{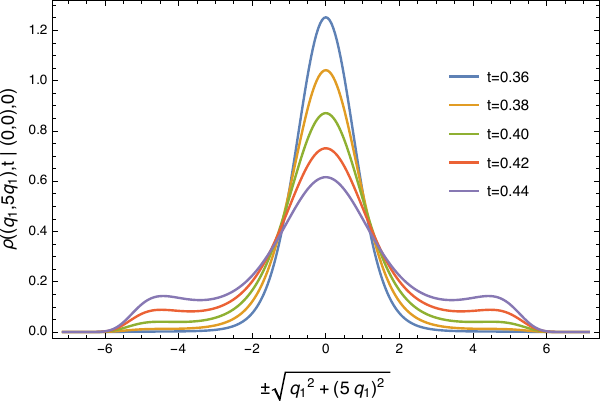}
\caption{Slices along direction $(q_1,q_2)=(q_1,5 q_1)$ of the transition probability $\rho\left((q_1,q_2),t \vert (0,0),0 \right)$ for the Van der Pol oscillator Eq.~(\ref{vdp}) with $\xi=3$, $\lambda=0.5$. Note that, despite the Laplace approximation, the resulting probability density is not necessarily Gaussian.}
    \label{fig:vdp}
\end{figure}

As a second application, we consider the Van der Pol oscillator driven by white noise \citep{NAESS}, that is described by the Langevin equation of motion for the coordinate $z$ as
\begin{equation}\label{vdp}
\ddot{z}(t)+2\xi[z(t)^2-1]\dot{z}(t)+z(t)=\sqrt{2 \lambda}f(t) \;,
\end{equation}
where $f(t)$ denotes a standard stationary Gaussian white noise, $\lambda>0$ represents the diffusion coefficient and $\xi>0$ the strength of the non-linearity. By defining the terms
\[
\bm{\Omega}=\begin{pmatrix}
       0 &\,\,\,-1  \\
      1  &\,\,\,-2\xi
     \end{pmatrix}\,\,\,\,\,\text{and}\,\,\,\,\,\bm{\nu}(\bm{q})=\begin{pmatrix}
       0  \\
      q_1^2q_2
     \end{pmatrix} \;,
\]
it is possible to write the stochastic equation of motion in phase space as a 2-dimensional Langevin equation in the form of Eq.~(\ref{lang}),
with $\bm{\sigma}=(0,\sqrt{2 \lambda})^T$. The associated Fokker-Planck equation has then coefficients
\[
\bm{\Sigma}=
\begin{pmatrix}
       0 &\,\,\,0  \\
      0  &\,\,\,2\lambda
     \end{pmatrix}\quad\text{and}\quad\bm{\mu}(\bm{q})= 
     -\bm{\Omega} \bm{q}-2\xi \bm{\nu}(\bm{q}) \;.
\]
Note that, in this example, the corresponding Onsager-Machlup function Eq.~(\ref{OM}) is no longer a quadratic function of $\bm{q}$ and $\dot{\bm{q}}$. Therefore, the Laplace approximation will lead to a result that is \textit{a priori} not exact. From the second-order expansion we nevertheless expect the result to be accurate for small values of diffusion $\lambda$ and final time $t$. Applying the method we presented, we obtain that the Hamilton and Jacobi equations for the system are respectively of the form
\begin{equation}\label{Hamvdp}
\displaystyle{\frac{\dd}{\dd\tau}\begin{pmatrix}
      \bm{q}  \\
     \bm{p}
     \end{pmatrix}
       =\begin{pmatrix}
       -\bm{\Omega} &\,\,\,\bm{\Sigma}  \\
      \mathbb{0}  &\,\,\,\bm{\Omega}^T
     \end{pmatrix}
     \begin{pmatrix}
       \bm{q}  \\
       \bm{p}
     \end{pmatrix}}-2\xi\bm{\psi}(\bm{q},\bm{p}) \;,
\end{equation}
\begin{equation}\label{Jacvdp}
\frac{\dd}{\dd\tau}\begin{pmatrix}
      \bm{H}  \\
     \bm{M}
     \end{pmatrix}
       =\begin{pmatrix}
       -\bm{\Omega} &\,\,\,\bm{\Sigma}  \\
      \mathbb{0}  &\,\,\,\bm{\Omega}^T
     \end{pmatrix}
     \begin{pmatrix}
      \bm{H}  \\
     \bm{M}
     \end{pmatrix}
     -2\xi\bm{\Psi}\begin{pmatrix}
      \bm{H}  \\
     \bm{M}
     \end{pmatrix} \;,
\end{equation}
with
\begin{equation}
\begin{split}
&\bm{\psi}(\bm{q},\bm{p})=(0,\,q_1^2 q_2,\,-2q_1 q_2 p_2,\,-q_1^2 p_2)^T\;,\\
& \\
&\bm{\Psi}=\begin{pmatrix}
       0 &\,\,\,0 &\,\,\,0 &\,\,\,0  \\
       2q_1 q_2 &\,\,\,q_1^2 &\,\,\,0 &\,\,\,0  \\
       -2q_2 p_2 &\,\,\,-2q_1 p_2 &\,\,\,0 &\,\,\,-2q_1 q_2  \\
       -2q_1 p_2 &\,\,\,0 &\,\,\,0 &\,\,\,-q_1^2  
     \end{pmatrix} \;.
\end{split}
\end{equation}
Solving numerically Eqs.~(\ref{Hamvdp},\ref{Jacvdp}), subject to the associated boundary conditions, we are able to obtain through Eq.~(\ref{fin2}) the Laplace approximation of the (non-Gaussian) transition probability solving the Fokker-Planck equation for the stochastic Van der Pol oscillator, see Fig. \ref{fig:vdp}.

\subsection{Investigating normalization}
As a final example, let us investigate in one dimension a system with action
\begin{equation}
    S = \int_{t_0=0}^t \left[ \dfrac{1}{2} \alpha \dot{q}(\tau)^2 - \beta \phi(q(\tau)) \right] \dd \tau \;,
\end{equation}
where $\alpha$, $\beta$ represent some physical parameters, $\phi(q(\tau))$ is a general potential that here we take of the form $\phi(q)=a q^2 + b q$, and let us define $\kappa=\sqrt{\frac{2\beta a}{\alpha}}$. From the Euler-Lagrange equations with the appropriate boundary conditions we obtain $q^f(\tau)$ and $\hat{q}(\tau)$, while from the Jacobi equations we obtain 
\begin{equation}
H^f(\tau)=\frac{\sin(\kappa(t-\tau))}{\alpha\kappa},\quad\hat{H}(\tau)=\cos(\kappa(t-\tau))\;.
\end{equation}
These results allow us to express the probability that $q(t)=q$, given that $q(0)=0$, as
\begin{equation}
\rho(q,t\vert 0,0) = e^{-\frac{1}{2}\frac{(q-\text{Av})^2}{\text{Var}}} /\sqrt{2\pi \text{Var}}\;, 
\end{equation}
which is a Gaussian probability distribution with mean and variance respectively
\begin{equation}
\text{Av}=\frac{b}{2a}\left(\frac{1-\cos(\kappa t)}{\cos(\kappa t)}\right),\quad\text{Var}=\frac{\tan(\kappa t)}{\alpha \kappa}\;.
\end{equation}
This example illustrates how $\mathcal{N}\neq 1$ in general, even if the cross-term $C=0$.  
\cleardoublepage
\part{Stochastic elastic rods}
\chapter{Looping probabilities}

\section{Introduction on elastic rods}
\subsection{Background on equilibria and stability}
\begin{figure*}
\captionsetup{justification=centering,margin=1cm}
\includegraphics[width=1\textwidth]{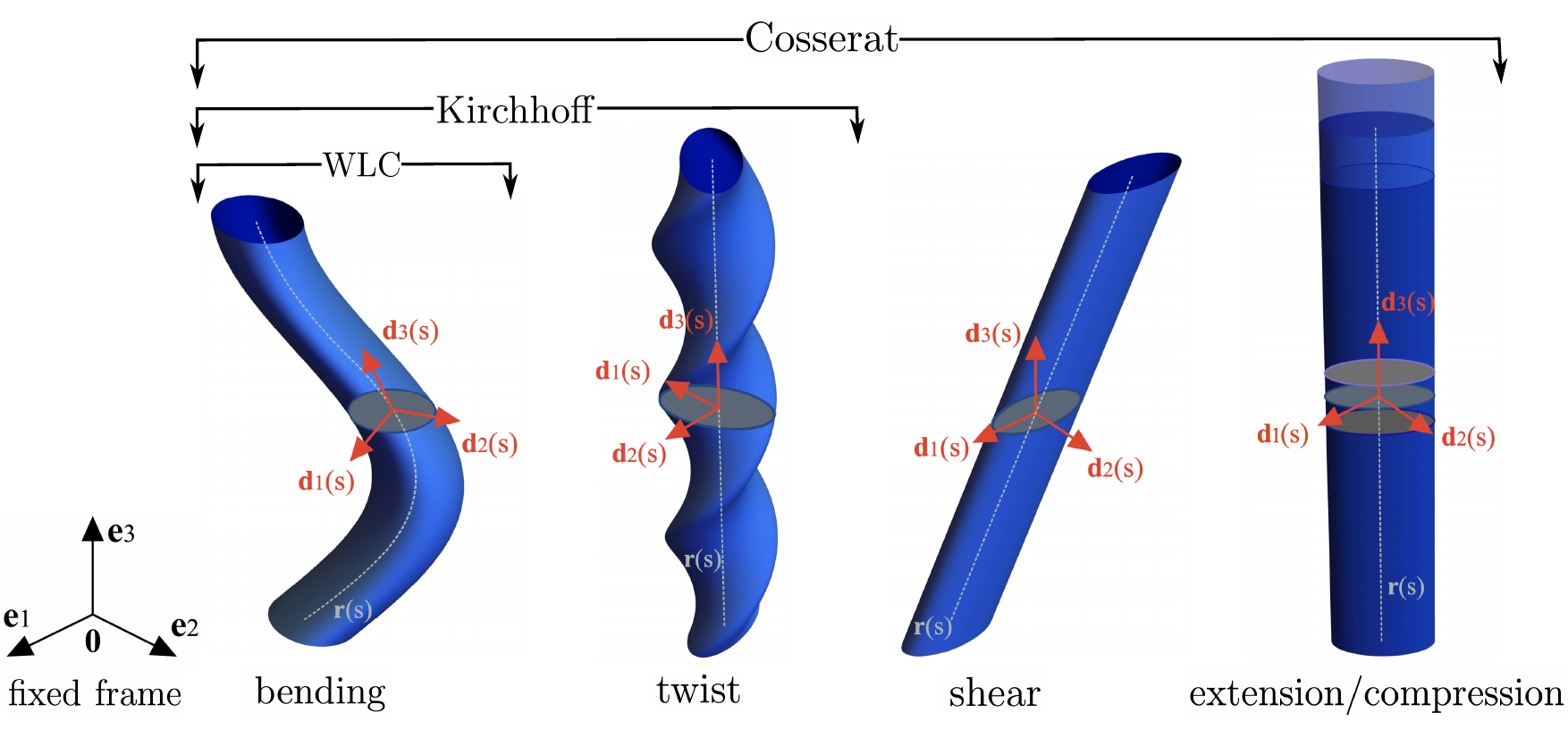}
\caption{Schematic representation of a Cosserat rod with an elliptical cross-section (non-isotropic), where bending, twist, shear and extension / compression are allowed deformations. For Kirchhoff rods only bending and twist are permissible. The standard WLC model takes only bending deformations into account and the rod is assumed to be intrinsically straight with circular cross-section (isotropic).}
\label{fig20}
\end{figure*}

The following can be regarded as a brief introduction to the topic of elastic rods, where we highlight the crucial results that will be used in the course of the chapter, mainly on two-point boundary value problems, stability, and special treatment of the inextensible-unshearable case. A comprehensive overview of the theory of elastic rods in the context of continuum mechanics can be found in \citep{ANT}. In particular, we follow the specific notation and Hamiltonian formulations introduced in \citep{HAM}. Briefly, a configuration of a Cosserat rod is a framed curve 
\begin{equation}
\bm{q}(s)=(\bm{R}(s),\bm{r}(s))\in SE(3)\quad\text{for each}\quad s\in [0,L]\;, 
\end{equation}
which may be bent, twisted, stretched or sheared. The vector $\bm{r}(s)\in\mathbb{R}^3$ and the matrix $\bm{R}(s)\in SO(3)$ model respectively the rod centerline and the orientation of the material in the rod cross-section via a triad of orthonormal directors $\lbrace\bm{d}_i(s)\rbrace_{i=1,2,3}$ attached to the rod centerline, with respect to a fixed frame $\lbrace\bm{e}_i\rbrace_{i=1,2,3}$. As a matter of notation, the columns of the matrix $\bm{R}(s)$ in coordinates are given by the components of the vectors ${\bm{d}_j(s)}$ in the fixed frame $\lbrace\bm{e}_i\rbrace$, namely $\bm{R}_{i,j}(s)=\bm{e}_i\cdot\bm{d}_j(s),\,\,i,j=1,..,3$. In Fig. \ref{fig20} we show a schematic representation of the the degrees of freedom allowed within the special Cosserat theory of rods in relation to other simpler models that will be outlined in the course of this section.

Strains are defined as $\bm{u}(s)$, $\bm{v}(s)$ in $\mathbb{R}^3$ where 
\begin{equation}
\bm{d}_i'(s)=\bm{u}(s)\times\bm{d}_i(s)\quad\text{for}\quad i=1,..,3\;,\quad\bm{r}'(s)=\bm{v}(s)\;, 
\end{equation}
with $\bm{u}$ the Darboux vector and the prime denoting the derivative with respect to $s$. Sans-serif font is used to denote components in the director basis (\eg $\mathsf{u}_i=\bm{u}\cdot\bm{d}_i$), and we write  $\bm{\mathsf{u}}=(\mathsf{u}_1,\mathsf{u}_2,\mathsf{u}_3)$, $\bm{\mathsf{v}}=(\mathsf{v}_1,\mathsf{v}_2,\mathsf{v}_3)$, etc. Physically, $\mathsf{u}_1$ and $\mathsf{u}_2$ represent the bending strains and $\mathsf{u}_3$ the twist strain. Analogously, $\mathsf{v}_1$ and $\mathsf{v}_2$ are associated with transverse shearing, whereas $\mathsf{v}_3$ with stretching or compression of the rod. In compact form, we have 
\begin{equation}
\bm{\mathsf{u}}^{\times}(s)=\bm{R}(s)^T \bm{R}'(s)\;,\quad\bm{\mathsf{v}}(s)=\bm{R}(s)^T \bm{r}'(s)\;,
\end{equation}
where $\bm{\mathsf{u}}^{\times}$ is the skew-symmetric matrix or cross product matrix of $\bm{\mathsf{u}}$:
\begin{equation}
\bm{\mathsf{u}}^{\times}=
\begin{pmatrix}
0 & -\mathsf{u}_3 & \mathsf{u}_2\\
\mathsf{u}_3 & 0 & -\mathsf{u}_1\\
-\mathsf{u}_2 & \mathsf{u}_1 & 0\\
\end{pmatrix}\;.
\end{equation}

The stresses $\bm{m}(s)$ and $\bm{n}(s)$ are defined as the resultant moment and force arising from averages of the stress field acting across the material cross-section at $\bm{r}(s)$. In the absence of any distributed loading, at equilibrium the stresses satisfy the balance laws 
\begin{equation}\label{EuL}
{\bm{{n}}}'=\bm{0}\;,\quad{\bm{{m}}}'+\bm{{r}}'\times\bm{{n}}=\bm{0}\;. 
\end{equation}
Equilibrium configurations can be found once constitutive relations are introduced, which we do in a way that facilitates the recovery of the inextensible, unshearable limit typically adopted in polymer physics.

Namely, we consider a pair of functions $W,\,W^*:\mathbb{R}^3\times\mathbb{R}^3\times [0,L]\rightarrow\mathbb{R}$ that (for each $s\in[0,L]$) are strictly convex, dual functions under Legendre transform in their first two arguments, and with $\bm{0}\in\mathbb{R}^6$ their unique global minimum. If $\hat{\bm{\mathsf{u}}}(s)$ and $\hat{\bm{\mathsf{v}}}(s)$ are the strains of the unique energy minimizing configuration $\hat{\bm{q}}(s)$, then, $\forall\epsilon>0$ we introduce the Hamiltonian function 
\begin{equation}
\mathcal{H}(\bm{\mathsf{m}},\bm{\mathsf{n}})=W^*(\bm{\mathsf{m}},\epsilon\bm{\mathsf{n}};s)+\bm{\mathsf{m}}\cdot\hat{\bm{\mathsf{u}}}+\bm{\mathsf{n}}\cdot\hat{\bm{\mathsf{v}}}\;, 
\end{equation}
and the constitutive relations are 
\begin{equation}
\bm{\mathsf{u}}=\frac{\partial \mathcal{H}}{\partial \bm{\mathsf{m}}}=W_1^*(\bm{\mathsf{m}},\epsilon\bm{\mathsf{n}};s)+\hat{\bm{\mathsf{u}}}\;,\quad\bm{\mathsf{v}}=\frac{\partial \mathcal{H}}{\partial\bm{\mathsf{n}}}=\epsilon W_2^*(\bm{\mathsf{m}},\epsilon\bm{\mathsf{n};s})+\hat{\bm{\mathsf{v}}}, 
\end{equation}
which can be inverted to obtain 
\begin{equation}
\bm{\mathsf{m}}=\frac{\partial \mathcal{L}}{\partial \bm{\mathsf{u}}}=W_1\left(\bm{\mathsf{u}}-\hat{\bm{\mathsf{u}}},\frac{\bm{\mathsf{v}}-\hat{\bm{\mathsf{v}}}}{\epsilon};s\right)\;,\quad\bm{\mathsf{n}}=\frac{\partial \mathcal{L}}{\partial \bm{\mathsf{v}}}=\frac{1}{\epsilon}W_2\left(\bm{\mathsf{u}}-\hat{\bm{\mathsf{u}}},\frac{\bm{\mathsf{v}}-\hat{\bm{\mathsf{v}}}}{\epsilon};s\right)\;, 
\end{equation}
where the Lagrangian 
\begin{equation}
\mathcal{L}(\bm{\mathsf{u}},\bm{\mathsf{v}})=W\left(\bm{\mathsf{u}}-\hat{\bm{\mathsf{u}}},\frac{\bm{\mathsf{v}}-\hat{\bm{\mathsf{v}}}}{\epsilon};s\right)
\end{equation}
defines the elastic potential energy of the system as 
\begin{equation}
E=\int_0^L{W\left(\bm{\mathsf{u}}-\hat{\bm{\mathsf{u}}},\frac{\bm{\mathsf{v}}-\hat{\bm{\mathsf{v}}}}{\epsilon};s\right)}\,\dd s\;. 
\end{equation}
Note the use of the subscripts to denote partial derivatives with respect to the first or second argument. The standard case of linear constitutive relations arises when
\begin{equation}
W^*(\bm{\mathsf{x}};s)=\frac{1}{2}\bm{\mathsf{x}}\cdot\bm{\mathcal{R}}(s)\bm{\mathsf{x}}\;,\quad W(\bm{\mathsf{y}};s)=\frac{1}{2}\bm{\mathsf{y}}\cdot\bm{\mathcal{P}}(s)\bm{\mathsf{y}}
\end{equation}
for $\bm{\mathsf{x}}$,$\,\bm{\mathsf{y}}\,\in\mathbb{R}^6$, where $\mathbb{R}^{6\times 6}\ni\bm{\mathcal{P}}^{-1}(s)=\bm{\mathcal{R}}(s)=\bm{\mathcal{R}}(s)^T>0$, with $\bm{\mathcal{P}}(s)$ a general non-uniform stiffness matrix and $\bm{\mathcal{R}}(s)$ the corresponding compliance matrix. For each $\epsilon>0$ and given $W$, $W^*$, we arrive at a well-defined Cosserat rod theory, where, \textit{e.g.}, the full potential energy of the system might include end-loading terms of the form $\bm{\lambda}\cdot(\bm{r}(L)-\bm{r}(0))$, $\bm{\lambda}\in\mathbb{R}^3$. 

The point of the above formulation is that the Hamiltonian and associated constitutive relations behave smoothly in the limit $\epsilon\rightarrow 0$, which imply the unshearability and inextensibility constraint on the strains ${\bm{\mathsf{v}}}(s)=\hat{\bm{\mathsf{v}}}(s)$, where $\hat{\bm{\mathsf{v}}}(s)$ are prescribed. This is precisely a Kirchhoff rod model. However, the $\epsilon\rightarrow 0$ limit of the Cosserat Lagrangian is not smooth; rather the potential energy density for the Kirchhoff rod is the Legendre transform of 
\begin{equation}
W^*(\bm{\mathsf{m}},\bm{0};s)+\bm{\mathsf{m}}\cdot\hat{\bm{\mathsf{u}}}+\bm{\mathsf{n}}\cdot\hat{\bm{\mathsf{v}}}
\end{equation}
with respect to $\bm{\mathsf{m}}\in\mathbb{R}^3$, or 
\begin{equation}
W^{(\mathtt{K})}({\bm{\mathsf{u}}}-\hat{\bm{\mathsf{u}}};s)-{\bm{\mathsf{n}}}\cdot\hat{\bm{\mathsf{v}}}\;.
\end{equation}
In the case of linear elasticity, for a Cosserat rod with 
\begin{equation}
\bm{\mathcal{P}}(s)=\begin{pmatrix}
       \bm{\mathcal{K}} & \bm{\mathcal{B}}  \\
       \bm{\mathcal{B}}^T & \bm{\mathcal{A}}
     \end{pmatrix}
\end{equation}
and $\bm{\mathcal{K}}(s)$, $\bm{\mathcal{B}}(s)$, $\bm{\mathcal{A}}(s)$ in $\mathbb{R}^{3\times 3}$, the $(1,1)$ block of the compliance matrix is $\bm{\mathcal{R}}_{1,1}=(\bm{\mathcal{K}}-\bm{\mathcal{B}}\bm{\mathcal{A}}^{-1}\bm{\mathcal{B}}^T)^{-1}$ and 
\begin{equation}
W^{(\mathtt{K})}({\bm{\mathsf{u}}}-\hat{\bm{\mathsf{u}}};s)=\frac{1}{2}(\bm{\mathsf{u}}-\hat{\bm{\mathsf{u}}})\cdot\bm{\mathcal{K}}^{(\mathtt{K})}(s)(\bm{\mathsf{u}}-\hat{\bm{\mathsf{u}}})\;,\quad\bm{\mathcal{K}}^{(\mathtt{K})}={\bm{\mathcal{K}}^{(\mathtt{K})}}^T=\bm{\mathcal{R}}_{1,1}^{-1}>0\;.
\end{equation}

Uniform helical WLC models are recovered in the case of a uniform Kirchhoff rod when $\hat{\bm{\mathsf{u}}}(s)$, $\hat{\bm{\mathsf{v}}}(s)$ and $\bm{\mathcal{K}}^{(\mathtt{K})}(s)$ are all taken to be constant. (For any uniform rod, Cosserat or Kirchhoff, the Hamiltonian function is constant along equilibria). Linearly elastic Kirchhoff rods are (transversely) isotropic when $\bm{\mathcal{K}}^{(\mathtt{K})}(s)=diag\{k_1(s),k_2(s),k_3(s)\}$ with $k_1=k_2$ and $\hat{{\mathsf{u}}}_1=\hat{{\mathsf{u}}}_2=\hat{{\mathsf{v}}}_1=\hat{{\mathsf{v}}}_2=0$. Then $\mathsf{m}_3$ is constant on equilibria, and 
\begin{equation}
W^{(\mathtt{K})}=\frac{1}{2}\left[k_1\kappa^2+k_3(\mathsf{u}_3-\hat{\mathsf{u}}_3)^2\right]
\end{equation}
reduces to a function of the square geometrical curvature $\kappa(s)$ of the curve (where it should be noted that $\mathsf{u}_3(s)$ is still the twist of the $\lbrace\bm{d}_i\rbrace$ frame which is not directly related to the geometrical torsion of the Frenet framing of the rod centerline). The WLC model arises when $k_1(s)$ is constant and the twist moment $\mathsf{m}_3$ vanishes.

There is an extensive literature concerning the study of equilibria of a given elastic rod. Numerically this involves the solution of a two-point boundary value problem, which can reasonably now be regarded as a straightforward well-understood procedure. Often coordinates on $SO(3)$ are introduced and the resulting system of second-order Euler Lagrange equations associated with the potential energy is solved numerically. We adopt an Euler parameters (or quaternions) parametrization of $SO(3)$, but solve the associated first-order canonical Hamiltonian system subject to appropriate (self-adjoint) two-point boundary conditions, so that the inextensible, unshearable Kirchhoff rod is a simple smooth limit of the extensible, shearable Cosserat case. 

In this chapter we are primarily interested in the two specific boundary value problems, denoted respectively by $(\mathtt{f})$ and $(\mathtt{m})$:
\begin{equation}\label{f}
(\mathtt{f})\quad\bm{r}(0)=\bm{0},\,\,\bm{R}(0)=\mathbb{1},\,\,\bm{r}(L)=\bm{r}_L,\,\,\bm{R}(L)=\bm{R}_L\;,
\end{equation}
\begin{equation}\label{m}
(\mathtt{m})\quad\bm{r}(0)=\bm{0},\,\,\,\bm{R}(0)=\mathbb{1},\,\,\,\bm{r}(L)=\bm{r}_L,\,\,\,\bm{m}(L)=\bm{0}\;.
\end{equation}
The boundary value problem $(\mathtt{f})$ arises in modelling looping in $SE(3)$ including the particular case of cyclization where $\bm{r}_L=\bm{0}$ and $\bm{R}_L=\mathbb{1}$. The boundary value problem $(\mathtt{m})$ arises in modelling looping in $\mathbb{R}^3$, where the value of $\bm{R}_L$ is a variable left free, over which one marginalises. In general, for rod two-point boundary value problems, equilibria with given boundary conditions are non-unique. For isotropic or uniform rods, and for specific choices of $\bm{r}_L$ and $\bm{R}_L$ in $(\mathtt{f})$ and $(\mathtt{m})$, equilibria can arise in continuous isoenergetic families \citep{RING}, a case of primary interest here. 

As we assume hyper-elastic constitutive relations with 
\begin{equation}\label{energy}
E(\bm{q})=\int_0^L{W(\bm{\mathsf{u}}-\hat{\bm{\mathsf{u}}},{\bm{\mathsf{v}}-\hat{\bm{\mathsf{v}}}};s)}\,\dd s\;,
\end{equation}
stability of rod equilibria can reasonably be discussed dependent on whether an equilibrium is a local minimum of the associated potential energy variational principle. For Cosserat rods classification of which equilibria are local minima has a standard and straightforward solution. The second variation $\delta^2E$ is a quadratic functional of the perturbation field $\bm{\mathsf{h}}=(\delta\bm{\mathsf{c}},\delta\bm{\mathsf{t}})$, where the sans-serif font $\bm{\mathsf{q}}(s)=(\bm{\mathsf{c}}(s),\bm{\mathsf{t}}(s))\in\mathbb{R}^6$ is a given parametrisation of $SE(3)$ for the configuration variable in the director basis which will be specified later in the chapter, and reads as
\begin{equation}\label{sec}
\delta ^2 E=\int_0^L{\left({\bm{\mathsf{h}}'}\cdot\bm{\mathsf{P}}\bm{\mathsf{h}}'+2{\bm{\mathsf{h}}'}\cdot{\bm{\mathsf{C}}}\bm{\mathsf{h}}+\bm{\mathsf{h}}\cdot{\bm{\mathsf{Q}}}\bm{\mathsf{h}}\right)}\,\dd s\;,
\end{equation}
where $\bm{\mathsf{P}}(s)$, ${\bm{\mathsf{C}}}(s)$ and ${\bm{\mathsf{Q}}}(s)$ are coefficient matrices in $\mathbb{R}^{6\times 6}$ computed at any equilibrium. The Jacobi equations are the (second-order) system of Euler-Lagrange equations for Eq.~(\ref{sec}), or equivalently the linearisation of the original Euler-Lagrange equations for the potential energy variational principle. One then solves a $6\times 6$ matrix valued system, namely an initial value problem for the Jacobi equations with initial conditions coinciding with the ones given later in the chapter when computing probability densities from Jacobi fields (shooting towards $s=0$, where in both $(\mathtt{f})$ and $(\mathtt{m})$ Dirichlet boundary conditions are present; the case with Neumann boundary conditions at both ends is more delicate \citep{NEU}). Provided that the determinant of the matrix solution does not vanish in $[0,L)$, then there is no conjugate point and the equilibrium is a local minimum \citep{JHMStab,ISO,HEL}. 

As described fully in \citep{JHMT}, the constrained case of Kirchhoff is more subtle and a theory dating back to Bolza for isoperimetrically constrained calculus of variations must be applied \citep{BOLZA}. However, the Hamiltonian version of the Jacobi equations for rods (just like the Hamiltonian version of the Euler-Lagrange equilibrium equations) has a smooth limit as $\epsilon\rightarrow 0$, and the limit corresponds to the Hamiltonian formulation of the Bolza conjugate point conditions as described in \citep{ISO}. The Jacobi equations in first-order Hamiltonian form are written as
\begin{equation}\label{JacHam}
\begin{pmatrix}
      {{\bm{\mathsf{H}}'}}  \\
     {{\bm{\mathsf{M}}'}}
     \end{pmatrix}
       =\bm{J}{{\bm{\mathsf{E}}}}\begin{pmatrix}
       {{\bm{\mathsf{H}}}}  \\
     {{\bm{\mathsf{M}}}}
     \end{pmatrix}\;,\quad\bm{J}=\begin{pmatrix} \mathbb{0} & \mathbb{1}\\ -\mathbb{1} & \mathbb{0} \end{pmatrix}\;,
\end{equation}
with the Hamiltonian skew-symmetric matrix $\bm{J}\in\mathbb{R}^{12\times 12}$, $\bm{\mathsf{E}}(s)$ the symmetric matrix driving the system which will be detailed later, and $\bm{\mathsf{M}}(s)\in\mathbb{R}^{6\times 6}$ the conjugate variable of the Jacobi fields $\bm{\mathsf{H}}(s)$ under the Legendre transform.

In the following, we assume the existence and stability of the minimizers of the elastic energy Eq.~(\ref{energy}) $\bm{q}^f$ and $\bm{q}^m$ satisfying the boundary conditions $(\mathtt{f})$ in Eq.~(\ref{f}) and $(\mathtt{m})$ in Eq.~(\ref{m}) respectively. Note that the intrinsic configuration of the rod $\hat{\bm{q}}$ is itself a minimizer (global) satisfying 
\begin{equation}\label{hat}
\bm{r}(0)=\bm{0}\;,\,\,\,\bm{R}(0)=\mathbb{1}\;,\,\,\,\bm{n}(L)=\bm{m}(L)=\bm{0}\;.
\end{equation}
Stability of equilibria is not the focus of this work, but we will show that the volume of certain Jacobi fields, \textit{i.e.}, the actual (positive) value of a Jacobi determinant, plays a central role in the evaluation formula for the quadratic path integrals that arise in our Laplace approximations to looping probabilities, according to the derivation presented in the previous chapter.

We remark that the connection between Jacobi fields and quadratic imaginary path integrals is well known in the case that the coefficient matrix ${\bm{\mathsf{C}}}(s)$ in the cross-terms in Eq.~(\ref{sec}) vanishes (or is symmetric and so can be integrated away). By contrast, for elastic rods a non-symmetric ${\bm{\mathsf{C}}}(s)$ is typically present and the approach of Papadopoulos \citep{PAP1} is required to evaluate the quadratic path integrals, and as described in \citep{LUDT, LUD} a further Riccati transformation for the Papadopoulos solution formula is necessary to recover a Jacobi fields expression. Moreover, in \citep{2020}, \ie in the previous chapter, the latter studies are generalised for different choices of boundary conditions on the paths, in particular for dealing with the partition function and solving the marginalised problem. 

The main contributions of this chapter are to demonstrate that the approach of \citep{LUDT, LUD} for conditional probability densities can be extended in two ways. First, isolated equilibria to boundary value problem $(\mathtt{m})$ can be treated, in addition to the case of isolated equilibria to boundary value problem $(\mathtt{f})$, and second, the case of non-isolated equilibria of both boundary value problems $(\mathtt{f})$ and $(\mathtt{m})$ (as arises for isotropic rods) can be handled by appropriately generalising a particular regularization procedure \citep{MCK, FAL} within Forman's theorem in the field of functional determinants \citep{FORM}. Furthermore, the underlying physical phenomena arising from the different cases are discussed and explained within some guiding examples. For a polymer, the questions we are trying to answer would be interpreted as follows: what is a good estimate of the probability of the end monomers coming into contact with each other? How is the latter value changing if we impose an orientation constraint on the binding site? How does the shape of the cross-section (isotropic or non-isotropic) affect the statistics? And finally, what happens if we deviate from the standard inextensible and unshearable model and incorporate shear and extension as possible deformations?

\subsection{A preview of the examples considered}
The method developed in the present chapter will be applied, as a fundamental example, to a linearly elastic, uniform, with diagonal stiffness matrix, intrinsically straight and untwisted rod 
\begin{equation}
\bm{\mathcal{P}}(s)=\bm{\mathcal{P}}=diag\lbrace k_1,k_2,k_3,a_1,a_2,a_3\rbrace\;,\quad\hat{\bm{\mathsf{u}}}=\bm{0}\;,\quad\hat{\bm{\mathsf{v}}}=(0,0,1)\;. 
\end{equation}
Neither intrinsic shear nor extension is present. Since we are primarily interested in ring-closure or cyclization probabilities, we look for minimizers of the energy satisfying the boundary conditions reported in Eq.~(\ref{f}), Eq.~(\ref{m}) with $\bm{r}_L=\bm{0}$ and $\bm{R}_L=\mathbb{1}$. 

\begin{figure*}
\captionsetup{justification=centering,margin=1cm}
\subfigure[]{\includegraphics[width=.441\textwidth]{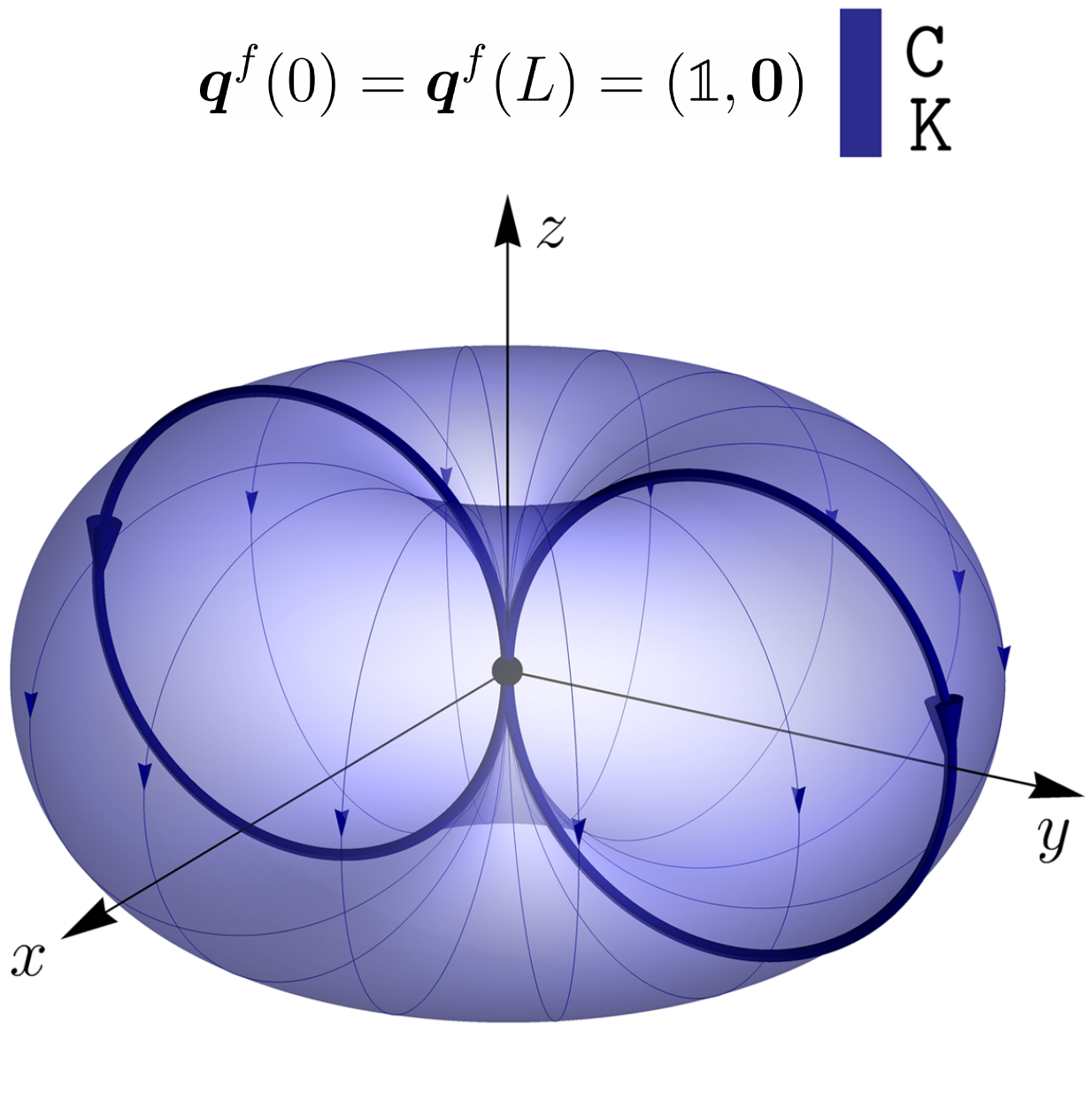}}\qquad\qquad
\subfigure[]{\includegraphics[width=.44\textwidth]{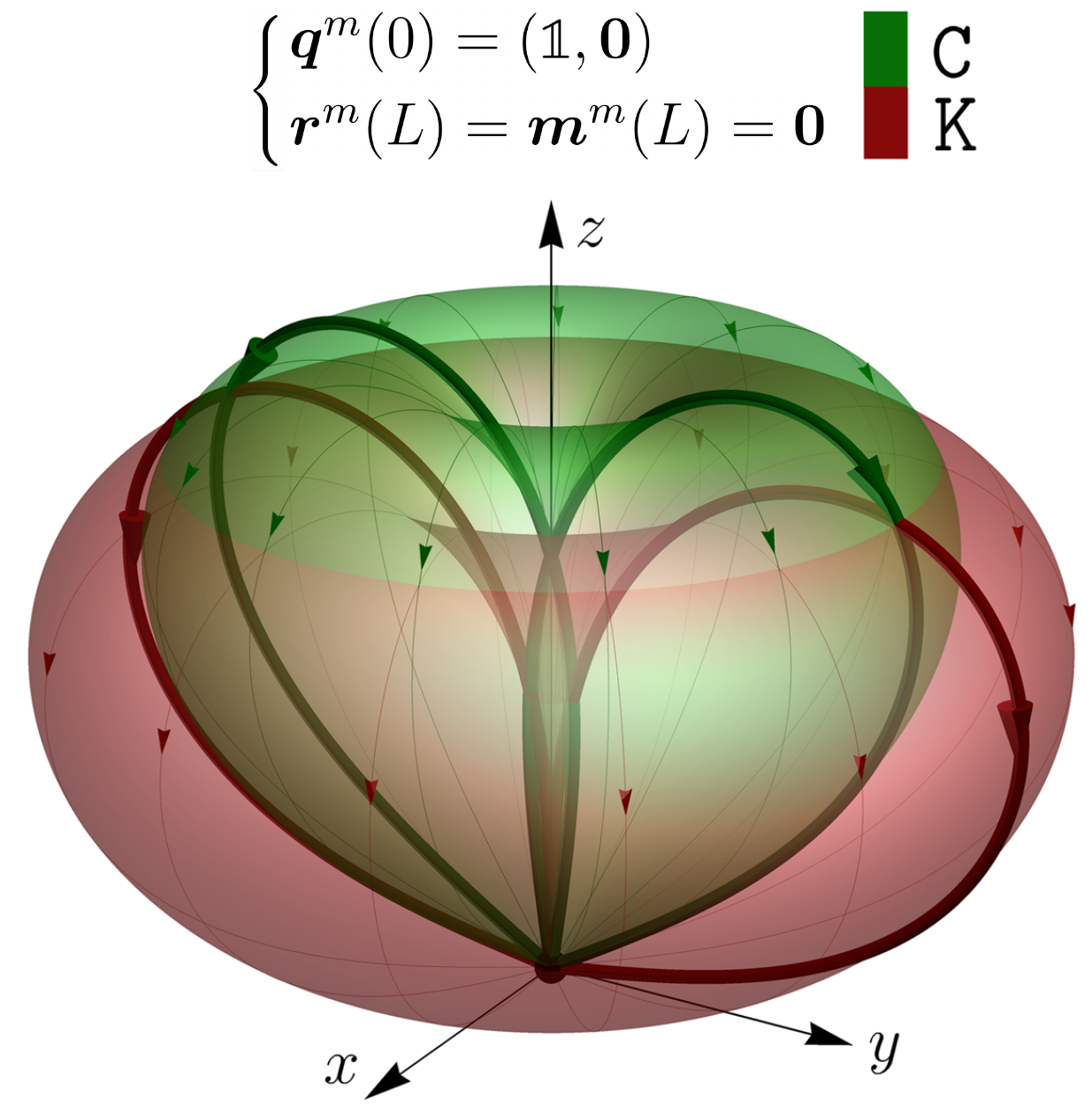}}
\caption{The thick lines represent the pairs of isolated minima for the non-isotropic case; the manifolds of minimizers for the isotropic case are displayed accordingly. In panel (a) the solutions for the full case are the same for Kirchhoff and Cosserat rods. In panel (b) we underline the effect of shear and extension for the marginal case, which modifies the red solutions (Kirchhoff) into the green ones (Cosserat).}
\label{fig1}
\end{figure*}

\begin{figure*}
\centering
\captionsetup{justification=centering,margin=1cm}
\subfigure[]{
\includegraphics[width=.444\textwidth]{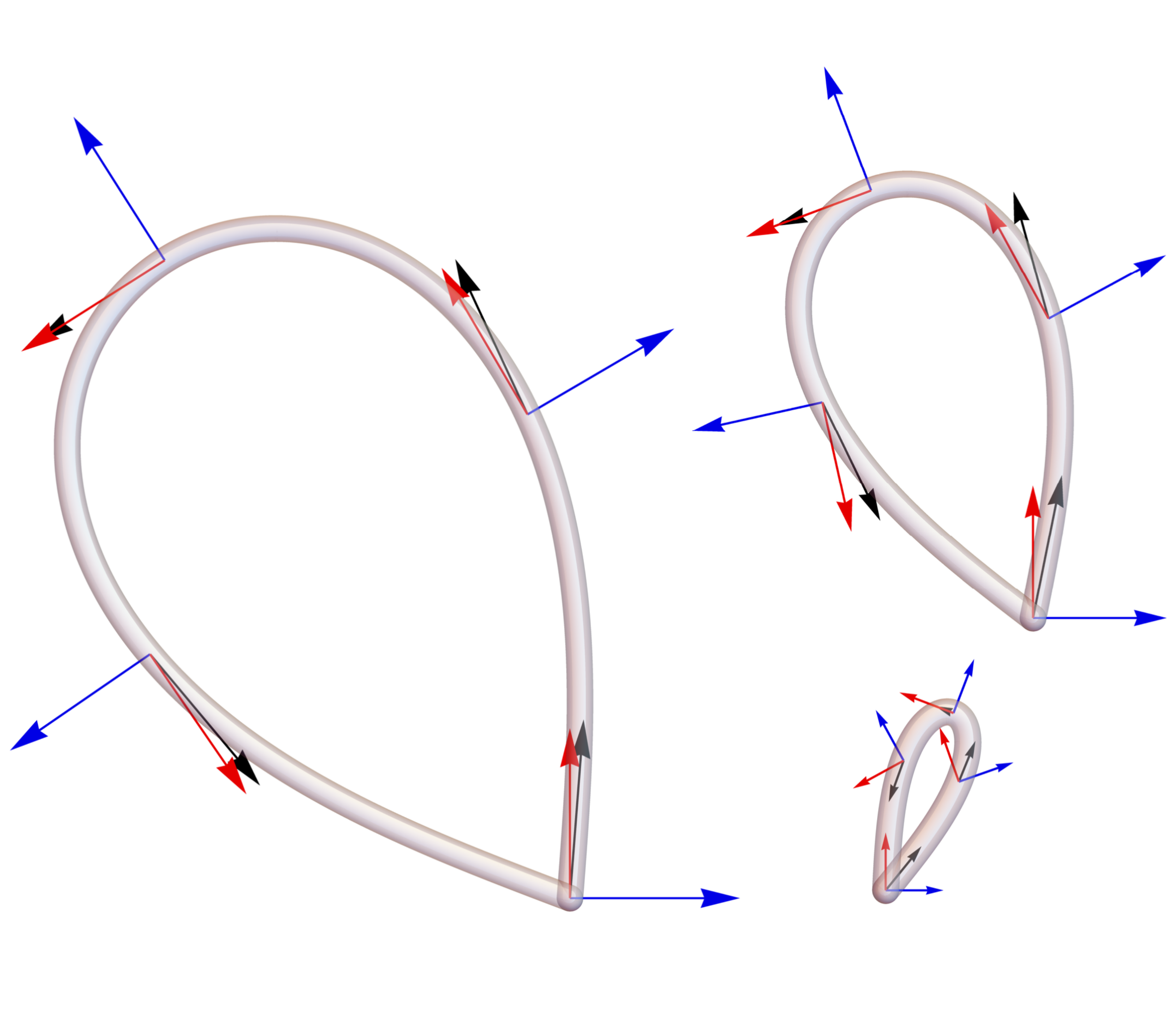}\qquad
}
\subfigure[]{
\includegraphics[width=.44\textwidth]{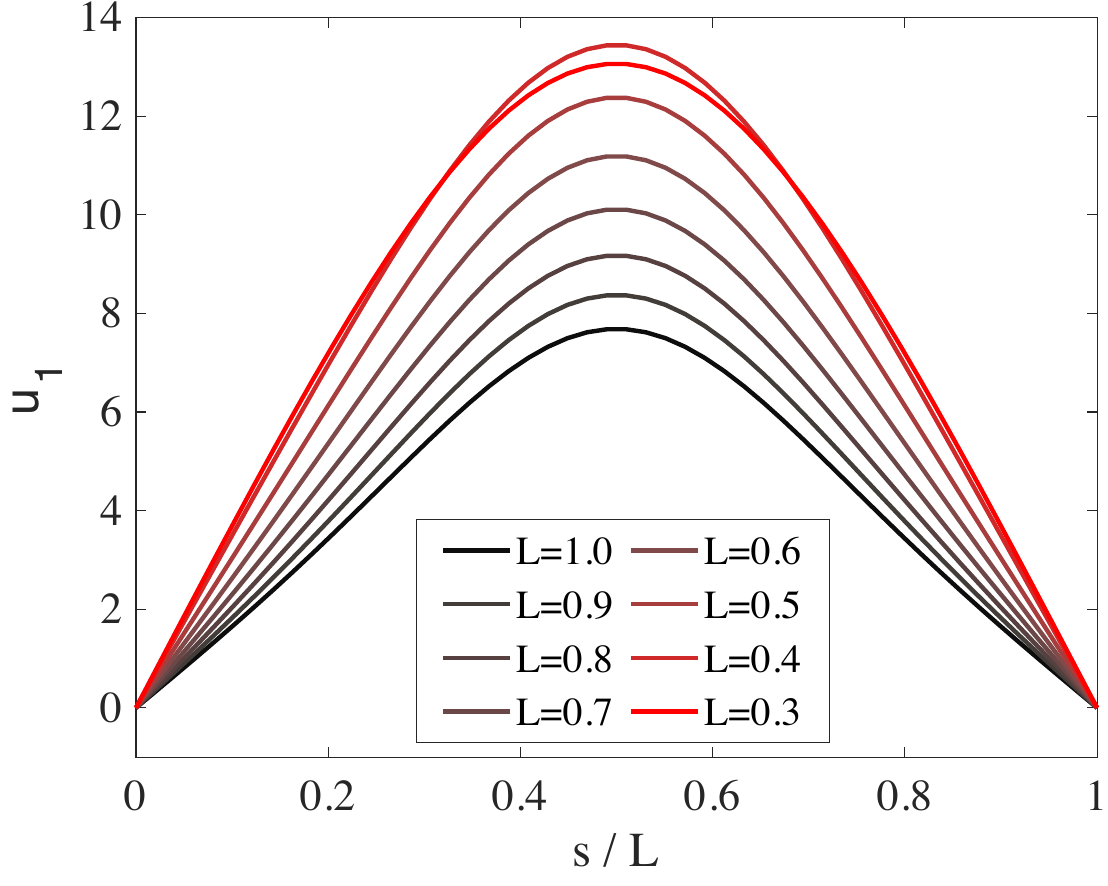}
}

\subfigure[]{
\includegraphics[width=.44\textwidth]{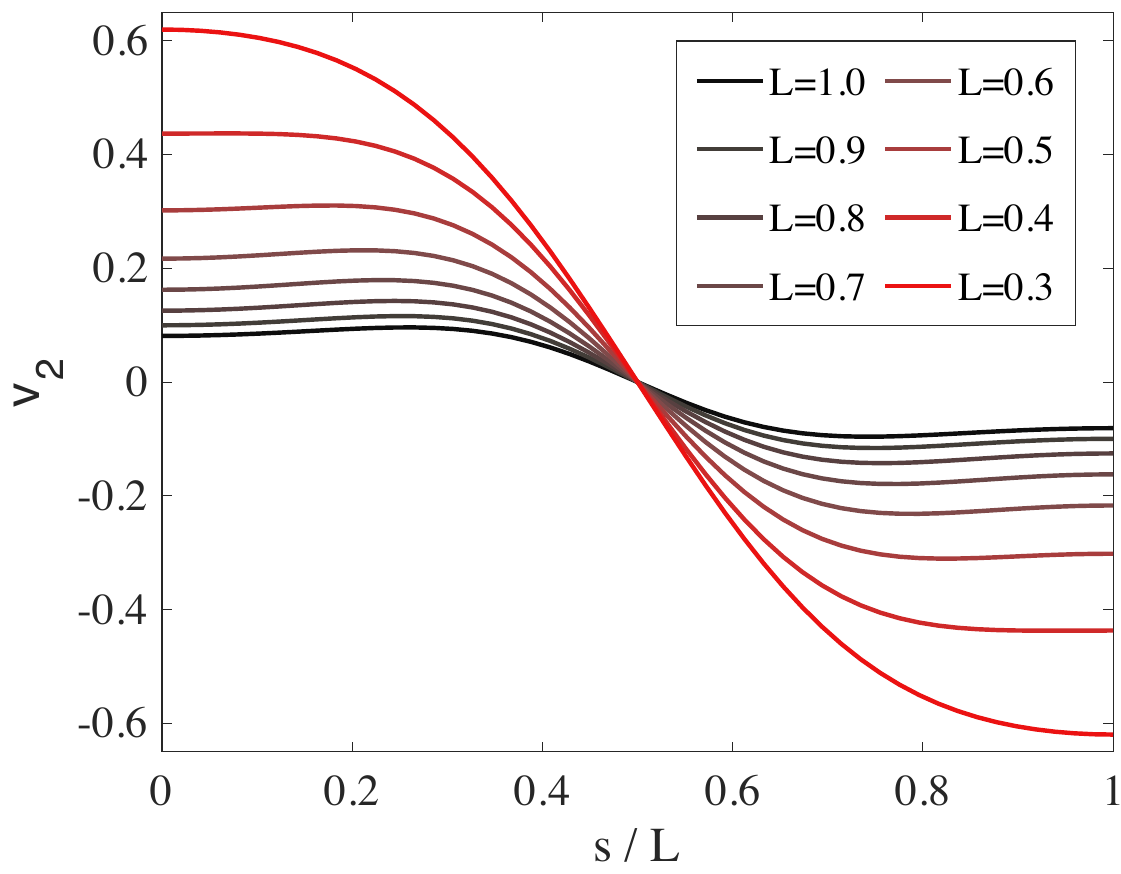}\qquad\,
}
\subfigure[]{
\includegraphics[width=.443\textwidth]{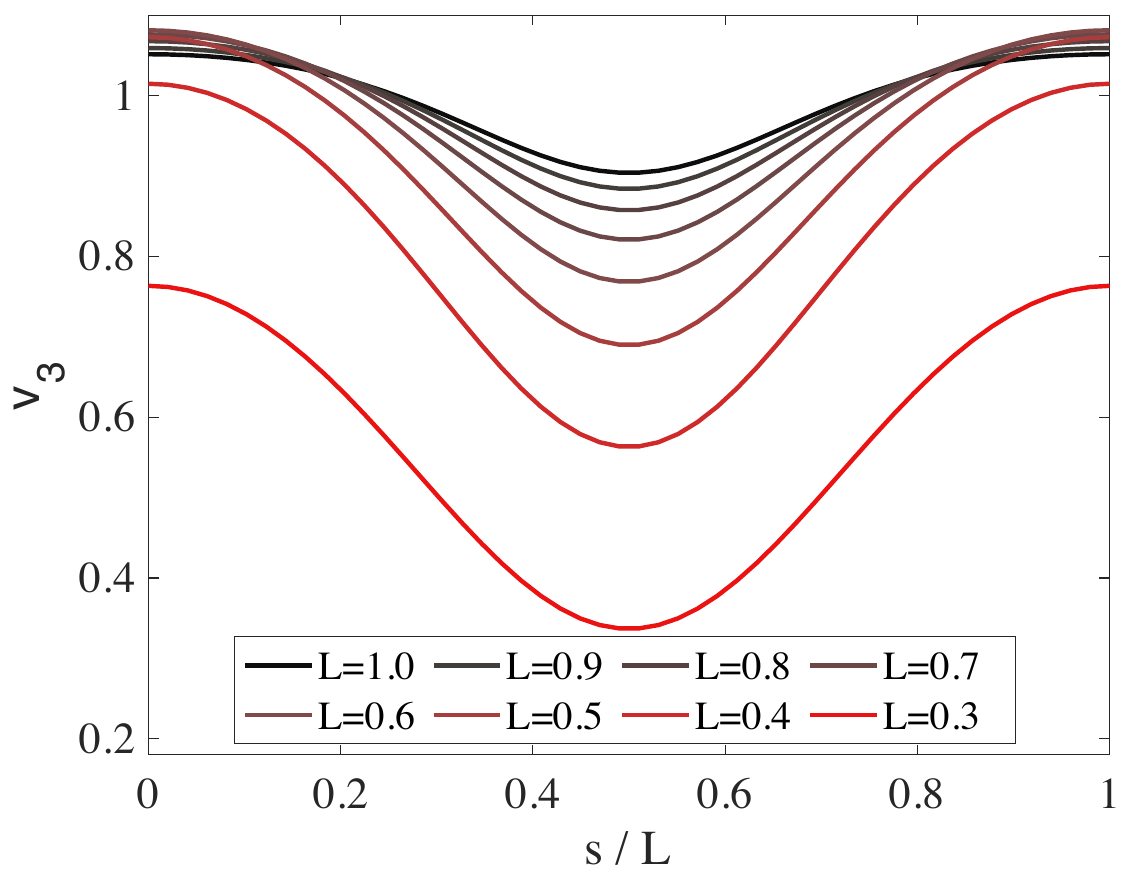}
}
\caption{Analysis of the $(\mathtt{m})$ cyclization problem for a non-isotropic Cosserat rod with $k_1=0.5$, $k_2=5$, $k_3=10$ and $a_1=a_2=a_3=100$. In panel (a) we report the shapes of the teardrop minimizers for $L=1$, $L=0.6$ and $L=0.3$. The tangent $\bm{r}'(s)$ and the vectors $\bm{d}_2(s)$, $\bm{d}_3(s)$ of the moving frame are displayed in black, blue and red respectively. In panels (b), (c) and (d) we plot the bending, shear and stretch components $\mathsf{u}_1^m(s)$, $\mathsf{v}_2^m(s)$ and $\mathsf{v}_3^m(s)$ respectively for a uniformly spaced set of undeformed lengths ranging from $L=1$ to $L=0.3$. The values are computed numerically.}
\label{fig2124}
\end{figure*}

First, we consider a non-isotropic rod ($k_1\neq k_2$), further assuming without loss of generality that $k_1<k_2$. For the case of full looping $(\mathtt{f})$, there exist two circular, untwisted, isolated minima $\bm{{q}}^f$ lying on the $y-z$ plane characterised by
\begin{equation}
\bm{\mathsf{u}}^f=\left(\pm\frac{2\pi}{L},0,0\right)\;,\quad\bm{\mathsf{v}}^f=(0,0,1)\;.
\end{equation}
In particular, the one having non-positive $y$ coordinate is given by 
\begin{equation}
\begin{split}
&\bm{r}^f(s)=\frac{L}{2\pi}\left(0,\cos{\left(\varphi^f(s)\right)}-1,\sin{\left(\varphi^f(s)\right)}\right)\;,\\
\\
&\bm{R}^f(s)=\begin{pmatrix}
1 & 0 & 0\\
0 & \cos{\left(\varphi^f(s)\right)} & -\sin{\left(\varphi^f(s)\right)}\\
0 & \sin{\left(\varphi^f(s)\right)} & \cos{\left(\varphi^f(s)\right)}
\end{pmatrix}\;,
\end{split}
\end{equation}
and the rotation matrix $\bm{R}^f(s)$ is a counter-clockwise planar rotation about the $x$ axis of an angle $\varphi^f(s)=\frac{2\pi s}{L}$, $s\in[0,L]$. Consequently, $\bm{\mathsf{u}}^f=(\frac{2\pi}{L},0,0)$, $\bm{\mathsf{m}}^f=(\frac{2\pi k_1}{L},0,0)$, $\bm{\mathsf{v}}^f=(0,0,1)$, $\bm{\mathsf{n}}^f=\bm{0}$ and the energy is simply computed as
\begin{equation}
E(\bm{{q}}^f)=\frac{2\pi^2 k_1}{L}\;. 
\end{equation}
We observe that these solutions are special for the fact of being the same for both Kirchhoff and Cosserat rods, which is not the case in general. By contrast, there are no simple analytical expressions for the two planar and untwisted teardrop shaped isolated minimizers $\bm{{q}}^m$ involved in the marginal looping problem $(\mathtt{m})$, and elliptic functions or numerics must be used. For example, in the Kirchhoff case, the rotation angle $\varphi^m(s)$ can be derived using elliptic functions in terms of the constant unknown force $\bm{n}^m=(0,n2,n3)$ \citep{AN1, AN2, AN3}. The qualitative shapes of the minimal energy configurations are reported in Fig. \ref{fig1}. It is important to underline that here in $(\mathtt{m})$ the solutions for Kirchhoff and Cosserat rods are different, since the latter are characterised by $\bm{\mathsf{v}}^m(s)=(0,\mathsf{v}_2^m(s)\neq 0,\mathsf{v}_3^m(s)\neq 1)$. More precisely, in Fig. \ref{fig2124} we provide a specific numerical analysis for the Cosserat teardrop solution varying the undeformed length of the rod $L$. We recall that the projection of the tangent $\bm{r}'$ on the director $\bm{d}_2$ is the component $\mathsf{v}_2$ of the shear strain, whereas the projection of the tangent on the director $\bm{d}_3$ is the component $\mathsf{v}_3$ of the stretch.  We observe that the bending and shear components $\mathsf{u}_1^m(s)$, $\mathsf{v}_2^m(s)$ are overall increasing (in the sense of departing from zero) when decreasing $L$, while the stretch $\mathsf{v}_3^m(s)$ decreases and increases (in the sense of departing from one) respectively in the interior and at the boundaries of the interval $[0,L]$. Namely, bending reaches its maximum at $s=L/2$ and vanishes at the boundaries; there is no shear at $s=L/2$ and it is maximized symmetrically within the intervals $[0,L/2)$ and $(L/2,L]$; compression is maximum for $s=L/2$ and slight extension can be observed close to the boundaries. Critical behaviours occur for small values of $L$, where compression dominates and bending starts to decrease.

To be precise, among the equilibria satisfying the boundary conditions for the full and marginal cases, there are also equilibria with figure-eight centerlines, but in the present study their contributions will be neglected because of their higher elastic energy. Moreover, a detailed stability analysis should be carried out, and it is possible to show that the circle and teardrop solutions are stable, with exceptions for the Cosserat rod in the limit of the undeformed length $L$ going to zero, where bifurcations occur. In particular, the ``compressed'' trivial solution $\bm{q}^c$, characterised by 
\begin{equation}
\bm{r}^c=\bm{0},\quad\bm{R}^c=\bm{\mathbb{1}},\quad\bm{\mathsf{u}}^c=\bm{0},\quad\bm{\mathsf{m}}^c=\bm{0},\quad\bm{\mathsf{v}}^c=\bm{0},\quad\bm{\mathsf{n}}^c=(0,0,-a_3),\quad E(\bm{{q}}^c)=\frac{a_3 L}{2},
\end{equation}
starts to play an important role (this is not mentioned in \citep{LUD}). We will show that for the full Cosserat case it exists $L^f>0$ such that the latter solution becomes stable and has lower energy than the circular minimizer $\bm{q}^f$ if $0<L<L^f$. In this regime the system will be mainly driven by the compressed solution (even if the circle remains stable). Moreover, for the marginal Cosserat case, it exists $L^m>0$ such that the stable teardrop solution $\bm{q}^m$ ceases to exist in the interval $0<L<L^m$, merging with the compressed solution which becomes stable. In both the cases, this observation will have a strong impact on the trend of the cyclization probability densities, that is confirmed by Monte Carlo simulations.

In addition to the above statements, the isotropic case requires a more detailed analysis for the presence of a continuous symmetry. Namely, for a general linearly elastic (transversely) isotropic Cosserat rod defined by 
\begin{equation}
\bm{\mathcal{P}}(s)=diag\{k_1,k_2,k_3,a_1,a_2,a_3\}\;\;\text{with}\;\; k_1=k_2,\quad a_1=a_2,\quad \hat{{\mathsf{u}}}_1=\hat{{\mathsf{u}}}_2=\hat{{\mathsf{v}}}_1=\hat{{\mathsf{v}}}_2=0, 
\end{equation}
it is known \citep{RING} that for cyclization boundary conditions $(\mathtt{f})$ in Eq.~(\ref{f}) and $(\mathtt{m})$ in Eq.~(\ref{m}) the equilibria are non-isolated and form a manifold obtained, starting from a known solution, by a rigid rotation of the rod of an angle $\theta$ about the $z$ axis and a subsequent rotation of the framing by an angle $-\theta$ about $\bm{d}_3(s)$, for $\theta\in[0,2\pi)$. As a consequence, in our particular examples, once selected \eg the non-isotropic solution lying in the $y-z$ plane, $y\leq 0$ and characterized by the configuration $(\bm{R}(s),\,\,\bm{r}(s))$, $s\in[0,L]$, then we get an entire family of minimizers 
\begin{equation}
\bm{R}(s;\theta)=\bm{Q}_{\theta}\bm{R}(s)\bm{Q}^T_{\theta}\;,\quad\bm{r}(s;\theta)=\bm{Q}_{\theta}\bm{r}(s)\;, 
\end{equation}
where $\bm{Q}_{\theta}$ is defined as the counter-clockwise planar rotation matrix about the $z$ axis of an angle $\theta\in[0,2\pi)$ (Fig. \ref{fig1}). As a side note for the $(\mathtt{f})$ example, being the circular solutions the same for Kirchhoff and Cosserat rods, the isotropy symmetry arises even if $a_1\neq a_2$.

Furthermore, for a general linearly elastic uniform rod, for which the stiffness matrix $\bm{\mathcal{P}}$ and the intrinsic strains $\hat{\bm{\mathsf{u}}}$, $\hat{\bm{\mathsf{v}}}$ are independent of $s$, another continuous symmetry is present for the cyclization boundary conditions full in Eq.~(\ref{f}). In fact, starting from a known solution characterized by the configuration $(\bm{R}(s),\,\,\bm{r}(s))$, $s\in[0,L]$, it is possible to obtain a family of equilibria parametrised by $s^*\in[0,L)$ in the following way: select $s^*\in[0,L)$, rigidly translate the rod by $-\bm{r}(s^*)$, reparametrise the rod using the parameter $t\in[0,L]$ such that $s=t+s^*$ $(\text{mod}\,\,L)$, rigidly rotate the rod about the origin by means of $\bm{R}^T(s^*)$. However, in our uniform examples, the symmetry of uniformity is not playing any role, due to the circular centerline of the minimizers which is a fixed point of the transformation and, in the marginal case, to the impossibility of satisfying the condition $\bm{m}^m(L)=\bm{0}$ after the application of the symmetry.

In the present paper we will deal with only one symmetry parameter, namely $\theta\in[0,2\pi)$ associated with isotropic rods. Nevertheless, the theory can be applied to the uniformity symmetry alone and generalised to cases in which isotropy and uniformity allow the coexistence of two non-degenerate symmetry parameters $(\theta,\,s^*)$ generating a manifold of equilibria isomorphic to a torus, as it is the case of figure-eight minimizers with full cyclization boundary conditions. Finally, note that in the following theory there is no assumption either of uniformity of the rod, nor, in general, of a straight intrinsic shape.

\section{Fluctuating elastic rods}
\subsection{The path integral approach}
Let us consider now an elastic rod at thermodynamic equilibrium with a heat bath in absence of external forces, assuming without loss of generality that $\bm{q}(0)=\bm{q}_0=(\mathbb{1},\bm{0})$. Then, given a prescribed $\bm{q}_L=(\bm{R}_L,\bm{r}_L)\in SE(3)$, we formulate the problem of computing a conditional probability density function for the other end of the rod to satisfy at $s=L$ either $\bm{q}(L)=\bm{q}_L$, or the weaker condition $\bm{r}(L)=\bm{r}_L$. The first case gives rise to a conditional probability density function $(\mathtt{f})$ over the space $SE(3)$, whereas the second one represents the $\mathbb{R}^3$-valued marginal $(\mathtt{m})$ over the final rotation variable, with no displacement constraint on $\bm{R}(L)$.

If a polymer interacts with a solvent heat bath, the induced thermal motion gives rise to a stochastic equilibrium that we model making use of a Boltzmann distribution on rod configurations satisfying $\bm{q}(0)=\bm{q}_0$ \citep{LUDT, LUD}, of the form 
\begin{equation}
\frac{e^{-\beta E(\bm{q}(s))}}{\mathcal{Z}}\;, 
\end{equation}
with $\beta$ the inverse temperature and $\mathcal{Z}$ the partition function of the system. A precise treatment of the previous expression requires the introduction of the path integral formalism \citep{BookFeynman, BookChaichian, BookSchulman, BookWiegel}. Namely, the $SE(3)$ and $\mathbb{R}^3$ densities $\rrf$ and $\rrm$ are respectively given as the ratios of infinite dimensional Wiener integrals \citep{2020}: 
\begin{equation}\label{dens}
\rrf(\bm{q}_L,L|\bm{q}_0,0)=\frac{\mathcal{K}_f}{\mathcal{Z}}\;,\quad\rrm(\bm{r}_L,L|\bm{q}_0,0)=\frac{\mathcal{K}_m}{\mathcal{Z}}\;,
\end{equation}
\begin{equation}\label{pathint_ch2}
{\mathcal{K}_f}=\int\limits_{\bm{q}(0)=\bm{q}_0}^{\bm{q}(L)=\bm{q}_L}{e^{-\beta E(\bm{q})}\,\mathcal{D}\bm{q}}\;,\,\,\,{\mathcal{K}_m}=\int\limits_{\bm{q}(0)=\bm{q}_0}^{\bm{r}(L)=\bm{r}_L}{e^{-\beta E(\bm{q})}\,\mathcal{D}\bm{q}}\;.
\end{equation}
The limits of integration are dictated by the boundary conditions Eq.~(\ref{f}) and Eq.~(\ref{m}) respectively  and $\mathcal{Z}$ is a path integral over all paths with boundary conditions given in Eq.~(\ref{hat}) that guarantees the normalisation condition:
\begin{equation}\label{normex_ch2}
\mathcal{Z} = \int\limits_{\bm{q}(0)=\bm{q}_0}{e^{-\beta E(\bm{q})}\,\mathcal{D}\bm{q}}\;,
\end{equation}
\begin{equation}
\int_{SE(3)}{\rrf(\bm{q}_L,L|\bm{q}_0,0)}\,\dd \bm{q}_L=\int_{\mathbb{R}^3}{\rrm(\bm{r}_L,L|\bm{q}_0,0)}\,\dd \bm{r}_L=1\;.
\end{equation}
The prescriptions $\bm{m}(L)=\bm{0}$ for $\mathcal{K}_m$ and $\bm{m}(L)=\bm{n}(L)=\bm{0}$ for $\mathcal{Z}$ account for Neumann natural boundary conditions at $s=L$ and concern the minimizers. We stress that it is key that at this stage the model is an extensible, shearable rod, namely with Cosserat energy Eq.~(\ref{energy}), otherwise the problem could not be expressed as simple boundary conditions at $s=0$ and $s=L$. Moreover, to apply all the path integral machinery, we first have to deal with the rotation group $SO(3)$, being part of the configuration variable $\bm{q}(s)=(\bm{R}(s),\bm{r}(s))$, which gives rise to a manifold structure that should be treated carefully in order to recover eventually a ``flat space" formulation.

\subsection{Formulation in SE(3): background}
Below is a revisited version of the set-up introduced in \citep{LUDT}, where a quaternion based analysis of the rotation group is exploited to relate the path integral expression to a known form. In addition, in this work we explicitly include the metric correction by means of appropriate ghost fields, a technique used for exponentiating the measure and formally show how it enters the Laplace expansion. For an explicit evaluation of the path integrals in Eq.~(\ref{pathint_ch2}) and Eq.~(\ref{normex_ch2}), we therefore introduce appropriate coordinates in $SE(3)$. As done originally by Feynman \citep{BookFeynman}, a path integral can be defined via a ``time-slicing'' procedure, or ``parameter-slicing'' in our case, which is to replace the infinite-dimensional integral $\mathcal{D}\bm{q}$ with the limit for $n\rightarrow\infty$ of $n$ iterated finite-dimensional integrals $\prod\limits_{j=1}^{n}{\dd \bm{q}_j}$. These have to be performed on the space of framed curves, whose measure can be chosen to be the product of the Lebesgue measure on the three-dimensional Euclidean space $E(3)$ and of the Haar bi-invariant measure on $SO(3)$, which may be uniquely defined up to a constant factor \citep{BookTung, BookSattinger}. 

In order to avoid difficulty that can arise from the non simple connectivity of $SO(3)$, it is often convenient to consider instead its universal (double) covering $SU(2)$. Any matrix in $SU(2)$ can be parametrized by a quadruple of real numbers 
\begin{equation}
\bm{\gamma}=(\gamma_1,\gamma_2,\gamma_3,\gamma_4)\quad\text{such that}\quad\bm{\gamma}\cdot\bm{\gamma}=1
\end{equation}
living on the unit sphere $S^3$ in $\mathbb{R}^4$. The latter quadruple is know as a unit quaternion or a set of Euler parameters \citep{HAM}. Recalling that by Euler's theorem each element of $SO(3)$ is equivalent to a rotation of an angle $\varphi$ about a unit vector $\bm{w}$, the Euler parameters are expressed as a function of $\varphi$ and $\bm{w}$ as 
\begin{equation}
\gamma_4=\cos{\left(\varphi/{2}\right)}\;,\quad\gamma_i=w_i\sin{\left({\varphi}/{2}\right)}\;,\quad i=1,2,3\;. 
\end{equation}
Hence $\bm{\gamma}$ and $-\bm{\gamma}$ encode the same rotation matrix and the correspondence from $SU(2)$ to $SO(3)$ is 2 to 1. 

Referring to \citep{LUDT}, for parametrising the group of proper rotations we restrict ourself to one hemisphere of the unit sphere $S^3$ in $\mathbb{R}^4$, and we introduce the matrices 
\begin{equation}\label{BMat}
     \bm{B}_1=\begin{pmatrix}
       0 & 0 & 0 & 1 \\
       0 & 0 & 1 & 0 \\
       0 & -1 & 0 & 0 \\
       -1 & 0 & 0 & 0 \\
     \end{pmatrix}\;, \,\,
    \bm{B}_2=\begin{pmatrix}
       0 & 0 & -1 & 0 \\
       0 & 0 & 0 & 1 \\
       1 & 0 & 0 & 0 \\
       0 & -1 & 0 & 0 \\
     \end{pmatrix}\;, \,\,
     \bm{B}_3=\begin{pmatrix}
       0 & 1 & 0 & 0 \\
       -1 & 0 & 0 & 0 \\
       0 & 0 & 0 & 1 \\
       0 & 0 & -1 & 0 \\
     \end{pmatrix}\;,
\end{equation}
satisfying the algebra 
\begin{equation}
\bm{B}_j\bm{B}_k=-\delta_{jk}\bm{\mathbb{1}}-\bm{\epsilon}_{ijk}\bm{B}_i\;,
\end{equation}
where $\bm{\epsilon}_{ijk}$ is the total antisymmetric or Levi-Civita tensor and summation over equal indices is intended. Furthermore, given a unit quaternion $\bar{\bm{\gamma}}$, $\lbrace \bm{B}_1\bar{\bm{\gamma}},\bm{B}_2\bar{\bm{\gamma}},\bm{B}_3\bar{\bm{\gamma}},\bar{\bm{\gamma}}\rbrace$ is an orthonormal basis of $\mathbb{\bm{R}}^4$ and each quadruple of Euler parameters $\bm{\gamma}$ (hence each rotation) can be expressed in coordinates with respect to the latter basis. In particular, for one hemisphere of $S^3$, we consider the new variable $\bm{\mathsf{b}}=(\mathsf{b}_1,\mathsf{b}_2,\mathsf{b}_3)\in {B}_1^3$ living in the open ball of $\mathbb{R}^3$ such that 
\begin{equation}
\bm{\gamma}(\bm{\mathsf{b}})=\sum\limits_{i=1}^3{\mathsf{b}_i\bm{B}_i\bar{\bm{\gamma}}}+\sqrt{1-\Vert \bm{\mathsf{b}}\Vert^2}\bar{\bm{\gamma}}\;.
\end{equation}
Therefore, $\bm{\gamma}(\bm{\mathsf{b}})$ defines a $1$-to-$1$ parametrisation of $SO(3)$, adapted to the rotation expressed by the unit quaternion $\bar{\bm{\gamma}}$, meaning that $\bm{\gamma}(\bm{\mathsf{b}}=\bm{0})=\bar{\bm{\gamma}}$. To be precise, we should remark that the image of such a parametrisation does not include the elements lying on a maximal circle (which depends on $\bar{\bm{\gamma}}$) of the unit sphere in $\mathbb{R}^4$, since $SO(3)$ is not simply connected and rotations about a generic axis of a fixed angle are inevitably neglected.

For Euler parameters, the infinitesimal measure is given by 
\begin{equation}
\dd \bm{q}_j=\delta\left(1-\Vert\bm{\gamma}_j\Vert^2\right)\dd \bm{\gamma}_j\,\dd \bm{r}_j\;, 
\end{equation}
so that the Haar volume measure on $SO(3)$ becomes a surface measure on $S^3$ \citep{BookTung}. Thus, the parametrisation $\bm{\phi}=\bm{\gamma}(\bm{\mathsf{b}}):{B}_1^3\subseteq\mathbb{R}^3\rightarrow\mathcal{M}\subseteq\mathbb{R}^4,$ with $\mathcal{M}$ an hemisphere of $S^3$, naturally induces a metric tensor $\bm{\mathsf{g}}$ on the tangent space at each point of $\mathcal{M}$. Denoting the coordinate vectors as $\bm{\phi}_i=\frac{\partial\bm{\phi}}{\partial\mathsf{b}_i}$, $i=1,2,3$, the components of the metric tensor are given by ${\mathsf{g}}_{i,k}=\bm{\phi}_i\cdot\bm{\phi}_k$, $i,k=1,2,3$, and we get 
\begin{equation}
\bm{\mathsf{g}}(\bm{\mathsf{b}})=\bm{\mathbb{1}}+\frac{\bm{\mathsf{b}}\otimes\bm{\mathsf{b}}}{{1-\Vert\bm{\mathsf{b}}\Vert^2}}\;,\quad\dd \bm{q}_j=\sqrt{\det{[\bm{\mathsf{g}}(\bm{\mathsf{b}}_j)]}}\,\dd \bm{\mathsf{b}}_j\,\dd \bm{r}_j\;,
\end{equation}
with the metric correction being equal to 
\begin{equation}
\sqrt{\det{[\bm{\mathsf{g}}(\bm{\mathsf{b}}_j)]}}=\frac{1}{\sqrt{1-\Vert\bm{\mathsf{b}}_j\Vert^2}}\;.
\end{equation}

Last, in order to deal with variables defined in the whole of $\mathbb{R}^3$, we introduce the Gibbs vector 
\begin{equation}
\bm{\mathsf{c}}=\frac{\bm{\mathsf{b}}}{\sqrt{1-\Vert \bm{\mathsf{b}}\Vert^2}}\;. 
\end{equation}
As a consequence, we have derived a $\bar{\bm{\gamma}}$-adapted parametrization of $SE(3)$ denoted by $\bm{\mathsf{q}}(s)=(\bm{\mathsf{c}}(s),\bm{\mathsf{t}}(s))\in\mathbb{R}^6$ as
\begin{equation}\label{par}
\bm{\gamma}(\bm{\mathsf{c}})=\frac{1}{\sqrt{1+\Vert \bm{\mathsf{c}}\Vert^2}}\left(\sum\limits_{i=1}^3{\mathsf{c}_i\bm{B}_i\bar{\bm{\gamma}}}+\bar{\bm{\gamma}}\right),\,\,\,\bm{\mathsf{t}}=\bm{R}(\bar{\bm{\gamma}})^T\bm{r}\;,
\end{equation}
with $\bm{\mathsf{c}}=(\mathsf{c}_1,\mathsf{c}_2,\mathsf{c}_3)\in\mathbb{\bm{R}}^3$ and $\bm{R}(\bar{\bm{\gamma}})$ the rotation matrix expressed by $\bar{\bm{\gamma}}$.
In particular, exploiting the Feynman discrete interpretation of the path integral measure, \citep{BookFeynman}
\begin{equation}\label{par2_ch2}
\begin{split}
\bm{\mathsf{g}}(\bm{\mathsf{c}})=\frac{\bm{\mathbb{1}}}{1+\Vert\bm{\mathsf{c}}\Vert^2}-\frac{\bm{\mathsf{c}}\otimes\bm{\mathsf{c}}}{\left(1+\Vert\bm{\mathsf{c}}\Vert^2\right)^2}\;,\qquad\qquad\\
\\
\dd \bm{q}_j=\sqrt{\det{[\bm{\mathsf{g}}(\bm{\mathsf{c}}_j)]}}\,\dd \bm{\mathsf{c}}_j\,\dd \bm{\mathsf{t}}_j=\frac{1}{\left(1+\Vert\bm{\mathsf{c}}_j\Vert^2\right)^2}\,\dd \bm{\mathsf{c}}_j\,\dd \bm{\mathsf{t}}_j\;.
\end{split}
\end{equation}

The latter results are implemented by choosing three different curves of unit quaternions $\bar{\bm{\gamma}}(s)$ to be the curves defined by the rotation component ${\bm{R}}(\bar{\bm{\gamma}})$ of the minimizers $\bm{q}^f$, $\bm{q}^m$ and $\hat{\bm{{q}}}$ respectively, which characterise the three different parametrisations involved in the computation of $\mathcal{K}_f$, $\mathcal{K}_m$ and $\mathcal{Z}$ in view of the Laplace approximation. Then, replacing the configuration variable $\bm{q}(s)\in SE(3)$ with the sans-serif fonts $\bm{\mathsf{q}}(s)\in \mathbb{R}^6$, we can formally write the integrand and measure in Eq.~(\ref{pathint_ch2}) and Eq.~(\ref{normex_ch2}) as 
\begin{equation}
e^{-\beta E(\bm{\mathsf{q}})}\sqrt{\det{[\bm{\mathsf{g}}(\bm{\mathsf{c}})]}}\,\mathcal{D}\bm{\mathsf{q}}\;. 
\end{equation}
The treatment of the metric factor relies on the introduction of real-valued ghost fields for exponentiating the measure, as can be found in \citep{GHO}. This means rewriting the factor as a Gaussian path integral in the ghost field $\bm{\mathsf{z}}(s)\in\mathbb{R}^3$ satisfying $\bm{\mathsf{z}}(0)=\bm{0}$ with energy 
\begin{equation}
\frac{1}{2}\int_0^L{\bm{\mathsf{z}}^T\bm{\mathsf{g}}^{-1}(\bm{\mathsf{c}})\bm{\mathsf{z}}\,\dd s}\;. 
\end{equation}
After that, we consider the path integral expressions in the joint variable $\bm{\mathsf{w}}=(\bm{\mathsf{q}},\bm{\mathsf{z}})$, \eg
\begin{equation}\label{curved}
\mathcal{K}_f=\int\limits_{\bm{\mathsf{w}}(0)=(\bm{\mathsf{q}}_0,\bm{0})}^{\bm{\mathsf{q}}(L)=\bm{\mathsf{q}}_L}{e^{-\beta\big[E(\bm{\mathsf{q}}(s))+\frac{1}{2}\int_0^L{\bm{\mathsf{z}}(s)^T\bm{\mathsf{g}}^{-1}(\bm{\mathsf{c}}(s))\bm{\mathsf{z}}(s)\,\dd s}\big]}\,\mathcal{D}\bm{\mathsf{w}}}\;.
\end{equation}

In the following, even if the theory could be given in principle for a general strain energy density $W$, in order to perform concrete computations we refer to the case of linear elasticity, where $W$ is a quadratic function of the shifted strains, driven by the stiffness matrix $\bm{\mathcal{P}}(s)$:
\begin{equation}\label{energylin}
E(\bm{\mathsf{q}})=\frac{1}{2}\int_0^L{     
\begin{pmatrix}
       \bm{\mathsf{u}}-\hat{\bm{\mathsf{u}}} \\
        {\bm{\mathsf{v}}-\hat{\bm{\mathsf{v}}}}
\end{pmatrix}^T 
\begin{pmatrix}
       \bm{\mathcal{K}} & \bm{\mathcal{B}}  \\
       \bm{\mathcal{B}}^T & \bm{\mathcal{A}}
\end{pmatrix}
\begin{pmatrix}
       \bm{\mathsf{u}}-\hat{\bm{\mathsf{u}}} \\
        {\bm{\mathsf{v}}-\hat{\bm{\mathsf{v}}}}
\end{pmatrix}}\dd s\;.
\end{equation}
Moreover, we also refer to the particular looping case of ring-closure or cyclization, evaluating $\rrf$ at $\bm{q}_L=\bm{q}_0$ and the marginal $\rrm$ at $\bm{r}_L=\bm{0}$; the same conditions apply to the minimizers.

\section{Looping probabilities in the case of isolated minimizers}
Since the elastic energy functional Eq.~(\ref{energylin}) is non quadratic in $\bm{\mathsf{q}}$, after the parametrisation we approximate $\mathcal{K}_f$, $\mathcal{K}_m$ and $\mathcal{Z}$ by means of a second-order expansion about a minimal energy configuration \citep{PAP1, MOR, BookWiegel, BookChaichian, BookSchulman, 2020}, known as the semi-classical method, or, in our real-valued context, Laplace expansion \citep{PIT}. The present work follows the set-up of \citep{2020}. We further recall that such an approximation holds when the energy required to deform the system is large with respect to the temperature of the heat bath, \ie in the short-length scale, or stiff, regimes.

First, note that there is no contribution to the result coming from the ghost energy when approximating path integrals of the kind of Eq.~(\ref{curved}) to second order in the joint variable $\bm{\mathsf{w}}$. This is a consequence of the structure of the metric tensor Eq.~(\ref{par2_ch2}), \ie 
\begin{equation}
\bm{\mathsf{g}}^{-1}(\bm{\mathsf{c}})=(1+\bm{\mathsf{c}}\cdot\bm{\mathsf{c}})(\bm{\mathbb{1}}+\bm{\mathsf{c}}\otimes\bm{\mathsf{c}})\;,
\end{equation}
and therefore we can consider only the elastic energy Eq.~(\ref{energylin}) in the variable $\bm{\mathsf{q}}$. In fact, the minima $\bm{q}^f$ and $\bm{q}^m$ (here assumed to be isolated) encoded within the associated adapted parametrisations lead to the minimizers $\qqf$ and $\qqm$ (denoted generically by $\qqa$, $\alpha$ standing for both $f$ and $m$) characterised by $\bm{\mathsf{c}}^{\alpha}=\bm{0}$. In particular, the Neumann natural boundary condition $\bm{m}(L)=\bm{0}$ for $\bm{q}^m$  translates into 
\begin{equation}\label{parmomM}
\frac{\partial W}{\partial\bm{\mathsf{c}}'}(L)=2\left[\frac{\bm{\mathbb{1}}+\bm{\mathsf{c}}^{\times}}{1+\Vert\bm{\mathsf{c}}\Vert^2}\bm{\mathsf{m}}\right](L)=\bm{0}
\end{equation}
for $\qqm$. Note that Eq.~(\ref{parmomM}) can be derived from the computation
\begin{equation}\label{DarEu}
\frac{\partial W}{\partial\bm{\mathsf{c}}'}=\frac{\partial W}{\partial\bm{\mathsf{u}}}\frac{\partial \bm{\mathsf{u}}}{\partial\bm{\mathsf{c}}'}\quad\text{with}\quad\bm{\mathsf{u}}=2\left(\sum\limits_{i=1}^3{\bm{e}_i\otimes\bm{B}_i{\bm{\gamma}}}\right)\bm{\gamma}'\;,
\end{equation}
and we observe from Eq.~(\ref{par}) that $\bm{\gamma}'$ is a function of $\bm{\mathsf{c}}'$, leading to 
\begin{equation}
\frac{\partial W}{\partial\bm{\mathsf{c}}'}=2\bm{\mathsf{m}}\left(\sum\limits_{i=1}^3{\bm{e}_i\otimes\bm{B}_i{\bm{\gamma}}}\right)\frac{\partial \bm{\gamma}}{\partial\bm{\mathsf{c}}}
\end{equation}
which, in column vector notation, gives Eq.~(\ref{parmomM}). In the Laplace approximation for $\mathcal{K}_f$ and $\mathcal{K}_m$ the energy is expanded about the associated $\qqa$ as 
\begin{equation}
E(\bm{\mathsf{q}})\sim E(\qqa)+\frac{1}{2}\delta^2E(\bm{\mathsf{h}};\qqa)\;,\quad\bm{\mathsf{q}}=\qqa+\bm{\mathsf{h}}\;, 
\end{equation}
being the first variation zero. The second variation $\delta^2E$ is reported in Eq.~(\ref{sec}), with $\bm{\mathsf{h}}=(\delta\bm{\mathsf{c}},\delta\bm{\mathsf{t}})$ the perturbation field describing fluctuations about the minimizer $\qqa$ and satisfying the linearised version of the parametrised boundary conditions, \ie 
\begin{equation}
\bm{\mathsf{h}}(0)=\bm{\mathsf{h}}(L)=\bm{0}\quad\text{for}\quad(\mathtt{f})\;,
\end{equation}
\begin{equation}
\bm{\mathsf{h}}(0)=\bm{0}\;,\quad\delta\bm{\mathsf{t}}(L)=\bm{0}\;,\quad\delta\frac{\partial W}{\partial\bm{\mathsf{c}}'}(L)=2[\delta\bm{\mathsf{m}}-\bm{\mathsf{m}}^m\times\delta\bm{\mathsf{c}}](L)=\bm{0}\quad\text{for}\quad(\mathtt{m})\;.
\end{equation}
Analogously, for $\mathcal{Z}$ the energy is expanded about $\qqh$, being 
\begin{equation}\label{parmomN}
\frac{\partial W}{\partial\bm{\mathsf{t}}'}(L)=\frac{\partial W}{\partial\bm{\mathsf{v}}}\frac{\partial \bm{\mathsf{v}}}{\partial\bm{\mathsf{t}}'}(L)=[\hat{\bm{R}}^T\bm{R}\,\bm{\mathsf{n}}](L)=\bm{0} 
\end{equation}
the associated Neumann natural boundary condition arising from $\bm{n}(L)=\bm{0}$ in $\hat{\bm{q}}$ (in addition to the boundary condition for the moment as described for $\qqm$). In this case, the linearised parametrised boundary conditions are given by 
\begin{equation}
\begin{split}
&\qquad\qquad\qquad\qquad\qquad\bm{\mathsf{h}}(0)=\bm{\mathsf{\mu}}(L)=\bm{0}\;, \\
\\
&\bm{\mathsf{\mu}}(L)=\left(\delta\frac{\partial W}{\partial\bm{\mathsf{c}}'},\delta\frac{\partial W}{\partial\bm{\mathsf{t}}'}\right)(L)=\left(2(\delta\bm{\mathsf{m}}-\hat{\bm{\mathsf{m}}}\times\delta\bm{\mathsf{c}}),\delta\bm{\mathsf{n}}-2\hat{\bm{\mathsf{n}}}\times\delta\bm{\mathsf{c}}\right)(L)\;.
\end{split}
\end{equation}

\subsection{Second variation of linearly elastic rods: background}
Computing the second variation for rods in the present case of linear elasticity is regarded as a standard procedure in the context of continuum mechanics. We will provide appropriate citations where the necessary computations are performed. Following the approach of \citep{LUDT, LUD}, the Jacobi fields for the inextensible and unshearable case are recovered as a smooth limit in Hamiltonian form, and we write a unified expression which encompasses both the Cosserat and Kirchhoff cases. 

The second variation Eq.~(\ref{sec}) is characterised by $\bm{\mathsf{P}}$, related to the stiffness matrix $\bm{\mathcal{P}}$, and ${\bm{\mathsf{C}}}$, ${\bm{\mathsf{Q}}}$ which can be computed as follows in terms of strains, forces and moments of the minimizer involved, generically denoted by $\bar{\bm{q}}=({\bm{R}}(\bar{\bm{\gamma}}),\bar{\bm{r}})$. In elastic rod theory, the natural parametrisation for the variation field around $\bar{\bm{q}}$ is directly provided by the Lie algebra $so(3)$ of the rotation group in the director frame, namely 
\begin{equation}
\delta\bm{R}={\bm{R}}(\bar{\bm{\gamma}})\delta\bm{\mathsf{\eta}}^{\times}\;, 
\end{equation}
where $\delta\bm{\mathsf{\eta}}^{\times}$ denotes the skew-symmetric matrix or cross product matrix of $\delta\bm{\mathsf{\eta}}\in\mathbb{R}^3$. In order to show the relation between $\delta\bm{\mathsf{\eta}}$ and the variation field $\delta\bm{\mathsf{c}}$, we use the formula 
\begin{equation}
\delta\bm{\mathsf{\eta}}=2\left(\sum\limits_{i=1}^3{\bm{e}_i\otimes\bm{B}_i\bar{\bm{\gamma}}}\right)\delta\bm{\gamma}
\end{equation}
(which is substantially the relation between the Darboux vector and Euler parameters Eq.~(\ref{DarEu}), see \eg \citep{HAM}) with 
\begin{equation}
\delta\bm{\gamma}=\frac{\partial\bm{\gamma}}{\partial\bm{\mathsf{c}}}\big|_{\bm{\mathsf{c}}=\bm{0}}\delta\bm{\mathsf{c}}=\sum\limits_{j=1}^3\bm{B}_j\bar{\bm{\gamma}}\delta\mathsf{c}_j
\end{equation}
referring to Eq.~(\ref{BMat}), Eq.~(\ref{par}) and we conclude that $\delta\bm{\mathsf{\eta}}(s)=2\,\delta\bm{\mathsf{c}}(s)$. 

With reference to \citep{NAD}, the second variation of the linear hyper-elastic energy Eq.~(\ref{energylin}) in the director variable $\bm{\mathsf{\omega}}=(\delta\bm{\mathsf{\eta}},\delta\bm{\mathsf{t}})$ is 
\begin{equation}
\delta ^2 E=\int_0^L{\left[{\bm{\mathsf{\omega}}'}^T\bm{\mathcal{P}}\bm{\mathsf{\omega}}'+2{\bm{\mathsf{\omega}}'}^T{\bm{\mathcal{C}}}\bm{\mathsf{\omega}}+\bm{\mathsf{\omega}}^T{\bm{\mathcal{Q}}}\bm{\mathsf{\omega}}\right]}\,\dd s\;,
\end{equation}
where $\bm{\mathcal{P}}$ is the stiffness matrix and $\bm{\mathcal{C}}$, $\bm{\mathcal{Q}}$ are respectively given in terms of strains, forces and moments by 
\begin{equation}\label{Cmat}
\begin{split}
\bm{\mathcal{C}}=\begin{pmatrix}
        \bm{\mathcal{K}}\bm{\mathsf{u}}^{\times}+\bm{\mathcal{B}}\bm{\mathsf{v}}^{\times}-\frac{1}{2}\bm{\mathsf{m}}^{\times}, &\,\,\,\, \bm{\mathcal{B}}\bm{\mathsf{u}}^{\times}  \\
       \\
       \bm{\mathcal{A}}\bm{\mathsf{v}}^{\times}+\bm{\mathcal{B}}^T\bm{\mathsf{u}}^{\times}-\bm{\mathsf{n}}^{\times}, &\,\,\,\, \bm{\mathcal{A}}\bm{\mathsf{u}}^{\times}
     \end{pmatrix}\;,\,\,\,
\bm{\mathcal{Q}}=\begin{pmatrix}
       \bm{\mathcal{Q}}_{1,1}, & \bm{\mathcal{Q}}_{1,2}  \\
       \\
       \bm{\mathcal{Q}}^T_{1,2}, & \bm{\mathcal{Q}}_{2,2}
     \end{pmatrix}\;,\,\,\,\text{with}
\end{split}
\end{equation}
\begin{small}
\begin{equation}\label{Qmat}
\begin{split}
&\bm{\mathcal{Q}}_{1,1}=\frac{1}{2}(\bm{\mathsf{n}}^{\times}\bm{\mathsf{v}}^{\times}+\bm{\mathsf{v}}^{\times}\bm{\mathsf{n}}^{\times})+\frac{1}{2}(\bm{\mathsf{m}}^{\times}\bm{\mathsf{u}}^{\times}+\bm{\mathsf{u}}^{\times}\bm{\mathsf{m}}^{\times})-\bm{\mathsf{u}}^{\times}\bm{\mathcal{K}}\bm{\mathsf{u}}^{\times}-\bm{\mathsf{v}}^{\times}\bm{\mathcal{A}}\bm{\mathsf{v}}^{\times}-\bm{\mathsf{u}}^{\times}\bm{\mathcal{B}}\bm{\mathsf{v}}^{\times}-\bm{\mathsf{v}}^{\times}\bm{\mathcal{B}}^T\bm{\mathsf{u}}^{\times}\;,\\
&\bm{\mathcal{Q}}_{1,2}=-\bm{\mathsf{u}}^{\times}\bm{\mathcal{B}}\bm{\mathsf{u}}^{\times}-\bm{\mathsf{v}}^{\times}\bm{\mathcal{A}}\bm{\mathsf{u}}^{\times}+\bm{\mathsf{n}}^{\times}\bm{\mathsf{u}}^{\times}\;,\,\,\,\,\bm{\mathcal{Q}}_{2,2}=-\bm{\mathsf{u}}^{\times}\bm{\mathcal{A}}\bm{\mathsf{u}}^{\times}\;.
\end{split}
\end{equation}
\end{small}
Finally, introducing the matrix 
\begin{equation}
\bm{\mathcal{D}}=\begin{pmatrix}
       2\mathbb{1} & \mathbb{0}  \\
       \mathbb{0} & \mathbb{1}
     \end{pmatrix}\in\mathbb{R}^{6\times 6}\;,
\end{equation}
we have that the second variation in the variable $\bm{\mathsf{h}}=(\delta\bm{\mathsf{c}},\delta\bm{\mathsf{t}})$ Eq.~(\ref{sec}) is given by $\delta ^2 E=\int_0^L{[{\bm{\mathsf{h}}'}^T\bm{\mathsf{P}}\bm{\mathsf{h}}'+2{\bm{\mathsf{h}}'}^T{\bm{\mathsf{C}}}\bm{\mathsf{h}}+\bm{\mathsf{h}}^T{\bm{\mathsf{Q}}}\bm{\mathsf{h}}]}\,\dd s$, with $\bm{\mathsf{P}}=\bm{\mathcal{D}}\bm{\mathcal{P}}\bm{\mathcal{D}}$, $\bm{\mathsf{C}}=\bm{\mathcal{D}}\bm{\mathcal{C}}\bm{\mathcal{D}}$ and $\bm{\mathsf{Q}}=\bm{\mathcal{D}}\bm{\mathcal{Q}}\bm{\mathcal{D}}$.

The Jacobi equations in first-order Hamiltonian form associated with the latter second variation functional are given in Eq.~(\ref{JacHam}) and are driven by the symmetric matrix 
\begin{equation}\label{E1}
\bm{\mathsf{E}}(s)=\begin{pmatrix}
     \bm{\mathsf{E}}_{1,1}={\bm{\mathsf{C}}}^T{\bm{\mathsf{P}}}^{-1}{\bm{\mathsf{C}}}-{\bm{\mathsf{Q}}}\,\,\,\,\, & \bm{\mathsf{E}}_{1,2}=-{\bm{\mathsf{C}}}^T{\bm{\mathsf{P}}}^{-1}\\
     \\
       \bm{\mathsf{E}}_{2,1}=\bm{\mathsf{E}}_{1,2}^{^T}\,\,\,\,\,\, & \bm{\mathsf{E}}_{2,2}={\bm{\mathsf{P}}}^{-1} 
       \end{pmatrix}\in\mathbb{R}^{12\times 12}\;.
\end{equation}
The Jacobi fields $\bm{\mathsf{H}}(s)\in\mathbb{R}^{6\times 6}$, together with the conjugate variable under the Legendre transform $\bm{\mathsf{M}}(s)\in\mathbb{R}^{6\times 6}$ represent the solutions of the Jacobi equations once prescribed appropriate initial conditions. The columns $\bm{\mathsf{h}}$ of $\bm{\mathsf{H}}$ and the ones $\bm{\mathsf{\mu}}$ of $\bm{\mathsf{M}}$ are related by $\bm{\mathsf{\mu}}=\bm{\mathsf{P}}\bm{\mathsf{h}}'+\bm{\mathsf{C}}\bm{\mathsf{h}}$.

Note that until now the formulation adopted is for the general Cosserat rod with extension, shear and hence an invertible stiffness matrix $\bm{\mathcal{P}}$. The constrained inextensible and unshearable Kirchhoff case requires the stiffness components $\bm{\mathcal{B}}$ and $\bm{\mathcal{A}}$ to diverge (as discussed in \citep{HAM, LUDT, LUD}), specifically as $\bm{\mathcal{B}}/{\epsilon}$ and $\bm{\mathcal{A}}/{\epsilon^2}$, for $\epsilon\rightarrow 0$. Switching to the Hamiltonian formulation, given a Cosserat rod the compliance matrix $\bm{\mathcal{R}}$ (which is the inverse of $\bm{\mathcal{P}}$) has a smooth limit for $\epsilon\rightarrow 0$. Namely, for a Kirchhoff rod we recover 
\begin{equation}
\bm{\mathcal{R}}(s)=\begin{pmatrix}
       \bm{\mathcal{R}}_{1,1} & \bm{\mathcal{R}}_{1,2}=\bm{\mathbb{0}}\\
        \bm{\mathcal{R}}_{2,1}=\bm{\mathbb{0}} &  \bm{\mathcal{R}}_{2,2}=\bm{\mathbb{0}}
     \end{pmatrix}\;,\quad\bm{\mathcal{R}}_{1,1}(s)=(\bm{\mathcal{K}}-\bm{\mathcal{B}}\bm{\mathcal{A}}^{-1}\bm{\mathcal{B}}^T)^{-1}\;.
\end{equation}
In conclusion, once prescribed a symmetric and positive definite matrix $\bm{\mathcal{K}}^{(\mathtt{K})}=\bm{\mathcal{R}}_{1,1}^{-1}$, there exists a sequence of positive definite and symmetric compliance matrices for the Cosserat case converging smoothly to the Kirchhoff case, implying that the expressions 
\begin{small}
\begin{equation}\label{E2}
\begin{split}
&\bm{\mathsf{E}}_{1,1}=\bm{\mathcal{D}}\begin{pmatrix}
* & \bm{\mathbb{0}}  \\
\bm{\mathbb{0}} & \bm{\mathbb{0}}
\end{pmatrix}\bm{\mathcal{D}}\;,\\
\\
&*=\frac{1}{2}(\bm{\mathsf{n}}^{\times}\bm{\mathsf{v}}^{\times}+\bm{\mathsf{v}}^{\times}\bm{\mathsf{n}}^{\times})-\frac{1}{4}\bm{\mathsf{m}}^{\times}\bm{\mathcal{R}}_{1,1}\bm{\mathsf{m}}^{\times}-\frac{1}{2}(\bm{\mathsf{m}}^{\times}\bm{\mathcal{R}}_{1,2}\bm{\mathsf{n}}^{\times}+\bm{\mathsf{n}}^{\times}\bm{\mathcal{R}}^T_{1,2}\bm{\mathsf{m}}^{\times})-\bm{\mathsf{n}}^{\times}\bm{\mathcal{R}}_{2,2}\bm{\mathsf{n}}^{\times}\;,
\end{split}
\end{equation}
\end{small}
\begin{equation}\label{E3}
\begin{split}
&\bm{\mathsf{E}}_{1,2}=\bm{\mathcal{D}}\begin{pmatrix}
\bm{\mathsf{u}}^{\times}-\frac{1}{2}\bm{\mathsf{m}}^{\times}\bm{\mathcal{R}}_{1,1}-\bm{\mathsf{n}}^{\times}\bm{\mathcal{R}}_{1,2}^T, &\,\,\,\, \bm{\mathsf{v}}^{\times}-\frac{1}{2}\bm{\mathsf{m}}^{\times}\bm{\mathcal{R}}_{1,2}-\bm{\mathsf{n}}^{\times}\bm{\mathcal{R}}_{2,2}  \\
 \\ \bm{\mathbb{0}}, &\,\,\,\, \bm{\mathsf{u}}^{\times}
\end{pmatrix}\bm{\mathcal{D}}^{-1}\;,\\
\\
&\bm{\mathsf{E}}_{2,2}=\bm{\mathcal{D}}^{-1}\bm{\mathcal{R}}\bm{\mathcal{D}}^{-1}\;,
\end{split}
\end{equation}
for the blocks of the matrix $\bm{\mathsf{E}}(s)$ Eq.~(\ref{E1}) hold for both Cosserat and Kirchhoff rods. We emphasise that for the Kirchhoff case $\delta\frac{\partial W}{\partial\bm{\mathsf{t}}'}$ is a basic unknown of the Jacobi equations and cannot be found using the relation $\bm{\mathsf{\mu}}=\bm{\mathsf{P}}\bm{\mathsf{h}}'+\bm{\mathsf{C}}\bm{\mathsf{h}}$, since the latter is not defined.

\subsection{Laplace approximation final formulas}
The results that we are going to show about the full looping conditional probability in the case of isolated minimizers are derived in \citep{LUDT}, where the Gaussian path integrals are carried out in the variables $\bm{\mathsf{\omega}}=(\delta\bm{\mathsf{\eta}},\delta\bm{\mathsf{t}})$ instead of $\bm{\mathsf{h}}=(\delta\bm{\mathsf{c}},\delta\bm{\mathsf{t}})$ as done here. As a consequence of the latter choice, in \citep{LUDT} all the formulas have a factor of $8$ in front corresponding to the Jacobian factor of the transformation (actually in the cited work a factor of $2$ is present, but it is a typographical error, it should be $8$). Furthermore, in \citep{LUDT} the evaluation of the partition function is performed by means of an appropriate change of variables. By contrast here we prefer to adopt the Laplace expansion approach, which can be easily extended to non-quadratic energies. The investigation regarding marginal looping probabilities is instead original of the present work; in the following we present all the cases together with the associated approximated formulas when the minimizers entering the Laplace method are isolated.

The resulting path integrals arising from the Laplace method are of the form, \eg
\begin{equation}\label{Gauss}
\mathcal{K}_f\approx e^{-\beta E(\qqf)}\int\limits_{\bm{\mathsf{h}}(0)=\bm{0}}^{\bm{\mathsf{h}}(L)=\bm{0}}{e^{-\frac{\beta}{2}\delta^2 E(\bm{\mathsf{h}};\qqf)}\,\mathcal{D}\bm{\mathsf{h}}}\;,
\end{equation}
and similarly for $\mathcal{K}_m$ and $\mathcal{Z}$ but considering the different minimizers and linearised boundary conditions. Then, applying the results derived in \citep{2020} for Gaussian path integrals, namely Eq.~(\ref{fin2}), Eq.~(\ref{fin3}), Eq.~(\ref{fin22}), Eq.~(\ref{fin33}), which are in turn extensions of the work of Papadopoulos \citep{PAP1}, the approximate form of the conditional probability density reads as
\begin{equation}\label{fin}
\rra\approx 
\left(\frac{\beta}{2\pi}\right)^{x(\alpha)}
\frac{e^{-\beta E\left(\qqa\right)}}{\sqrt{\det{[\hha(0)]}}}\;,
\end{equation}
with $x(f)=3$, $x(m)=3/2$, and we are interested in the cyclization values\\ $\rrf(\bm{q}_0,L|\bm{q}_0,0)$, $\rrm(\bm{0},L|\bm{q}_0,0)$. We denote by $\hha(s)$ the Jacobi fields computed at $\qqa$, solutions of the associated Jacobi equations Eq.~(\ref{JacHam}) with $\bm{\mathsf{E}}(s)$ reported in Eq.~(\ref{E2}), Eq.~(\ref{E3}) and initial conditions given at $s=L$ as
\begin{equation}\label{inFM}
\begin{split}
\hhf(L)=\mathbb{0}\;,\,\,\mmf(L)=-\mathbb{1}\;;\quad\qquad\qquad\\
\\
\hhm(L)=
\begin{pmatrix}
     \mathbb{1}_{3\times 3} & \mathbb{0}_{3\times 3}\\
      \mathbb{0}_{3\times 3} & \mathbb{0}_{3\times 3}
       \end{pmatrix}\;,\,\,\mmm(L)=\begin{pmatrix}
    \mathbb{0}_{3\times 3} & \mathbb{0}_{3\times 3}\\
      \mathbb{0}_{3\times 3} & -\mathbb{1}_{3\times 3}
       \end{pmatrix}\;.
\end{split}
\end{equation}

In principle, denoting by $\hhh(s)$ $\in\mathbb{R}^{6\times 6}$ the Jacobi fields computed at $\hat{\bm{\mathsf{q}}}$ subject to the initial conditions 
\begin{equation}
\hhh(L)=\mathbb{1}\;,\quad\mmh(L)=\mathbb{0}\;, 
\end{equation}
the numerator and denominator in Eq.~(\ref{fin}) should be respectively 
\begin{equation}
e^{-\beta\left(E\left(\qqa\right)-E\left(\qqh\right)\right)}\quad\text{and}\quad\sqrt{\det{\left[\hha\hhh^{-1}(0)\right]}}\;, 
\end{equation}
in order to include the contribution coming from the evaluation of the partition function $\mathcal{Z}$. However, the result simplifies since $E(\hat{\bm{\mathsf{q}}})=0$, being $\hat{\bm{\mathsf{q}}}$ the intrinsic configuration of the rod. At the same time $\bm{\mathsf{E}}_{1,1}$ is the zero matrix for this case, which implies $\mmh(s)=\bm{\mathbb{0}}\,\,\forall s$ (according to the initial conditions $\mmh(L)=\bm{\mathbb{0}}$) and consequently $\hhh(s)$ must satisfy a linear system whose matrix has zero trace. Thus, by application of the generalized Abel's identity or Liouville's formula, $\forall s$ we have that 
\begin{equation}
\det[{{\hhh}}(s)]=\det[{{\hhh}}(L)]=\det[\bm{\mathbb{1}}]=1\;. 
\end{equation}
Furthermore, it is worth to mention that here the partition function computation is not affected by approximations, even if it apparently undergoes the Laplace expansion. In fact, there exists a change of variables presented in \citep{LUDT, LUD} which allows an equivalent exact computation exploiting the specific boundary conditions involved in $\mathcal{Z}$. In general, the latter change of variables is not applicable and the present method must be used, \eg for non-linear elasticity or in the case of a linearly elastic polymer subject to external end-loadings, for which the shape of the energy leads to a non-trivial contribution of the partition function that must be approximated.

\section{Looping in the case of non-isolated minimizers}
In this section we consider non-isolated minimizers arising as a consequence of continuous symmetries of the problem. In particular, we provide a theory for one symmetry parameter, namely $\theta\in[0,2\pi)$ (as we want to deal with isotropic rods), but the same scheme can be suitably generalised to more symmetry parameters. The presence of a family of minimizers denoted by $\qqa(s;\theta)$ translates into a zero mode 
\begin{equation}
\bm{\mathsf{\psi}}^{\alpha}(s;\theta)=\frac{\partial}{\partial\theta}\qqa(s;\theta)
\end{equation}
\citep{MOR} of the self-adjoint operator 
\begin{equation}\label{secOp}
\bm{\mathsf{S}}=-\bm{\mathsf{P}}\frac{d^2}{d s^2}+(\bm{\mathsf{C}}^T-\bm{\mathsf{C}}-\bm{\mathsf{P}}')\frac{\dd}{\dd  s}+\bm{\mathsf{Q}}-\bm{\mathsf{C}}'
\end{equation}
associated with the second variation Eq.~(\ref{sec}), namely 
\begin{equation}
\delta ^2 E=(\bm{\mathsf{h}},\bm{\mathsf{S}}\bm{\mathsf{h}})\;, 
\end{equation}
where $(\cdot,\cdot)$ is the scalar product in the space of square-integrable functions $L^2([0,L];\mathbb{R}^6)$. Consequently, we cannot proceed as before, for otherwise expression Eq.~(\ref{fin}) will diverge for the existence of a conjugate point at $s=0$. This is because the zero mode solves the Jacobi equation $\bm{\mathsf{S}}\bm{\mathsf{\psi}}^{\alpha}(s;\theta)=\bm{0}$. Moreover, by definition it satisfies the linearised boundary condition $\bm{\mathsf{\psi}}^{\alpha}(0;\theta)=\bm{0}$ and it is compatible with the initial conditions Eq.~(\ref{inFM}), which imply that the columns of $\hha(0)$ are not linearly independent and therefore the determinant vanishes.

Thus, in evaluating expression Eq.~(\ref{pathint_ch2}) for $\mathcal{K}_f$ and $\mathcal{K}_m$, we adapt the parametrization to the minimizer corresponding to $\theta=0$, our choice of the gauge in applying the collective coordinates method, which amounts to a Faddeev-Popov-type procedure \citep{FADPOP}, widely used in the context of quantum mechanics for solitons or instantons \citep{POLYA, COL1, BER, JAR}, of inserting the Dirac delta transformation identity
\begin{equation}\label{Fad-Pop}
1=\left\lvert\frac{\partial}{\partial\theta}F\right\rvert_{\theta=0}\int{\delta(F(\theta))}\,\dd \theta,\,\,\,F(\theta)=\left(\bm{\mathsf{q}}-\qqa,\frac{\bm{\mathsf{\psi}}^{\alpha}}{\Vert\bm{\mathsf{\psi}}^{\alpha}\Vert}\right)
\end{equation}
within the path integral, in order to integrate over variations which are orthogonal to the zero mode. In fact, for a smooth function $g(x)$ having a zero at $x_0$, the generalised scaling property for the Dirac delta can be written as 
\begin{equation}
\delta(g(x))=\frac{\delta(x-x_0)}{\vert g'(x_0)\vert}\;.
\end{equation}
Once performed the Laplace expansion as before about $\qqa$, exchanged the order of integration $\mathcal{D}\bm{\mathsf{h}}\leftrightarrow\text{d}\theta$ to get a contribution of $2\pi$, and having approximated to leading order both the metric tensor and the factor 
\begin{equation}
\left\lvert\frac{\partial}{\partial\theta}F\right\rvert_{\theta=0}\approx \Vert\bm{\mathsf{\psi}}^{\alpha}(s;0)\Vert\;,
\end{equation}
we are left with the computation of a ratio of Gaussian path integrals
\begin{equation}\label{Isopathint}
\frac{1}{\mathcal{Z}_g}\int{e^{-\pi(\bm{\mathsf{h}},\frac{\beta}{2\pi}\bm{\mathsf{S}}\bm{\mathsf{h}})}\delta\left[\left(\bm{\mathsf{h}},\frac{\bm{\mathsf{\psi}}^{\alpha}}{\Vert\bm{\mathsf{\psi}}^{\alpha}\Vert}\right)\right]\,\mathcal{D}\bm{\mathsf{h}}}\;,
\end{equation}
for the linearised parametrised boundary conditions associated with Eq.~(\ref{f}) and Eq.~(\ref{m}) respectively. For notation simplicity, throughout this section $\bm{\mathsf{S}}$ stands for $\frac{\beta}{2\pi}\bm{\mathsf{S}}$ and $\hat{\bm{\mathsf{S}}}$ for $\frac{\beta}{2\pi}\hat{\bm{\mathsf{S}}}$, the latter operator driving the Gaussian path integral $\mathcal{Z}_g$ arising from the partition function $\mathcal{Z}$, in which the minimizer $\qqh$ is isolated. Note that, since the argument of the delta distribution must vanish for $\theta=0$ according to Eq.~(\ref{Fad-Pop}), then the integration for the numerator is performed on the minimizer $\qqa(s;0)$ with associated zero mode $\bm{\mathsf{\psi}}^{\alpha}(s;0)$; in the following they will be both denoted simply by $\qqa$ and $\bm{\mathsf{\psi}}^{\alpha}$.

\subsection{Functional determinants and Forman's theorem}
Interpreting Eq.~(\ref{Isopathint}) as 
\begin{equation}
\sqrt{\frac{\text{Det}(\hat{\bm{\mathsf{S}}})}{{\text{Det}^{\star}({\bm{\mathsf{S}}})}}}\;, 
\end{equation}
\textit{i.e.}, the square root of the ratio of the functional determinants for the operators $\hat{\bm{\mathsf{S}}}$ and ${\bm{\mathsf{S}}}$, the latter with removed zero eigenvalue (thus the $^{\star}$ sign) \citep{MCK, FAL}, we consider the following general strategy for its evaluation. Given the second variation operator ${\bm{\mathsf{S}}}$ acting on $\bm{\mathsf{h}}(s)\in\mathbb{R}^6$, with $s\in[0,L]$ and boundary conditions determined by the square matrices $\bm{\mathsf{T}}_0$ and $\bm{\mathsf{T}}_L$ in $\mathbb{R}^{12\times 12}$ as 
\begin{equation}
\bm{\mathsf{T}}_0\begin{pmatrix}
      \bm{\mathsf{h}}(0) \\
       \bm{\mathsf{\mu}}(0)
     \end{pmatrix}+\bm{\mathsf{T}}_L\begin{pmatrix}
      \bm{\mathsf{h}}(L) \\
       \bm{\mathsf{\mu}}(L)
     \end{pmatrix}=\bm{0}\;, 
\end{equation}
we state Forman's theorem \citep{FORM} in Hamiltonian form as 
\begin{equation}\label{For}
\frac{\text{Det}({\bm{\mathsf{S}}})}{\text{Det}(\hat{\bm{\mathsf{S}}})}=\frac{\det{[\bm{\mathsf{T}}_0\bm{\mathsf{W}}(0)+\bm{\mathsf{T}}_L\bm{\mathsf{W}}(L)]}}{\det[{\bm{\mathsf{W}}(L)}]}\;,
\end{equation}
for $\bm{\mathsf{W}}(s)\in\mathbb{R}^{12\times 12}$ whose columns $
(\bm{\mathsf{h}},\bm{\mathsf{\mu}})^T$ solve the homogeneous problem ${\bm{\mathsf{S}}}\bm{\mathsf{h}}=\bm{0}$ (\textit{i.e.}, the Jacobi equations Eq.~(\ref{JacHam}) with the extra $\frac{\beta}{2\pi}$ factor, completed as $\bm{\mathsf{W}}'=\bm{J}\bm{\mathsf{E}}\bm{\mathsf{W}}$), and the trivial partition function contribution has already been evaluated. It is important to note the freedom of choosing $\bm{\mathsf{W}}(0)$, $\bm{\mathsf{W}}(L)$ consistently; the latter statements are justified by the following considerations.

Given two matrix differential operators 
\begin{equation}
\bm{\mathsf{\Omega}}=\bm{\mathsf{G}}_0(s)\frac{\dd^2}{\dd s^2}+\bm{\mathsf{G}}_1(s)\frac{\dd}{\dd s}+\bm{\mathsf{G}}_2(s)\quad\text{and}\quad\hat{\bm{\mathsf{\Omega}}}=\bm{\mathsf{G}}_0(s)\frac{\dd^2}{\dd s^2}+\hat{\bm{\mathsf{G}}}_1(s)\frac{\dd}{\dd s}+\hat{\bm{\mathsf{G}}}_2(s)
\end{equation}
with non-zero eigenvalues (with respect to the boundary conditions), acting on $\bm{\mathsf{h}}(s)\in\mathbb{R}^d$, where $\bm{\mathsf{G}}_0$, $\bm{\mathsf{G}}_1$, $\hat{\bm{\mathsf{G}}}_1$, $\bm{\mathsf{G}}_2$, $\hat{\bm{\mathsf{G}}_2}\in\mathbb{R}^{d\times d}$, $\bm{\mathsf{G}}_0$ is invertible and $s\in[a,b]$, the results of Forman \citep{FORM} provide a simple way of computing the ratio of functional determinants $\text{Det}(\bm{\mathsf{\Omega}})/\text{Det}(\hat{\bm{\mathsf{\Omega}}})$, once prescribed the boundary conditions 
\begin{equation}
\bm{\mathsf{I}}_a\begin{pmatrix}
      \bm{\mathsf{h}}(a) \\
      \bm{\mathsf{h}}'(a)
     \end{pmatrix}+\bm{\mathsf{I}}_b\begin{pmatrix}
      \bm{\mathsf{h}}(b) \\
       \bm{\mathsf{h}}'(b)
     \end{pmatrix}=\bm{0}\quad\text{for}\quad\bm{\mathsf{\Omega}}\quad\text{and}
\end{equation}
\begin{equation}
\hat{\bm{\mathsf{I}}}_a\begin{pmatrix}
      \bm{\mathsf{h}}(a) \\
      \bm{\mathsf{h}}'(a)
     \end{pmatrix}+\hat{\bm{\mathsf{I}}}_b\begin{pmatrix}
      \bm{\mathsf{h}}(b) \\
       \bm{\mathsf{h}}'(b)
     \end{pmatrix}=\bm{0}\quad\text{for}\quad\hat{\bm{\mathsf{\Omega}}}\;, 
\end{equation}
being $\bm{\mathsf{I}}_a$, $\bm{\mathsf{I}}_b$, $\hat{\bm{\mathsf{I}}}_a$, $\hat{\bm{\mathsf{I}}}_b\in\mathbb{R}^{2d\times 2d}$. Namely
\begin{equation}\label{For1}
\frac{\text{Det}(\bm{\mathsf{\Omega}})}{\text{Det}(\hat{\bm{\mathsf{\Omega}}})}=\frac{\det{[\bm{\mathsf{I}}_a+\bm{\mathsf{I}}_b\bm{\mathsf{F}}(b)]}}{\sqrt{\det[{\bm{\mathsf{F}}(b)}]}}\frac{\sqrt{\det[{\hat{\bm{\mathsf{F}}}(b)}]}}{\det{[\hat{\bm{\mathsf{I}}}_a+\hat{\bm{\mathsf{I}}}_b\hat{\bm{\mathsf{F}}}(b)]}}\;,
\end{equation}
with $\bm{\mathsf{F}}(s)$ ($\hat{\bm{\mathsf{F}}}(s)$) in $\mathbb{R}^{2d\times 2d}$ the fundamental solution of the linear differential system 
\begin{equation}
\bm{\mathsf{F}}'=\bm{\mathsf{\Gamma}}\bm{\mathsf{F}}\;,\quad\bm{\mathsf{F}}(a)=\bm{\mathbb{1}}\quad(\hat{\bm{\mathsf{F}}}'=\hat{\bm{\mathsf{\Gamma}}}\hat{\bm{\mathsf{F}}}\;,\quad\hat{\bm{\mathsf{F}}}(a)=\bm{\mathbb{1}}) 
\end{equation}
associated with the homogeneous problem $\bm{\mathsf{\Omega}}\bm{\mathsf{h}}=\bm{0}$ ($\hat{\bm{\mathsf{\Omega}}}\bm{\mathsf{h}}=\bm{0}$) and $\bm{\mathsf{\Gamma}}$ ($\hat{\bm{\mathsf{\Gamma}}}$) the matrix of first-order reduction interpreting $\bm{\mathsf{h}}'$ as an independent variable \citep{MCK, FAL}. 

In particular, we specialize to general second variation operators for $s\in[0,L]$, $d=6$, and we make the choice 
\begin{equation}
\bm{\mathsf{\Omega}}={\bm{\mathsf{S}}}=-{\bm{\mathsf{P}}}(s)\frac{\dd^2}{\dd s^2}+({\bm{\mathsf{C}}}^T(s)-{\bm{\mathsf{C}}}(s)-{\bm{\mathsf{P}}}'(s))\frac{\dd}{\dd s}+{\bm{\mathsf{Q}}}(s)-{\bm{\mathsf{C}}}'(s)
\end{equation}
computed in either $\bm{\mathsf{q}}^f$ or $\bm{\mathsf{q}}^m$ and $\hat{\bm{\mathsf{\Omega}}}={\hat{\bm{\mathsf{S}}}}$ computed in $\hat{\bm{\mathsf{q}}}$. Note that, for notation convenience, throughout this section $\bm{\mathsf{P}}$, $\bm{\mathsf{C}}$ and $\bm{\mathsf{Q}}$ stand for $\frac{\beta}{2\pi}\bm{\mathsf{P}}$, $\frac{\beta}{2\pi}\bm{\mathsf{C}}$ and $\frac{\beta}{2\pi}\bm{\mathsf{Q}}$. Moreover, defining $\bm{\mathsf{Y}}(s)=\bm{\mathsf{F}}(s)\bm{\mathsf{Y}}(0)$ for a given non-singular matrix $\bm{\mathsf{Y}}(0)$, changing variables in Hamiltonian form by means of 
\begin{equation}
\bm{\mathsf{Y}}=\bm{\mathsf{O}}\bm{\mathsf{W}}\;,\quad\bm{\mathsf{O}}(s)=\begin{pmatrix}
      \bm{\mathbb{1}} &  \bm{\mathbb{0}} \\
     -{\bm{\mathsf{P}}}^{-1}{\bm{\mathsf{C}}} & {\bm{\mathsf{P}}}^{-1}
     \end{pmatrix}
\end{equation}
being $\bm{\mathsf{W}}$ partitioned in $6$ by $6$ blocks as 
\begin{equation}
\bm{\mathsf{W}}(s)=\begin{pmatrix}
      \bm{\mathsf{H}} &  \bm{\mathsf{H}}^* \\
    \bm{\mathsf{M}} & \bm{\mathsf{M}}^*
     \end{pmatrix}\;,
\end{equation}
and doing the same in terms of $\hat{\bm{\mathsf{F}}}$, it is easily shown that Forman's theorem Eq.~(\ref{For1}) for $\bm{\mathsf{S}}$, $\hat{\bm{\mathsf{S}}}$ becomes Eq.~(\ref{For}) multiplied by 
\begin{equation}
\frac{\det[{\hat{\bm{\mathsf{W}}}(L)}]}{\det{[\hat{\bm{\mathsf{T}}}_0\hat{\bm{\mathsf{W}}}(0)+\hat{\bm{\mathsf{T}}}_L\hat{\bm{\mathsf{W}}}(L)]}}\;,
\end{equation}
with the Hamiltonian version of the boundary conditions being equal to 
\begin{equation}
\bm{\mathsf{T}}_0=\bm{\mathsf{I}}_0\bm{\mathsf{O}}(0)\;,\quad\bm{\mathsf{T}}_L=\bm{\mathsf{I}}_L\bm{\mathsf{O}}(L)
\end{equation}
and $\bm{\mathsf{W}}'=\bm{J}\bm{\mathsf{E}}\bm{\mathsf{W}}$ (the same is done for the ``hat'' term). Since the trace of $\bm{J}\bm{\mathsf{E}}$ is always zero, the so-called generalized Abel's identity or Liouville’s formula implies that $\det[{\bm{\mathsf{W}}}]$ ($\det[{\hat{\bm{\mathsf{W}}}}]$) is constant. We further observe that for ${\hat{\bm{\mathsf{S}}}}$ the boundary conditions on the paths (being the ones entering the path integral for the partition function) must be given by the matrices 
\begin{equation}
\hat{\bm{\mathsf{T}}}_0=\begin{pmatrix}
      \bm{\mathbb{1}} &  \bm{\mathbb{0}} \\
     \bm{\mathbb{0}} & \bm{\mathbb{0}}
     \end{pmatrix}\;,\quad\hat{\bm{\mathsf{T}}}_L=\begin{pmatrix}
      \bm{\mathbb{0}} &  \bm{\mathbb{0}} \\
     \bm{\mathbb{0}} & \bm{\mathbb{1}}
     \end{pmatrix}\;,
\end{equation}
and choosing ${\hat{\bm{\mathsf{H}}}}(L)=\bm{\mathbb{1}}$, ${\hat{\bm{\mathsf{M}}}}(L)= \bm{\mathbb{0}}$ within $\hat{\bm{\mathsf{W}}}(L)$, the ``hat'' contribution reduces to $\text{det}[\hat{\bm{\mathsf{H}}}(0)]$, which is equal to $1$ by direct inspection (see the previous section).

\subsection{Laplace approximation final formulas}
The idea is now to compute expression Eq.~(\ref{For}) for the operator ${\bm{\mathsf{S}}}$ subject to carefully chosen perturbed boundary conditions $\bm{\mathsf{T}}^{(\varepsilon)}_0$, in order to avoid the zero mode. This gives rise to a quasi-zero eigenvalue that can be found analytically using our extension to general second variation operators (including cross-terms) of the trick introduced in \citep{MCK}. Finally, by taking the limit for $\varepsilon\rightarrow 0$ in the ratio of the regularized expression Eq.~(\ref{For}) to the regularized quasi-zero eigenvalue, we recover the desired quantity ${\text{Det}^{\star}({\bm{\mathsf{S}}})}/{\text{Det}(\hat{\bm{\mathsf{S}}})}$. We anticipate here the results for the approximation formulas of the probability densities in the case of non-isolated minimizers, valid also for Kirchhoff rods as detailed in the previous section (note that the factor $\Vert\bm{\mathsf{\psi}}^{\alpha}\Vert$ simplifies out within the regularization procedure)
\begin{equation}\label{finiso}
\rra\approx2\pi\,e^{-\beta E(\qqa)}\sqrt{\frac{[\bm{\mathsf{\mu}}_{{\bm{\mathsf{\psi}}}^{\alpha}}(0)]_i}{]{\hha(0)}[_{i,i}}}\;,
\end{equation}
and we are interested in the cyclization values $\rrf(\bm{q}_0,L|\bm{q}_0,0)$, $\rrm(\bm{0},L|\bm{q}_0,0)$. In particular, $\bm{\mathsf{\mu}}_{{\bm{\mathsf{\psi}}}^{\alpha}}\in\mathbb{R}^6$ and $\hha\in\mathbb{R}^{6\times 6}$ are respectively the conjugate momentum of the zero mode and the Jacobi fields associated with ${\bm{\mathsf{S}}}^{\alpha}$, both computed by means of Eq.~(\ref{JacHam}) but recalling the contribution of $\frac{\beta}{2\pi}$. Moreover, here we denote with $[\cdot]_i$ the $i$-th component of a vector, with $]\cdot[_{i,i}$ the principal minor of a square matrix removing the $i$-th row and the $i$-th column, and the index $i$ depends on the choice of the boundary regularization, based on the non-zero components of $\bm{\mathsf{\mu}}_{{\bm{\mathsf{\psi}}}^{\alpha}}$. The appropriate initial conditions for $\hha$ are given at $s=L$ as: 
\begin{equation}\label{inFMiso}
\begin{split}
\hhf(L)=\bm{\mathbb{0}}\;,\,\,\mmf(L)=\bm{\mathsf{\chi}}\;;\qquad\qquad\quad\\
\\
\hhm(L)=\begin{pmatrix}
      \bm{\mathsf{X}}_{1,1} &   \bm{\mathsf{X}}_{1,2} \\
     \bm{\mathbb{0}} & \bm{\mathbb{0}}
     \end{pmatrix}\;,\,\,\mmm(L)=\begin{pmatrix}
      \bm{\mathbb{0}} &  \bm{\mathbb{0}} \\
      \bm{\mathsf{X}}_{2,1} &  \bm{\mathsf{X}}_{2,2}
     \end{pmatrix}\;,\\
\\
\end{split}
\end{equation}
where $\bm{\mathsf{\chi}}$ is an arbitrary matrix with unit determinant such that the $i$-th column corresponds to $\bm{\mathsf{\mu}}_{{\bm{\mathsf{\psi}}}^f}(L)$ and 
\begin{equation}
\bm{\mathsf{X}}=\begin{pmatrix}
      \bm{\mathsf{X}}_{1,1} &   \bm{\mathsf{X}}_{1,2} \\
     \bm{\mathsf{X}}_{2,1} & \bm{\mathsf{X}}_{2,2}
     \end{pmatrix}\in\mathbb{R}^{6\times 6}\;, 
\end{equation}
partitioned in $3$ by $3$ blocks, is an arbitrary matrix with determinant equal to $-1$ such that the $i$-th column corresponds to $([{\bm{\mathsf{\psi}}}^m]_{1:3},[\bm{\mathsf{\mu}}_{{\bm{\mathsf{\psi}}}^m}]_{4:6})^T(L)$.

\subsection{The regularization procedure}
We are now ready to explain how to regularize the functional determinants for ${\bm{\mathsf{S}}}^f$ and ${\bm{\mathsf{S}}}^m$ respectively, in order to get rid of the zero eigenvalue. Starting from the pure Dirichlet case, the boundary conditions are given as 
\begin{equation}\label{BCLD}
{\bm{\mathsf{T}}}_0^{(\varepsilon)}=\begin{pmatrix}
      \bm{\mathbb{1}} &  \bm{\mathcal{E}} \\
     \bm{\mathbb{0}} & \bm{\mathbb{0}}
     \end{pmatrix}\;,\quad{\bm{\mathsf{T}}}_L=\begin{pmatrix}
      \bm{\mathbb{0}} &  \bm{\mathbb{0}} \\
     \bm{\mathbb{1}} & \bm{\mathbb{0}}
     \end{pmatrix}\;,
\end{equation}
with $\bm{\mathcal{E}}$ the zero matrix with a non-zero diagonal entry $\varepsilon$ in position $i,i$ serving as a perturbation to avoid the zero mode. Then, choosing ${\bm{\mathsf{H}}}^f(L)$, ${\bm{\mathsf{M}}}^f(L)$ as given in Eq.~(\ref{inFMiso}), and applying the formula for the determinant of a block matrix
\begin{equation}\label{fblock}
\det\begin{pmatrix}
      \bm{A} &  \bm{B} \\
     \bm{C} & \bm{D}
     \end{pmatrix}=\det{(\bm{D})}\det{(\bm{A}-\bm{B}\bm{D}^{-1}\bm{C})}\;,\quad\bm{\mathsf{A}},\,\,\bm{\mathsf{B}},\,\,\bm{\mathsf{C}},\,\,\bm{\mathsf{D}}\in\mathbb{R}^{n\times n}
\end{equation}
for $n=6$, from Eq.~(\ref{For}) we get 
\begin{equation}\label{regepsD}
\frac{\text{Det}^{(\varepsilon)}({\bm{\mathsf{S}}^f})}{\text{Det}(\hat{\bm{\mathsf{S}}})}=\frac{\det{[\bm{\mathsf{H}}^f(0)+\bm{\mathcal{E}}\bm{\mathsf{M}}^f(0)]}}{\det[{\bm{\mathsf{M}}^f(L)}]}=\varepsilon [\bm{\mathsf{\mu}}_{{\bm{\mathsf{\psi}}}^f}(0)]_i\,]{{\bm{\mathsf{H}}}^f(0)}[_{i,i}\;. 
\end{equation}
By construction the zero mode represents the $i$-th column of $\bm{\mathsf{H}}^f(s)$ and satisfies the linearised boundary condition ${{\bm{\mathsf{\psi}}}^f}(0)=\bm{0}$, hence the last equality.

On the other hand, for the marginalized case, the boundary conditions are given by ${\bm{\mathsf{T}}}_0^{(\varepsilon)}$ as before and 
\begin{equation}\label{BCLDN}
{\bm{\mathsf{T}}}_L=\begin{pmatrix}
      \bm{\mathbb{0}} &  \bm{\mathbb{0}} \\
     \bm{\mathbb{1}}^{\bm{\mathbb{0}}} & \bm{\mathbb{1}}_{\bm{\mathbb{0}}}
     \end{pmatrix}\;,\quad\text{being}\quad\bm{\mathbb{1}}^{\bm{\mathbb{0}}}=\begin{pmatrix}
      \bm{\mathbb{0}} &  \bm{\mathbb{0}} \\
     \bm{\mathbb{0}} & \bm{\mathbb{1}}
     \end{pmatrix}\;,\quad\bm{\mathbb{1}}_{\bm{\mathbb{0}}}=\begin{pmatrix}
      \bm{\mathbb{1}} &  \bm{\mathbb{0}} \\
     {\bm{\mathbb{0}}} & {\bm{\mathbb{0}}}
     \end{pmatrix}
\end{equation}
partitioned in $3$ by $3$ blocks. Then, choosing ${\bm{\mathsf{H}}}^m(L)$, ${\bm{\mathsf{M}}}^m(L)$ as given in Eq.~(\ref{inFMiso}), from Eq.~(\ref{For}) we get 
\begin{equation}\label{regepsDN}
\frac{\text{Det}^{(\varepsilon)}({\bm{\mathsf{S}}^m})}{\text{Det}(\hat{\bm{\mathsf{S}}})}=-\frac{\det{[\bm{\mathsf{H}}^m(0)+\bm{\mathcal{E}}\bm{\mathsf{M}}^m(0)]}}{\det[{\bm{\mathsf{X}}}]}=\varepsilon [\bm{\mathsf{\mu}}_{{\bm{\mathsf{\psi}}}^m}(0)]_i\,]{{\bm{\mathsf{H}}}^m(0)}[_{i,i}\;. 
\end{equation}
By construction the zero mode represents the $i$-th column of $\bm{\mathsf{H}}^m(s)$ and satisfies the linearised boundary condition ${{\bm{\mathsf{\psi}}}^m}(0)=\bm{0}$, hence the last equality. In addition, when computing $\text{det}[\bm{\mathsf{W}}^m(L)]$ in Eq.~(\ref{For}), we have applied the formula for the determinant of a block matrix Eq.~(\ref{fblock}) for $n=6$ and the following determinant identity for $n=3$:
\begin{equation}\label{Id}
\begin{split}
\text{det}&\left[\begin{pmatrix}
      \bm{\mathsf{X}}_{1,1} &   \bm{\mathsf{X}}_{1,2} \\
     \bm{\mathbb{0}} & \bm{\mathbb{0}}
     \end{pmatrix}\begin{pmatrix}
      \bm{\mathsf{A}} &   \bm{\mathsf{B}} \\
     \bm{\mathsf{C}} & \bm{\mathsf{D}}
     \end{pmatrix}-\begin{pmatrix}
      \bm{\mathsf{\alpha}} &   \bm{\mathsf{\beta}} \\
     \bm{\mathsf{\gamma}} & \bm{\mathsf{\delta}}
     \end{pmatrix}\begin{pmatrix}
      \bm{\mathsf{A}} &   \bm{\mathsf{B}} \\
     \bm{\mathsf{C}} & \bm{\mathsf{D}}
     \end{pmatrix}^{-1}\begin{pmatrix}
      \bm{\mathbb{0}} &   \bm{\mathbb{0}} \\
     \bm{\mathsf{X}}_{2,1} & \bm{\mathsf{X}}_{2,2}
     \end{pmatrix}\begin{pmatrix}
      \bm{\mathsf{A}} &   \bm{\mathsf{B}} \\
     \bm{\mathsf{C}} & \bm{\mathsf{D}}
     \end{pmatrix}\right]=\\
     \\
     &=(-1)^n\text{det}\left[\begin{pmatrix}
      \bm{\mathsf{X}}_{1,1} &   \bm{\mathsf{X}}_{1,2} \\
     \bm{\mathsf{X}}_{2,1} & \bm{\mathsf{X}}_{2,2}
     \end{pmatrix}\begin{pmatrix}
      \bm{\mathsf{A}} &   \bm{\mathsf{B}} \\
     \bm{\mathsf{\gamma}} & \bm{\mathsf{\delta}}
     \end{pmatrix}\right]\;,
     \end{split}
\end{equation}
for given matrices $\bm{\mathsf{X}}_{1,1}$, $\bm{\mathsf{X}}_{1,2}$, $\bm{\mathsf{X}}_{2,1}$, $\bm{\mathsf{X}}_{2,2}$, $\bm{\mathsf{A}}$, $\bm{\mathsf{B}}$, $\bm{\mathsf{C}}$, $\bm{\mathsf{D}}$, $\bm{\mathsf{\alpha}}$, $\bm{\mathsf{\beta}}$, $\bm{\mathsf{\gamma}}$, $\bm{\mathsf{\delta}}\in\mathbb{R}^{n\times n}$, which can be proven to be true by direct computation. Namely, the matrix inside the square brackets on the left-hand side of Eq.~(\ref{Id}) is also equal to a block matrix having $(1,1)$, $(1,2)$, $(2,1)$ and $(2,2)$ blocks respectively equal to
\begin{equation}
\begin{split}
&(1,1)\qquad\bm{\mathsf{X}}_{1,1}\bm{\mathsf{A}}+\bm{\mathsf{X}}_{1,2}\bm{\mathsf{C}}-(\bm{\mathsf{\beta}}-\bm{\mathsf{\alpha}}\bm{\mathsf{A}}^{-1}\bm{\mathsf{B}})(\bm{\mathsf{D}}-\bm{\mathsf{C}}\bm{\mathsf{A}}^{-1}\bm{\mathsf{B}})^{-1}(\bm{\mathsf{X}}_{2,1}\bm{\mathsf{A}}+\bm{\mathsf{X}}_{2,2}\bm{\mathsf{C}})\;,\\
\\
&(1,2)\qquad\bm{\mathsf{X}}_{1,1}\bm{\mathsf{B}}+\bm{\mathsf{X}}_{1,2}\bm{\mathsf{D}}-(\bm{\mathsf{\beta}}-\bm{\mathsf{\alpha}}\bm{\mathsf{A}}^{-1}\bm{\mathsf{B}})(\bm{\mathsf{D}}-\bm{\mathsf{C}}\bm{\mathsf{A}}^{-1}\bm{\mathsf{B}})^{-1}(\bm{\mathsf{X}}_{2,1}\bm{\mathsf{B}}+\bm{\mathsf{X}}_{2,2}\bm{\mathsf{D}})\;,\\
\\
&(2,1)\qquad -(\bm{\mathsf{\delta}}-\bm{\mathsf{\gamma}}\bm{\mathsf{A}}^{-1}\bm{\mathsf{B}})(\bm{\mathsf{D}}-\bm{\mathsf{C}}\bm{\mathsf{A}}^{-1}\bm{\mathsf{B}})^{-1}(\bm{\mathsf{X}}_{2,1}\bm{\mathsf{A}}+\bm{\mathsf{X}}_{2,2}\bm{\mathsf{C}})\;,\\
\\
&(2,2)\qquad -(\bm{\mathsf{\delta}}-\bm{\mathsf{\gamma}}\bm{\mathsf{A}}^{-1}\bm{\mathsf{B}})(\bm{\mathsf{D}}-\bm{\mathsf{C}}\bm{\mathsf{A}}^{-1}\bm{\mathsf{B}})^{-1}(\bm{\mathsf{X}}_{2,1}\bm{\mathsf{B}}+\bm{\mathsf{X}}_{2,2}\bm{\mathsf{D}})\;,
\end{split}
\end{equation}
whose determinant can be computed using Eq.~(\ref{fblock}) and simplified in order to obtain
\begin{equation}
\begin{split}
&(-1)^n\left[\det{(\bm{\mathsf{A}})}\det{(\bm{\mathsf{\delta}}-\bm{\mathsf{\gamma}}\bm{\mathsf{A}}^{-1}\bm{\mathsf{B}})}\right]\left[\det{(\bm{\mathsf{D}}-\bm{\mathsf{C}}\bm{\mathsf{A}}^{-1}\bm{\mathsf{B}})}^{-1}\det{(\bm{\mathsf{A}})}^{-1}\right]\Big[\det{(\bm{\mathsf{X}}_{2,1}\bm{\mathsf{B}}+\bm{\mathsf{X}}_{2,2}\bm{\mathsf{D}})}\\
&\quad\det{\left((\bm{\mathsf{X}}_{1,1}\bm{\mathsf{A}}+\bm{\mathsf{X}}_{1,2}\bm{\mathsf{C}})-(\bm{\mathsf{X}}_{1,1}\bm{\mathsf{B}}+\bm{\mathsf{X}}_{1,2}\bm{\mathsf{D}})(\bm{\mathsf{X}}_{2,1}\bm{\mathsf{B}}+\bm{\mathsf{X}}_{2,2}\bm{\mathsf{D}})^{-1}(\bm{\mathsf{X}}_{2,1}\bm{\mathsf{A}}+\bm{\mathsf{X}}_{2,2}\bm{\mathsf{C}})\right)}\Big]\;,\\
\\
&=(-1)^n\text{det}\begin{pmatrix}
      \bm{\mathsf{A}} &   \bm{\mathsf{B}} \\
     \bm{\mathsf{\gamma}} & \bm{\mathsf{\delta}}
     \end{pmatrix}\det \begin{pmatrix}
      \bm{\mathsf{A}} &   \bm{\mathsf{B}} \\
     \bm{\mathsf{C}} & \bm{\mathsf{D}}
     \end{pmatrix}^{-1}\det \begin{pmatrix}
      \bm{\mathsf{X}}_{1,1}\bm{\mathsf{A}}+\bm{\mathsf{X}}_{1,2}\bm{\mathsf{C}} &\quad   \bm{\mathsf{X}}_{1,1}\bm{\mathsf{B}}+\bm{\mathsf{X}}_{1,2}\bm{\mathsf{D}} \\
      \\
     \bm{\mathsf{X}}_{2,1}\bm{\mathsf{A}}+\bm{\mathsf{X}}_{2,2}\bm{\mathsf{C}} &\quad \bm{\mathsf{X}}_{2,1}\bm{\mathsf{B}}+\bm{\mathsf{X}}_{2,2}\bm{\mathsf{D}}
     \end{pmatrix}\;,
\end{split}
\end{equation}
where we have used also the other formula for the determinant of a block matrix
\begin{equation}\label{fblock2}
\det\begin{pmatrix}
      \bm{A} &  \bm{B} \\
     \bm{C} & \bm{D}
     \end{pmatrix}=\det{(\bm{A})}\det{(\bm{D}-\bm{C}\bm{A}^{-1}\bm{B})}\;.
\end{equation}
Since
\begin{equation}
\begin{pmatrix}
      \bm{\mathsf{X}}_{1,1}\bm{\mathsf{A}}+\bm{\mathsf{X}}_{1,2}\bm{\mathsf{C}} &\quad   \bm{\mathsf{X}}_{1,1}\bm{\mathsf{B}}+\bm{\mathsf{X}}_{1,2}\bm{\mathsf{D}} \\
      \\
     \bm{\mathsf{X}}_{2,1}\bm{\mathsf{A}}+\bm{\mathsf{X}}_{2,2}\bm{\mathsf{C}} &\quad \bm{\mathsf{X}}_{2,1}\bm{\mathsf{B}}+\bm{\mathsf{X}}_{2,2}\bm{\mathsf{D}}
     \end{pmatrix}=\begin{pmatrix}
      \bm{\mathsf{X}}_{1,1} &  \bm{\mathsf{X}}_{1,2} \\
     \bm{\mathsf{X}}_{2,1} & \bm{\mathsf{X}}_{2,2}
     \end{pmatrix}\begin{pmatrix}
     \bm{\mathsf{A}} &  \bm{\mathsf{B}}\\
     \bm{\mathsf{C}} & \bm{\mathsf{D}}
     \end{pmatrix}\;,
\end{equation}
then we recover the result Eq.~(\ref{Id}).

The last step consists of finding the non-zero eigenvalue $\lambda^{\alpha}_{(\varepsilon)}$ associated with the eigenfunction $\bm{\mathsf{\psi}}^{\alpha}_{(\varepsilon)}$ (arising from the zero mode $\bm{\mathsf{\psi}}^{\alpha}$) of the operator $\bm{\mathsf{S}}^{\alpha}$ with perturbed boundary conditions. First, we have that
\begin{equation}
(\bm{\mathsf{\psi}}^{\alpha},\bm{\mathsf{S}}^{\alpha}\bm{\mathsf{\psi}}^{\alpha}_{(\varepsilon)})=\lambda^{\alpha}_{(\varepsilon)}(\bm{\mathsf{\psi}}^{\alpha},\bm{\mathsf{\psi}}^{\alpha}_{(\varepsilon)})\;, 
\end{equation}
and the left-hand side can be rewritten as 
\begin{equation}
\begin{split}
(\bm{\mathsf{\psi}}^{\alpha},\bm{\mathsf{S}}^{\alpha}\bm{\mathsf{\psi}}^{\alpha}_{(\varepsilon)})&=(\bm{\mathsf{\psi}}^{\alpha},\bm{\mathsf{S}}^{\alpha}\bm{\mathsf{\psi}}^{\alpha}_{(\varepsilon)})-(\bm{\mathsf{S}}^{\alpha}\bm{\mathsf{\psi}}^{\alpha},\bm{\mathsf{\psi}}^{\alpha}_{(\varepsilon)})\;,\\
&=[\bm{\mathsf{\mu}}_{\bm{\mathsf{\psi}}^{\alpha}}\cdot\bm{\mathsf{\psi}}^{\alpha}_{(\varepsilon)}]_0^L-[\bm{\mathsf{\mu}}_{\bm{\mathsf{\psi}}^{\alpha}_{(\varepsilon)}}\cdot\bm{\mathsf{\psi}}^{\alpha}]_0^L\;,\\
&=-(\bm{\mathsf{\mu}}_{\bm{\mathsf{\psi}}^{\alpha}}\cdot\bm{\mathsf{\psi}}^{\alpha}_{(\varepsilon)})(0)\;,\\
&=\varepsilon [\bm{\mathsf{\mu}}_{\bm{\mathsf{\psi}}^{\alpha}}(0)]_i [\bm{\mathsf{\mu}}_{\bm{\mathsf{\psi}}^{\alpha}_{(\varepsilon)}}(0)]_i\;, 
\end{split}
\end{equation}
where the second equality comes after integration by parts and the third and fourth ones are a consequence of the boundary conditions Eq.~(\ref{BCLD}) and Eq.~(\ref{BCLDN}). More precisely, for the second equality we use Eq.~(\ref{secOp}) and write (omitting the $\alpha$ superscript)
\begin{equation}
\begin{split}
(\bm{\mathsf{\psi}},\bm{\mathsf{S}}\bm{\mathsf{\psi}}_{(\varepsilon)})-(\bm{\mathsf{S}}\bm{\mathsf{\psi}},\bm{\mathsf{\psi}}_{(\varepsilon)})&=\left(\bm{\mathsf{\psi}},-\bm{\mathsf{P}}\bm{\mathsf{\psi}}''_{(\varepsilon)}+(\bm{\mathsf{C}}^T-\bm{\mathsf{C}}-\bm{\mathsf{P}}')\bm{\mathsf{\psi}}'_{(\varepsilon)}+(\bm{\mathsf{Q}}-\bm{\mathsf{C}}')\bm{\mathsf{\psi}}_{(\varepsilon)}\right)\\
&\quad+\left(\bm{\mathsf{P}}\bm{\mathsf{\psi}}''-(\bm{\mathsf{C}}^T-\bm{\mathsf{C}}-\bm{\mathsf{P}}')\bm{\mathsf{\psi}}'-(\bm{\mathsf{Q}}-\bm{\mathsf{C}}')\bm{\mathsf{\psi}},\bm{\mathsf{\psi}}_{(\varepsilon)}\right)\;,\\
&=-\left((\bm{\mathsf{P}}\bm{\mathsf{\psi}}'_{(\varepsilon)}+\bm{\mathsf{C}}\bm{\mathsf{\psi}}_{(\varepsilon)})',\bm{\mathsf{\psi}}\right)+\left(\bm{\mathsf{C}}\bm{\mathsf{\psi}},\bm{\mathsf{\psi}}'_{(\varepsilon)}\right)\\
&\quad+\left((\bm{\mathsf{P}}\bm{\mathsf{\psi}}'+\bm{\mathsf{C}}\bm{\mathsf{\psi}})',\bm{\mathsf{\psi}}_{(\varepsilon)}\right)-\left(\bm{\mathsf{C}}\bm{\mathsf{\psi}}_{(\varepsilon)},\bm{\mathsf{\psi}}'\right)\;,\\
&=-[\bm{\mathsf{\mu}}_{\bm{\mathsf{\psi}}_{(\varepsilon)}}\cdot\bm{\mathsf{\psi}}]_0^L+\left(\bm{\mathsf{P}}\bm{\mathsf{\psi}}'_{(\varepsilon)}+\bm{\mathsf{C}}\bm{\mathsf{\psi}}_{(\varepsilon)},\bm{\mathsf{\psi}}'\right)+\left(\bm{\mathsf{C}}\bm{\mathsf{\psi}},\bm{\mathsf{\psi}}'_{(\varepsilon)}\right)\\
&\quad+[\bm{\mathsf{\mu}}_{\bm{\mathsf{\psi}}}\cdot\bm{\mathsf{\psi}}_{(\varepsilon)}]_0^L-\left(\bm{\mathsf{P}}\bm{\mathsf{\psi}}'+\bm{\mathsf{C}}\bm{\mathsf{\psi}},\bm{\mathsf{\psi}}'_{(\varepsilon)}\right)-\left(\bm{\mathsf{C}}\bm{\mathsf{\psi}}_{(\varepsilon)},\bm{\mathsf{\psi}}'\right)\;,\\
&=[\bm{\mathsf{\mu}}_{\bm{\mathsf{\psi}}}\cdot\bm{\mathsf{\psi}}_{(\varepsilon)}]_0^L-[\bm{\mathsf{\mu}}_{\bm{\mathsf{\psi}}_{(\varepsilon)}}\cdot\bm{\mathsf{\psi}}]_0^L\;,
\end{split}
\end{equation}
where we have exploited the symmetry of $\bm{\mathsf{P}}$ and $\bm{\mathsf{Q}}$.

Finally, being 
\begin{equation}
\lambda^{\alpha}_{(\varepsilon)}=\frac{\varepsilon [\bm{\mathsf{\mu}}_{\bm{\mathsf{\psi}}^{\alpha}}(0)]_i [\bm{\mathsf{\mu}}_{\bm{\mathsf{\psi}}^{\alpha}_{(\varepsilon)}}(0)]_i}{(\bm{\mathsf{\psi}}^{\alpha},\bm{\mathsf{\psi}}^{\alpha}_{(\varepsilon)})}\;, 
\end{equation}
and recalling Eq.~(\ref{regepsD}), Eq.~(\ref{regepsDN}), then
\begin{equation}
\frac{\text{Det}^{\star}(\bm{\mathsf{S}}^{\alpha})}{\text{Det}({\hat{\bm{\mathsf{S}}}})}=\lim_{\varepsilon\rightarrow 0}(\lambda^{\alpha}_{(\varepsilon)})^{-1}\frac{\text{Det}^{(\varepsilon)}(\bm{\mathsf{S}}^{\alpha})}{\text{Det}({\hat{\bm{\mathsf{S}}}})}=\Vert\bm{\mathsf{\psi}}^{\alpha}\Vert^2\,\frac{]{{{\bm{\mathsf{H}}^{\alpha}}}(0)}[_{i,i}}{[\bm{\mathsf{\mu}}_{\bm{\mathsf{\psi}}^{\alpha}}(0)]_i}\;,
\end{equation}
therefore obtaining the result Eq.~(\ref{finiso}).

We conclude with a technical remark. We observe that \textit{a priori} the solution formulas for isolated minimizers could be recovered by applying Forman's theorem in the framework of functional determinants (as done here for the non-isolated case); however, there we exploit the insightful connection with the more standard theory of path integrals via ``time-slicing''. Exploring the latter possibility not only allows us to gain a deeper understanding of the subject, but is crucial to developing the right ideas for using Forman's formalism in a more general setting. In this regard, it would be appropriate in future studies to investigate in depth the underlying connections between the ``time-slicing'' approach and Forman's results.

\section{A Monte Carlo algorithm for stochastic elastic rods}\label{MC}
In this section we refer to the approach of \citep{MCDNA, ALEX2, MCD} for DNA Monte Carlo simulations of $J$-factors, using the ``half-molecule'' technique \citep{ALEX1} for enhancing the efficiency. Namely we give a Monte Carlo sampling algorithm for fluctuating linearly elastic rods according to the Boltzmann distribution having partition function Eq.~(\ref{normex_ch2}), \ie 
\begin{equation}
\mathcal{Z}=\int_{\bm{q}(0)=\bm{q}_0}{e^{-\beta E(\bm{q})}\,\mathcal{D}\bm{q}}
\end{equation}
with energy Eq.~(\ref{energylin}), and we use the compact notation $\bm{\mathsf{u}}_{\Delta}=\bm{\mathsf{u}}-\hat{\bm{\mathsf{u}}}$, $\bm{\mathsf{v}}_{\Delta}=\bm{\mathsf{v}}-\hat{\bm{\mathsf{v}}}$ for the shifted strains. First of all, we need to rewrite the infinite-dimensional problem as a finite-dimensional one by means of a ``parameter-slicing method". This is achieved, after parametrizing the configuration variable as $\bm{\mathsf{q}}(s)=(\bm{\mathsf{c}}(s),\bm{\mathsf{t}}(s))\in\mathbb{R}^6$, setting $\epsilon=\frac{L}{n}$ with $n$ a large positive integer and $s_j=j\epsilon$ for $j=0,...,n$. Moreover, by exploiting the change of variables $(\bm{\mathsf{c}}_j,\bm{\mathsf{t}}_j)\rightarrow(\bm{\mathsf{u}}_j,\bm{\mathsf{v}}_j)$ as presented in \citep{LUDT}, we get the following equality up to a constant factor for the discrete version of the partition function $\mathcal{Z}$
\begin{equation}\label{partfin}
\int{e^{-\beta\epsilon\sum\limits_{j=0}^{n}{W(\bm{\mathsf{c}},\bm{\mathsf{t}})_j}}}\prod\limits_{j=1}^n\left(1+\Vert\bm{\mathsf{c}}_j\Vert^2\right)^{-2}\dd\bm{\mathsf{c}}_j\dd\bm{\mathsf{t}}_j\sim\int{e^{-{\beta\epsilon}\sum\limits_{j=0}^{n-1}{W({\bm{\mathsf{u}}_{\Delta}},{\bm{\mathsf{v}}_{\Delta}})_j}}}\prod\limits_{j=0}^{n-1}\mathcal{J}(\bm{\mathsf{u}}_j)\dd\bm{\mathsf{u}}_j\dd\bm{\mathsf{v}}_j\;,
\end{equation}
with 
\begin{equation}
W_j=\frac{1}{2}\left[{\bm{\mathsf{u}}_{\Delta}}_j^T \bm{\mathcal{K}}_j{\bm{\mathsf{u}}_{\Delta}}_j+2{\bm{\mathsf{u}}_{\Delta}}_j^T \bm{\mathcal{B}}_j{\bm{\mathsf{v}}_{\Delta}}_j+{\bm{\mathsf{v}}_{\Delta}}_j^T \bm{\mathcal{A}}_j{\bm{\mathsf{v}}_{\Delta}}_j\right]\;,\quad\mathcal{J}(\bm{\mathsf{u}}_j)=\left(1- \frac{\epsilon^2}{4}\Vert\bm{\mathsf{u}}_j\Vert^2\right)^{-\frac{1}{2}}
\end{equation}
and the subscript $j$ indicates that the associated term is evaluated in $s_j$. We observe that the Jacobian factor $\mathcal{J}$ can be neglected, as discussed in \citep{PL}, leading to the Gaussian distribution 
\begin{equation}\label{GausDisc}
\rho_{\mathcal{Z}}=\frac{e^{-{\beta\epsilon}\sum\limits_{j=0}^{n-1}{W({\bm{\mathsf{u}}_{\Delta}},{\bm{\mathsf{v}}_{\Delta}})_j}}}{\displaystyle{\int e^{-{\beta\epsilon}\sum\limits_{j=0}^{n-1}{W({\bm{\mathsf{u}}_{\Delta}},{\bm{\mathsf{v}}_{\Delta}})_j}}\,\,\prod\limits_{j=0}^{n-1}\dd\bm{\mathsf{u}}_j\dd\bm{\mathsf{v}}_j}} 
\end{equation}
which can be easily sampled by a direct Monte Carlo method in order to get random instances of $\bm{\mathsf{u}}_j$, $\bm{\mathsf{v}}_j$, $j=0,...,n-1$, associated with a random framed curve with initial data $\bm{q}_0=(\bm{\mathbb{1}},\bm{0})$. Note that, in the proposed uniform example with diagonal stiffness matrix, the Gaussian factorises and the sampling is simply performed componentwise in terms of independent univariate Gaussians. 

Since the conditional probability density is a function of the variables $\bm{R}_L$, $\bm{r}_L$, we need to reconstruct $\bm{R}_n$, $\bm{r}_n$ from the sampled strains by discretization of the differential equations 
\begin{equation}
\bm{\gamma}'(s)=\frac{1}{2}\sum\limits_{i=1}^3 [\bm{\mathsf{u}}(s)]_i\bm{B}_i\bm{\gamma}(s)\;,\quad\bm{r}'(s)=\bm{R}(\bm{\gamma}(s))\bm{\mathsf{v}}(s), 
\end{equation}
with $[\bm{\mathsf{u}}(s)]_i$ the $i$th component of $\bm{\mathsf{u}}$ and $\bm{R}(\bm{\gamma})$ the rotation matrix associated with the quaternion $\bm{\gamma}$. This is achieved, \eg, by application of the scalar factor method, derived in \citep{QUAT1} and discussed in \citep{QUAT2}, which is an efficient and precise one-step method for integrating the Darboux vector $\bm{\mathsf{u}}$, preserving the unit norm of the quaternion. Defining 
\begin{equation}\label{recon1}
{\delta_{\bm{\gamma}}}_j=\frac{\epsilon}{2}\sum\limits_{i=1}^3[\bm{\mathsf{u}}_j]_i\bm{B}_i\bm{\gamma}_j\;, 
\end{equation}
then we have that 
\begin{equation}\label{recon2}
\begin{cases}
\bm{\gamma}_{j+1}=\left(\bm{\gamma}_{j}+\frac{\tan{(\Vert{\delta_{\bm{\gamma}}}_j\Vert)}}{\Vert{\delta_{\bm{\gamma}}}_j\Vert}{\delta_{\bm{\gamma}}}_j\right)\cos{(\Vert{\delta_{\bm{\gamma}}}_j\Vert)}\;,\\
\bm{\gamma}_0=(0,0,0,1)\;,
\end{cases}
\end{equation}
and consequently 
\begin{equation}\label{recon3}
\begin{cases}
\bm{r}_{j+1}=\bm{r}_{j}+\epsilon\bm{R}(\bm{\gamma}_j)\bm{\mathsf{v}}_j\;,\\
\bm{r}_0=\bm{0}\;. 
\end{cases}
\end{equation}

In the spirit of \citep{MCD} for computing cyclization densities, we are now able to generate Monte Carlo trajectories and assess whether or not $\bm{q}_n=(\bm{R}_n=\bm{R}(\bm{\gamma}_n),\bm{r}_n)$ is falling inside the given small region $\mathcal{R}_{\zeta,\xi}$ of $SE(3)$ centred in $(\bm{\mathbb{1}},\bm{0})$ parametrized as the Cartesian product $\mathcal{B}_{\zeta}\times\mathcal{B}_{\xi}$ of two open balls in $\mathbb{R}^3$, centred in $\bm{0}$, of radius $\zeta,\xi>0$ respectively. Namely, 
\begin{equation}
(\bm{R}_n,\bm{r}_n)\in\mathcal{R}_{\zeta,\xi}\quad\text{if and only if}\quad\Vert\bm{\mathsf{c}}(\bm{\gamma}_n)\Vert<\zeta\quad\text{and}\quad\Vert\bm{r}_n\Vert<\xi\;, 
\end{equation}
with $\bm{\mathsf{c}}\in\mathbb{R}^3$ the same parametrization of $SO(3)$ presented above, adapted to $\bar{\bm{\gamma}}=(0,0,0,1)$. Note that, since 
\begin{equation}
\bm{\mathsf{c}}(\bm{\gamma}_n)=\frac{([\bm{\gamma}_n]_1,[\bm{\gamma}_n]_2,[\bm{\gamma}_n]_3)}{[\bm{\gamma}_n]_4}\quad\text{and}\quad\Vert\bm{\gamma}_n\Vert=1\;, 
\end{equation}
the condition $\Vert\bm{\mathsf{c}}(\bm{\gamma}_n)\Vert<\zeta$ is equivalent to 
\begin{equation}
\sqrt{[\bm{\gamma}_n]_4^{-2}-1}<\zeta\;. 
\end{equation}

Moreover, we have the following link between the probability of the set $\mathcal{R}_{\zeta,\xi}$ ($ \mathbb{P}(\mathcal{R}_{\zeta,\xi})$) computed using Monte Carlo simulations and the conditional probability density defined in the theoretical framework
\begin{equation}\label{link}
\begin{split}
\frac{|\lbrace\text{samples:}\,(\bm{R}_n,\bm{r}_n)\in\mathcal{R}_{\zeta,\xi}\rbrace|}{|\lbrace\text{all samples}\rbrace|}&\approx \mathbb{P}(\mathcal{R}_{\zeta,\xi})=\int_{\mathcal{R}_{\zeta,\xi}}{\rho_f(\bm{q}_L,L|\bm{q}_0,0)}\dd\bm{q}_L\;,\\
\\
&\approx |\mathcal{R}_{\zeta,\xi}|\,\rho_f(\bm{q}_0,L|\bm{q}_0,0)\;,
\end{split}
\end{equation}
where the notation $|\cdot |$ stands for the number of elements of a discrete set or the measure of a continuous set, and the accuracy of the approximation increases with 
\begin{equation}
n\rightarrow\infty\;,\quad|\lbrace\text{all samples}\rbrace |\rightarrow\infty\;,\quad\zeta\rightarrow 0\;,\quad\xi\rightarrow 0\;. 
\end{equation}
The set $\mathcal{R}_{\zeta,\xi}$ is measured by means of the product of the Haar measure and the Lebesgue measure for the $SO(3)$ and the $E(3)$ components. Thus, making use of the parametrisation, 
\begin{equation}\label{volume}
\begin{split}
|\mathcal{R}_{\zeta,\xi}|&=\int_{\mathcal{R}_{\zeta,\xi}}{\dd\bm{q}_n}=\int_{\mathcal{B}_{\zeta}\times\mathcal{B}_{\xi}}{\left(1+\Vert\bm{\mathsf{c}}_n\Vert^2\right)^{-2}\,\dd \bm{\mathsf{c}}_n\,\dd\bm{r}_n}\;,\\
\\
&=\frac{8\pi^2\xi^3}{3}\left(\arctan{(\zeta)}-\frac{\zeta}{1+\zeta^2}\right)\;.
\end{split}
\end{equation}
Regarding the marginal $\rho_m(\bm{r}_0,L|\bm{q}_0,0)$, the method is applied only considering the condition on $\bm{r}_n$ for being inside the open ball $\mathcal{B}_{\xi}$ with measure $| \mathcal{B}_{\xi} |=\frac{4\pi\xi^3}{3}$, and neglecting all the details concerning the rotation component. 

More specifically, in order to enhance the efficiency of the algorithm, we refer to the approach adopted in \citep{MCDNA, ALEX2, MCD} for DNA Monte Carlo simulations, using the ``half-molecule'' technique as developed by Alexandrowicz \citep{ALEX1}. In this technique, one computes $M$ random instances each of the first and second halves of the framed curve and then considers all first-half-second-half pairs in order to generate $M^2$ random curves, allowing a large sample size contributing for each density data point and providing the necessary accuracy to the estimation. In particular, we give here the specifications for the simulations reported in the following section. For the $(\mathtt{f})$ computations, $\sim 10^{15}$ samples were produced for each data point, choosing $n=200$ and $\zeta$, $\xi$ ranging from $2.5$ \% to $6.6$ \% of the parameter $L$. The estimated density value corresponds to the mean taken over $81$ ``boxes'', along with the standard deviation for these boxes defining the range of the bar for each Monte Carlo data point. For the $(\mathtt{m})$ cases, $\sim 10^{13}$ samples were produced for each data point, choosing $n=200$ and $\xi$ ranging from $0.1$ \% to $4$ \% of the parameter $L$; $40$ different ``boxes'' were used for the final estimation.

\section{Results and discussion for the examples considered}
\subsection{A preliminary stability analysis}
\begin{figure*}
\centering
\captionsetup{justification=centering,margin=1cm}
\subfigure[]{
\includegraphics[width=.454\textwidth]{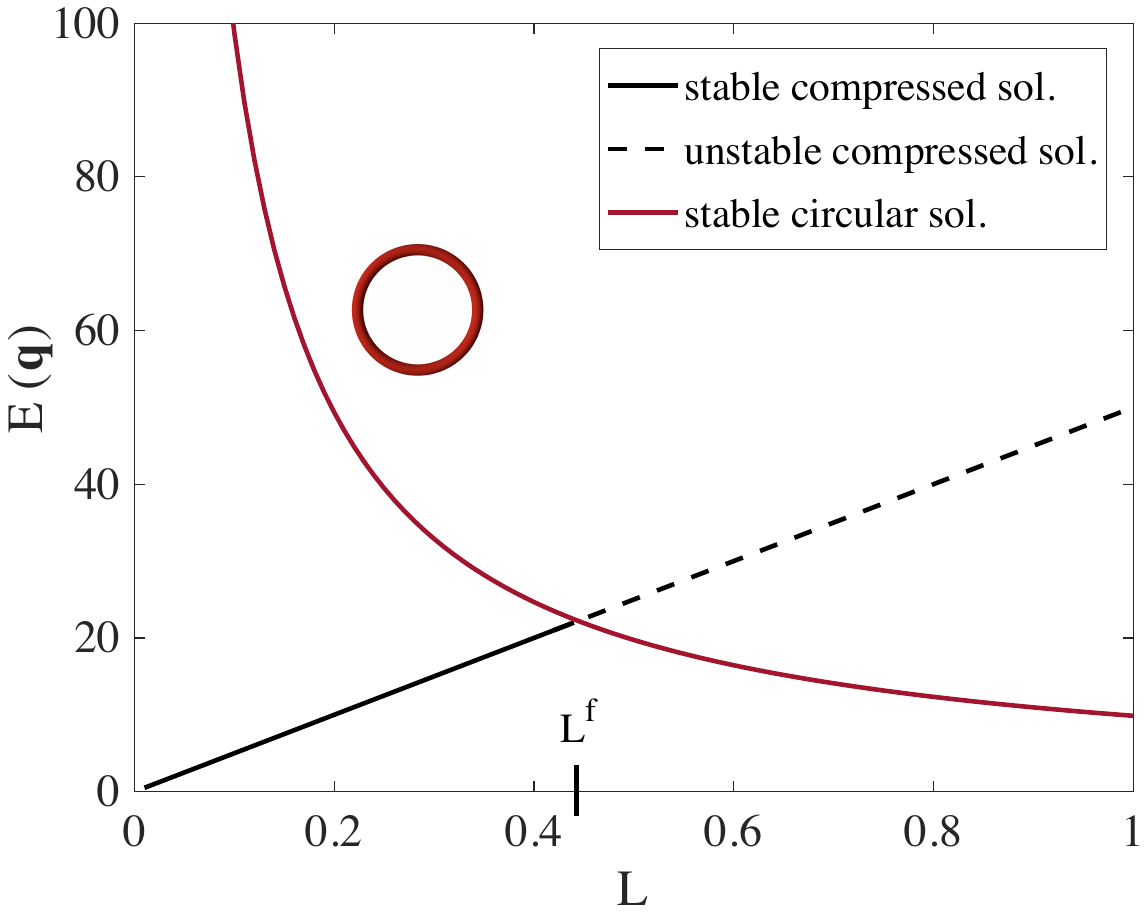}\qquad
}
\subfigure[]{
\includegraphics[width=.45\textwidth]{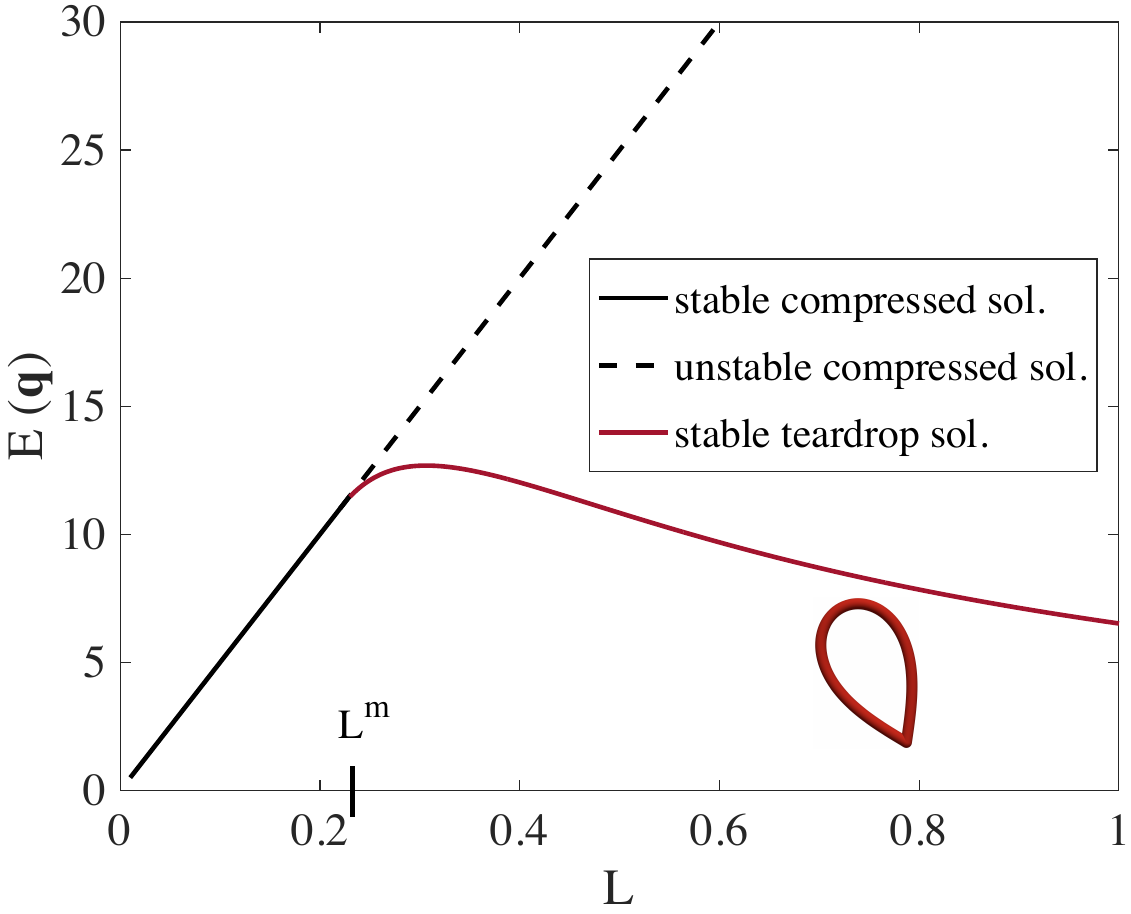}
}

\subfigure[]{
\includegraphics[width=.457\textwidth]{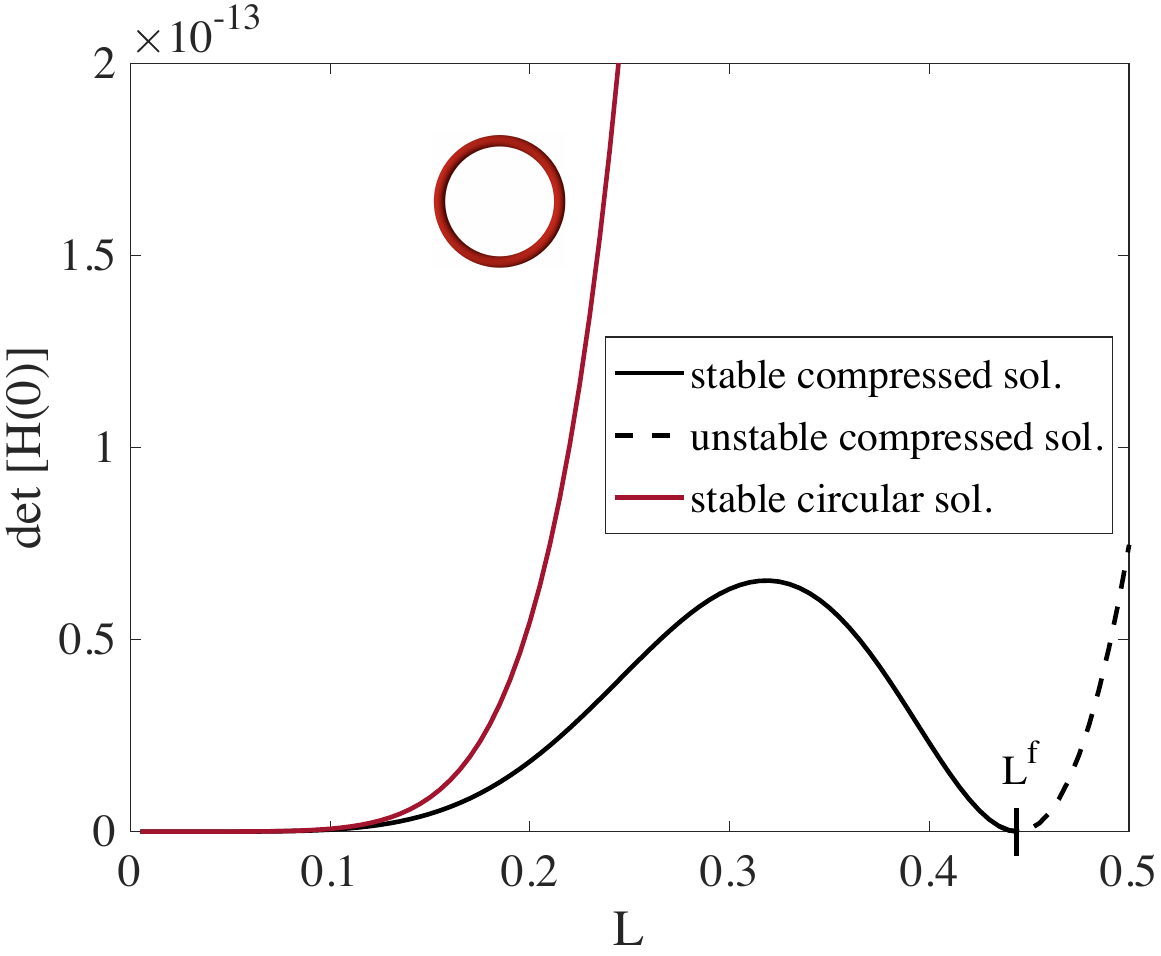}\qquad
}
\subfigure[]{
\includegraphics[width=.45\textwidth]{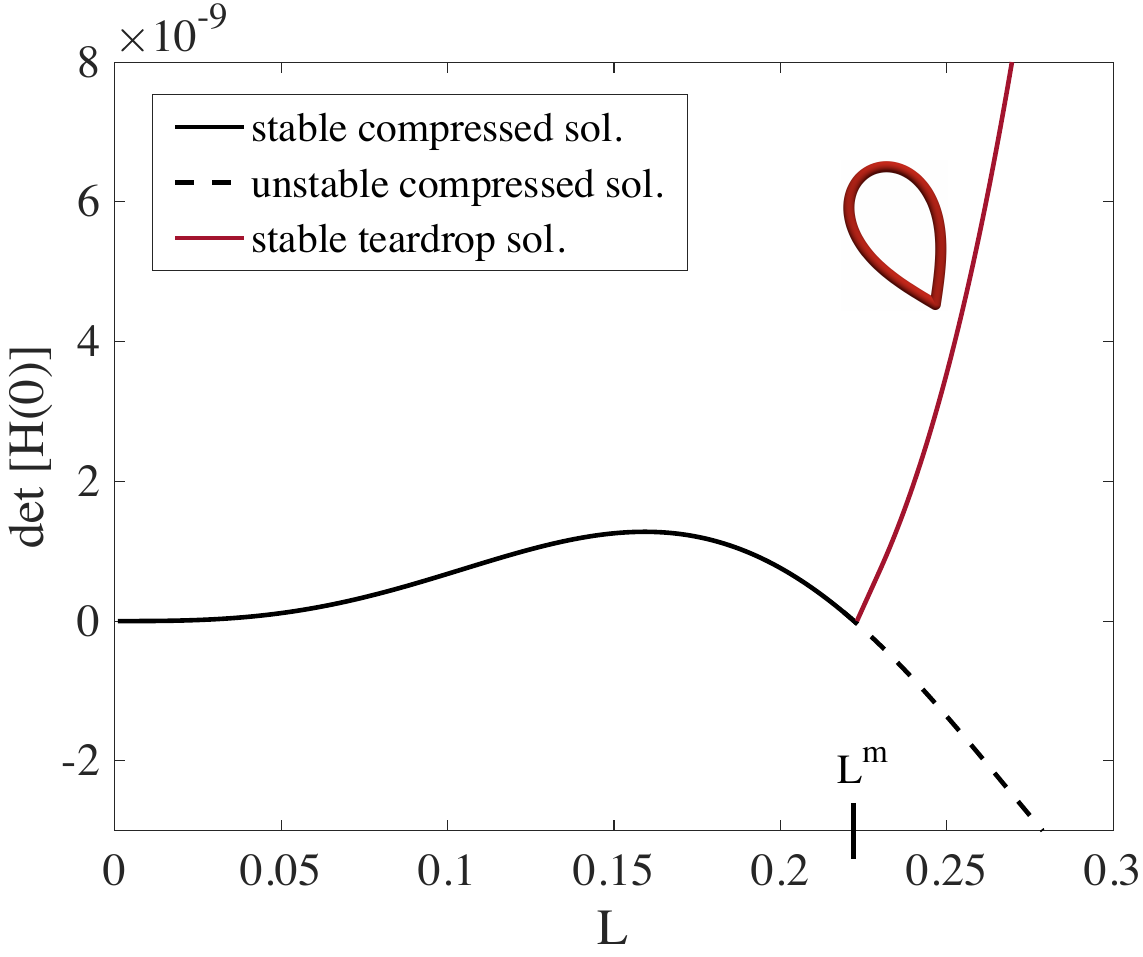}
}
\caption{Stability analysis for a non-isotropic Cosserat rod with $k_1=0.5$, $k_2=5$, $k_3=10$ and $a_1=a_2=a_3=100$. Continuous lines represent quantities associated with stable solutions, dashed lines to unstable ones. In panels (a) and (b) the energies for the circular and teardrop equilibria are displayed in red, together with the compressed solution, in black, which becomes unstable after the bifurcation point $L^f$ or $L^m$. In panels (c) and (d) we report the values of $\det{[\bm{\mathsf{H}}(0)]}$  computed on the associated solutions for the full and marginal cases respectively, with conjugate points arising when the curves hit zero.}
\label{fig2345}
\end{figure*} 

We refer to a linearly elastic, uniform, with diagonal stiffness matrix, intrinsically straight and untwisted rod ($\bm{\mathcal{P}}(s)=\bm{\mathcal{P}}=diag\lbrace k_1,k_2,k_3,a_1,a_2,a_3\rbrace$, $\hat{\bm{\mathsf{u}}}=\bm{0}$, $\hat{\bm{\mathsf{v}}}=(0,0,1)$), as presented above. For Cosserat rods, cyclization problems full and marginal always admit a ``compressed'' trivial solution $\bm{q}^c$, characterised by $\bm{r}^c=\bm{0}$, $\bm{R}^c=\bm{\mathbb{1}}$, $\bm{\mathsf{u}}^c=\bm{0}$, $\bm{\mathsf{m}}^c=\bm{0}$, $\bm{\mathsf{v}}^c=\bm{0}$, $\bm{\mathsf{n}}^c=(0,0,-a_3)$, with energy $E(\bm{{q}}^c)=\frac{a_3 L}{2}$. Analysing the determinant of the associated Jacobi fields with initial conditions Eq.~(\ref{inFM}) by means of conjugate point theory, we observe that the latter solution is stable (\ie a minimizer of the energy) in the range $0<L<L^f$ for the full case, and in $0<L<L^m$ for the marginal case, where 
\begin{equation}
L^f=\frac{2\pi}{a_3}min\{\sqrt{k_1\,a_2},\,\,\sqrt{k_2\,a_1}\}\;,\quad L^m=\frac{L^f}{2}\;. 
\end{equation}
In fact
\begin{equation}
\det{[\hhf_c(s)]}=\frac{1}{ 4{a_3}^5  {k_3}}(L-s)^2 \sin ^2\left(\frac{ {a_3}}{2 \sqrt{ {k_1}{a_2}}}(L-s)\right) \sin ^2\left(\frac{ {a_3}}{2 \sqrt{ {k_2}{a_1}}}(L-s)\right)\;,
\end{equation}
\begin{equation}
\det{[\hhm_c(s)]}=\frac{1}{{a_3}^3}\sqrt{\frac{ {k_1}  {k_2}}{  {a_1}  {a_2} }}(L-s) \sin \left(\frac{ {a_3} }{\sqrt{ {k_1}{a_2}}}(L-s)\right) \sin \left(\frac{ {a_3} }{\sqrt{ {k_2}{a_1}}}(L-s)\right)\;.
\end{equation}
Moreover, as already mentioned, for full looping $(\mathtt{f})$ there exist also circular solutions $\bm{q}^f$, which are stable for all $L>0$, with energy $\frac{2\pi^2 k_1}{L}$. (This is true except for 
\begin{equation}\label{except}
k_1>k_3\;,\quad L<2\pi\sqrt{\frac{k_1-k_3}{a_1}}\;, 
\end{equation}
but in the present work we will not treat such an instability of the circular solution). Note that if $k1\leq k2$ and $a_1=a_2=a_3$, then 
\begin{equation}
E(\bm{q}^c)=E(\bm{q}^f)\quad\text{at}\quad L^f={2\pi}\sqrt{\frac{k_1}{a_3}}\quad\text{and}\quad E(\bm{q}^c)<E(\bm{q}^f)\quad\text{for}\quad 0<L<L^f\;. 
\end{equation}
For marginal looping $(\mathtt{m})$, the stable teardrop solution $\bm{q}^m$ ceases to exist in the interval $0<L<L^m$, merging with the compressed solution which becomes stable. We show the bifurcation diagrams in Fig. \ref{fig2345} for a non-isotropic Cosserat rod. The isotropic case is totally analogous, except from the fact that an entire family of minimizers is involved and a conjugate point is always present due to the continuous symmetry. Observe that $E(\bm{q}^m)$ does not explode for small lengths, but instead reaches a maximum and decreases towards $E(\bm{q}^c)$. By contrast, for a Kirchhoff rod the circular and teardrop solutions exist and are stable for all $L>0$, with energy diverging approaching $L=0$, and no compressed solution is present.

Since in the Cosserat case the compressed (isolated) solution is a minimizer for the short-length scale regimes, we evaluate analytically its contribution $\rho^c_{\alpha}$ to the cyclization probability density $(\mathtt{f})$ and $(\mathtt{m})$ for $0<L<L^f$ and $0<L<L^m$ respectively. Making use of Eq.~(\ref{fin}) with initial conditions Eq.~(\ref{inFM}) and setting the non-dimensional length $\tilde{L}=\frac{L}{l_p}$ for a given $l_p>0$, we get 
\begin{equation}\label{compr}
\rho^c_{\alpha}\approx e^{-E_p\tilde{L}}\frac{1}{{l_p}^{3}{\tilde{L}}^{\frac{1}{x}}}\sqrt{y\,\csc(x\,\vartheta_1\,\tilde{L})^{\frac{2}{x}}\csc(x\,\vartheta_2\,\tilde{L})^{\frac{2}{x}}}\;,
\end{equation}
where $x=x(\alpha)$ with $x(f)=1$, $x(m)=2$ and 
\begin{equation}\label{nd1}
E_p=\beta\frac{l_p\,a_3}{2}\;,\quad\vartheta_1=\frac{l_p\,a_3}{2\sqrt{k_1\,a_2}}\;,\quad\vartheta_2=\frac{l_p\,a_3}{2\sqrt{k_2\,a_1}}\;, 
\end{equation}
$y=y(\alpha)$ with 
\begin{equation}\label{nd2}
y(f)=\frac{\beta^2}{\pi^6}k_3\,a_3\,E_p^4\;,\quad y(m)=\frac{l_p^2}{\pi^3}\sqrt{\frac{a_1\,a_2}{k_1\,k_2}}\,E_p^3\;. 
\end{equation}
Since $\beta=[E]^{-1}$, $k_i={[E]}{[l]}$ and $a_i=\frac{[E]}{[l]}$ for $i=1,...,3$, where $E$ stands for energy and $l$ for length, the quantities defined in Eq.~(\ref{nd1}) and Eq.~(\ref{nd2}) are dimensionless. The latter formula is valid for both isotropic (setting $k_1=k_2$, $a_1=a_2$) and non-isotropic rods. In the following we focus on the contribution $\rho_{\alpha}$ to the cyclization probability density full and marginal coming from the circular and teardrop minimizers respectively.

\subsection{Non-isotropic polymers}
The uniform and intrinsically straight example leading to isolated circular minimizers in the non-isotropic case has been previously studied in \citep{LUDT} and for planar rods in \citep{LUD}. In the cited works one can find the approximation formulas for full looping derived by solving analytically the associated Jacobi systems, to which one is referred for detailed calculations, including the treatment of in-plane and out-of-plane fluctuations. We note that, in the case of Cosserat rods, the investigation found in \citep{LUDT, LUD} do not consider the analysis arising from the presence of a stable compressed solution, which leads to a different behaviour in the limit of zero length. The diverging behaviour that we are going to show for the conditional probability density at zero length is a consequence of the linearly elastic hypothesis on the energy functional and cannot be regarded as a physical behaviour in the context of polymers made of elementary units. However, we show the existence of a length-range, not affected by a compressed stable solution, where high looping probabilities occur due to an energy relaxation of the teardrop minimizer achieved by exploiting the degrees of freedom associated with extension and shear (this will occur also for isotropic polymers). We believe this mechanism to be relevant and to be shared also by general problems with full boundary conditions in the context of looping probabilities of stochastic Cosserat polymers. In the following we justify our statements and compare the derived results with extensive Monte Carlo simulations.

First, we consider a non-isotropic rod ($k_1\neq k_2$), further assuming without loss of generality that $k_1<k_2$. For the case of full looping $(\mathtt{f})$, there exist two circular, untwisted, isolated minimizers $\bm{{q}}^f$ lying on the $y-z$ plane with energy $\frac{2\pi^2 k_1}{L}$. The existence of a couple of reflected minima simply translates into a factor of 2 in front of Eq.~(\ref{fin}) and the Laplace expansion is performed about one of them (\eg about the one having non-positive $y$ coordinate). For this case Eq.~(\ref{JacHam}) is a constant coefficients Jacobi system, that we solve analytically together with the first set of initial conditions in Eq.~(\ref{inFM}), in order to obtain the approximated formula for the cyclization probability density $\rho_f(\bm{q}_0,L|\bm{q}_0,0)$ for both Cosserat and Kirchhoff rods. Namely,
\begin{equation}
\rrf\approx 2\,e^{-\frac{2\beta\pi^2 k_1}{L}}\frac{1}{L^6}\sqrt{\frac{2^9 \beta^6 \pi^2 k_1^5(k_3-k_1)}{a^2 b(1-\cos\lambda)}}\;,
\end{equation}
with 
\begin{equation}\label{quant1}
\lambda=2\pi\sqrt{\frac{(k_3-k_1)(k_2-k_1)}{k_2 k_3}}\;,\quad a=1+\left(\frac{2\pi}{L}\right)^2\left(\frac{1}{a_2}+\frac{1}{a_3}\right)k_1\;,\quad b=1+\left(\frac{2\pi}{L}\right)^2\frac{k_3-k_1}{a_1}\;. 
\end{equation}
Setting the length scale $l_p=2\beta k_1$, which corresponds to the planar tangent-tangent persistence length for the same rod but constrained in two dimensions \citep{PL}, and the non-dimensional length $\tilde{L}=\frac{L}{l_p}$, we get 
\begin{equation}
\rrf\approx 2\,e^{-\frac{\pi^2}{\tilde{L}}} h_{I}\,h_{O}\;, 
\end{equation}
where $h_{I}$ and $h_{O}$ are the in-plane (of the minimizer) and out-of-plane contributions
\begin{equation}\label{Pop}
    {h_{I}}=\frac{1}{{l_p}^{2}{\tilde{L}}^{7/2}}\sqrt{\frac{\pi}{a^2}}\;,\quad {h_{O}}_{(\mathtt{non-iso})} = \frac{1}{l_p{\tilde{L}}^{5/2}} \sqrt{\frac{8 \pi (1-\nu_3)}{b\,\nu_3(1-\cos\lambda)}}\;,
\end{equation}
\begin{equation}\label{An1}
{\rho_f}_{(\mathtt{non-iso})} \approx 2\,e^{-\frac{\pi^2}{\tilde{L}}}\frac{1}{{l_p}^{3}{\tilde{L}}^{6}}\sqrt{\frac{8 \pi^2 (1-\nu_3)}{a^2\,b\,\nu_3(1-\cos\lambda)}}\;,
\end{equation}
with 
\begin{equation}\label{quant2}
\lambda=2\pi\sqrt{(1-\nu_2)(1-\nu_3)}\;,\quad a = 1 + \left(\frac{2\pi}{{\tilde{L}}}\right)^2 (\eta_2+\eta_3)\;,\quad b = 1 + \left(\frac{2\pi}{{\tilde{L}}}\right)^2 (\omega_1-\eta_1)\;, 
\end{equation}
\begin{equation}\label{quant3}
\nu_2 = \frac{k_1}{k_2}\;,\quad \nu_3 = \frac{k_1}{k_3}\;,\quad \eta_1 = \frac{k_1}{a_1\,l_p^2}\;,\quad \eta_2 = \frac{k_1}{a_2\,l_p^2}\;,\quad \eta_3 = \frac{k_1}{a_3 \,l_p^2}\;,\quad \omega_1 = \frac{k_3}{a_1\,l_p^2}\;. 
\end{equation}
The Kirchhoff case is recovered setting $a=b=1$ (and condition Eq.~(\ref{except}) arises when $b=0$). Moreover, the density obtained disregarding the factor ${h_{O}}$ coincides with the cyclization probability density for planar rods given in \citep{LUD}. Note that the in-plane and out-of-plane contributions are computed by performing two separated Gaussian path integrals for the in-plane and out-of-plane variation fields, exploiting the decomposition of the second variation in two distinguished terms \citep{LUDT}. Moreover, the expressions in Eq.~(\ref{Pop}), Eq.~(\ref{An1}) are valid under stability assumptions for $k_1<k_2$, $k_1\neq k_3$ and equal to the limit $k_3\rightarrow k_1$, \ie $\nu_3\rightarrow 1$, if $k_3=k_1$. We further underline that Eq.~(\ref{An1}) diverges in the isotropic limit $k_2\rightarrow k_1$, \ie $\nu_2\rightarrow 1$.

In general, for computing the density $\rrm(\bm{0},L|\bm{q}_0,0)$ from Eq.~(\ref{fin}) together with the second set of initial conditions in Eq.~(\ref{inFM}), numerics must be used. In fact, for the case of marginal looping $(\mathtt{m})$, there are no simple analytical expressions for the two planar ($y-z$ plane) and untwisted teardrop shaped isolated minimizers $\bm{{q}}^m$.
However, in the Kirchhoff case there exists a scaling argument in the variable $L$, which allows one to provide a qualitative expression. Namely, given the fact that we can compute numerically a Kirchhoff equilibrium $\bm{{q}}^m_p$ for a given rod length $l_p$, characterised by 
\begin{equation}
\bm{r}_p(s_p)\;,\quad \bm{R}_p(s_p)\;,\quad \bm{\mathsf{v}}_p(s_p)=\hat{\bm{\mathsf{v}}}\;,\quad \bm{\mathsf{u}}_p(s_p)\;,\quad \bm{\mathsf{n}}_p(s_p)\;,\quad \bm{\mathsf{m}}_p(s_p)\quad\text{for}\quad s_p\in[0,l_p]\;, 
\end{equation}
then for each $L>0$ it can be easily checked that 
\begin{equation}
\begin{split}
\bm{r}(s)=\tilde{L}\,\bm{r}_p\left(\frac{s}{\tilde{L}}\right)\;,\quad \bm{R}(s)=\bm{R}_p\left(\frac{s}{\tilde{L}}\right)\;,\quad \bm{\mathsf{v}}(s)=\,\bm{\mathsf{v}}_p\left(\frac{s}{\tilde{L}}\right)=\hat{\bm{\mathsf{v}}}\;,\quad \bm{\mathsf{u}}(s)=\frac{1}{\tilde{L}}\,\bm{\mathsf{u}}_p\left(\frac{s}{\tilde{L}}\right)\;,\\
\\
\bm{\mathsf{n}}(s)=\frac{1}{\tilde{L}^2}\,\bm{\mathsf{n}}_p\left(\frac{s}{\tilde{L}}\right)\;,\quad \bm{\mathsf{m}}(s)=\frac{1}{\tilde{L}}\,\bm{\mathsf{m}}_p\left(\frac{s}{\tilde{L}}\right)\;,\quad \tilde{L}=\frac{L}{l_p}\qquad\qquad\qquad\quad
\end{split}
\end{equation}
define a Kirchhoff equilibrium $\bm{{q}}^m$ for $s\in[0,L]$. This immediately implies that
\begin{equation}
E(\bm{{q}}^m)=\frac{1}{\tilde{L}}\,E(\bm{{q}}^m_p)\;. 
\end{equation}
Moreover, since the matrix $\bm{\mathsf{E}}(s)$ is given in terms of strains, forces and moments at the equilibrium by means of Eq.~(\ref{E2}), Eq.~(\ref{E3}), it is possible to obtain the scaling for the Jacobi fields as 
\begin{equation}
{\text{det}[\hhm(0)]}=\tilde{L}^{9}\,{\text{det}[\hhm_p(0)]}\;. 
\end{equation}
Finally, defining 
\begin{equation}
E_p=\beta E(\bm{{q}}^m_p)\quad\text{and}\quad h_p=\left(\frac{\beta}{2\pi}\right)^\frac{3}{2}\frac{l_p^{3}}{\sqrt{\text{det}[\hhm_p(0)]}}\;, 
\end{equation}
we get 
\begin{equation}\label{SAn1}
{\rrm}_{(\mathtt{non-iso})}\approx 2\,e^{-\frac{E_p}{\tilde{L}}}\frac{h_p}{{l_p}^{3}{\tilde{L}}^{\frac{9}{2}}}\;,
\end{equation}
where $E_p$ and $h_p$ have to be computed numerically, and the factor $2$ accounts for the contribution of both the minimizers. By contrast, a simple scaling argument is not present for a Cosserat rod, therefore allowing for more complex behaviours.

\begin{figure*}
\captionsetup{justification=centering,margin=1cm}
\subfigure[]{\includegraphics[width=.487\textwidth]{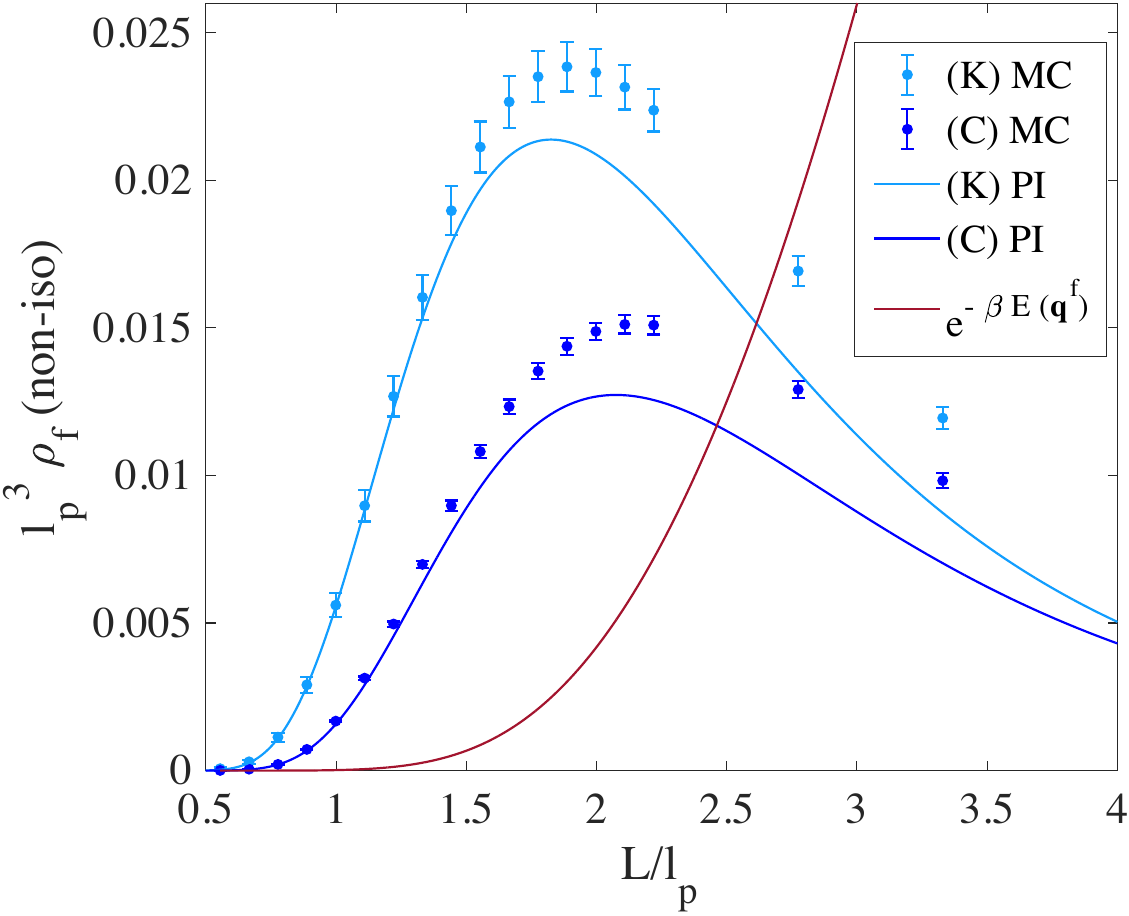}}\quad
\subfigure[]{\includegraphics[width=.49\textwidth]{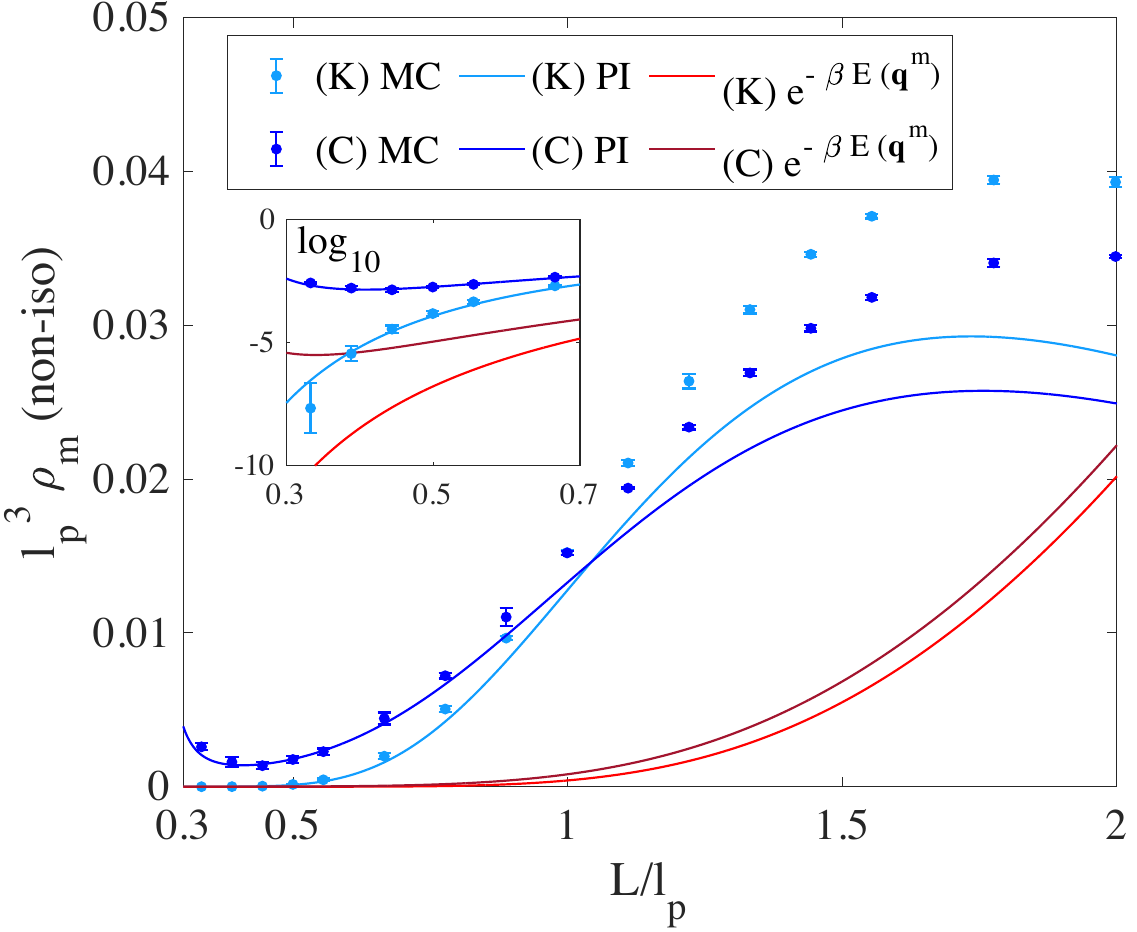}}
\caption{Comparison of cyclization densities between the path integral (PI) Laplace approximation (continuous lines) and Monte Carlo (MC) (discrete points with standard deviation error bars) for a non-isotropic rod. For Kirchhoff $(\mathtt{K})$ we set $\beta=1$, $k_1=0.5$, $k_2=5$, $k_3=10$; for the Cosserat $(\mathtt{C})$ case, we also set $a_1=a_2=a_3=100$. The quantities are reported in non-dimensional form. In particular, the undeformed length of the rod is expressed in units of real persistence length $l_p\approx 0.9$, the harmonic average of $k_1$ and $k_2$. In panel (a) we address the full case, reporting the values for $\rrf$ and displaying in red the zero-order contribution. In panel (b) the results for the marginal density $\rrm$ are reported, with a zoom window in $\log_{10}$ scale in order to underline the peculiar small length trend. In this case, Kirchhoff and Cosserat rods differ in zero-order contribution of the energy, and two different red curves are displayed. Note that the error bar ``$\mu\pm\sigma$'' expressed in logarithmic scale is correctly given by ``$\log_{10}(\mu)\pm\frac{\sigma}{\ln(10)\,\mu}$'' since we can make the approximation $\dd(\log_{10}(\mu))\approx\frac{\dd\mu}{\ln(10)\,\mu}$, and the absolute error becomes the relative error in logarithmic scale.}
\label{fig67}
\end{figure*}

We show the results in Fig. \ref{fig67} for a specific choice of the parameters, in the range $L>L^f$ and $L>L^m$ respectively for full and marginal, so that the only accounted minimizers for the computation of the cyclization probability densities are the circular and the teardrop solutions, and we can apply Eq.~(\ref{An1}) and Eq.~(\ref{SAn1}). The simulations show good agreement between the Laplace approximation and Monte Carlo in the target small length domain. Even though the second-order expansion loses its quantitative power for larger lengths, the qualitative behaviour is captured and the error does not explode. We recall that looping is a rare event and Monte Carlo simulations are usually expensive and unfeasible; by contrast, the method proposed in the present chapter is performing successfully with much higher efficiency. It is also important to underline that for the specific example considered the difference in $\rrf$ between Kirchhoff and Cosserat rods is due only to Jacobi fields, since the energy factor is the same, the circular minima having no extension and no shear deformations. The marginal case $(\mathtt{m})$ is more representative of the general behaviour where Kirchhoff and Cosserat minimizers are distinct solutions, which is true also for $(\mathtt{f})$ boundary conditions for arbitrary (non-uniform, with non-straight intrinsic shape) elastic rods. In fact, in the short-length scale regimes, the possibility to exploit the additional degrees of freedom associated with extension and shear is crucial for minimizing the overall elastic energy, in the face of an increasingly penalizing bending contribution. This phenomenon allows the probability density to be remarkably higher than the Kirchhoff case below the persistence length, remaining almost constant and even increasing in the range where for the Kirchhoff rod (and therefore also for the WLC model) is exponentially vanishing. By contrast, for large lengths extension and shear become negligible. In addition, as a general statement, the Jacobi factor is fundamental to determine the peak of the density, in a domain where the energy is monotonically decreasing with length. On the other hand, the energy contribution dominates the system for smaller lengths. Finally, we clearly observe overall higher values for the marginal density compared to the full case because of the less restrictive boundary conditions.

\subsection{Isotropic polymers}
Now we consider the isotropic case, \ie, $k_1=k_2$, $a_1=a_2$ and a one-parameter family of non-isolated circular or teardrop minima arises. Given the minimizer in the $y-z$ plane with $y\leq 0$ represented by $\bm{r}(s)=(0,r_2(s),r_3(s))$ and $\bm{R}(s)$ a counter-clockwise planar rotation about the $x$ axis of an angle $\varphi(s)$, the one-parameter family of minimizers can be expressed as 
\begin{equation}
\bm{r}(s;\theta)=\bm{Q}_{\theta}\bm{r}(s)\;,\quad\bm{R}(s;\theta)=\bm{Q}_{\theta}\bm{R}(s)\bm{Q}_{\theta}^T\;, 
\end{equation}
where $\bm{Q}_{\theta}$ is defined as the counter-clockwise planar rotation about the $z$ axis of an angle $\theta\in [0,2\pi)$. Thus, taking the derivative of such minimizers with respect to $\theta$ and finally setting $\theta=0$, the zero mode can be easily recovered in the chosen parametrisation to be 
\begin{equation}
{\bm{\mathsf{\psi}}}(s)=\left(0,\frac{1}{2}\sin{(\varphi)},\frac{1}{2}(\cos{(\varphi)}-1),-r_2,0,0\right)\;. 
\end{equation}
Moreover, the conjugate momentum of ${\bm{\mathsf{\psi}}}$ is derived in general (for both Cosserat and Kirchhoff rods) substituting the zero mode itself and its to-be-found moment as the unknowns of the Jacobi equations in Hamiltonian form Eq.~(\ref{JacHam}) computed on the minimizer associated with $\theta=0$ (recalling to multiply $\bm{\mathsf{E}}_{1,1}$ by $\frac{\beta}{2\pi}$ and $\bm{\mathsf{E}}_{2,2}$ by $\frac{2\pi}{\beta}$), and reads as 
\begin{equation}
\bm{\mathsf{\mu}}_{{\bm{\mathsf{\psi}}}}(s)=\left(0,\frac{\beta k_1}{2\pi}(\cos{(\varphi)}+1)\varphi ',-\frac{\beta k_1}{2\pi}\sin{(\varphi)}\varphi ',-\frac{\beta}{2\pi} n_2,0,0\right)\;. 
\end{equation}

At this point it is straightforward to apply the theory developed above for non-isolated minimizers, choosing $\bm{\mathsf{\chi}}$ to be a matrix with unit determinant such that the second column ($i=2$) corresponds to $\bm{\mathsf{\mu}}_{{\bm{\mathsf{\psi}}}^f}(L)$ and $\bm{\mathsf{X}}$ a matrix with determinant equal to $-1$ such that the fourth column ($i=4$) corresponds to $([{\bm{\mathsf{\psi}}}^m]_{1:3},[\bm{\mathsf{\mu}}_{{\bm{\mathsf{\psi}}}^m}]_{4:6})^T(L)$, according to Eq.~(\ref{inFMiso}). Consequently, the initial conditions for the Jacobi equations are well defined, the energy is computed, \textit{e.g.}, for the minimizer corresponding to $\theta=0$ as before, and Eq.~(\ref{finiso}) is analytical for $\rho_f(\bm{q}_0,L|\bm{q}_0,0)$
\begin{equation}
\rho_f\approx 2\pi\,e^{-\frac{2\beta\pi^2 k_1}{L}}\frac{1}{L^{\frac{13}{2}}}\sqrt{\frac{2^9 \beta^{7}\pi k_1^6 k_3}{a^2\,b}}\;,
\end{equation}
with 
\begin{equation}
a=1+\left(\frac{2\pi}{L}\right)^2\left(\frac{1}{a_1}+\frac{1}{a_3}\right)k_1
\end{equation}
and $b$ is the same defined for non-isotropic polymers in Eq.~(\ref{quant1}). More specifically, and in non-dimensional form, we have
\begin{equation}
\rrf\approx 2\pi e^{-\frac{\pi^2}{\tilde{L}}} h_{I}\,h_{O}\;, 
\end{equation}
where
\begin{equation}\label{An2}
{h_{O}}_{(\mathtt{iso})} = \frac{1}{l_p{\tilde{L}}^{3}} \sqrt{\frac{4}{b\,\nu_3}},\quad{\rho_f} _{(\mathtt{iso})}\approx2\pi e^{-\frac{\pi^2}{ \tilde{L}}}\frac{1}{{l_p}^{3}{\tilde{L}}^{\frac{13}{2}}}\sqrt{\frac{4 \pi}{a^2 b\,\nu_3}}\;,
\end{equation}
with 
\begin{equation}
a = 1 + \left(\frac{2\pi}{\tilde{L}}\right)^2 (\eta_1+\eta_3)
\end{equation}
and all the other quantities have been defined previously in Eq.~(\ref{quant2}), Eq.~(\ref{quant3}), with $l_p=2\beta k_1$ and $\tilde{L}=\frac{L}{l_p}$. In particular, $h_{I}$ is the same as for the non-isotropic case Eq.~(\ref{Pop}) and therefore the zero mode arises for the out-of-plane factor for which the above regularization is applied. The Kirchhoff limit is recovered as before setting $a=b=1$.

For the marginal density $\rrm$ numerics must be used, but in the Kirchhoff case we can carry on the scaling argument in the variable $L$ as before, obtaining
\begin{equation}\label{SAn2}
{\rrm}_{(\mathtt{iso})}\approx 2\pi\,e^{-\frac{E_p}{\tilde{L}}}\frac{h_p}{{l_p}^{3}{\tilde{L}}^{5}}\;,
\end{equation}
for given
\begin{equation}
E_p=\beta E(\qqm_p)\;,\quad h_p=l_p^{3}\sqrt{\frac{[\bm{\mathsf{\mu}}_{{{\bm{\mathsf{\psi}}}_{ p}^{m}}}(0)]_i}{]{\hhm_p(0)}[_{i,i}}}
\end{equation}
computed numerically.

\begin{figure*}
\centering
\captionsetup{justification=centering,margin=1cm}
\subfigure[]{\includegraphics[width=.47\textwidth]{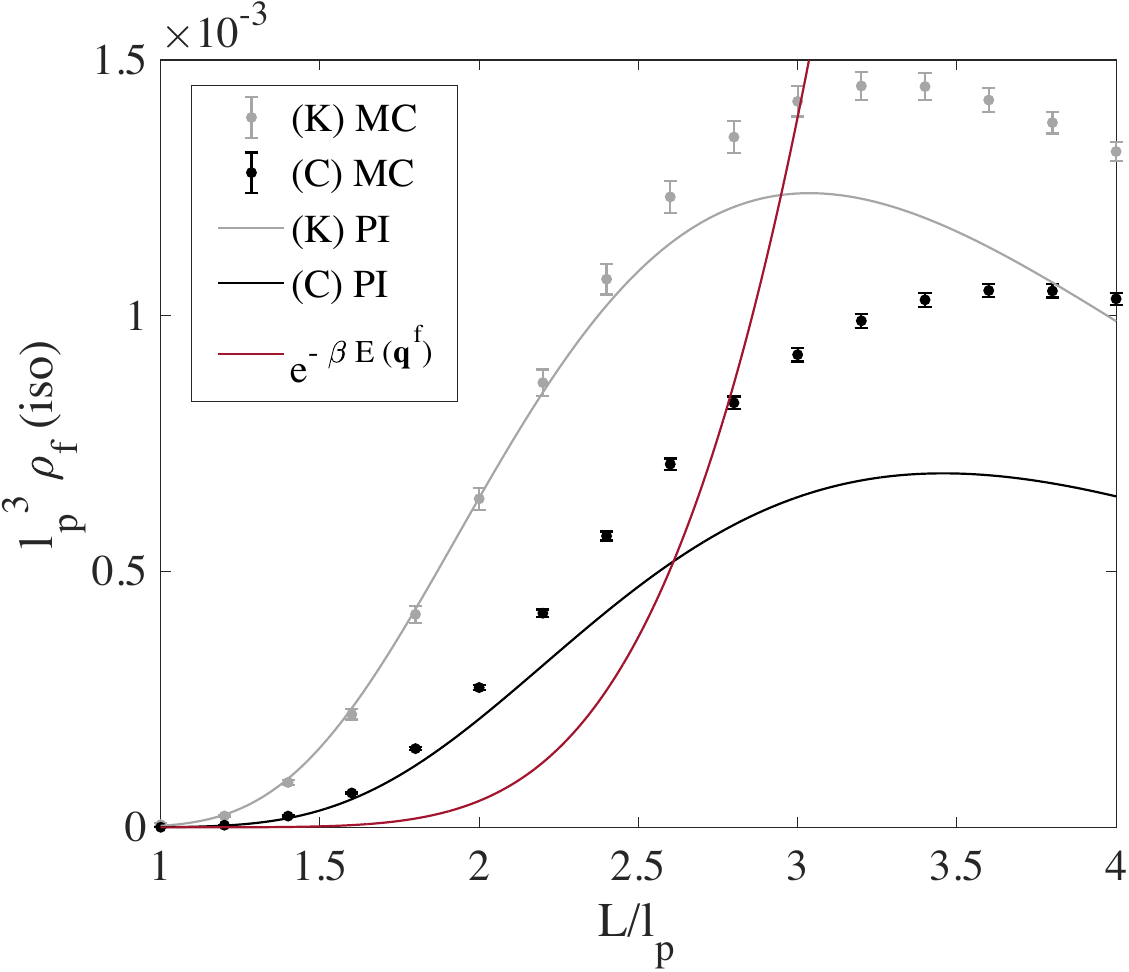}}\qquad
\subfigure[]{\includegraphics[width=.46\textwidth]{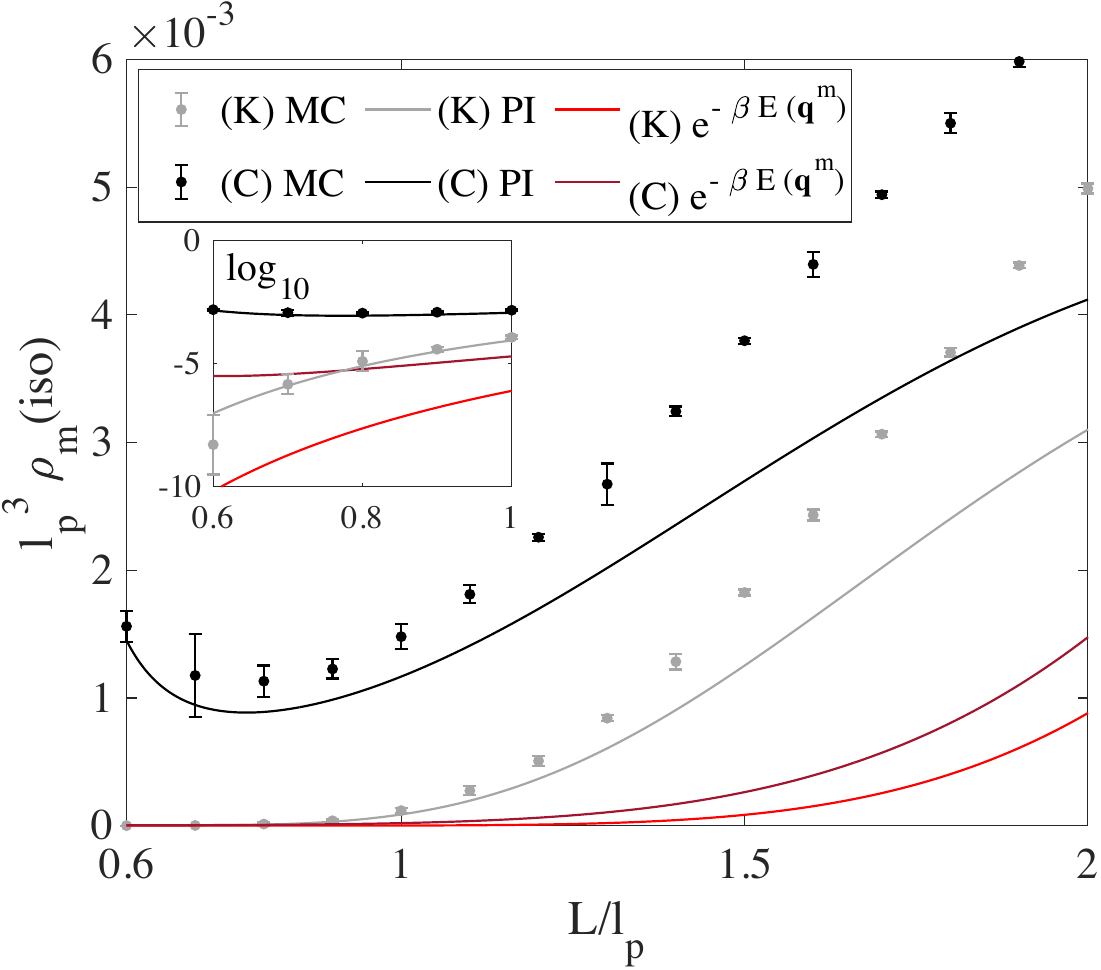}}

\subfigure[]{\includegraphics[width=.47\textwidth]{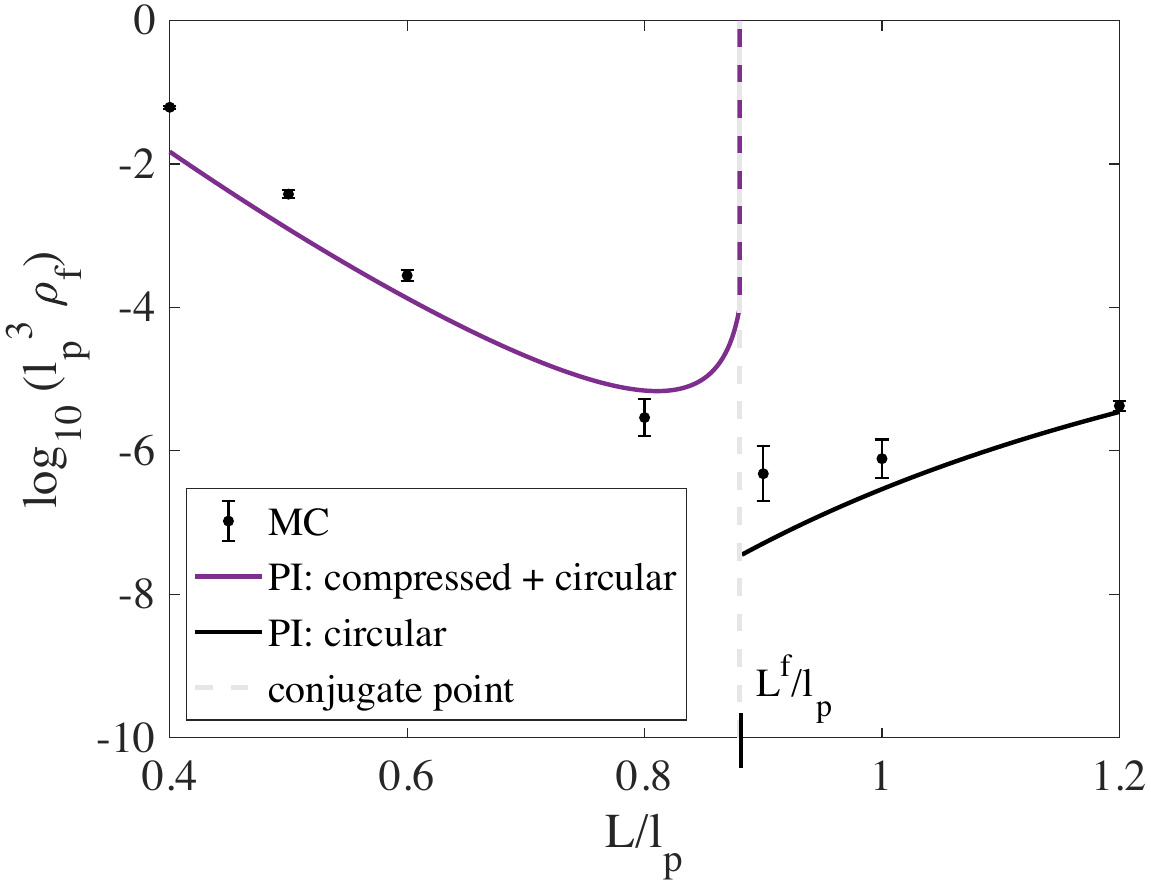}}\qquad
\subfigure[]{\includegraphics[width=.47\textwidth]{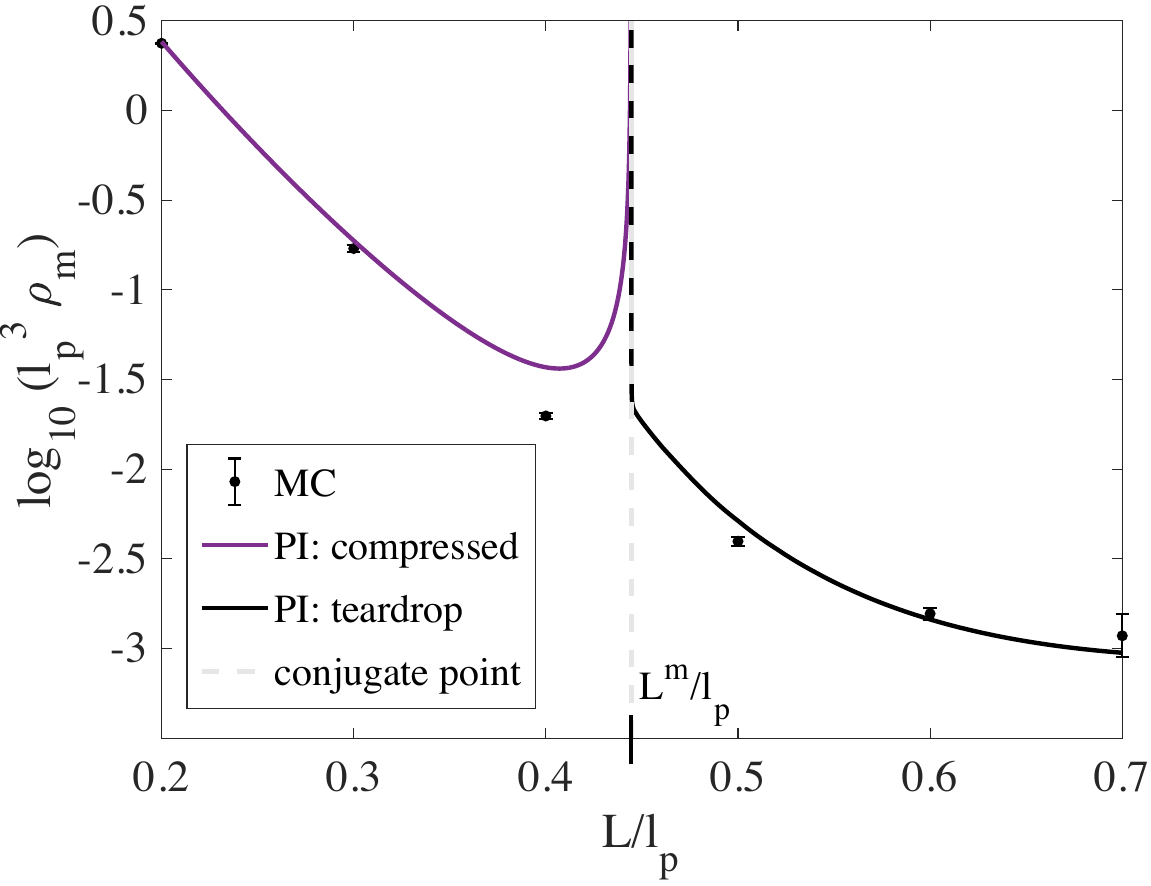}}
\caption{Comparison of cyclization densities between the path integral (PI) approximation and Monte Carlo (MC) for an isotropic rod. For Kirchhoff $(\mathtt{K})$ we set $\beta=1$, $k_1=k_2=0.5$, $k_3=10$; for the Cosserat $(\mathtt{C})$ case, we also set $a_1=a_2=a_3=100$. The quantities are reported in non-dimensional form and the undeformed length of the rod is expressed in units of real persistence length $l_p\approx 0.5$. In panels (a) and (c) we address the full case, reporting the values for $\rrf$ and displaying in red the zero-order contribution. The behaviour for Cosserat in the small-length regime is shown in (c). In panels (b) and (d) the results for the marginal density $\rrm$ are reported, with a zoom window in $\log_{10}$ scale; the two different zero-order contributions for Kirchhoff and Cosserat rods are displayed in red. The behaviour for Cosserat in the small-length regime is shown in panel (d).}
\label{fig891011}
\end{figure*}

It is interesting to note that formulas Eq.~(\ref{An1}), Eq.~(\ref{SAn1}), Eq.~(\ref{An2}) and Eq.~(\ref{SAn2}) scale differently with length as far as the second-order correction term is concerned. The latter scalings naturally arise from the ones observed within simpler WLC models, see Chap. $7$ in \citep{YAM00}. The comparison between Laplace and Monte Carlo simulations for isotropic polymers is shown in Fig. \ref{fig891011}, (a) and (b), for the same parameters addressed in the non-isotropic case, but now sending $k_2\rightarrow k_1$. Once more time we only consider the contributions of the manifolds made of circular and teardrop minimizers, setting $L>L^f$ and $L>L^m$. The fact that now $k_2$ is ten times smaller than the same parameter adopted in Fig. \ref{fig67} implies that the overall trend of the density is shifted to the right in units of persistence length, allowing large effects of shear and extension compared to the more standard inextensible and unshearable models, as already discussed. We further observe that the approximation error is generally higher for Cosserat rods and for marginal looping, which is a consequence of the Laplace expansion that depends on the stiffness values and boundary conditions. For the simple examples considered, there clearly exist more accurate formulas for the Kirchhoff case in the literature, \textit{e.g.}, Eq.~(\ref{SAn2}) can be related to the WLC formula (7.68) (p. 266 in \citep{YAM00}). However, the power of the method explained above lies in its generality and ability to easily provide approximation formulas for a wide range of potentially realistic and complex problems in the short length scale regimes. By contrast, since the Cosserat case represents itself a novelty, we believe that basic examples are still important to understand the underlying physical behaviour.

It is natural to ask what happens for $L\leq L^f$ and $L\leq L^m$, respectively in full and marginal, for Cosserat rods (for Kirchhoff the former analysis based only on circular and teardrop solutions is valid for all lengths). Due to the presence of the stable compressed solution in this range, the density diverges for vanishing length, and this is true for both isotropic and non-isotropic rods. In particular, for $(\mathtt{f})$ here we sum up the contributions coming from the compressed solution Eq.~(\ref{compr}) and the manifold of circular minimizers Eq.~(\ref{An2}); for $(\mathtt{m})$ only the compressed solution is present and we apply Eq.~(\ref{compr}). At the critical lengths $L^f$ and $L^m$ a conjugate point arises for the compressed solution (in $(\mathtt{m})$ the conjugate point arises also in the teardrop minimizer) and the Jacobi fields are singular, leading to an incorrect explosion of the probability density, which should be regularised. We do not address such regularisations, but in Fig. \ref{fig891011}, (c) and (d), we report the results for this length regime, together with Monte Carlo simulations which connect our approximation formulas valid on the left and on the right of the singularities.

\begin{figure*}
\centering
\captionsetup{justification=centering,margin=1cm}
\subfigure[]{\includegraphics[width=.451\textwidth]{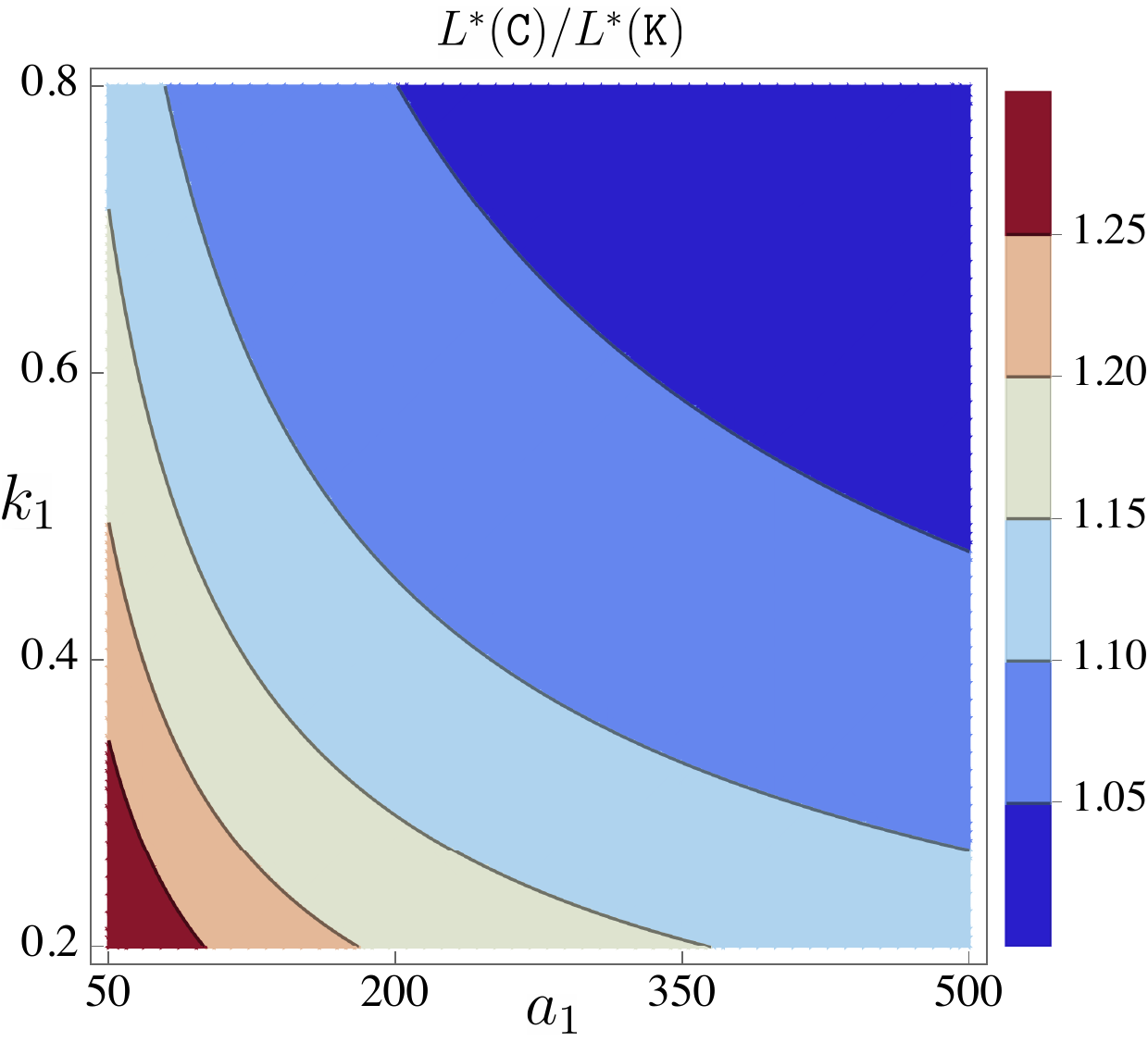}}\quad\qquad
\subfigure[]{\includegraphics[width=.45\textwidth]{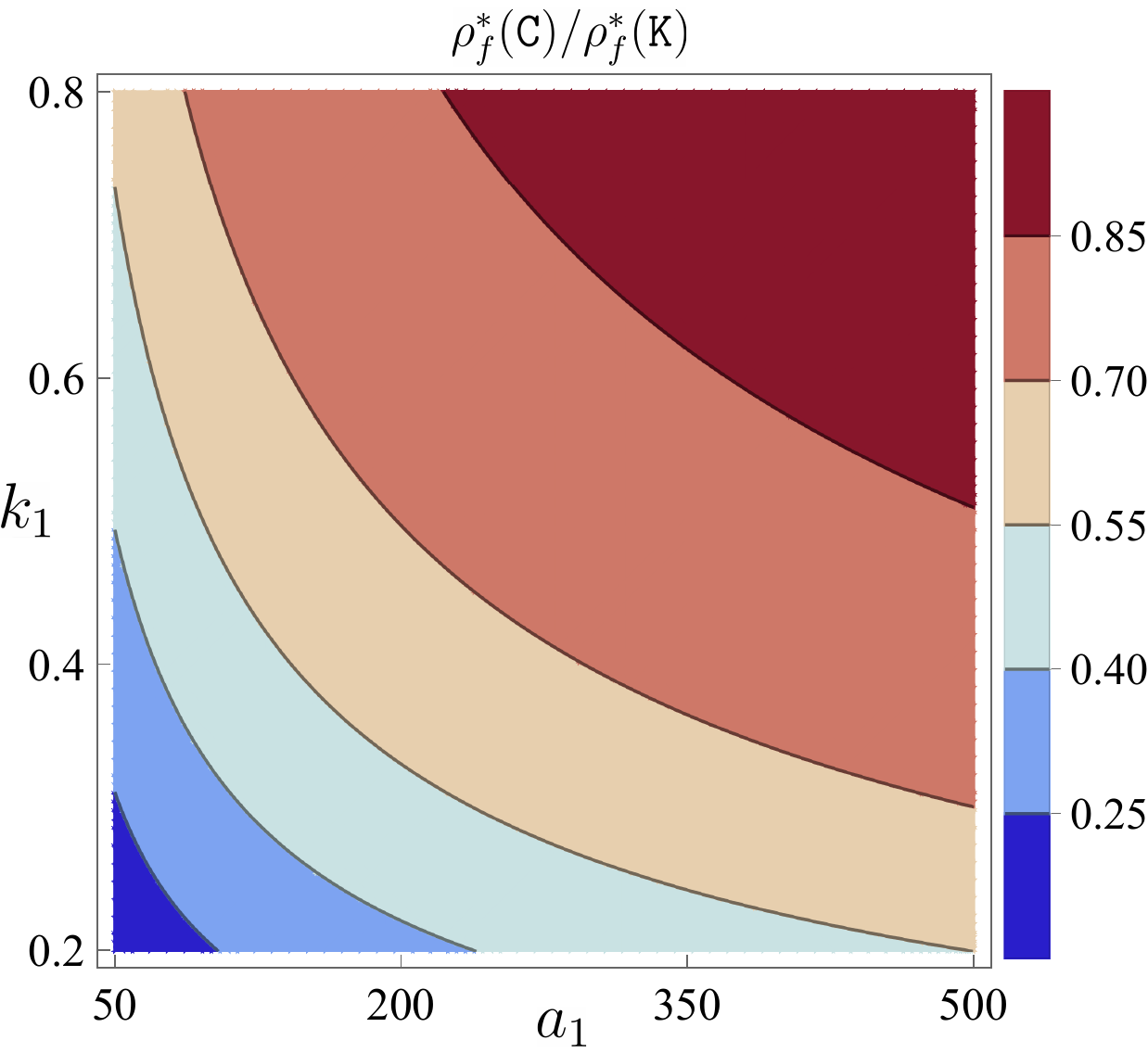}}
\caption{Contour plots for an isotropic rod (with the same parameters already used) showing the ratio of Cosserat to Kirchhoff for the length $L^*$ at which the maximum of $\rrf$ occurs ($L^*(\mathtt{C})/L^*(\mathtt{K})$ in panel (a)) and for the value $\rrf^*$ at the maximum ($\rrf^*(\mathtt{C})/\rrf^*(\mathtt{K})$ in panel (b)) as a function of $k_1=k_2$ and $a_1=a_2=a_3$.}
\label{fig1213}
\end{figure*}

Finally, in order to highlight the effect of shear and extension for larger lengths, in Fig. \ref{fig1213} we compare the Kirchhoff and the Cosserat cases in terms of the length and the value of the probability density at which the maximum of $\rrf$ occurs, the first increasing and the second decreasing in presence of extension and shear. 

%\begin{equation}\label{eqn:rate_eqns}
%\frac{\textrm{d}}{\textrm{d}t}\left[
%\begin{array}{l}
%P_{\textit{0}} \\
%P_{\textit{1}} \\
%P_{\textit{T}}
%\end{array}
%\right] =
%\left[
%\begin{array}{l}
%\frac{P_{\textit{1}}}{\tau_{\textit{10}}} + \frac{P_{\textit{T}}}{\tau_{\textit{T}}} - \frac{P_{\textit{0}}}{\tau_{\textit{ex}}} \\
%- \frac{P_{\textit{1}}}{\tau_{\textit{10}}} - \frac{P_{\textit{1}}}{\tau_{isc}} + \frac{P_{\textit{0}}}{\tau_{\textit{ex}}} \\
%\frac{P_{\textit{1}}}{\tau_{isc}} -  \frac{P_{\textit{T}}}{\tau_{\textit{T}}}
%\end{array}
%\right]
%\end{equation}
%
%
%\begin{equation}\label{eqn:avgfluorescence}
%\bar{I_{f}}(\vec{r})
%	= \gamma(\vec{r}) \left(1 - \frac{\tau_{\textit{T}} P_{\textit{T}}^{{eq}}\left(1-\exp \left(-\frac{(T_p - t_p)}{\tau_{\textit{T}}}\right)\right)}{1-\exp\left(-\frac{(T_p - t_p)}{\tau_{\textit{T}}} + k_{\textit{2}} t_p\right)} \times \frac{\left(\exp\left(k_{\textit{2}} t_p\right)-1\right)}{t_p} \right)
%\end{equation}
\chapter{End-to-end probabilities with end-loading}
The following work can be regarded as an extension of what done in the previous chapter and it is not present in \citep{2020, 2022}. The focus is again a path integral formulation for fluctuating elastic rods, but now the elastic chain is subject to a prescribed deterministic external end-loading, in addition to the stochastic forcing from the heat bath itself. The very new obstacle when dealing with an external end-loading, compared to the work done in \citep{LUDT, LUD}, is the evaluation of the partition function. In \citep{LUDT, LUD}, only the elastic internal energy was accounted and its quadratic structure allowed an exact Gaussian computation in the space of strains $\bm{\mathsf{u}}$-$\bm{\mathsf{v}}$ based on a change of variables, showing that the partition function was not providing any non-trivial contribution to the result, as also remarked in the present paper. However, an end-loading translates into an additional non-linear, non-local function of the strains in the energy and the mentioned strategy cannot be applied. Nevertheless, the Laplace approximation for the partition function developed in the previous chapters is still valid and will play a fundamental normalizing role. Hence the target is to demonstrate how to derive an approximate conditional probability density function governing the relative location and orientation of the two ends, when the fluctuating rod is subject to a prescribed external end-loading, and to illustrate the results with suitable examples.

We consider the linearly elastic energy defined in Eq.~(\ref{energylin}), \ie
\begin{equation}\label{energylin2}
E(\bm{q})=\frac{1}{2}\int_0^L{     
\begin{pmatrix}
       \bm{\mathsf{u}}-\hat{\bm{\mathsf{u}}} \\
        {\bm{\mathsf{v}}-\hat{\bm{\mathsf{v}}}}
\end{pmatrix}^T 
\begin{pmatrix}
       \bm{\mathcal{K}} & \bm{\mathcal{B}}  \\
       \bm{\mathcal{B}}^T & \bm{\mathcal{A}}
\end{pmatrix}
\begin{pmatrix}
       \bm{\mathsf{u}}-\hat{\bm{\mathsf{u}}} \\
        {\bm{\mathsf{v}}-\hat{\bm{\mathsf{v}}}}
\end{pmatrix}}\dd s\;,
\end{equation}
and we add the contribution of an external force applied on $\bm{r}(L)$, expressed in coordinates with respect to the fixed frame $\lbrace \bm{e}_i\rbrace_{i=1,2,3}$ and denoted by $\bm{\lambda}\in\mathbb{R}^3$. Thus we write the energy for the new loading term as
\begin{equation}\label{energyfor}
E_l(\bm{q};\bm{\lambda})=-\bm{\lambda}\cdot\left(\bm{r}(L)-\bm{r}(0)\right)=-\bm{\lambda}\cdot\int_0^L{\bm{r}'(s)}\dd s\;,
\end{equation}
where we use conventionally the minus sign, and the total potential energy of the system is given by 
\begin{equation}\label{energylinfor}
E_T(\bm{q};\bm{\lambda})=\int_0^L{     
W_T}\,\,\dd s\;,\quad W_T=\frac{1}{2}\begin{pmatrix}
       \bm{\mathsf{u}}-\hat{\bm{\mathsf{u}}} \\
        {\bm{\mathsf{v}}-\hat{\bm{\mathsf{v}}}}
\end{pmatrix}^T 
\begin{pmatrix}
       \bm{\mathcal{K}} & \bm{\mathcal{B}}  \\
       \bm{\mathcal{B}}^T & \bm{\mathcal{A}}
\end{pmatrix}
\begin{pmatrix}
       \bm{\mathsf{u}}-\hat{\bm{\mathsf{u}}} \\
        {\bm{\mathsf{v}}-\hat{\bm{\mathsf{v}}}}
\end{pmatrix}-\bm{\lambda}\cdot\bm{r}'(s)\;.
\end{equation}
In this chapter we are interested in determining the minimizers $\bm{q}^f$ and $\bm{q}^l$ of energy Eq.~(\ref{energylinfor}) satisfying respectively the Dirichlet-Dirichlet boundary conditions 
\begin{equation}\label{ffor}
(\mathtt{f})\quad\bm{r}(0)=\bm{0},\,\,\bm{R}(0)=\mathbb{1},\,\,\bm{r}(L)=\bm{r}_L,\,\,\bm{R}(L)=\bm{R}_L\;,
\end{equation}
and the Dirichlet-Neumann natural boundary conditions
\begin{equation}\label{lfor}
(\mathtt{l})\quad\bm{r}(0)=\bm{0}\;,\,\,\,\bm{R}(0)=\mathbb{1}\;,\,\,\,\bm{n}(L)=\bm{\lambda}\;,\,\,\,\bm{m}(L)=\bm{0}\;.
\end{equation}
The boundary condition $\bm{n}(L)=\bm{\lambda}$ in Eq.~(\ref{lfor}) is justified by the fact that the first variation of the loading energy Eq.~(\ref{energyfor}) reduces to a boundary term
\begin{equation}
\delta E_l=-\bm{\lambda}\cdot\int_0^L{\delta\bm{r}'(s)}\dd s=-\left[\bm{\lambda}\cdot\delta\bm{r}(s)\right]_0^L\;,
\end{equation}
which comes together with the boundary term $\left[\bm{n}\cdot\delta\bm{r}(s)\right]_0^L$
arising from the first variation of the elastic energy Eq.~(\ref{energylin2}). The first-order necessary condition for a minimizer of energy Eq.~(\ref{energylinfor}) requires the first variation to vanish. Thus, in absence of Dirichlet constraint at $s=L$,
\begin{equation}
(\bm{n}(L)-\bm{\lambda})\cdot\delta\bm{r}(L)=0\,\,\Rightarrow\,\,\bm{n}(L)=\bm{\lambda}\;.
\end{equation}
From the above statements it is then clear that $\bm{q}^f$ and $\bm{q}^l$ are actually minimizers of the elastic energy Eq.~(\ref{energylin2}) determined by the Euler-Lagrange equations Eq.~(\ref{EuL}) 
\begin{equation}
{\bm{{n}}}'=\bm{0}\;,\quad{\bm{{m}}}'+\bm{{r}}'\times\bm{{n}}=\bm{0}\;,
\end{equation}
and satisfying the boundary conditions Eq.~(\ref{ffor}) and Eq.~(\ref{lfor}) respectively.  In the following we assume stability of the minimizers and we will not consider explicitly neither the marginal case, nor the problem of non-isolation of minima, even though the following presentation applies to the latter cases from the considerations reported in the previous chapter. Finally, note that the reference configuration of the rod $\hat{\bm{q}}$ satisfying the Neumann natural boundary conditions at $s=L$  for the elastic energy Eq.~(\ref{energylin2}) is in general different from the minimizer $\bm{q}^l$ satisfying the Neumann natural boundary conditions at $s=L$  for the total potential energy Eq.~(\ref{energylinfor}).

\section{The partition function}
We start the section recalling the underlying path integral formulation, starting from an assumed Boltzmann distribution on rod configurations satisfying $\bm{q}(0)=\bm{q}_0=(\mathbb{1},\bm{0})$, of the form 
\begin{equation}
\frac{e^{-\beta E_T(\bm{q}(s);\bm{\lambda})}}{\mathcal{Z}}\;, 
\end{equation}
with $\beta$ the inverse temperature and $\mathcal{Z}$ the partition function of the system. The conditional probability density for the end of the rod subject to end-loading to be in configuration $\bm{q}_L=(\bm{R}_L,\bm{r}_L)$ at $s=L$, given that the other end satisfies $\bm{q}(0)=\bm{q}_0$ at $s=0$, is provided by the ratio of infinite dimensional Wiener integrals: 
\begin{equation}\label{denfor}
\rrf(\bm{q}_L,L|\bm{q}_0,0)=\frac{\mathcal{K}_f}{\mathcal{Z}}\;,\quad \int_{SE(3)}{\rrf(\bm{q}_L,L|\bm{q}_0,0)}\,\dd \bm{q}_L=1\;,
\end{equation}
\begin{equation}\label{pifor}
{\mathcal{K}_f}=\int\limits_{\bm{q}(0)=\bm{q}_0}^{\bm{q}(L)=\bm{q}_L}{e^{-\beta E_T(\bm{q};\bm{\lambda})}\,\mathcal{D}\bm{q}}\;,\quad\mathcal{Z} = \int\limits_{\bm{q}(0)=\bm{q}_0}{e^{-\beta E_T(\bm{q};\bm{\lambda})}\,\mathcal{D}\bm{q}}\;.
\end{equation}
In order to evaluate the expressions in Eq.~(\ref{pifor}), we parametrize the configuration variable $\bm{q}(s)=(\bm{R}(s),\bm{r}(s))\in SE(3)$ by means of $\bm{\mathsf{q}}(s)=(\bm{\mathsf{c}}(s),\bm{\mathsf{t}}(s))\in\mathbb{R}^6$ as done in Eq.~(\ref{par}). In particular, we choose two different curves of unit quaternions $\bar{\bm{\gamma}}(s)$ to be the curves defined by the rotation component ${\bm{R}}(\bar{\bm{\gamma}})$ of the minimizers $\bm{q}^f$ and $\bm{q}^l$ respectively, which characterise the two different parametrisations involved in the computation of $\mathcal{K}_f$ and $\mathcal{Z}$ in view of the Laplace approximation.
Neglecting the Jacobian factor of the parametrization as justified in the previous chapter, we can write
\begin{equation}\label{piforz}
\mathcal{Z}\approx\int\limits_{\bm{\mathsf{q}}(0)=\bm{\mathsf{q}}_0}{e^{-\beta\int_0^L{\left[W-\bm{\lambda}\cdot(\bm{R}^l\bm{\mathsf{t}})'\right]\,\dd s}}\,\mathcal{D}\bm{\mathsf{q}}}\;,\quad\bm{\mathsf{t}}={\bm{R}^l}^T\bm{r}\;,
\end{equation}
and an analogous expression holds for $\mathcal{K}_f$. In particular, using Eq.~(\ref{parmomM}) and Eq.~(\ref{parmomN}), the Neumann natural boundary conditions $\bm{m}(L)=\bm{0}$ and $\bm{n}(L)=\bm{\lambda}$ for $\bm{q}^l$ translate into 
\begin{equation}
\frac{\partial W_T}{\partial\bm{\mathsf{c}}'}(L)=2\left[\frac{\bm{\mathbb{1}}+\bm{\mathsf{c}}^{\times}}{1+\Vert\bm{\mathsf{c}}\Vert^2}\bm{\mathsf{m}}\right](L)=\bm{0}\;,
\end{equation}
\begin{equation}
\frac{\partial W_T}{\partial\bm{\mathsf{t}}'}(L)=\left[{\bm{R}^l}^T(\bm{n}-\bm{\lambda})\right](L)=\bm{0}\;,
\end{equation} 
for $\bm{\mathsf{q}}^l$, and our path integral computation methods smoothly apply.

In the Laplace approximation for $\mathcal{K}_f$ and $\mathcal{Z}$ the total energy is expanded about the associated minimizer $\bar{\bm{\mathsf{q}}}$ as 
\begin{equation}
E_T(\bm{\mathsf{q}})\sim E_T(\bar{\bm{\mathsf{q}}})+\frac{1}{2}\delta^2 E_T(\bm{\mathsf{h}};\bar{\bm{\mathsf{q}}})\;,\quad\bm{\mathsf{q}}=\bar{\bm{\mathsf{q}}}+\bm{\mathsf{h}}\;, 
\end{equation}
being the first variation zero. Furthermore, we observe that the shape of the second variation $\delta^2 E_T$ coincides with the one of $\delta^2 E$ reported in Eq.~(\ref{sec}), with $\bm{\mathsf{h}}=(\delta\bm{\mathsf{c}},\delta\bm{\mathsf{t}})$. This is because the second variation of the loading energy Eq.~(\ref{energyfor}) trivially vanishes. Then, the perturbation field describing fluctuations about the minimizer $\bar{\bm{\mathsf{q}}}$ satisfies the linearised version of the parametrised boundary conditions, \ie 
\begin{equation}
\bm{\mathsf{h}}(0)=\bm{\mathsf{h}}(L)=\bm{0}\quad\text{for}\quad(\mathtt{f})\;,
\end{equation}
\begin{equation}
\begin{cases}
\bm{\mathsf{h}}(0)=\bm{0}\quad\text{for}\quad(\mathtt{l})\;,\\
\\
\bm{\mathsf{\mu}}(L)=\left(\delta\frac{\partial W_T}{\partial\bm{\mathsf{c}}'},\delta\frac{\partial W_T}{\partial\bm{\mathsf{t}}'}\right)(L)=\left(2(\delta\bm{\mathsf{m}}-{\bm{\mathsf{m}}}^l\times\delta\bm{\mathsf{c}}),\delta\bm{\mathsf{n}}-2{\bm{\mathsf{n}}}^l\times\delta\bm{\mathsf{c}}\right)(L)=\bm{0}\quad\text{for}\quad(\mathtt{l})\;.
\end{cases}
\end{equation}
Thus, according to Eq.~(\ref{fin2}), the approximate conditional probability density is given by 
\begin{equation}\label{finfor}
\rho_f(\bm{q}_L,L \vert \bm{q}_0,0)\approx e^{\beta\left[E_T({\bm{\mathsf{q}}}^l)-E_T(\bm{\mathsf{q}}^f)\right]}\sqrt{\det\left[\frac{\beta}{2\pi}\frac{{{\bm{\mathsf{H}}}^l}}{{\bm{\mathsf{H}}^{f}}}(0)\right]} \;,
\end{equation}
where
\begin{equation}\label{jacfor}
\begin{cases}
             \begin{pmatrix}
      {{\bm{\mathsf{H}}^{f}}}  \\
     {{\bm{\mathsf{M}}^{f}}}
     \end{pmatrix}'
       =\bm{J}{{\bm{\mathsf{E}}^{f}}}\begin{pmatrix}
       {{\bm{\mathsf{H}}^{f}}}  \\
     {{\bm{\mathsf{M}}^{f}}}
     \end{pmatrix}
   \\
   \\
\begin{pmatrix}
      {{\bm{\mathsf{H}}^{f}}}  \\
     {{\bm{\mathsf{M}}^{f}}}
     \end{pmatrix}
     (L)
       =\bm{J}\begin{pmatrix}
       {\mathbb{1}} \\
     \mathbb{0}
     \end{pmatrix}
       \end{cases}\;, \qquad
\begin{cases}
          \begin{pmatrix}
      {{\bm{\mathsf{H}}^{l}}}  \\
     {{\bm{\mathsf{M}}^{l}}}
     \end{pmatrix}'
       =\bm{J}{{{\bm{\mathsf{E}}}^l}}\begin{pmatrix}
       {{\bm{\mathsf{H}}^{l}}}  \\
     {{\bm{\mathsf{M}}^{l}}}
     \end{pmatrix}
   \\
   \\
\begin{pmatrix}
      {{\bm{\mathsf{H}}^{l}}}  \\
     {{\bm{\mathsf{M}}^{l}}}
     \end{pmatrix}
     (L)
       =\bm{J}\begin{pmatrix}
     \mathbb{0} \\
     {\mathbb{1}} 
     \end{pmatrix}
       \end{cases}\;,
\end{equation} 
with ${\bm{\mathsf{E}}}\in\mathbb{R}^{12\times 12}$ given in Eq.~(\ref{E2}), Eq.~(\ref{E3}) in terms of strains, forces and moments, valid for both Kirchhoff and Cosserat rods, and
\begin{equation}
\bm{J}=\begin{pmatrix} \mathbb{0} &\,\, \mathbb{1}\\ -\mathbb{1} &\,\, \mathbb{0}\end{pmatrix}\;.
\end{equation}
Note that the $(\mathtt{f})$ contribution in Eq.~(\ref{finfor}), arising from the evaluation of the numerator $\mathcal{K}_f$ in Eq.~(\ref{denfor}), is the same as the one in Eq.~(\ref{fin}), with the only difference of the loading term in the energy. By contrast, the $(\mathtt{l})$ contribution in Eq.~(\ref{finfor}) from the evaluation of the partition function is not equal to $1$, as was true for Eq.~(\ref{fin}). In fact, in presence of an external end-loading, the second-order expansion cannot be performed about the unstressed reference configuration of the rod because $\bm{q}^l\neq\hat{\bm{q}}$. Therefore, the non trivial minimizer $\bm{q}^l$ enters the expression of the energy and determines the matrix $\bm{\mathsf{E}}^l$ for the computation of the Jacobi fields ${{\bm{\mathsf{H}}^{l}}}$ in Eq.~(\ref{jacfor}), leading to a new system of equations.

\section{Importance sampling Monte Carlo algorithm}
In this section we extend the Monte Carlo algorithm for stochastic elastic rods presented in Sec. \ref{MC} of the previous chapter. In absence of an external end-loading, a suitable change of variables allowed to write the discrete path integral distribution as a  $6\,n$-dimensional Gaussian $\rho_{\mathcal{Z}}$ in the variable $\bm{\mathsf{w}}=(\bm{\mathsf{u}}_0,...,\bm{\mathsf{u}}_{n-1},\bm{\mathsf{v}}_0,...,\bm{\mathsf{v}}_{n-1})$ by means of expression Eq.~(\ref{GausDisc}). However, in presence of end-loading, the new term in the energy Eq.~(\ref{energyfor}) is a non-local and non-linear function of the strains, and the Gaussian property is lost. Namely, according to the Boltzmann distribution having partition function given in Eq.~(\ref{pifor}), \ie 
\begin{equation}
\mathcal{Z}=\int_{\bm{q}(0)=\bm{q}_0}{e^{-\beta E_T(\bm{q};\bm{\lambda})}\,\mathcal{D}\bm{q}}
\end{equation}
with total energy Eq.~(\ref{energylinfor}), we rewrite the infinite-dimensional problem as a finite-dimensional one by means of a ``parameter-slicing method". As already seen before, this is achieved after parametrizing the configuration variable as $\bm{\mathsf{q}}(s)=(\bm{\mathsf{c}}(s),\bm{\mathsf{t}}(s))\in\mathbb{R}^6$ by setting $\epsilon=\frac{L}{n}$ with $n$ a large positive integer and $s_j=j\epsilon$ for $j=0,...,n$. Moreover, the change of variables $(\bm{\mathsf{c}}_1,...,\bm{\mathsf{c}}_n,\bm{\mathsf{t}}_1,...,\bm{\mathsf{t}}_n)\rightarrow(\bm{\mathsf{u}}_0,...,\bm{\mathsf{u}}_{n-1},\bm{\mathsf{v}}_0,...,\bm{\mathsf{v}}_{n-1})$ with Jacobian
\begin{equation}
\prod\limits_{j=0}^{n-1}\frac{\epsilon^6}{8}\left(1- \frac{\epsilon^2}{4}\Vert\bm{\mathsf{u}}_j\Vert^2\right)^{-\frac{1}{2}}\left(1+\Vert\bm{\mathsf{c}}_{j+1}\Vert^2\right)^{2}\;,
\end{equation}
as presented in \citep{LUDT}, leads to the following equality up to a constant factor for the discrete version of the partition function $\mathcal{Z}$
\begin{equation}\label{partfinfor}
\begin{split}
&\int{e^{-\beta\epsilon\sum\limits_{j=0}^{n}{W_T(\bm{\mathsf{c}},\bm{\mathsf{t}})_j}}}\prod\limits_{j=1}^n\left(1+\Vert\bm{\mathsf{c}}_j\Vert^2\right)^{-2}\dd\bm{\mathsf{c}}_j\dd\bm{\mathsf{t}}_j\\
\\
&\quad\sim \mathcal{Z}_n=\int{\left(\prod\limits_{j=0}^{n-1}\mathcal{J}(\bm{\mathsf{u}}_j)\right)\,\,e^{{\beta\epsilon}\bm{\lambda}\cdot\left(\bm{\mathsf{v}}_0+\sum\limits_{j=1}^{n-1}{\bm{R}_j\bm{\mathsf{v}}_j}\right)}\,\,e^{-{\beta\epsilon}\sum\limits_{j=0}^{n-1}{W({\bm{\mathsf{u}}_{\Delta}},{\bm{\mathsf{v}}_{\Delta}})_j}}}\prod\limits_{j=0}^{n-1}\dd\bm{\mathsf{u}}_j\dd\bm{\mathsf{v}}_j\;,
\end{split}
\end{equation}
with 
\begin{equation}
\bm{\mathsf{u}}_{\Delta}=\bm{\mathsf{u}}-\hat{\bm{\mathsf{u}}}\;,\quad \bm{\mathsf{v}}_{\Delta}=\bm{\mathsf{v}}-\hat{\bm{\mathsf{v}}}\;,
\end{equation}
\begin{equation}
\bm{R}_j= \bm{R}(\bm{\gamma}_j)\;,\quad \bm{\gamma}_j=\bm{\gamma}_j(\bm{\mathsf{u}}_0,...,\bm{\mathsf{u}}_{j-1})\,\,\text{according to Eq.~(\ref{recon1}) and Eq.~(\ref{recon2})}\;,
\end{equation}
\begin{equation}
W_j=\frac{1}{2}\left[{\bm{\mathsf{u}}_{\Delta}}_j^T \bm{\mathcal{K}}_j{\bm{\mathsf{u}}_{\Delta}}_j+2{\bm{\mathsf{u}}_{\Delta}}_j^T \bm{\mathcal{B}}_j{\bm{\mathsf{v}}_{\Delta}}_j+{\bm{\mathsf{v}}_{\Delta}}_j^T \bm{\mathcal{A}}_j{\bm{\mathsf{v}}_{\Delta}}_j\right]\;,\quad\mathcal{J}(\bm{\mathsf{u}}_j)=\left(1- \frac{\epsilon^2}{4}\Vert\bm{\mathsf{u}}_j\Vert^2\right)^{-\frac{1}{2}}\;,
\end{equation}
and the subscript $j$ indicates that the associated term is evaluated in $s_j$. Finally, defining the Gaussian
\begin{equation}
g_n(\bm{\mathsf{w}})=\frac{e^{-{\beta\epsilon}\sum\limits_{j=0}^{n-1}{W({\bm{\mathsf{u}}_{\Delta}},{\bm{\mathsf{v}}_{\Delta}})_j}}}{\mathcal{G}_n} \;,\quad \mathcal{G}_n=\displaystyle{\int e^{-{\beta\epsilon}\sum\limits_{j=0}^{n-1}{W({\bm{\mathsf{u}}_{\Delta}},{\bm{\mathsf{v}}_{\Delta}})_j}}\,\,\prod\limits_{j=0}^{n-1}\dd\bm{\mathsf{u}}_j\dd\bm{\mathsf{v}}_j}
\end{equation}
and the weighting function
\begin{equation}
f_n(\bm{\mathsf{w}})=\left(\prod\limits_{j=0}^{n-1}\mathcal{J}(\bm{\mathsf{u}}_j)\right)\,\,e^{{\beta\epsilon}\bm{\lambda}\cdot\left(\bm{\mathsf{v}}_0+\sum\limits_{j=1}^{n-1}{\bm{R}_j\bm{\mathsf{v}}_j}\right)}\;,
\end{equation}
the new target distribution has the form 
\begin{equation}\label{nonGaus}
\rho_{\mathcal{Z}}=\frac{\mathcal{G}_n}{\mathcal{Z}_n}f_n(\bm{\mathsf{w}})g_n(\bm{\mathsf{w}})\;,
\end{equation}
which cannot be sampled as before by a ``naive'' direct Monte Carlo method, even if we neglect the Jacobian term $\mathcal{J}(\bm{\mathsf{u}}_j)$.

Before giving the algorithm, we recall that the conditional probability density is a function of the variables $\bm{R}_L$, $\bm{r}_L$ and we reconstruct $\bm{R}_n$, $\bm{r}_n$ from the strains as described in Eq.~(\ref{recon1}), Eq.~(\ref{recon2}) and Eq.~(\ref{recon3}). Moreover, we are able to assess whether or not $\bm{q}_n=(\bm{R}_n=\bm{R}(\bm{\gamma}_n),\bm{r}_n)$ is falling inside the given small region $\mathcal{R}_{\zeta,\xi}$ of $SE(3)$ centred in $(\bm{R}_L(\bm{\gamma}_L),\bm{r}_L)$ and parametrized as the Cartesian product $\mathcal{B}_{\zeta}\times\mathcal{B}_{\xi}$ of two open balls in $\mathbb{R}^3$ of radius $\zeta,\xi>0$ respectively. Namely, 
\begin{equation}
(\bm{R}_n,\bm{r}_n)\in\mathcal{R}_{\zeta,\xi}\quad\text{if and only if}\quad\Vert\bm{\mathsf{c}}(\bm{\gamma}_n)\Vert<\zeta\quad\text{and}\quad\Vert\bm{r}_n-\bm{r}_L\Vert<\xi\;, 
\end{equation}
with $\bm{\mathsf{c}}\in\mathbb{R}^3$ the same parametrization of $SO(3)$ presented in Eq.~(\ref{par}) adapted to $\bar{\bm{\gamma}}=\bm{\gamma}_L$, \ie
\begin{equation}
\bm{\gamma}_n(\bm{\mathsf{c}})=\frac{1}{\sqrt{1+\Vert \bm{\mathsf{c}}\Vert^2}}\left(\sum\limits_{i=1}^3{\mathsf{c}_i\bm{B}_i\bm{\gamma}_L}+\bm{\gamma}_L\right)\;.
\end{equation}
Since we easily recover the relation
\begin{equation}
\bm{\mathsf{c}}(\bm{\gamma}_n)=\frac{(\bm{\gamma}_n\cdot\bm{B}_1\bm{\gamma}_L,\bm{\gamma}_n\cdot\bm{B}_2\bm{\gamma}_L,\bm{\gamma}_n\cdot\bm{B}_3\bm{\gamma}_L)}{\bm{\gamma}_n\cdot\bm{\gamma}_L}\;, 
\end{equation}
where the matrices $\bm{B}_1$, $\bm{B}_2$, $\bm{B}_3$ are given in Eq.~(\ref{BMat}), then the condition $\Vert\bm{\mathsf{c}}(\bm{\gamma}_n)\Vert<\zeta$ is equivalent to 
\begin{equation}
\sqrt{(\bm{\gamma}_n\cdot\bm{\gamma}_L)^{-2}-1}<\zeta\;,
\end{equation}
and the volume of $\mathcal{R}_{\zeta,\xi}$ in terms of the product of the Haar measure and the Lebesgue measure is still the same as in Eq.~(\ref{volume})
\begin{equation}
|\mathcal{R}_{\zeta,\xi}|=\frac{8\pi^2\xi^3}{3}\left(\arctan{(\zeta)}-\frac{\zeta}{1+\zeta^2}\right)\;.
\end{equation}

In the light of the above statements, we observe that instead of sampling from the distribution $\rho_{\mathcal{Z}}$ in Eq.~(\ref{nonGaus}), for us it is sufficient to compute an expected value of a given function $a(\bm{\mathsf{w}})$ of the form 
\begin{equation}
\int{a(\bm{\mathsf{w}})}\rho_{\mathcal{Z}}(\bm{\mathsf{w}})\,\dd \bm{\mathsf{w}}=\frac{\mathcal{G}_n}{\mathcal{Z}_n}\int{a(\bm{\mathsf{w}})f_n(\bm{\mathsf{w}})g_n(\bm{\mathsf{w}})}\,\dd \bm{\mathsf{w}}\;.
\end{equation}
First of all, denoting with $\bm{\mathsf{w}}_i$, $i=1,...,N$, a sample from the Gaussian distribution $g_n(\bm{\mathsf{w}})$ and with $N$ the total number of samples generated, we have
\begin{equation}
\frac{\mathcal{Z}_n}{\mathcal{G}_n}=\int{f_n(\bm{\mathsf{w}})g_n(\bm{\mathsf{w}})}\,\dd \bm{\mathsf{w}}\approx \frac{\sum\limits_{i=1}^N{f_n(\bm{\mathsf{w}}_i)}}{N}\;,\quad \int{a(\bm{\mathsf{w}})f_n(\bm{\mathsf{w}})g_n(\bm{\mathsf{w}})}\,\dd \bm{\mathsf{w}}\approx \frac{\sum\limits_{i=1}^N{a(\bm{\mathsf{w}}_i)f_n(\bm{\mathsf{w}}_i)}}{N}\;,
\end{equation}
so that
\begin{equation}
\int{a(\bm{\mathsf{w}})}\rho_{\mathcal{Z}}(\bm{\mathsf{w}})\,\dd \bm{\mathsf{w}}\approx \frac{\sum\limits_{i=1}^N{a(\bm{\mathsf{w}}_i)f_n(\bm{\mathsf{w}}_i)}}{\sum\limits_{i=1}^N{f_n(\bm{\mathsf{w}}_i)}}\;.
\end{equation}
Since our target is the computation of the probability of the set $\mathcal{R}_{\zeta,\xi}$ with respect to $\rho_f(\bm{q},L|\bm{q}_0,0)$, then $a(\bm{\mathsf{w}})=\bm{1}_{\mathcal{R}_{\zeta,\xi}}(\bm{q}_n(\bm{\mathsf{w}}))$, \ie the indicator or characteristic function of the set $\mathcal{R}_{\zeta,\xi}$ which equals $1$ if $\bm{q}_n(\bm{\mathsf{w}})\in\mathcal{R}_{\zeta,\xi}$ and $0$ otherwise. Thus
\begin{equation}
 \mathbb{P}(\mathcal{R}_{\zeta,\xi})\approx \int{\bm{1}_{\mathcal{R}_{\zeta,\xi}}(\bm{q}_n(\bm{\mathsf{w}}))}\rho_{\mathcal{Z}}(\bm{\mathsf{w}})\,\dd \bm{\mathsf{w}}\approx \frac{\sum\limits_{i=1}^N{\bm{1}_{\mathcal{R}_{\zeta,\xi}}(\bm{q}_n(\bm{\mathsf{w}}_i))f_n(\bm{\mathsf{w}}_i)}}{\sum\limits_{i=1}^N{f_n(\bm{\mathsf{w}}_i)}}\;.
\end{equation}
On the other hand,
\begin{equation}
\mathbb{P}(\mathcal{R}_{\zeta,\xi})=\int_{\mathcal{R}_{\zeta,\xi}}{\rho_f(\bm{q},L|\bm{q}_0,0)}\dd\bm{q}\approx |\mathcal{R}_{\zeta,\xi}|\,\rho_f(\bm{q}_L,L|\bm{q}_0,0)\;,
\end{equation}
and we finally get the estimation algorithm
\begin{equation}
\rho_f(\bm{q}_L,L|\bm{q}_0,0)\approx\frac{\sum\limits_{i=1}^N{\bm{1}_{\mathcal{R}_{\zeta,\xi}}(\bm{q}_n(\bm{\mathsf{w}}_i))f_n(\bm{\mathsf{w}}_i)}}{\sum\limits_{i=1}^N{f_n(\bm{\mathsf{w}}_i)}|\mathcal{R}_{\zeta,\xi}|}\;.
\end{equation}
Note that when $f_n=1$ we recover the direct Monte Carlo sampling method and the formula given in Eq.~(\ref{link}), namely
\begin{equation}
\rho_f(\bm{q}_L,L|\bm{q}_0,0)\approx\frac{|\lbrace\text{samples:}\,(\bm{R}_n,\bm{r}_n)\in\mathcal{R}_{\zeta,\xi}\rbrace|}{|\lbrace\text{all samples}\rbrace|\,\,|\mathcal{R}_{\zeta,\xi}|}\;,
\end{equation}
where the notation $|\cdot |$ stands for the number of elements of a discrete set or the measure of a continuous set, and the accuracy of the approximation increases with 
\begin{equation}
n\rightarrow\infty\;,\quad N\rightarrow\infty\;,\quad\zeta\rightarrow 0\;,\quad\xi\rightarrow 0\;. 
\end{equation}

\section{A cyclic loading example}
The following example is designed to underline the role played by the different quantities involved in Eq.~(\ref{fin}) and Eq.~(\ref{finfor}). We consider a Cosserat elastic rod clamped at $\bm{q}(0)=\bm{q}_0=(\mathbb{1},\bm{0})$ with diagonal and constant stiffness matrix $\bm{\mathcal{P}}(s)=\bm{\mathcal{P}}$, having constant intrinsic strains $\hat{\bm{\mathsf{u}}}(s)=\hat{\bm{\mathsf{u}}}$ and $\hat{\bm{\mathsf{v}}}(s)=(0,0,1)$, so that the intrinsic configuration is a helix and no intrinsic shear nor extension is present. Moreover, we define a family of cyclic loadings in the $x-y$ plane to be applied at $\bm{r}(L)$ as
\begin{equation}\label{cyc}
\bm{\lambda}_i=r\begin{pmatrix}
\sin{(\theta_i)}\\
\cos{(\theta_i)}\\
0
\end{pmatrix}\;,\quad r\in\mathbb{R}_{>0}\;,\quad \theta_i\in[0,2\pi] \;,\quad i=1,...,p\;.
\end{equation}
Defining $\hat{\bm{q}}(L)$ to be the intrinsic configuration of the rod at $s=L$, the aim is to compare 2 different situations. 
\begin{enumerate}
\item The first problem, that we call problem $(1)$, consists of computing the conditional probability densities $\rho_i^{(1)}(\bm{q}_L,L \vert \bm{q}_0,0)$ evaluated at ${\bm{q}}_L=\hat{\bm{q}}(L)$ for our fluctuating elastic rod clamped at $\bm{q}_0=(\mathbb{1},\bm{0})$ and subject to the external end-loading $\bm{\lambda}_i$, $i=1,...,p$. As a consequence, the Dirichlet-Dirichlet minimizer $\bm{q}^f(s)$ in Eq.~(\ref{finfor}) coincides with $\hat{\bm{q}}(s)$ and $E_T(\bm{{q}}^f;\bm{\lambda}_i)=-\bm{\lambda}_i\cdot\hat{\bm{r}}(L)$. Moreover, in principle ${{\bm{\mathsf{H}}^{f}}}(0)$ appearing in Eq.~(\ref{finfor}) can be computed analytically since the Jacobi system in this case has constant coefficients. On the other hand, the problem is characterised also by $p$ minimizers $\bm{q}^l_i(s)$ satisfying the boundary conditions Eq.~(\ref{lfor}) to be found numerically, with $E_T(\bm{{q}}^l_i)=E(\bm{{q}}^l_i)-\bm{\lambda}_i\cdot\bm{{r}}^l_i(L)$ given in Eq.~(\ref{energylinfor}) and ${{\bm{\mathsf{H}}^{l}_i}}(0)$ computed by means of the second system in Eq.~(\ref{jacfor}). Thus we have 
\begin{equation}\label{enfl1}
\begin{split}
\rho_i^{(1)}(\hat{\bm{q}}(L),L \vert \bm{q}_0,0)&\approx \frac{e^{\beta\bm{\lambda}_i\cdot(\hat{\bm{r}}-\bm{{r}}^l_i)(L)}}{\sqrt{\det\left[{\bm{\mathsf{H}}}^f(0)\right]}}\,\,\left(\frac{\beta}{2\pi}\right)^3 e^{\beta E(\bm{{q}}^l_i)}\sqrt{\det\left[{{\bm{\mathsf{H}}}^l_i}(0)\right]} \;,\\
&=e^{\beta\left[E(\bm{{q}}^l_i)+\bm{\lambda}_i\cdot(\hat{\bm{r}}-\bm{{r}}^l_i)(L)\right]}\sqrt{\det\left[\frac{\beta}{2\pi}\frac{{\bm{\mathsf{H}}}^l_i}{{\bm{\mathsf{H}}}^f}(0)\right]}\;,
\end{split}
\end{equation}
and we call the exponential part and the square root part in the latter expression respectively energy factor and fluctuation factor for problem $(1)$.
\item A dual problem of $(1)$, that we call problem  $(2)$, consists of computing the conditional probability densities $\rho_i^{(2)}(\bm{q}_L,L \vert \bm{q}_0,0)$ evaluated at ${\bm{q}}_L=\bm{q}^l_i(L)$ for the same rod but without an external end-loading, where $\bm{q}^l_i(s)$ are the minimizers presented in problem $(1)$. As noted in the previous chapter in deriving Eq.~(\ref{fin}), there is no contribution to the probability density from the intrinsic configuration minimizer $\hat{\bm{q}}(s)$, since $E(\hat{\bm{q}})=0$ and $\det{[\hat{\bm{\mathsf{H}}}(0)]}=1$. On the other hand, for Eq.~(\ref{fin}) here we have $p$ Dirichlet-Dirichlet minimizers $\bm{q}^f_i(s)$ coinciding with $\bm{q}^l_i(s)$ having elastic energy $E(\bm{{q}}^f_i)=E(\bm{q}^l_i)$ and ${{\bm{\mathsf{H}}^{f}_i}}(0)$ computed by means of the first system in Eq.~(\ref{jacfor}). Thus we have
\begin{equation}\label{enfl2}
\begin{split}
\rho_i^{(2)}(\bm{q}^l_i(L),L \vert \bm{q}_0,0)&\approx \left(\frac{\beta}{2\pi}\right)^3 \left[e^{\beta E(\bm{q}^l_i)}\sqrt{\det\left[{{\bm{\mathsf{H}}}^f_i}(0)\right]}\right]^{-1} \;\\
&=e^{-\beta E(\bm{q}^l_i)}\sqrt{\det\left[\frac{\beta}{2\pi}{\bm{\mathsf{H}}^f_i(0)}^{-1}\right]}\;,
\end{split}
\end{equation}
and we call the exponential part and the square root part in the latter expression respectively energy factor and fluctuation factor for problem $(2)$.
\end{enumerate}
Since $\bm{q}^l_i(s)=\bm{q}^f_i(s)$ then we also have ${\bm{\mathsf{E}}}_i={\bm{\mathsf{E}}}^l_i={\bm{\mathsf{E}}}^f_i$ and the two systems of equations for ${{\bm{\mathsf{H}}^{l}_i}}$ and ${{\bm{\mathsf{H}}^{f}_i}}$ can be written as a single one 
\begin{equation}
\begin{cases}
             \begin{pmatrix}
      \bm{\mathsf{H}}^{f}_i &  \bm{\mathsf{H}}^{l}_i\\
     \bm{\mathsf{M}}^{f}_i  &  \bm{\mathsf{M}}^{l}_i
     \end{pmatrix}'
       =\bm{J}{\bm{\mathsf{E}}}_i\begin{pmatrix}
      \bm{\mathsf{H}}^{f}_i &  \bm{\mathsf{H}}^{l}_i\\
     \bm{\mathsf{M}}^{f}_i  &  \bm{\mathsf{M}}^{l}_i
     \end{pmatrix}
   \\
   \\
\begin{pmatrix}
      \bm{\mathsf{H}}^{f}_i &  \bm{\mathsf{H}}^{l}_i\\
     \bm{\mathsf{M}}^{f}_i  &  \bm{\mathsf{M}}^{l}_i
     \end{pmatrix}
     (L)
       =\bm{J}
       \end{cases} \;.
\end{equation}
We further observe that the latter linear system is driven by a matrix whose trace is zero ($\bm{J}{\bm{\mathsf{E}}}_i$ is the product of a skew-symmetric matrix and a symmetric matrix) and by application of the generalized Abel's identity or Liouville’s formula we have the following relation
\begin{equation}
\det{\begin{pmatrix}
      \bm{\mathsf{H}}^{f}_i &  \bm{\mathsf{H}}^{l}_i\\
     \bm{\mathsf{M}}^{f}_i  &  \bm{\mathsf{M}}^{l}_i
     \end{pmatrix}
     (0)}=\det{\begin{pmatrix}
      \bm{\mathsf{H}}^{f}_i &  \bm{\mathsf{H}}^{l}_i\\
     \bm{\mathsf{M}}^{f}_i  &  \bm{\mathsf{M}}^{l}_i
     \end{pmatrix}
     (L)}=\det{(\bm{J})}=1\;.
\end{equation}

\begin{figure*}
\centering
\captionsetup{justification=centering,margin=1cm}
\subfigure[]{
\qquad\qquad\includegraphics[width=.34\textwidth]{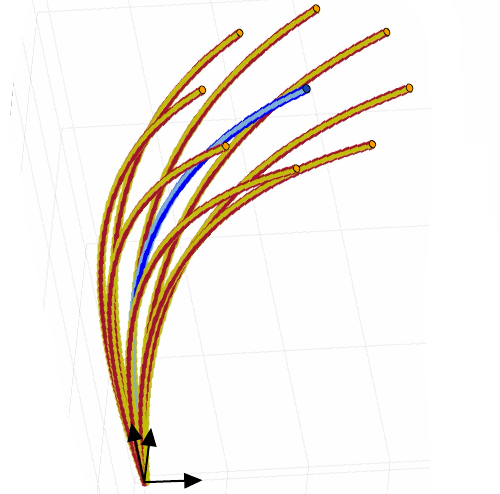}\quad\,
}
\subfigure[]{
\includegraphics[width=.448\textwidth]{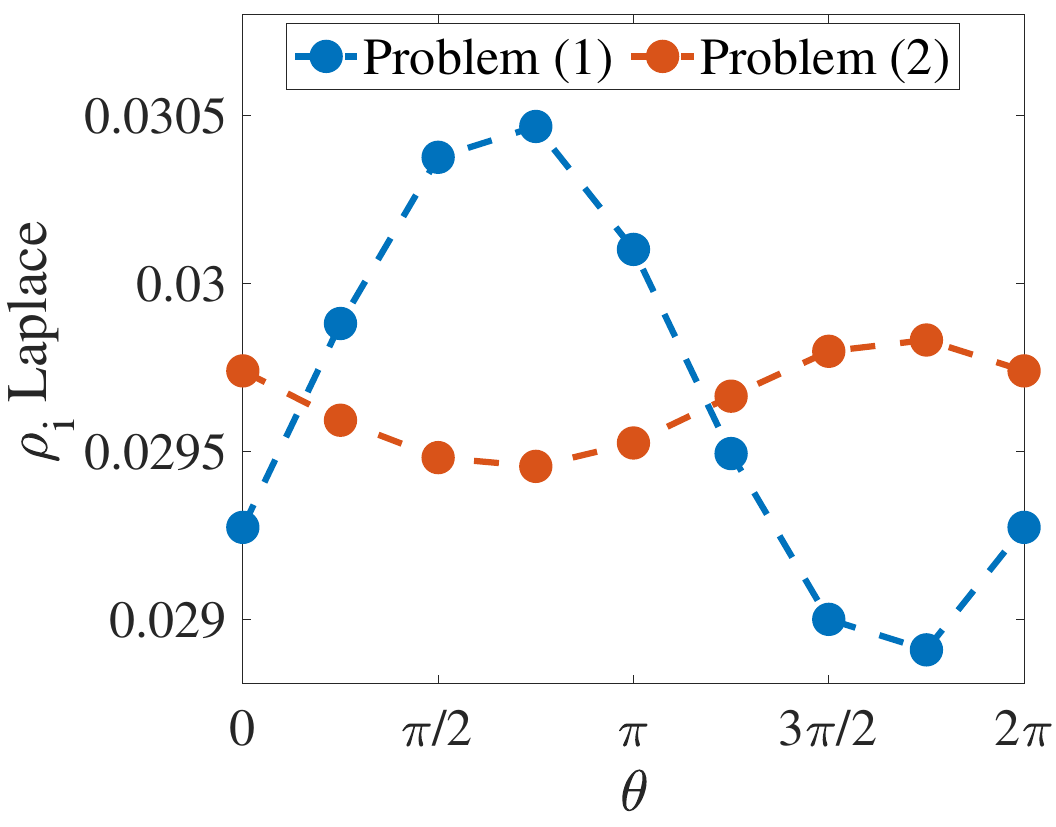}
}

\subfigure[]{
\includegraphics[width=.44\textwidth]{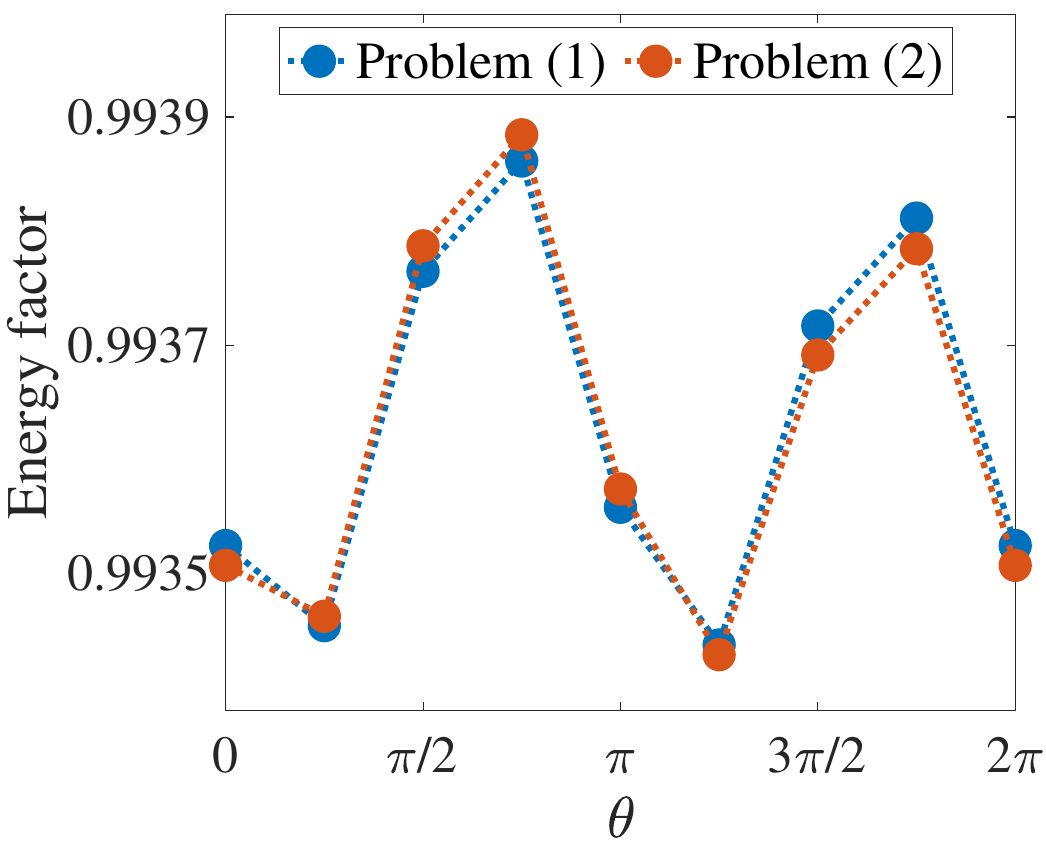}\qquad
}
\subfigure[]{
\includegraphics[width=.4418\textwidth]{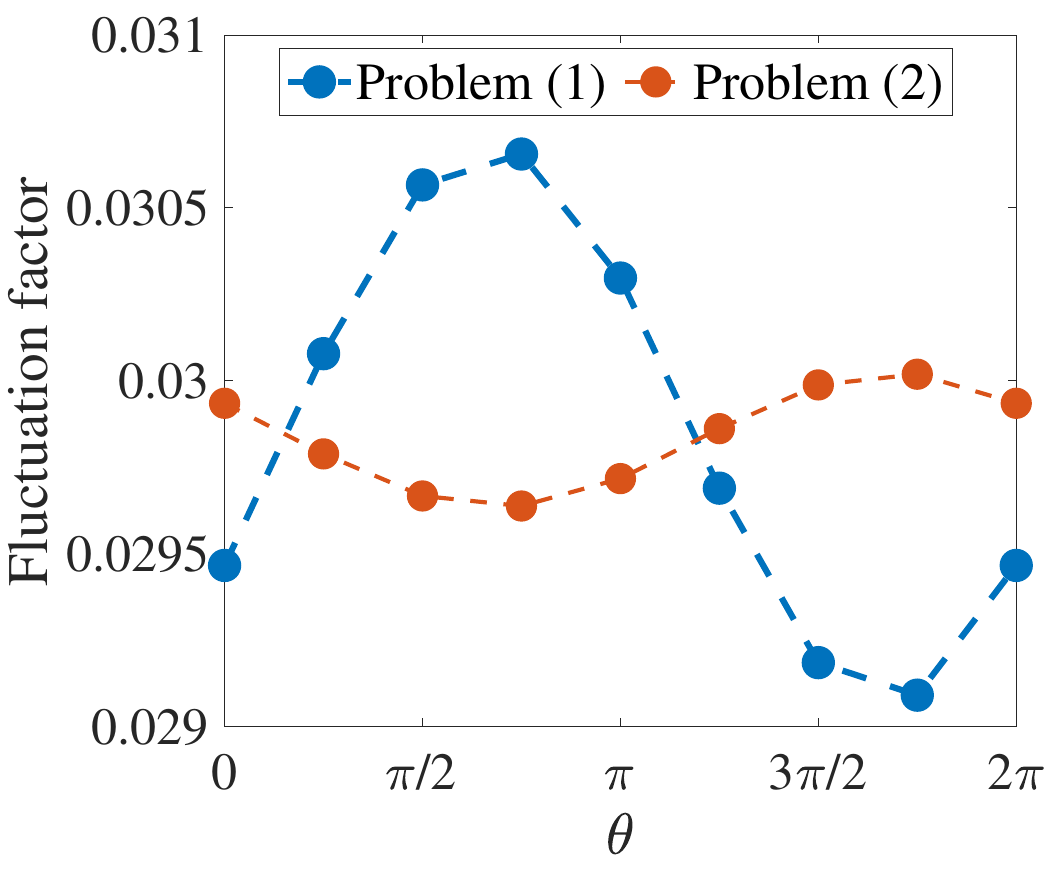}
}

\subfigure[]{
\includegraphics[width=.451\textwidth]{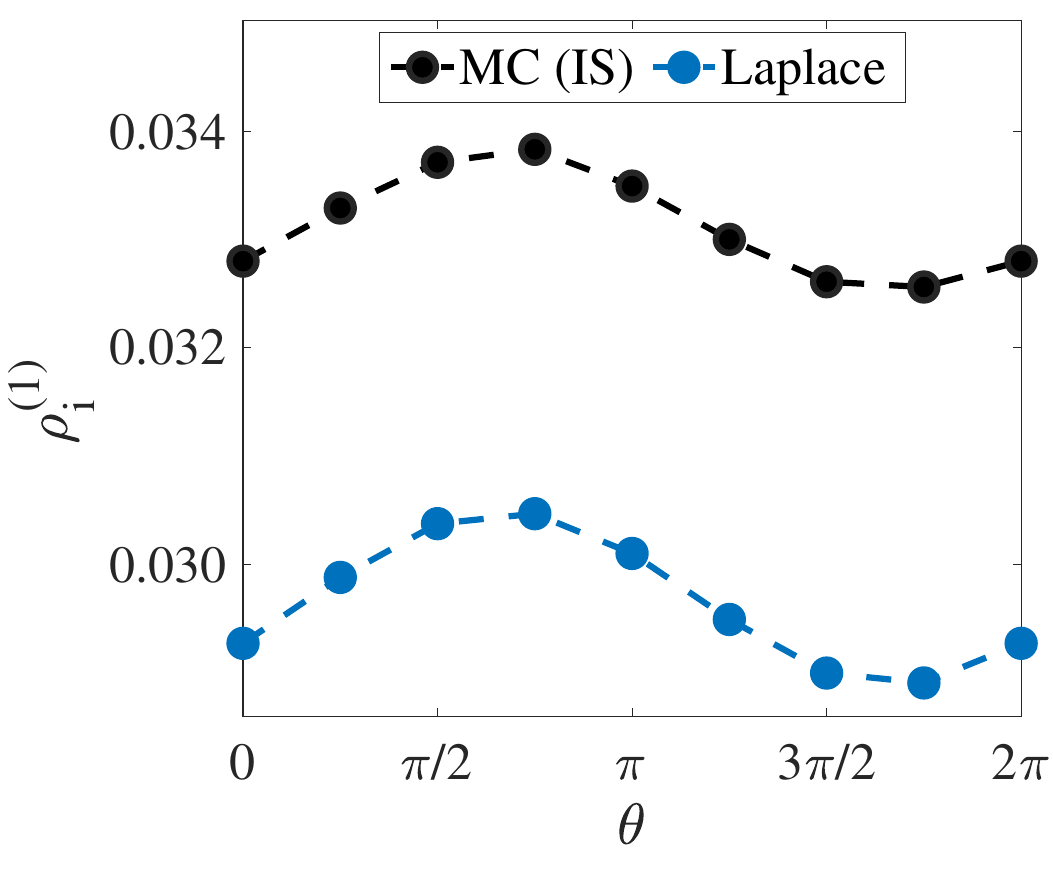}\quad\,\,
}
\subfigure[]{
\includegraphics[width=.441\textwidth]{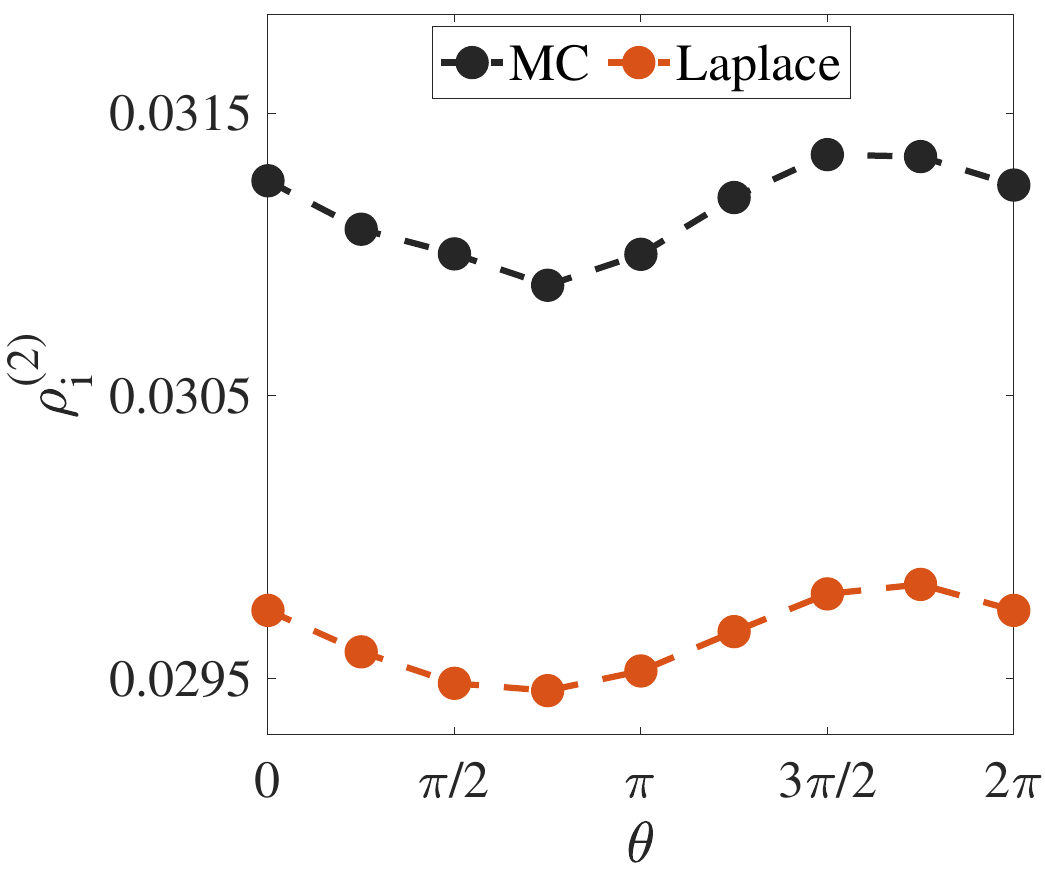}
}
\caption{A schematic representation of the system is given in panel (a): in blue the intrinsic configuration $\hat{\bm{q}}(s)$ and in orange the $8$ minimizers $\bm{q}^l_i(s)$. In (b) we show the Laplace approximation of the conditional probability densities for both problems (1) and (2) as a function of the loading parameter $\theta$. The energy and fluctuation contributions to the final result as defined in Eq.~(\ref{enfl1}), Eq.~(\ref{enfl2}) are displayed in panels (c) and (d). Finally, in (e), (f) we report the comparison with importance sampling Monte Carlo for problem (1) and direct Monte Carlo sampling for problem (2).}
\label{fig1419}
\end{figure*} 

The results are shown in Fig. \ref{fig1419} for $\beta=1$ and a Cosserat elastic rod having the stiffness matrix $\bm{\mathcal{P}}$ equal to the identity matrix, $\hat{\bm{\mathsf{u}}}=(\frac{\pi}{5},\frac{\pi}{5},\frac{\pi}{5})$ and $L=1$, with a family of cyclic loadings as in Eq.~(\ref{cyc}) characterized by $r=\frac{1}{10}$ and $\theta_i=\frac{\pi}{4}i$, $i=0,1,...,8$. Observe that in both problems we are ``travelling'' back and forth from the same end configurations and the ``trip'' is performed with or without the presence of an external end-loading. Finally we can answer the question: which is the most or least probable situation in this roulette made of elastic rods for problem $(1)$ and $(2)$ respectively? Remarkably the oscillating trend is coming from the Jacobi fields, whereas the energy factor can be regarded as constant as a function of the loading parameter $\theta$. Therefore, in this example the fluctuation factor is dominating the system and this highlights the importance of understanding Gaussian path integrals under different boundary conditions by means of the work done throughout this document. In addition, we note that problem $(2)$ can be solved using the theory presented in \citep{LUDT} since the external loading is not present, and $\det{[\hat{\bm{\mathsf{H}}}(0)]}$ equals one. By contrary, in problem $(1)$ the same determinant for the partition function is fully providing the qualitative trend of the system, playing a fundamental role in the computation of the conditional probability density. The comparison with Monte Carlo simulations (where sampling is clearly less demanding here than for looping) shows a quantitative and qualitative success of the approximation method. For both problems (1) and (2) we make use of the algorithm derived in the previous section, without including the contribution of the Jacobian factor, which is negligible. On one hand, for problem (1), we adopt an importance sampling Monte Carlo algorithm for dealing with the external end-loading. On the other hand, setting $\bm{\lambda}=\bm{0}$, the same algorithm reduces to direct Monte Carlo sampling for problem (2). We conclude this chapter by noting that the relative error between Monte Carlo and our approximated results observed in panels (e) and (f) of Fig. \ref{fig1419} is quantitatively different between problem (1) and problem (2). Namely, the relative error is higher for the first problem and this is due to the fact that the partition function is approximated. As we have already discussed several times in the course of this work, the computation of the partition function in absence of external end-loading undergoes the Laplace expansion but the result is exact and hence the smaller error for problem (2).

\nnfootnote{The content of this chapter will be integrated in a publication, currently in preparation.}

\cleardoublepage
\cleardoublepage
\chapter*{Conclusions}
\markboth{Conclusions}{Conclusions}
\addcontentsline{toc}{chapter}{Conclusions}

In this work we presented a consistent approach, based on ratios of Wiener path integrals, to compute conditional (or transition) probabilities in the Laplace approximation. In the case of isolated minimizers of the energy, our method can be regarded as a generalization of work by Papadopoulos \citep{PAP1} and \citep{LUDT, LUD}, which allowed us to express the solutions of both conditional and unconditional Gaussian path integrals from the solutions of suitable matrix initial value problems for the Jacobi equations. In the framework of a general second variation operator with cross-terms acting on vector-valued curves, in addition to the previously known Dirichlet-Dirichlet case, the path integrals were solved for Dirichlet-Neumann boundary conditions and Dirichlet-Mixed (Dirichlet, Neumann) boundary conditions, that in our theory enter respectively the evaluation of the partition function (with a normalizing role) and of the numerator for marginal conditional probability densities. Moreover, the computation for the Neumann-Neumann case was performed. Our results are strongly connected with the conjugate point theory of stability in the calculus of variations, and it is possible to appreciate the relation between the interpretation of path integrals as functional determinants where the progressive product of eigenvalues is encoded by the determinant of appropriate Jacobi fields evolving in $[\sigma, t]$, and the conjugate points interpreted as zero eigenvalues of the second variation operator restricted to the interval $[\sigma, t]$, $\sigma\geq t_0$.

We phrased the latter formulation in the context of Fokker-Planck dynamics as a tool to investigate transition probabilities of stochastic processes. We observed that the accuracy of the method is only dependent on the accuracy of the Laplace approximation and, when the Lagrangian of the system is quadratic in position and velocity, then the results are exact. In this regard, the multidimensional Ornstein–Uhlenbeck process was solved exactly, whereas the stochastic Van der Pol oscillator led to an approximate solution, reasonably accurate in the small-time scale transient regime. As a side note, we discussed what is the effect of choosing different discretization prescriptions for continuous paths, and mentioned under which circumstances this can be arbitrary. 

The same computational framework was adopted in the stationary case of fluctuating Cosserat elastic rods at thermodynamic equilibrium with a heat bath. In particular, we addressed the problem of computing looping probabilities from a continuum perspective, for different choices of boundary conditions, with particular emphasis on extensible and shearable polymers, which are not generally treated in the standard literature of WLC-type models. Thus we have highlighted the role of extension and shear not only in modifying the expression of the quadratic Jacobi fluctuation term compared to inextensible and unshearable models, but also in minimizing the overall elastic energy and allowing ``relaxation'' phenomena to occur at small lengths with consequent higher values for the probability density. Note that the Kirchhoff inextensible and unshearable case was then recovered as a smooth limit in Hamiltonian form as done in \citep{LUDT, LUD}. Furthermore, we supplemented the theory employed for deriving general looping formulas with concrete examples in the case of ring closure or cyclization, the results of which were compared with extensive Monte Carlo simulations, showing overall good agreement. More precisely, the analysis was based on \citep{LUDT, LUD}, where the details concerning the choice of coordinates in $SO(3)$ are given, together with the treatment of the full looping problem in presence of isolated minimizers. Here we have extended the results to marginal looping probabilities, where no constraint on the end-orientation of the polymer is prescribed, which was made possible by the above studies on Gaussian path integrals under more general boundary conditions. 

In the following paragraph we take the opportunity to make an observation. In a first approximation DNA fits the WLC hypothesis of inextensibility and unshearability. However, contradictory results have been reported for DNA below the persistence length since the studies of Cloutier and Widom \citep{CW}, actually showing enhanced cyclization of short DNA molecules not explainable by WLC-type models. In a recent study \citep{BIO5} the authors conclude that ``determining whether the high bendability of DNA at short length scales comes from transient kinks or bubbles or stems from anharmonic elasticity of DNA requires improved computational methods and further studies''. We therefore observe that the Cosserat framework in the presence of extension and shear is a candidate for gaining a better understanding of the enhanced cyclization of short DNA molecules and adding a piece to the puzzle. In other words, even if DNA proves to have, under most circumstances, mechanical properties and thus parameters compatible with an inextensible and unshearable model, some energy-relaxation mechanisms could still fit into the observations reported here, especially in length regimes and boundary conditions where energy expressed simply as the square of the curvature cannot be of sufficient explanation. However, a basis for such an observation will require specific studies on the subject, which are outside the scope of the present work. Furthermore, birod models \citep{BIR, PRA} with sequence-dependent parameters are more accurate in capturing DNA conformations, but the theory devised here is comprehensive and can be applied analogously to this level of complexity, allowing the computation of different ring-closure probabilities without involving expensive MC simulations.

In the case of non-isolated minimizers instead, we applied an important result of Forman \citep{FORM} that we adapted and rewrote in Hamiltonian formulation, together with techniques which are typical of the field of quantum mechanics. Moreover, we generalized a regularization procedure using boundary perturbations described in \citep{MCK} in order to circumvent the zero mode problem and explicitly evaluate the singular path integrals. The latter analysis was capable of providing regularized cyclization probability densities for both full and marginal boundary conditions under the continuous symmetry associated with isotropic elastic rods. We mention that the symmetry of isotropy and uniformity can be present simultaneously (figure $8$ minimizers); in future, an analogous analysis in the presence of multiple zero modes could be performed, which is interesting from both a mathematical and a physical generalization perspective. The presence of a conjugate point which breaks down the standard Laplace method is not only a peculiarity of variational symmetries, but can arise in other situations, \eg we showed a bifurcation diagram varying the undeformed length for Cosserat rods. Having a systematic way of regularizing the path integrals in these circumstances is certainly a desirable task to be accomplished within future studies. In addition, one might be interested in investigating deeply the connection between Papadopoulos' evaluation method of Gaussian path integrals and Forman's analysis of functional determinants, both of which lead to equivalent results but exploit very different techniques.

When a linearly elastic energy functional is chosen, the partition function has the special property to be integrable exactly. This is because the energy is quadratic in the shifted strains and a suitable change of variables is sufficient to recover Gaussian integrals in configuration space \citep{LUDT, LUD}. The maneuver is allowed by the specific boundary conditions characterising the partition function, where the Dirichlet constraint is present only at one end. We pose the question: is it possible to exploit the quadratic structure of the linearly elastic energy to improve the computation and the level of accuracy of the Laplace approximation for the numerator, where the boundary conditions do not allow an easy way out? The answer could be enlightening to understand the behaviour of the cyclization probability density for longer lengths, and to clarify why the error does not explode but stabilises, making the approximation curve follow the same qualitative trend as the Monte Carlo estimate. We then conjecture the possibility of deriving a correction factor to be systematically included in the approximate formulae, exploiting the peculiar structure of the error resulting from the Laplace approximation of an energy that maintains a quadratic structure in certain variables.

Driven by the possibility of extending the range of applications of the theory, we also considered the effect of an external deterministic end-loading acting on fluctuating elastic rods in a thermal bath. The quadratic structure of the energy in some variables is then broken by an additional non-local term and the partition function was evaluated by means of a Laplace expansion in presence of Dirichlet-Neumann boundary conditions, leading to an approximate result. We designed an example to illustrate the effect of a non-straight intrinsic shape and a cyclic set of loadings in determining the profile of the probability density, which appears to be in counter-phase with respect to the same set-up but removing the external force contribution. The fluctuation factor was understood to fully characterize the behaviour of the latter system and our findings were benchmarked against Monte Carlo simulations by means of an appropriate algorithm for perturbed Gaussian distributions. In the future, it would be interesting to formulate the Euler's buckling problem from a stochastic perspective. Buckling of a Kirchhoff elastic rod occurs at a very specific critical compression load and is characterised by an instability of the intrinsically straight configuration which suddenly bends. We expect the study of the problem in the presence of stochastic fluctuations to be important in showing and understanding a dramatic change in the probability density function before and after the buckling load is reached.

In conclusion, this work touches on several seemingly distinct topics whose common thread is a random component that can be studied through the theory of path integrals. Both mathematics and physics can benefit from the analyses and observations reported in this work, where the new ideas proposed could lead to further developments in the fields of research considered.

%----------------------------------------------------------------------------------------
%	BIBLIOGRAPHY
%----------------------------------------------------------------------------------------

\printbibliography[heading=bibintoc]

%----------------------------------------------------------------------------------------

\end{document}